\documentclass[11pt,a4paper]{article}
\usepackage{iheppub}
\usepackage[bitstream-charter]{mathdesign}
\usepackage{XCharter}
\usepackage[dvipsnames,x11names]{xcolor}
\usepackage[T1]{fontenc}
\usepackage[utf8]{inputenc}
\usepackage[british]{babel}
\usepackage{amsmath}
\usepackage{tikz}
\usepackage{slashed}
\usepackage{placeins}
\usepackage[compat=1.1.0]{tikz-feynhand}
\usepackage{comment}
\usepackage{xfrac}
\usepackage{color}
\usepackage{xcolor,colortbl}
\usepackage{xspace}
\usepackage{orcidlink}
\allowdisplaybreaks

\usepackage{url}

\newcommand{\eq}[1]{\eqref{eq:#1}}
\newcommand{\fig}[1]{Fig.~\ref{fig:#1}}

\newcommand{\Sec}[1]{Sec.~\ref{sec:#1}}

\definecolor{blue}{rgb}{0,0.396,0.741}
\definecolor{MathBlue}{rgb}{0.368417, 0.506779, 0.709798}
\definecolor{MathYellow}{rgb}{0.880722, 0.611041, 0.142051}
\definecolor{MathGreen}{rgb}{0.560181, 0.691569, 0.194885}
\definecolor{MathRed}{rgb}{0.922526, 0.385626, 0.209179}
\definecolor{MathViolet}{rgb}{0.528488, 0.470624, 0.701351}

\newcommand{\yestag}{\tag{\stepcounter{equation}\theequation}}

\DeclareSymbolFont{usualmathcal}{OMS}{cmsy}{m}{n}
\DeclareSymbolFontAlphabet{\mathcal}{usualmathcal}

\begin{document}

\title{\boldmath Four Loop Renormalisation Group Equations in general Gross-Neveu-Yukawa Theories }%
\abstract{
We compute template formulae of all four-loop $\beta$-functions and anomalous dimensions of arbitrary renormalisable quantum field theories with fermions and scalar fields in the $\overline{\text{MS}}$ scheme. Using these results, novel contributions to the $\beta$-functions of the Higgs and top coupling are extracted. We discuss how four loops present a challenge to dimensional continuation from four to three dimensions and propose a solution. 
}

\author[a]{Tom Steudtner\,\orcidlink{0000-0003-1935-0417},} 
\affiliation[a]{
	Fakultät für Physik, TU Dortmund, Otto-Hahn-Str. 4, D-44221 Dortmund, Germany
}
\emailAdd{tom2.steudtner@tu-dortmund.de}

\maketitle

\section{Introduction}

Renormalisation Group Equations (RGEs) are a ubiquitous tool for precision calculations in quantum field theories (QFTs).
They connect the physics at different length or energy scales, predict critical phenomena and are the backbone of resummation techniques. As such, obtaining high-precision RGEs is of strategic interest for many areas of quantum physics. However, this task becomes formidable with increasingly high orders and more complex QFTs. 

Fortunately, the calculation can be factorised into two pieces. The technically challenging part consists of spinor algebra and loop integration, is rather generic for any renormalisable QFT. It can be conducted in a model independent manner, leading to intermediate results -- the template RGEs. Once determined, these templates can be utilised again and again to compute RGEs in a simple mapping procedure. 

Over the last four decades, the availability of template RGEs has been pushed to four-loop order for gauge interactions, three loops in the Yukawa- and scalar sectors, as well as two-loop order for vacuum expectation values~\cite{Machacek:1983tz,Machacek:1983fi,Machacek:1984zw,Jack:1984vj,Pickering:2001aq,Luo:2002ti,Chetyrkin:2012rz,Mihaila:2012pz,Bednyakov:2012en,Bednyakov:2012rb,Chetyrkin:2013wya,Bednyakov:2013eba,Bednyakov:2013cpa,Sperling:2013eva,Sperling:2013xqa,Mihaila:2014caa,Zoller:2015tha,Chetyrkin:2016ruf,Schienbein:2018fsw,Poole:2019kcm,Poole:2019txl,Bednyakov:2021qxa,Davies:2021mnc,Steudtner:2021fzs,Jack:2024sjr,Steudtner:2024teg}.
There are also many results which are higher order but slightly less general, for instance six-loop templates in purely scalar QFTs~\cite{ Chetyrkin:1981jq,Gorishnii:1983gp,Kleinert:1991rg,Calabrese:2003ww,Kleinert:1994td,Batkovich:2016jus,Kompaniets:2017yct,Adzhemyan:2019gvv,Kompaniets:2019xez, Bednyakov:2021ojn}, or five-loop results in simple gauge theories~\cite{Baikov:2016tgj,Luthe:2016ima,Herzog:2017ohr,Luthe:2017ttg}.

The evaluation of these template RGEs is technically straightforward, doing so by hand is ill-advised due to the condensed notation and overall complexity of the expressions~\cite{Molgaard:2014hpa,Roy:2019jqs}.
Thus, it is not surprising that progress with these template RGEs has been intertwined with software packages implementing them~\cite{Fonseca:2011sy,Staub:2013tta,Litim:2020jvl,Sartore:2020gou,Thomsen:2021ncy,Steudtner:FoRGEr}.

While the template formalism has been shaped through intensive model-building efforts in high-energy physics, the same technological setup has also proven beneficial for condensed matter systems. Here, second-order phase transitions can be related to critical points of continuum field theories in three dimensions. A large set of such transitions is captured by Wilson-Fisher-like fixed points, which can be approached in dimensional continuation from $d=4-\varepsilon$ to $d=3$~\cite{Wilson:1971dc}. This ansatz has reached a considerable precision leveraging scalar templates at six-loop order~\cite{Bednyakov:2021ojn} as well as advanced resummation techniques in $\varepsilon$~\cite{Kleinert:2001ax}.

More recently, there has been interest in Dirac materials (see e.g.~\cite{Herbut:2006cs,Boyack:2020xpe,Herbut:2023xgz}), where critical phenomena are described via Gross-Neveu-Yukawa theories.
These QFTs are more complicated than purely scalar ones as they also feature fermions,
but much simpler than general ones as gauge interactions are absent.
Unfortunately this middle ground is not reflected in the availability of template RGEs. While some specific four- and five-loop results exist~\cite{Zerf:2017zqi,Gracey:2025aoj}, general expressions are only available at three-loop~\cite{Jack:2024sjr}. The aim of this work is to improve the situation by computing template RGEs for gaugeless QFTs at four-loop order. This improves the precision for arbitrary Gross-Neveu-Yukawa theories and represents an intermediate step towards full general four-loop RGEs.

In \Sec{Notation}, we introduce notation and computation details of this endeavour. The results are summarised and cross-checks are conducted in \Sec{Results}. A brief conclusion is found in \Sec{End}. Due to their length, the explicit main results are listed in subsequent appendices.

\section{Computation}\label{sec:Notation}
In this work, we consider a QFT given by the general Lagrangian 
\begin{equation}\label{eq:master-template}
  \begin{aligned}
    \mathcal{L} =
     & \  \tfrac{1}{2} \partial^{\mu} \phi_a \partial_\mu \phi_a + \tfrac{i}{2} \psi^i (\tilde{\sigma}^\mu)^{\phantom{i}j}_i \partial_\mu \psi_j    - \tfrac{1}{2} y^{ajk} \,\phi_a(\psi_j \varepsilon \psi_k)   - \tfrac{1}{24} \lambda^{abcd} \,\phi_a \phi_b \phi_c \phi_d\,,
  \end{aligned}
  \end{equation}
following the notation of~\cite{Steudtner:2021fzs,Jack:2024sjr,Steudtner:2024teg}. Here $\phi_a$ is a vector containing real scalar field components which are counted via the indices $a,b,c, ..\,$. 
Meanwhile,  Weyl fermions as well as their complex conjugates are enumerated by $i,j,k,..\,$. This does not include spinor indices, which are kept implicit and conjugated via the Levi-Civita symbol $\varepsilon$.
Thus $\psi_i = (\psi^i)^*$ is pseudo-real and contains both left- and right handed parts. Hence $\tilde{\sigma}^\mu$ is either $\sigma^\mu$ or $\bar{\sigma}^\mu$ depending on which fermion couples to it. 
The Lagrangian~\eq{master-template} features a Yukawa interaction $y^{a ij}$ symmetric in its fermionic indices as well as a scalar quartic coupling tensor $\lambda^{abcd}$ which is totally symmetric overall.
In the following, we suppress all fermion indices and abbreviate products of Yukawa matrices $y^{aij}$ as
\begin{equation}
  y^{a ij} y^b_{jk} y^{c kl} \dots = y^{abc \dots}\,.
\end{equation}
For convenience, we make use of the symmetrisation operator $\mathcal{S}_n$ in order to hint at all $n$ distinct permutations of an expression with respect to its external indices, both scalar and fermionic. For instance, 
\begin{equation}
  \mathcal{S}_6 \,\mathrm{tr} \left(y^{abecde}\right) = \mathrm{tr} \left(y^{abecde} + y^{abedce} + y^{acebde}+ y^{acedbe} + y^{adebce} + y^{adecbe}\right)\,.
\end{equation}

The general shape of \eq{master-template} allows to define a Yukawa sector which contains interactions with $\gamma_5$-matrices when written in terms of four-component Dirac spinors.
As there is no na\"ive generalisation of $\gamma_5$ away from four spacetime dimensions, an ambiguity definition its treatment arises in dimensional regularisation~\cite{Jegerlehner:2000dz}. 
A consistent treatment of $\gamma_5$ is available~\cite{tHooft:1972tcz,Breitenlohner:1977hr} but 
would render the computation much more involved, see~\cite{Belusca-Maito:2023wah} for a recent review.
In the language of template RGEs, each term affected by the $\gamma_5$-ambiguity contains contractions of the tensor $\chi^{ij}$, which simply introduces factors $\pm 1$ depending on the chirality of each fermion field labelled as $j$~\cite{Poole:2019kcm}.

Fortunately, the ambiguity is absent in all gaugeless four-loop RGEs, which can be seen by following the arguments developed in~\cite{Chetyrkin:2012rz,Poole:2019kcm}. In $d=4-2\varepsilon$ dimensions, the ambiguity manifests in any traces with $\gamma_5$ and 4 or more other Dirac matrices. These yield totally antisymmetric tensors $\tilde{\epsilon}$, i.e.
\begin{equation}
  \mathrm{tr} \left(\gamma^\mu \gamma^\nu \gamma^\rho \gamma^\sigma \gamma_5\right) = 4 i \tilde{\epsilon}^{\mu\nu\rho\sigma}.
\end{equation}
Such objects are related to the rank-four Levi-Civita tensors $\epsilon$ up to unknown evanescent terms 
\begin{equation}
  \tilde{\epsilon}^{\mu\nu\rho\sigma} = \epsilon^{\bar{\mu} \bar{\nu} \bar{\rho} \bar{\sigma}} + \mathcal{O}(\varepsilon)\,,
\end{equation}
where the bared indices are restricted to the subdomain $d=4$.
Due to the antisymmetry of $\tilde{\epsilon}$, such ambiguous contributions often vanish. However, the ambiguity is present if two contracted $\tilde{\epsilon}$ tensors remain after loop integrations, or if a single one is contracted with an open line.
A non-vanishing $\tilde{\epsilon}$-tensor requires closed fermion lines with at least 4 propagators, and four independent momenta to yield contractions like $\mathrm{tr}(\slashed{k}_1\slashed{k}_2\slashed{k}_3\slashed{k}_4 \gamma_5)$.~\footnote{In the absence of gauge fields, there are no external Lorentz indices or direct contractions between two traces (before integration) to consider.}
For each closed fermion line, such a structure must remain after integration over its fermionic loop momentum. 

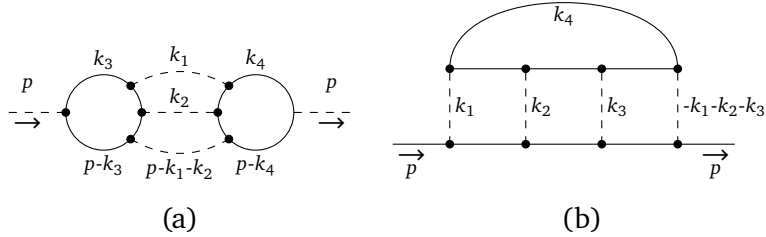
\begin{figure}[ht!]
\centering
\begin{tabular}{cc}
\begin{tikzpicture}
  \draw (.5,0) circle (.5);
  \draw (2.5,0) circle (.5);
  \draw[dashed] (-.75,0) -- (0,0);
  \draw[dashed] (3,0) -- (3.75,0);
  \filldraw[black] (0,0) circle (.05);
  \filldraw[black] (1,0) circle (.05);
  \filldraw[black] ({.5+.25*sqrt(2)},{+.25*sqrt(2)}) circle (.05);
  \filldraw[black] ({.5+.25*sqrt(2)},{-.25*sqrt(2)}) circle (.05);
  \filldraw[black] ({2.5-.25*sqrt(2)},{+.25*sqrt(2)}) circle (.05);
  \filldraw[black] ({2.5-.25*sqrt(2)},{-.25*sqrt(2)}) circle (.05);
  \filldraw[black] (2,0) circle (.05);
  \draw[dashed] (1,0) -- (2,0);
  \draw[dashed]  ({.5+.25*sqrt(2)},{+.25*sqrt(2)}) to [bend left] ({2.5-.25*sqrt(2)},{+.25*sqrt(2)});
  \draw[dashed]  ({.5+.25*sqrt(2)},{-.25*sqrt(2)}) to [bend right] ({2.5-.25*sqrt(2)},{-.25*sqrt(2)});
  \node[label=90:{\scriptsize $p$}] at (-.5,0) {};
  \node[label=90:{\scriptsize $p$}] at (+3.5,0) {};
  \node[label=90:{\scriptsize $k_3$}] at (.5,.3) {};
  \node[label=-90:{\scriptsize $p\text{-}k_3$}] at (0.5,-.3) {};
  \node[label=90:{\scriptsize $k_4$}] at (2.5,.3) {};
  \node[label=-90:{\scriptsize $p\text{-}k_4$}] at (2.5,-.3) {};
  \node[label=90:{\scriptsize $k_1$}] at (1.5,.35) {};
  \node[label=90:{\scriptsize $k_2$}] at (1.5,-.2) {};
  \node[label=-90:{\scriptsize $p\text{-}k_1\text{-}k_2$}] at (1.5,-.35) {};
  \node[label=-90:$\rightarrow$] at (-.5,.2) {};
  \node[label=-90:$\rightarrow$] at (+3.5,.2) {};
\end{tikzpicture}&
\begin{tikzpicture}
  \draw (-.75,0) -- (3.75,0);
  \draw (0,1) -- (3,1) to [bend right=90] (0,1);
  \draw[dashed] (0,0) -- (0,1);
  \draw[dashed] (1,0) -- (1,1);
  \draw[dashed] (2,0) -- (2,1);
  \draw[dashed] (3,0) -- (3,1);
  \filldraw[black] (0,0) circle (.05);
  \filldraw[black] (1,0) circle (.05);
  \filldraw[black] (2,0) circle (.05);
  \filldraw[black] (3,0) circle (.05);
  \filldraw[black] (0,1) circle (.05);
  \filldraw[black] (1,1) circle (.05);
  \filldraw[black] (2,1) circle (.05);
  \filldraw[black] (3,1) circle (.05);
  \node[label=90:{\scriptsize $k_4$}] at (1.5,1.3) {};
  \node[label=0:{\scriptsize $k_1$}] at (-.2,.5) {};
  \node[label=0:{\scriptsize $k_2$}] at (1-.2,.5) {};
  \node[label=0:{\scriptsize $k_3$}] at (2-.2,.5) {};
  \node[label=0:{\scriptsize $\text{-}k_1\text{-}k_2\text{-}k_3$}] at (3-.2,.5) {};
  \node[label=-90:{\scriptsize $p$}] at (-.5,0) {};
  \node[label=-90:{\scriptsize $p$}] at (3.5,0) {};
  \node[label=-90:$\rightarrow$] at (-.5,.2) {};
  \node[label=-90:$\rightarrow$] at (3.5,.2) {};
\end{tikzpicture}
\\
(a) & (b)
\end{tabular}
\caption{Sample of four-loop diagrams contributing to (a) the scalar and (b) fermion anomalous dimension with external momentum $p$ and loop momenta $k_{1..4}$.}
\label{fig:g5}
\end{figure}

Scalar field anomalous dimensions are of order $\propto y^8$, which means only graphs with two fermion loops $\propto \mathrm{tr}(y^4) \times \mathrm{tr}(y^4)$ are problematic. However, after integrating over each fermion loop, only two integration momenta and one external momentum can be exchanged between the trace structures, which causes all $\tilde{\varepsilon}$ tensors to vanish.
An example is given in \fig{g5}(a): after integration over momenta $k_{3,4}$, each of the fermion loops may yield a $\tilde{\varepsilon}$ tensor which is fully contracted with a combination of the momenta $k_{1,2}$ and $p$, and thus vanishes.

For fermion field anomalous dimension only graphs  $\propto y^4 \times \mathrm{tr}(y^4)$ and  $\propto y^2 \times \mathrm{tr}(y^6)$ are relevant. The former is examplified in \fig{g5}(b). The external momentum is routed along the open line, and integrating the closed fermion loops ($k_4$ in \fig{g5}(b)) yields a contraction of at most three loop momenta ($k_{1,2,3}$ in the example) and hence no $\tilde{\varepsilon}$ tensor survives.

The UV poles of both Yukawa and quartic vertex corrections are extracted for vanishing external momenta. Relevant diagrams for the Yukawa has at least one, and for the quartic at least two closed fermion loops, meaning only three or two loop momenta remain after integrating those loops. This can be seen in the graphs (a) and (b) of \fig{g5}, if one external scalar field is attached to each fermion loop, and all external momenta are set to zero, in particular $p=0$. For both cases, the number of loop momenta are insufficient for non-vanishing $\tilde{\varepsilon}$ terms.
In consequence, none of the RGEs here can produce a $\gamma_5$-ambiguity at four-loop order.

As the results were are interested in are independent of the ambiguity, we may choose a theory not containing any $\gamma_5$ for computational convenience. Instead of working directly in~\eq{master-template}, we select the QFT 
\begin{equation}\label{eq:test-theory}
  \begin{aligned}
    \mathcal{L}' =
     & \  \tfrac{1}{2} \partial^{\mu} \phi_a \partial_\mu \phi_a + \tfrac{i}{2} \bar{\Psi} \gamma^\mu \partial_\mu \Psi    - \phi_a \bar{\Psi} Y^a \Psi   - \tfrac{1}{24} \lambda^{abcd} \,\phi_a \phi_b \phi_c \phi_d\,,
  \end{aligned}
  \end{equation}
in terms of Dirac fermions and hermitian Yukawa matrices $Y^a$. 
We compute all renormalisation group equations at four-loop order in dimensional regularisation~\cite{Bollini:1972bi,Bollini:1972ui} and the $\overline{\text{MS}}$ scheme~\cite{tHooft:1973mfk,Bardeen:1978yd}. This is achieved with a custom version of the \texttt{MaRTIn} framework~\cite{Brod:2024zaz}, which uses \texttt{QGRAF}~\cite{Nogueira:1991ex} for diagram generation and \texttt{FORM}~\cite{Kuipers:2012rf} for all other manipulations. UV poles are computed via infrared rearrangement with a common mass parameter~\cite{Chetyrkin:1997fm}. Moreover, four-loop integrals are reduced using \texttt{FMFT}~\cite{Pikelner:2017tgv} to a set of masters which are taken from~\cite{Czakon:2004bu,Lee:2010hs}. 
Using the software tool \texttt{FoRGEr}~\cite{Steudtner:FoRGEr}, we translate all RGEs back into the more general language of~\eq{master-template}. As a result, complete template expressions are obtained.

\section{Results}\label{sec:Results}

We compute the scalar- and fermion field anomalous dimensions 
\begin{equation}
  \gamma_\phi^{ab} = - \left( \frac{\mathrm{d} \sqrt{Z_\phi}}{\mathrm{d} \log \mu}Z_\phi^{-1/2}\right)^{ab} = \sum_{\ell=1}^\infty \frac{\gamma_\phi^{(\ell) ab}}{(4\pi)^{2\ell}} \,,\qquad \gamma_\psi^{ij} = - \left( \frac{\mathrm{d} \sqrt{Z_\psi}}{\mathrm{d} \log \mu}Z_\psi^{-1/2}\right)^{ij}= \sum_{\ell=1}^\infty \frac{\gamma_\psi^{(\ell) ij}}{(4\pi)^{2\ell}}
\end{equation}
where $\phi_\text{bare}^a = \sqrt{Z_\phi}^a_{\phantom{a}b} \phi^b$ and $\psi_\text{bare}^i = \sqrt{Z_\psi}^i_{\phantom{i}j} \psi^j$, respectively. 
We choose the scalar (fermion) field renormalisations and hence the corresponding anomalous dimensions to be symmetric (hermitian).  
The Yukawa and scalar quartic $\beta$-functions 
\begin{align}
  \beta_\lambda^{abcd} &= \frac{\mathrm{d} \lambda^{abcd}}{\mathrm{d} \log \mu} = {\gamma_\phi}{}^{a}_{\phantom{a}e} \lambda^{ebcd} + {\gamma_\phi}{}^{b}_{\phantom{b}e} \lambda^{eacd}\label{eq:beta-lam}
  + {\gamma_\phi}{}^{c}_{\phantom{c}e} \lambda^{eabd} + {\gamma_\phi}{}^{d}_{\phantom{d}e} \lambda^{eabc} + \hat{\beta}_\lambda^{abcd} \,,\\
  \beta_y^{a} &= \frac{\mathrm{d} y^{a}}{\mathrm{d} \log \mu} = {\gamma_\phi}{}^{a}_{\phantom{a}b} y^b +  {\gamma_\psi} y^a +  y^a \gamma_\psi + \hat{\beta}_y^{a} \label{eq:beta-yuk}
\end{align}
consist of leg corrections according to the anomalous dimensions of the corresponding fields, as well as vertex corrections with the respective loop expansions
\begin{equation}
  \hat{\beta}_\lambda^{abcd} = \sum_{\ell=1}^\infty\frac{ \hat{\beta}_\lambda^{(\ell)\, abcd}}{(4\pi)^{2\ell}}\,,\qquad   \hat{\beta}_y^{a} = \sum_{\ell=1}^\infty\frac{ \hat{\beta}_y^{(\ell)\, a}}{(4\pi)^{2\ell}}\,.
\end{equation}
Up to three loops, the RGEs are listed in~\cite{Jack:2024sjr}. 
The explicit results at four-loop order are quite long and relegated to App.~\ref{app:gammaS}-\ref{app:betaQ}. Schematically, they read 
\begin{align*}
    \gamma_\phi^{(4)} 
    &= \lambda^4 \ \left[\text{4 terms}\right] &
    &\,+ \lambda^3 \,\mathrm{tr}( y^2) \ \left[\text{3 terms}\right] &
    &\,+ \lambda^2 \,\mathrm{tr}^2(y^2)  \ \left[\text{2 terms}\right] &
    \\ 
    &\,+ \lambda^2 \,\mathrm{tr}( y^4) \ \left[\text{8 terms}\right] &
    &\,+ \lambda \,\mathrm{tr}( y^4) \,\mathrm{tr}( y^2) \ \left[\text{4 terms}\right]&
    &\,+ \lambda \,\mathrm{tr}( y^6)  \ \left[\text{9 terms}\right] &
    \\ 
    &\,+  \mathrm{tr}( y^4) \,\mathrm{tr}^2( y^2)  \ \left[\text{2 terms}\right] &
    &\,+  \mathrm{tr}^2( y^4) \,  \ \left[\text{6 terms}\right] &
    &\,+  \mathrm{tr}( y^6) \,\mathrm{tr}( y^2)  \ \left[\text{11 terms}\right] &
    \\ 
    &\,+  \mathrm{tr}( y^8)  \ \left[\text{39 terms}\right] &
    \\ &= \ \left[\text{88 terms}\right]\,, \yestag{} \\[.5em]
    \gamma_\psi^{(4)} 
    &= y^2 \lambda^3  \ \left[\text{2 terms}\right] &
  &\,+  y^2 \, \mathrm{tr}(y^2) \,\lambda^2  \ \left[\text{3 terms}\right] &
  &\,+  y^4 \lambda^2  \ \left[\text{6 terms}\right] &
   \\ 
   &\,+  y^2 \, \mathrm{tr}^2(y^2) \,\lambda  \ \left[\text{2 terms}\right] &
   &\,+  y^2 \, \mathrm{tr}(y^4) \,\lambda  \ \left[\text{3 terms}\right] &
   &\,+  y^4 \, \mathrm{tr}(y^2) \,\lambda  \ \left[\text{5 terms}\right]&
   \\  
   &\,+  y^6  \lambda  \ \left[\text{8 terms}\right]&
   &\,+  y^2 \, \mathrm{tr}^3(y^2)   \ \left[\text{1 term}\right] &
   &\,+  y^2 \, \mathrm{tr}(y^4) \, \mathrm{tr}(y^2)   \ \left[\text{4 terms}\right] &
  \\ 
  &\,+  y^2 \, \mathrm{tr}(y^6)   \ \left[\text{8 terms}\right]&
  &\,+  y^4 \, \mathrm{tr}^2(y^2)   \ \left[\text{5 terms}\right] &
  &\,+  y^4 \, \mathrm{tr}(y^4)   \ \left[\text{9 terms}\right] &
  \\
  &\,+  y^6 \, \mathrm{tr}(y^2)   \ \left[\text{19 terms}\right] &
  &\,+  y^8    \ \left[\text{47 terms}\right] & 
  \\
  &= \ \left[\text{122 terms}\right]\,, \yestag{} \\[.5em]
 \hat{\beta}_{y}^{(4)} 
  &= y^3 \lambda^3  \ \left[\text{10 terms}\right] &
    &\,+  y^3 \, \mathrm{tr}(y^2) \,\lambda^2  \ \left[\text{12 terms}\right] &
    &\,+  y^5 \, \lambda^2  \ \left[\text{32 terms}\right] &
  \\ 
  &\,+  y^3 \, \mathrm{tr}^2(y^2) \,\lambda  \ \left[\text{4 terms}\right]&
  &\,+  y^3 \, \mathrm{tr}(y^4) \,\lambda  \ \left[\text{13 terms}\right] &
  &\,+  y^5 \, \mathrm{tr}(y^2) \,\lambda  \ \left[\text{24 terms}\right] &
   \\ 
   &\,+  y^7 \,\lambda  \ \left[\text{79 terms}\right]&
   &\,+  y^3 \, \mathrm{tr}^3(y^2)  \ \left[\text{1 term}\right]&
   &\,+  y^3 \, \mathrm{tr}(y^4) \, \mathrm{tr}(y^2)  \ \left[\text{9 terms}\right]&
   \\ 
   &\,+  y^3 \, \mathrm{tr}(y^6)   \ \left[\text{24 terms}\right]&
   &\,+  y^5 \, \mathrm{tr}^2(y^2)  \ \left[\text{11 terms}\right]&
   &\,+  y^5 \, \mathrm{tr}(y^4)  \ \left[\text{33 terms}\right]&
   \\ 
   &\,+  y^7 \, \mathrm{tr}(y^2)  \ \left[\text{79 terms}\right]&
   &\,+  y^9  \ \left[\text{269 terms}\right] &
   \\
   &= \ \left[\text{600 terms}\right] \,, \yestag{} \\[.5em]
   \hat{\beta}_{\lambda}^{(4)} 
   &= \lambda^5    \ \left[\text{26 terms}\right]&
   &\,+  \lambda^4\, \mathrm{tr}(y^2)   \ \left[\text{20 terms}\right]&
   &\,+  \lambda^3\, \mathrm{tr}^2(y^2)   \ \left[\text{8 terms}\right]&
      \\ 
   &\,+  \lambda^3\, \mathrm{tr}(y^4)   \ \left[\text{41 terms}\right]&
   &\,+  \lambda^2\, \mathrm{tr}^3(y^2)   \ \left[\text{2 terms}\right]&
   &\,+  \lambda^2\, \mathrm{tr}(y^4)\, \mathrm{tr}(y^2)   \ \left[\text{23 terms}\right]&
   \\ 
   &\,+  \lambda^2\, \mathrm{tr}(y^6)   \ \left[\text{83 terms}\right]&
   &\,+  \lambda\, \mathrm{tr}(y^4)\, \mathrm{tr}^2(y^2)   \ \left[\text{4 terms}\right]&
   &\,+  \lambda\, \mathrm{tr}^2(y^4)   \ \left[\text{19 terms}\right]&
   \\ 
   &\,+  \lambda\, \mathrm{tr}(y^6)\, \mathrm{tr}(y^2)    \ \left[\text{32 terms}\right]&
   &\,+  \lambda\, \mathrm{tr}(y^8)    \ \left[\text{142 terms}\right]&
   &\,+  \mathrm{tr}(y^6)\, \mathrm{tr}^2(y^2)    \ \left[\text{3 terms}\right]&
    \\ 
    &\,+  \mathrm{tr}(y^6)\, \mathrm{tr}(y^4)    \ \left[\text{67 terms}\right]&
    &\,+  \mathrm{tr}^2(y^4)\, \mathrm{tr}(y^2)    \ \left[\text{3 terms}\right]&
    &\,+  \mathrm{tr}(y^{10})    \ \left[\text{190 terms}\right] & \\
    &= \ \left[\text{663 terms}\right] \,,\yestag{}
\end{align*} 
where the number of terms counts over distinct diagrams, but does not yet include permutations of external indices. 

Using the dummy field technique~\cite{Martin:1993zk,Luo:2002ti,Schienbein:2018fsw}, $\beta$-functions for fermion and scalar masses as well as scalar cubic couplings follow from $\beta_y$ and $\beta_\lambda$. Moreover, $\beta$-functions for vacuum expectation values are simply related to scalar field anomalous dimension~\cite{Steudtner:2024teg}.

In theories where several scalars or fermions are not distinguished by global symmetries, a rotation between these fields is always possible. Sometimes these symmetry transformations are hidden as they also require a transformation of some couplings. These theories exhibit ambiguities taking the shape of antisymmetric (antihermitian) contributions to scalar (fermion) field anomalous dimensions~\cite{Herren:2021yur}, which occur at three-loop order and higher. Such ambiguities have been encountered previously in e.g.~\cite{Bednyakov:2012en,Fortin:2012hn,Herren:2017uxn,Steudtner:2024teg} and are also present in this work, both for scalar and fermion fields.
In their wake, zeros of $\beta$-functions do not faithfully predict conformal symmetry. 
That effect can be compensated by shifting the scalar (fermion) anomalous dimensions by a antisymmetric (antihermitian) corrections $\upsilon_{\psi}^{ab}$ $(\upsilon_{\phi}^{ij})$. Applying this procedure to the leg corrections of each $\beta$-function restores conformality at the zeros~\cite{Fortin:2012hn,Herren:2021yur}.
At three-loop order, $\upsilon_{\phi,\psi}$ are given in~\cite{Davies:2021mnc} including gauge interactions. A four-loop basis of $\upsilon_{\phi,\psi}$ can be readily extracted from $\gamma_{\phi,\psi}^{(4)}$ in App.~\ref{app:gammaS} and~\ref{app:gammaF}, respectively, by picking all terms with symmetrisations $\mathcal{S}_2$ and replacing them with antisymmetrisations.
However, computing the coefficients for such a basis is beyond the scope of this work.

Let us now turn towards the results of App.~\ref{app:gammaS}-\ref{app:betaQ}.
The pure scalar structures of $\gamma_\phi^{(4)}$ and $\hat{\beta}_\lambda^{(4)}$ agree with the literature~\cite{Kazakov:1979ik,Jack:1990eb,Jack:2018oec,Steudtner:2020tzo}. Moreover, all RGEs are compatible with the direct computation of the chiral-Ising, XY- and Heisenberg model available in~\cite{Zerf:2017zqi}.
Furthermore, we have cross-checked our findings with the $\mathcal{N}=1$ supersymmetric Wess-Zumino model featuring the superpotential 
\begin{equation}\label{eq:suppot}
  W = \tfrac16 Y^{ABC} \,\Phi_{A} \Phi_B \Phi_C\,.
\end{equation}
Here, $\Phi_A$ are chiral superfields containing a Weyl fermion $\psi_A$ and complex scalar $\phi_A$. After integrating superspace coordinates and auxiliary fields, one obtains an interaction Lagrangian
\begin{equation}\label{eq:SUSY1-lag}
  - \mathcal{L}_\text{int} = \tfrac12 Y^{ABC}\, \phi_{A} \psi_B \psi_{C} + \tfrac12 Y_{ABC} \, \phi^{A} \psi^B \psi^{C} + \tfrac1{4} Y^{ABE} Y_{CDE} \,\phi_{A} \phi_B \phi^C \phi^{D}  
\end{equation}
with raising and lowering of indices being understood as complex conjugation.
At four-loop order, this implies the following relations
\begin{align}
  \left(\gamma^{(4)}_\phi\right){}^A_{\phantom{A}B} &=\left(\gamma^{(4)}_\psi\right){}^A_{\phantom{A}B}  = \left(\gamma^{(4)}_\Phi\right){}^A_{\phantom{A}B}\,,\label{eq:superAD}\\
  \hat{\beta}^{(4)\,{ABC}}_{Y} &= 0\,, \label{eq:non-ren}\\
  \left(\hat{\beta}^{(4)}_{\lambda}\right){}^{AB}_{\phantom{AB}CD} &= 2\,Y^{ABE} \left(\gamma_\Phi^{(4)}\right){}^{F}_{\phantom{F}E} Y_{CDF}\,,\label{eq:susy-quartic}
\end{align}
where the first is a consequence of scalars and fermions forming a superfield,
the second follows from the $\mathcal{N}=1$ non-renormalisation theorem~\cite{Grisaru:1979wc} and the third one owns to the identification of the quartic by the super-Yukawa tensors in~\eq{SUSY1-lag}.
We have explicitly verified that \eq{superAD}--\eq{susy-quartic} hold in the $\overline{\text{MS}}$ scheme.
For a general scheme, the supersymmetry relations are collected in App.~\ref{app:SUSY1}. 
Note that $\gamma_\Phi^{(4)}$ refers to four-loop superfield anomalous dimensions which have been computed in~\cite{Ferreira:1996az,Ferreira:1997bi,Gracey:2021yvb}.

In three spacetime dimensions, $\mathcal{N}=1$ supersymmetry is implemented on real superfields $\Phi_A$. The superpotential~\eq{suppot} leads to the interaction Lagrangian
\begin{equation}\label{eq:susy-lag}
  - \mathcal{L}_\text{int} = \tfrac12 Y^{ABC}\, \varphi_{A} \chi_B \chi_{C}  + \tfrac1{24} Y^{ABE} Y^{CDE} \,\varphi_{A} \varphi_B \varphi_C \varphi_{D}  
\end{equation}
similar to the four-dimensional case, but consisting of real scalars $\varphi_A$ and real two-component spinors $\chi_A$. Continuation to four spacetime dimensions yields a case sometimes called $\mathcal{N}=\tfrac12$ supersymmetry, featuring totally symmetric Yukawa tensors $Y^{ABC}$ and $\lambda^{ABCD} = \mathcal{S}_3\,Y^{ABE} Y^{ECD}$. Scalar and fermion anomalous dimensions concur, but vertex corrections $\hat{\beta}^{(4)}_{Y}$ do not vanish as opposed to $\mathcal{N}=1$.   
At four-loop order this stipulates
\begin{align}
  \left(\gamma^{(4)}_\varphi\right){}^{AB} &=\left(\gamma^{(4)}_\chi\right){}^{AB} \,, \label{eq:N=1/2-AD}\\
  \hat{\beta}^{(4)\,{ABC}}_{Y} &= \hat{\beta}^{(4)\,{(ABC)}}_{Y} \,,\label{eq:N=1/2-yuk}\\
  \hat{\beta}^{(4)}_{\lambda^{ABCD}} &= 2\,\mathcal{S}_3\,Y^{ABE} \left(\gamma_\Phi^{(4)}\right)^{EF} Y^{CDF} + \mathcal{S}_6\,\hat{\beta}_{Y}^{(4)\,{ABE}} Y^{ECD}\,. \label{eq:N=1/2-quartic}
\end{align}
While similar conditions hold at one-, two- and three-loop order~\cite{Jack:2024sjr}, we find~\eq{N=1/2-yuk} and~\eq{N=1/2-quartic} to be violated at four-loop in the $\overline{\text{MS}}$ scheme. This is in accordance with the observations of~\cite{Zerf:2017zqi}, where the violation is attributed to differences in the spinor algebra between three and four dimensions. 
In three dimensions, odd $\gamma$-traces do not vanish, but give rise to terms proportional to Levi-Civita tensors, e.g. 
\begin{equation}\label{eq:3-gamma}
  \mathrm{tr}(\gamma^\mu \gamma^\nu \gamma^\rho) = -4 i \,\epsilon^{\mu\nu\rho}\,, 
\end{equation}
and similar for higher odd powers of $\gamma$.
This leads to new tensor contraction in the template expression for RGEs, consisting of odd traces of Yukawa matrices $\propto \mathrm{tr}(y^3)$ and higher powers, which are absent in four dimensions. 
While such terms vanish naturally in some scenarios -- for instance in ($d=4$) $\mathcal{N}=1$ supersymmetry due to the holomorphic structure of the superpotential-- this is not the general case.
However, as the terms in question are connected to Levi-Civita tensors, they only start appearing at four-loop order. This can be seen by following arguments somewhat analogous ato the discussion of the $\gamma_5$-ambiguity in \Sec{Notation}.

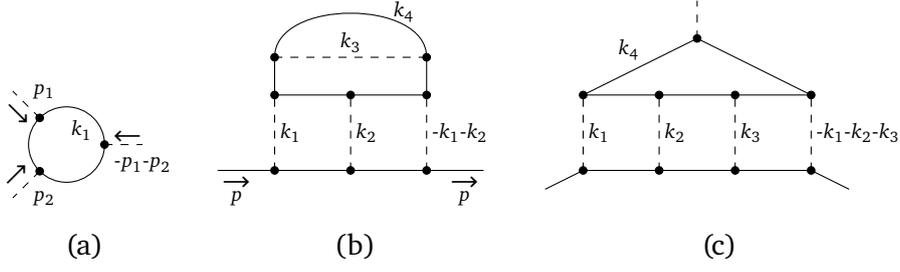
\begin{figure}[ht!]
\centering
\begin{tabular}{ccc}
\begin{tikzpicture}
  \draw (0,0) circle (.5);
  \draw[dashed] ({-.25*sqrt(2)},{+.25*sqrt(2)}) -- ({-.5*sqrt(2)},{+.5*sqrt(2)});
  \draw[dashed] ({-.25*sqrt(2)},{-.25*sqrt(2)}) -- ({-.5*sqrt(2)},{-.5*sqrt(2)});
  \draw[dashed] (.5,0)  -- (1,0) ;
  \node[label=0:{\scriptsize $p_1$}] at ({-.5*sqrt(2)},{+.5*sqrt(2)})  {};
  \node[label=0:{\scriptsize $p_2$}] at ({-.5*sqrt(2)},{-.5*sqrt(2)})  {};
  \node[label=-90:{\scriptsize $\text{-}p_1\text{-}p_2$}] at (1,0.1)  {};
  \node[label=45:{\scriptsize $k_1$}] at ({-.15*sqrt(2)},{-.15*sqrt(2)})  {};
  \node at (.8,.1)  {$\leftarrow$};
  \node[rotate=45] at (-.65,-.4)  {$\rightarrow$};
  \node[rotate=-45] at (-.65,+.4)  {$\rightarrow$};
  \filldraw[black] ({-.25*sqrt(2)},{+.25*sqrt(2)}) circle (.05);
  \filldraw[black] ({-.25*sqrt(2)},{-.25*sqrt(2)}) circle (.05);
  \filldraw[black] (.5,0) circle (.05);
\end{tikzpicture}&
\begin{tikzpicture}
  \draw (-.75,0) -- (2.75,0);
  \draw (0,1.5) -- (0,1) -- (2,1) -- (2,1.5) to [bend right=90] (0,1.5) ;
  \draw[dashed] (2,1.5) to [bend left=0] (0,1.5) ;
  \draw[dashed] (0,0) -- (0,1);
  \draw[dashed] (1,0) -- (1,1);
  \draw[dashed] (2,0) -- (2,1);
  \filldraw[black] (0,0) circle (.05);
  \filldraw[black] (1,0) circle (.05);
  \filldraw[black] (2,0) circle (.05);
  \filldraw[black] (0,1) circle (.05);
  \filldraw[black] (1,1) circle (.05);
  \filldraw[black] (2,1) circle (.05);
  \filldraw[black] (2,1.5) circle (.05);
  \filldraw[black] (0,1.5) circle (.05);
  \node[label=90:{\scriptsize $k_4$}] at (1.7,1.7) {};
  \node[label=90:{\scriptsize $k_3$}] at (1,1.3) {};
  \node[label=0:{\scriptsize $k_1$}] at (-.2,.5) {};
  \node[label=0:{\scriptsize $k_2$}] at (1-.2,.5) {};
  \node[label=0:{\scriptsize $\text{-}k_1\text{-}k_2$}] at (2-.2,.5) {};
  \node[label=-90:{\scriptsize $p$}] at (-.5,0) {};
  \node[label=-90:$\rightarrow$] at (-.5,.2) {};
  \node[label=-90:{\scriptsize $p$}] at (2.5,0) {};
  \node[label=-90:$\rightarrow$] at (2.5,.2) {};
\end{tikzpicture}
&
\begin{tikzpicture}
  \draw (-.5,-.25) -- (0,0) -- (3,0) -- (3.5,-.25);
  \draw (0,1) -- (3,1)  -- (1.5,1.75) --  (0,1);
  \draw[dashed] (0,0) -- (0,1);
  \draw[dashed] (1,0) -- (1,1);
  \draw[dashed] (2,0) -- (2,1);
  \draw[dashed] (3,0) -- (3,1);
  \draw[dashed] (1.5,1.75)  -- (1.5,2.25) ;
  \filldraw[black] (0,0) circle (.05);
  \filldraw[black] (1,0) circle (.05);
  \filldraw[black] (2,0) circle (.05);
  \filldraw[black] (3,0) circle (.05);
  \filldraw[black] (0,1) circle (.05);
  \filldraw[black] (1,1) circle (.05);
  \filldraw[black] (2,1) circle (.05);
  \filldraw[black] (3,1) circle (.05);
  \filldraw[black] (1.5,1.75) circle (.05);
  \node[label=90:{\scriptsize $k_4$}] at (0.6,1.2) {};
  \node[label=0:{\scriptsize $k_1$}] at (-.2,.5) {};
  \node[label=0:{\scriptsize $k_2$}] at (1-.2,.5) {};
  \node[label=0:{\scriptsize $k_3$}] at (2-.2,.5) {};
  \node[label=0:{\scriptsize $\text{-}k_1\text{-}k_2\text{-}k_3$}] at (3-.2,.5) {};
  \node[label=-90:\phantom{\scriptsize $p$}] at (-.5,0) {};
  \node[label=-90:\phantom{\scriptsize $p$}] at (3.5,0) {};
\end{tikzpicture}
\\
(a) & (b) & (c)
\end{tabular}
\caption{(a) Vanishing one-loop subdiagram $\propto \mathrm{tr}(y^3)$. (b) Vanising four-loop contribution to the fermion anomalous dimensions. (c) Non-vanishing Yukawa vertex correction. See main text for further explanations.}
\label{fig:levi3}
\end{figure}

In consequence, one must expect that dimensional continuation from $d=4-\varepsilon$ to $d=3$ systematically fails at four-loop order and higher. In the following, we will propose a solution to this issue. To do so, we first identify which terms absent in our basis of tensor structures are generated by odd $\gamma$-traces. 
In order to have these Levi-Civita terms give rise to non-vanishing contributions, an odd $\gamma$-trace needs to be paired up with either an open fermion line or another such trace. More so, odd $\gamma$-traces need to be contracted with three independent momenta in order to not vanish because of the antisymmetry of the Levi-Civita. Both loop and external momenta are admissible, though the structure must survive even after integrating the fermion loop to which the odd $\gamma$-trace corresponds.
Thus the structure $\propto \mathrm{tr}(y^3)$ may actually not occur: it originates from a fermion loop with three external scalar legs, as depicted in \fig{levi3}(a). After integration over the fermion loop momentum $k_1$, the expression only retains two independent external momenta $p_{1,2}$ and thus vanishes.

We will now argue where Levi-Civita terms enter for each of the four-loop RGEs.
Quartic vertex corrections exhibit UV poles at vanishing external momentum, which leaves only the four integration momenta. As they do not have open fermion lines, only diagrams with two traces may generate a Levi-Civita contribution. After integrating over both traces, only two momenta remain. Thus all Levi-Civita tensors vanish. 
Scalar leg corrections also require two traces, but are sensitive to an external momentum, which means three independent momenta are available. At four-loop order the structure of possible diagrams is either $\propto \mathrm{tr}(y^3) \times \mathrm{tr}(y^5)$ or $\propto \mathrm{tr}(y^3) \times\mathrm{tr}(y^3) \times\mathrm{tr}(y^2)$, both of which must vanish as $\mathrm{tr}(y^3)$ cannot generate Levi-Civita terms. 
Fermion leg corrections have an open line, so only a single odd trace is required. With the external momentum routed along the open line, three independent loop momenta remain after integrating the closed fermion loop. Only diagrams of the shape $\propto y^3\,\mathrm{tr}(y^5)$ are admissible. This requires a scalar exchange between two Yukawa vertices in the fermion loop, as for instance shown in \fig{levi3}(b). However, this scalar subloop also factorises and after integrating both $k_4$ and $k_3$, only two independent momenta $k_{1,2}$ remain. A similar pictures emerges for all other graphs contributing $\propto y^3\,\mathrm{tr}(y^5)$.
Thus, fermion anomalous dimensions cannot receive four-loop contributions due to Levi-Civitas.
Analysing the four-loop Yukawa vertex diagrams, all Levi-Civitas vanish except in a single family $\propto y^4 \mathrm{tr}(y^5)$, as already identified in~\cite{Zerf:2017zqi}. An example is shown in \fig{levi3}(c): after integration of $k_4$, three independent momenta $k_{1,2,3}$ are contracted with the Levi-Civita tensor.
In full generality we find 8 suitable terms
\begin{align*}
    \tilde{\beta}_y^a &=  y^{bcde}\,\mathrm{tr} \left( \Delta_1\, y^{abcde} + \Delta_2\, y^{abdce} +  \Delta_3\,y^{acbed} +  \Delta_4\, y^{acebd} \right) \\
    &\phantom{= }  + \mathcal{S}_2\,y^{bcde} \, \mathrm{tr} \left(\Delta_5\,\, y^{abecd} +  \Delta_6\, y^{abedc}  + \Delta_7\,y^{abced} +  \Delta_8\, y^{abdec}\right) \,. \yestag{}
  \end{align*}
We fix the coefficient $\Delta_{1\dots 8}$ by imposing $\mathcal{N}=\tfrac12$ supersymmetry via the relations~\eq{N=1/2-AD}--\eq{N=1/2-quartic}. Meanwhile, other coefficients of the RGEs remain fixed in the $\overline{\text{MS}}$ scheme. As noted in~\cite{Zerf:2017zqi}, this amounts to a modified regularisation procedure, giving rise to a renormalisation scheme we shall dub $\overline{\text{MS}}_3$ and 
\begin{equation}
  \gamma_{\phi,\psi}^{(4), \overline{\text{MS}}_3} = \gamma_{\phi,\psi}^{(4), \overline{\text{MS}}} \,,\quad \hat{\beta}_{\lambda}^{(4), \overline{\text{MS}}_3} = \hat{\beta}_{\lambda}^{(4), \overline{\text{MS}}}\,, \quad \hat{\beta}_{y}^{(4), \overline{\text{MS}}_3} = \hat{\beta}_{y}^{(4), \overline{\text{MS}}} + \tilde{\beta}_y\,.
\end{equation}
Due to the generality of our template expression, we expect that this procedure allows for dimensional continuation at four-loops not just in supersymmetric, but any renormalisable and gaugeless QFT. 
We explicitly find in the $\overline{\text{MS}}_3$ scheme
\begin{align*}
  \Delta_1 &= 2 ( 1 + 10\,\zeta_3 - 10\,\zeta_5 )\,,\quad
  &\Delta_2 &= \Delta_6 = \Delta_7 = 2(7\,\zeta_3 - 10\,\zeta_5)\,,\\
  \Delta_3 &= \Delta_5 = \Delta_8 =20 (\zeta_3 - \zeta_5)\,,\quad
  &\Delta_4 &= 6 (4 \,\zeta_3 - 5\,\zeta_5) \yestag{}\,,
\end{align*}
which exactly restore the conditions \eq{N=1/2-yuk} and \eq{N=1/2-quartic} violated in the conventional $\overline{\text{MS}}$ scheme.
Note that \eq{N=1/2-AD}--\eq{N=1/2-quartic} imply an enormous amount of conditions for the coefficients of an arbitrary renormalisation scheme. We will refrain from listing these here for the sake of brevity. Their extraction requires efficient tensor canonisation algorithms. Here we have used \texttt{FORM}~\cite{Kuipers:2012rf} with the option \texttt{renumber,1} as well as \texttt{TensorReduction} in \texttt{Mathematica}~\cite{Mathematica} for more advanced cases.

As a direct application of our results, we ameliorate the four-loop top coupling and Higgs quartic $\beta$-functions in the Standard Model of particle physics.
Using the notation $\alpha_i = g_i^2/(4\pi)^2$, $\alpha_t = y_t^2/(4\pi)^2$ and $\alpha_\lambda = \lambda/(4\pi)^2$, we find 
\begin{align}
  \frac{\mathrm{d} \alpha_t}{\mathrm{d} \log \mu} \bigg|_{\alpha^5}  &= {\color{gray}-\left[\tfrac{1379027}{81} + \tfrac{224}3 \pi^4  - \tfrac{759104}{27} \zeta_3 + \tfrac{20800}{3} \zeta_5\right]\alpha_3^4 \alpha_t } \notag \\
  &\quad - \left[\tfrac{567}{2} - \tfrac{27}{20} \pi^4 + \tfrac{1863}{2}\zeta_3 + 270 \zeta_5\right] \alpha_t^5 - \left[\tfrac{6897}{2} + 1512 \zeta_3\right] \alpha_t^4 \alpha_\lambda \notag \\
   &\quad - \left[2784 + 1116\zeta_3\right] \alpha_t^3 \alpha_\lambda^2 - \left[4848 - 1440 \zeta_3\right] \alpha_t^2 \alpha_\lambda^3 + 2340\, \alpha_t \alpha_\lambda^4 \\[.2em]
  \frac{\mathrm{d} \alpha_\lambda}{\mathrm{d} \log \mu}\bigg|_{\alpha^5}  &={\color{gray}  \left[\tfrac{104776}{27} + 32 \pi^4 + \tfrac{39616}{3} \zeta_3  - \tfrac{126400}{9}\zeta_5\right] \alpha_t^2 \alpha_3^3} \notag \\
  &\quad + \left[\tfrac{8601}{8} + \tfrac{27}{10} \pi^4 + 648 \zeta_3 + 4140 \zeta_5 \right] \alpha_t^5 \notag \\
  &\quad +\left[\tfrac{369039}{16} - \tfrac{357}{5} \pi^4 + 11163 \zeta_3 - 9480 \zeta_5\right] \alpha_t^4 \alpha_\lambda \notag \\
  &\quad - \left[\tfrac{43773}{2} - \tfrac{762}{5} \pi^4 + 16200 \zeta_3 - 28080 \zeta_5\right] \alpha_t^3 \alpha_\lambda^2\notag \\
  &\quad - \left[92664 + \tfrac{1344}{5} \pi^4 + 22608 \zeta_3 + 80640 \zeta_5\right] \alpha_t^2 \alpha_\lambda^3\notag \\
  &\quad - \left[31500 - \tfrac{2016}{5} \pi^4 + 69120 \zeta_3\right] \alpha_t \alpha_\lambda^4\notag \\
  &\quad - \left[206424 - \tfrac{2688}{5} \pi^4 + 204672 \zeta_3 + 280320 \zeta_5\right] \alpha_\lambda^5
\end{align}
where the grey terms have been computed previously in~\cite{Martin:2015eia,Chetyrkin:2016ruf}. Note that further contributions $\propto \alpha_{1,2,3}$ are not available but might be of similar size.

\section{Summary}\label{sec:End}

In this work, we have provided general four-loop expressions for all $\beta$-functions and anomalous dimensions in any renormalisable theory without gauge interactions. We have obtained and cross-checked $\overline{\text{MS}}$ results which are detailed in App.~\ref{app:gammaS}--\ref{app:betaQ}.
Not only are these expressions independent of the $\gamma_5$-ambiguity, but also provide a skeleton for other renormalisation schemes, corresponding to different coefficients.

As the technological burden for any four-loop computation is formidable, our results represent an attractive alternative to a direct calculation. This is in particular true for theories with many fields and couplings. To facilitate this strategy, we have implemented the RGEs in \texttt{FoRGEr}~\cite{Steudtner:FoRGEr} and also attached an electronic version of the expressions to this publication. 

With our findings at hand, we are able to calculate hitherto unknown four-loop contributions to the top coupling and Higgs self-interaction $\beta$-functions in the Standard Model. 
Furthermore, our template expressions can be utilised to investigate critical phenomena of Gross-Neveu-Yukawa theories in three spacetime dimensions via dimensional continuation from $d=4-\varepsilon$ and approaching $\varepsilon \to 1$. Starting at four-loop order, differences in the spinor algebra between four and three dimensions spoil this strategy. However, we have countered this phenomenon by the construction presented in Sec.~\ref{sec:Results}, enabling dimensional continuation at four-loop precision.

\section*{Acknowledgements}
The author is indebted to Sandra Kvedarait\.e and York Schröder for comments on the manuscript as well as Max Uetrecht for his support in the development of four-loop routines for \texttt{MaRTIn}~\cite{Brod:2024zaz}.

\appendix
\section{Scalar Anomalous Dimension}\label{app:gammaS}
The four-loop scalar field anomalous dimensions can be written as 
\begin{equation}\label{eq:gamS-terms}
  \gamma_\phi^{(4)} = \gamma_{\phi,\lambda^4}^{(4)} + \gamma_{\phi,\lambda^3 y^2}^{(4)} + \gamma_{\phi,\lambda^2 y^4}^{(4)} + \gamma_{\phi,\lambda y^6}^{(4)} + \gamma_{\phi,y^4 y^2 y^2}^{(4)} + \gamma_{\phi,y^4 y^4}^{(4)}  + \gamma_{\phi,y^6 y^2 }^{(4)}  + \gamma_{\phi,y^8}^{(4)}\,.
\end{equation}
The first term is the pure scalar contribution
\begin{equation}
  \begin{aligned}
    \gamma_{\phi,\lambda^4}^{(4)\, ab} &= 
\gamma_{\phi,1}\ \lambda^{acde}\,\lambda^{defg}\,\lambda^{fghi}\,\lambda^{bchi}
+\gamma_{\phi,2}\ \lambda^{acde}\,\lambda^{eghi}\,\lambda^{fghi}\,\lambda^{bcdf}\\
&\quad
+\gamma_{\phi,3}\ \lambda^{acde}\,\lambda^{dfgh}\,\lambda^{eghi}\,\lambda^{bcfi}
+\gamma_{\phi,4}\ \lambda^{acde}\,\lambda^{cdfg}\,\lambda^{eghi}\,\lambda^{bfhi}
  \end{aligned}
\end{equation}
which is known for some time~\cite{Kazakov:1979ik,Jack:1990eb,Jack:2018oec}. The other terms read
\begin{align*}
    \gamma_{\phi,\lambda^3 y^2}^{(4)\, ab} &= 
      \gamma_{\phi,5}\ \lambda^{acde}\,\lambda^{cdfg}\,\lambda^{bfgh}\,\mathrm{tr} \left(y^{e h}\right) 
      + \gamma_{\phi,6}\ \lambda^{acde}\,\lambda^{bcdf}\,\lambda^{efgh}\,\mathrm{tr} \left(y^{g h}\right)
      \\
      &\quad
      + \gamma_{\phi,7}\ \mathcal{S}_2\,\lambda^{acde}\,\lambda^{bcfg}\,\lambda^{defh}\,\mathrm{tr} \left(y^{g h}\right)\,,\yestag{}\\[.5em]
    \gamma_{\phi,\lambda^2 y^4}^{(4)\, ab} &= 
 \gamma_{\phi,8}\  \lambda^{acde}\,\lambda^{bcdf}\,\mathrm{tr} \left(y^{e g}\right)\,\mathrm{tr} \left(y^{g f}\right)
 + \gamma_{\phi,9}\ \lambda^{acde}\,\lambda^{bcfg}\,\mathrm{tr} \left(y^{d f}\right)\,\mathrm{tr} \left(y^{e g}\right)
 \\&\phantom{=}
+ \gamma_{\phi,10}\ \lambda^{acde}\,\lambda^{bcfg}\,\mathrm{tr} \left(y^{d e f g}\right)
 + \gamma_{\phi,11}\ \lambda^{acde}\,\lambda^{bcfg}\,\mathrm{tr} \left(y^{d f e g}\right) 
\\&\phantom{=}
+ \gamma_{\phi,12}\ \lambda^{acde}\,\lambda^{bcdf}\,\mathrm{tr} \left(y^{e f g g}\right)
 + \gamma_{\phi,13}\ \lambda^{acde}\,\lambda^{bcdf}\,\mathrm{tr} \left(y^{e g f g}\right)
 \\&\phantom{=}
+ \gamma_{\phi,14}\ \mathrm{tr} \left(y^{a b c d}\right)\,\lambda^{cefg}\,\lambda^{defg}
 + \gamma_{\phi,15}\ \mathrm{tr} \left(y^{a c b d}\right)\,\lambda^{cefg}\,\lambda^{defg}
  \\&\phantom{=}
+ \gamma_{\phi,16}\ \mathcal{S}_2\,\lambda^{acde}\,\lambda^{cdfg}\,\mathrm{tr} \left(y^{b e f g}\right)
 + \gamma_{\phi,17}\ \mathcal{S}_2\,\lambda^{acde}\,\lambda^{cdfg}\,\mathrm{tr} \left(y^{b f e g}\right)\,,\yestag{}\\[.5em]
    \gamma_{\phi,\lambda y^6}^{(4)\, ab} &= 
    \gamma_{\phi,18}\  \mathrm{tr} \left(y^{a b c d}\right)\,\lambda^{cdef}\,\mathrm{tr} \left(y^{e f}\right)
    + \gamma_{\phi,19}\  \mathrm{tr} \left(y^{a c b d}\right)\,\lambda^{cdef}\,\mathrm{tr} \left(y^{e f}\right)
    \\&\phantom{=}
    + \gamma_{\phi,20}\ \mathcal{S}_2\,\lambda^{acde}\,\mathrm{tr} \left(y^{e f}\right)\,\mathrm{tr} \left(y^{b c d f}\right)
    + \gamma_{\phi,21}\ \mathcal{S}_2\,\lambda^{acde}\,\mathrm{tr} \left(y^{e f}\right)\,\mathrm{tr} \left(y^{b c f d}\right)
    \\&\phantom{=}
    + \gamma_{\phi,22}\  \mathrm{tr} \left(y^{a b c d e f}\right)\,\lambda^{cdef}
    + \gamma_{\phi,23}\  \mathrm{tr} \left(y^{a c b d e f}\right)\,\lambda^{cdef}
    \\&\phantom{=}
    + \gamma_{\phi,24}\  \mathrm{tr} \left(y^{a c d b e f}\right)\,\lambda^{cdef}
    + \gamma_{\phi,25}\ \mathcal{S}_2\,\lambda^{acde}\,\mathrm{tr} \left(y^{b f f c d e}\right)
    \\&\phantom{=}
    + \gamma_{\phi,26}\ \mathcal{S}_2\,\lambda^{acde}\,\mathrm{tr} \left(y^{b f c f d e}\right)
    + \gamma_{\phi,27}\ \mathcal{S}_2\,\lambda^{acde}\,\mathrm{tr} \left(y^{b f c d f e}\right)
    \\&\phantom{=}
    + \gamma_{\phi,28}\ \mathcal{S}_2\,\lambda^{acde}\,\mathrm{tr} \left(y^{b f c d e f}\right)
    + \gamma_{\phi,29}\ \mathcal{S}_2\,\lambda^{acde}\,\mathrm{tr} \left(y^{b c f f d e}\right)
    \\&\phantom{=}
    + \gamma_{\phi,30}\ \mathcal{S}_2\,\lambda^{acde}\,\mathrm{tr} \left(y^{b c f d f e}\right)\,,\yestag{}\\[.5em]
    \gamma_{\phi,y^4 y^2 y^2}^{(4)\, ab} &= 
    \gamma_{\phi,31}\  \mathrm{tr} \left(y^{a b c d}\right)\,\mathrm{tr} \left(y^{c e}\right)\,\mathrm{tr} \left(y^{e d}\right)
    + \gamma_{\phi,32}\  \mathrm{tr} \left(y^{a c b d}\right)\,\mathrm{tr} \left(y^{c e}\right)\,\mathrm{tr} \left(y^{e d}\right)\,,\yestag{}\\[.5em]
    \gamma_{\phi, y^4 y^4}^{(4)\, ab} &= 
    \gamma_{\phi,33}\   \mathrm{tr} \left(y^{a b c d}\right)\,\mathrm{tr} \left(y^{c d e e}\right)
    + \gamma_{\phi,34}\   \mathrm{tr} \left(y^{a b c d}\right)\,\mathrm{tr} \left(y^{c e d e}\right)
    \\&\phantom{=}
    + \gamma_{\phi,35}\   \mathrm{tr} \left(y^{a c b d}\right)\,\mathrm{tr} \left(y^{c d e e}\right)
    + \gamma_{\phi,36}\   \mathrm{tr} \left(y^{a c b d}\right)\,\mathrm{tr} \left(y^{c e d e}\right)
    \\&\phantom{=}
    + \gamma_{\phi,37}\   \mathrm{tr} \left(y^{a c d e}\right)\,\mathrm{tr} \left(y^{b c d e}\right)
    + \gamma_{\phi,38}\   \mathrm{tr} \left(y^{a c d e}\right)\,\mathrm{tr} \left(y^{b d c e}\right)\,,\yestag{}\\[.5em]
    \gamma_{\phi, y^6 y^2}^{(4)\, ab} &= 
    \gamma_{\phi,39}\   \mathrm{tr} \left(y^{a b c e e d}\right)\,\mathrm{tr} \left(y^{c d}\right)
    + \gamma_{\phi,40}\  \mathrm{tr} \left(y^{a b e c d e}\right)\,\mathrm{tr} \left(y^{c d}\right)
    \\&\phantom{=}
    + \gamma_{\phi,41}\   \mathcal{S}_2\,\mathrm{tr} \left(y^{a b c e d e}\right)\,\mathrm{tr} \left(y^{c d}\right)
    + \gamma_{\phi,42}\   \mathcal{S}_2\,\mathrm{tr} \left(y^{b a c c d e}\right)\,\mathrm{tr} \left(y^{d e}\right)
    \\&\phantom{=}
    + \gamma_{\phi,43}\   \mathrm{tr} \left(y^{a e b c e d}\right)\,\mathrm{tr} \left(y^{c d}\right)
    + \gamma_{\phi,44}\   \mathrm{tr} \left(y^{a c b e d e}\right)\,\mathrm{tr} \left(y^{c d}\right)
    \\&\phantom{=}
    + \gamma_{\phi,45}\   \mathcal{S}_2\,\mathrm{tr} \left(y^{a c b d e e}\right)\,\mathrm{tr} \left(y^{c d}\right)
    + \gamma_{\phi,46}\   \mathcal{S}_2\,\mathrm{tr} \left(y^{a e b e c d}\right)\,\mathrm{tr} \left(y^{c d}\right)
    \\&\phantom{=}
    + \gamma_{\phi,47}\   \mathrm{tr} \left(y^{a c d b e e}\right)\,\mathrm{tr} \left(y^{c d}\right)
    + \gamma_{\phi,48}\   \mathcal{S}_2\,\mathrm{tr} \left(y^{a c e b e d}\right)\,\mathrm{tr} \left(y^{c d}\right)
    \\&\phantom{=}
    + \gamma_{\phi,49}\   \mathcal{S}_2\,\mathrm{tr} \left(y^{a c e b d e}\right)\,\mathrm{tr} \left(y^{c d}\right)\,,\yestag{}\\[.5em]
    \gamma_{\phi, y^8}^{(4)\, ab} &= 
    \gamma_{\phi,50}\   \mathrm{tr} \left(y^{a b c d e e d c}\right)
    + \gamma_{\phi,51}\  \mathrm{tr} \left(y^{a b c d e d e c}\right)
    +  \gamma_{\phi,52}\    \mathcal{S}_2\,\mathrm{tr} \left(y^{a b c d e d c e}\right)
    \\&\phantom{=}
    + \gamma_{\phi,53}\   \mathcal{S}_2\,\mathrm{tr} \left(y^{a b c d d e c e}\right)
        +  \gamma_{\phi,54}\   \mathrm{tr} \left(y^{a b c e d d c e}\right)
    + \gamma_{\phi,55}\  \mathrm{tr} \left(y^{a b c d e c d e}\right)
    \\&\phantom{=}
    +  \gamma_{\phi,56}\   \mathrm{tr} \left(y^{a b c d c e d e}\right)
    + \gamma_{\phi,57}\  \mathrm{tr} \left(y^{a b c d d e e c}\right)
    +  \gamma_{\phi,58}\   \mathrm{tr} \left(y^{a b c c d d e e}\right)
     \\&\phantom{=}
    + \gamma_{\phi,59}\   \mathcal{S}_2\,\mathrm{tr} \left(y^{a b c c d e d e}\right)
    +  \gamma_{\phi,60}\    \mathcal{S}_2\,\mathrm{tr} \left(y^{a b c c d e e d}\right)
    + \gamma_{\phi,61}\   \mathcal{S}_2\,\mathrm{tr} \left(y^{a c b c d d e e}\right)
     \\&\phantom{=}
    +  \gamma_{\phi,62}\    \mathcal{S}_2\,\mathrm{tr} \left(y^{a c b c d e d e}\right)
    + \gamma_{\phi,63}\   \mathcal{S}_2\,\mathrm{tr} \left(y^{a c b c d e e d}\right)
    +  \gamma_{\phi,64}\    \mathcal{S}_2\,\mathrm{tr} \left(y^{a c b d c e e d}\right)
    \\&\phantom{=}
    + \gamma_{\phi,66}\   \mathcal{S}_2\,\mathrm{tr} \left(y^{a c b d c e d e}\right)
    +  \gamma_{\phi,66}\    \mathcal{S}_2\,\mathrm{tr} \left(y^{a c b d c d e e}\right)
    + \gamma_{\phi,67}\  \mathrm{tr} \left(y^{a c b d d c e e}\right)
     \\&\phantom{=}
    +  \gamma_{\phi,68}\   \mathrm{tr} \left(y^{a c b d e c e d}\right)
    + \gamma_{\phi,69}\  \mathrm{tr} \left(y^{a c b d e c d e}\right)
    +  \gamma_{\phi,70}\   \mathrm{tr} \left(y^{a c c b d d e e}\right)
        \\&\phantom{=}
    + \gamma_{\phi,71}\  \mathrm{tr} \left(y^{a c c b d e e d}\right)
    +  \gamma_{\phi,72}\   \mathrm{tr} \left(y^{a c c b d e d e}\right)
    + \gamma_{\phi,73}\   \mathcal{S}_2\,\mathrm{tr} \left(y^{a c d b c d e e}\right)
     \\&\phantom{=}
    +  \gamma_{\phi,74}\    \mathcal{S}_2\,\mathrm{tr} \left(y^{a c d b d c e e}\right)
    + \gamma_{\phi,75}\   \mathcal{S}_2\,\mathrm{tr} \left(y^{a c d b c e d e}\right)
    +  \gamma_{\phi,76}\    \mathcal{S}_2\,\mathrm{tr} \left(y^{a c d b d e c e}\right)
    \\&\phantom{=}
    + \gamma_{\phi,77}\  \mathrm{tr} \left(y^{a c d b e c d e}\right)
    +  \gamma_{\phi,78}\   \mathrm{tr} \left(y^{a c d b e d c e}\right)
    + \gamma_{\phi,79}\ \mathrm{tr} \left(y^{a c d b c e e d}\right)
    \\&\phantom{=}
    +  \gamma_{\phi,80}\   \mathrm{tr} \left(y^{a c d b d e e c}\right)
    + \gamma_{\phi,81}\   \mathcal{S}_2\, \mathrm{tr} \left(y^{a c c d b d e e}\right)
    +  \gamma_{\phi,82}\    \mathcal{S}_2\, \mathrm{tr} \left(y^{a c c d b e d e}\right)
    \\&\phantom{=}
    + \gamma_{\phi,83}\  \mathrm{tr} \left(y^{a c c d b e e d}\right)
    +  \gamma_{\phi,84}\   \mathrm{tr} \left(y^{a c d c b e d e}\right)
    + \gamma_{\phi,85}\  \mathrm{tr} \left(y^{a c d e b e d c}\right)
     \\&\phantom{=}
    +  \gamma_{\phi,86}\   \mathcal{S}_2\,\mathrm{tr} \left(y^{a c d e b e c d}\right)
    + \gamma_{\phi,87}\   \mathcal{S}_2\,\mathrm{tr} \left(y^{a c d e b d c e}\right)
    + \gamma_{\phi,88}\  \mathrm{tr} \left(y^{a c d e b c d e}\right)\,.\yestag{}
  \end{align*}
The coefficients in the $\overline{\text{MS}}$ scheme are found to be
\begin{align*}
\gamma_{\phi,1} & = -\tfrac{5}{64}\,,&
\gamma_{\phi,2} & = -\tfrac{5}{96}\,,&
\gamma_{\phi,3} & = \tfrac{13}{96}\,,\\
\gamma_{\phi,4} & = \tfrac{1}{3}\,,&
\gamma_{\phi,5} & = -\tfrac{1}{32}\,,&
\gamma_{\phi,6} & = 0\,,\\
\gamma_{\phi,7} & = \tfrac{7}{192}\,,&
\gamma_{\phi,8} & = -\tfrac{1}{192}\,,&
\gamma_{\phi,9} & =  -\tfrac{1}{192}\,\\
\gamma_{\phi,10} & = \tfrac{17}{24}\,,&
\gamma_{\phi,11} & = \tfrac{3 \zeta_3}{4}-\tfrac{13}{48}\,,&
\gamma_{\phi,12} & = \tfrac{95}{192}\,,\\
\gamma_{\phi,13} & = \tfrac{29}{96}\,,&
\gamma_{\phi,14} & = \tfrac{65}{192}\,,&
\gamma_{\phi,15} & = \tfrac{85}{288}\,,\\
\gamma_{\phi,16} & = -\tfrac{47}{48}\,,&
\gamma_{\phi,17} & = \tfrac{3 \zeta_3}{4}-\tfrac{39}{32}\,,&
\gamma_{\phi,18} & = 0\,,\\
\gamma_{\phi,20} & = 0\,,&
\gamma_{\phi,21} & = -\tfrac{1}{6}\,,&
\gamma_{\phi,19} & = -\tfrac{5}{8}\,,\\
\gamma_{\phi,22} & = -\tfrac{35}{24}\,,&
\gamma_{\phi,23} & = -\tfrac{37}{12}\,,&
\gamma_{\phi,24} & = -\tfrac{13}{6}\,,\\
\gamma_{\phi,25} & = -\tfrac{5}{12}\,,&
\gamma_{\phi,26} & = \tfrac{25}{24}\,,&
\gamma_{\phi,27} & = \zeta_3-\tfrac{35}{12}\,,\\
\gamma_{\phi,28} & = \tfrac{\zeta_3}{2}-\tfrac{121}{48}\,,&
\gamma_{\phi,29} & = -\tfrac{9}{16}\,,&
\gamma_{\phi,30} & = \tfrac{\zeta_3}{2}-\tfrac{79}{48}\,,\\
\gamma_{\phi,31} & = \tfrac{1}{64}-\tfrac{\zeta_3}{4}\,,&
\gamma_{\phi,32} & = \tfrac{49}{96}-\tfrac{\zeta_3}{2}\,,&
\gamma_{\phi,33} & = -\tfrac{105}{32}\,,\\
\gamma_{\phi,34} & = -\tfrac{\zeta_3}{2}-2\,,&
\gamma_{\phi,35} & = -\tfrac{119}{48}\,,&
\gamma_{\phi,36} & = -\zeta_3-\tfrac{3}{4}\,,\\
\gamma_{\phi,37} & = \tfrac{19}{12}-2 \zeta_3\,,&
\gamma_{\phi,38} & = -7 \zeta_3+\tfrac{65}{6}-\tfrac{\pi ^4}{30}\,,&
\gamma_{\phi,39} & = \tfrac{41}{192}\,,\\
\gamma_{\phi,40} & = -\tfrac{47}{64}\,,&
\gamma_{\phi,41} & = \tfrac{37}{64}\,,&
\gamma_{\phi,42} & = \tfrac{7}{32}\,,\\
\gamma_{\phi,43} & = \zeta_3-\tfrac{73}{48}\,,&
\gamma_{\phi,44} & = \tfrac{65}{24}-2 \zeta_3\,,&
\gamma_{\phi,45} & = \tfrac{43}{192}\,,\\
\gamma_{\phi,46} & = -\tfrac{7}{12}\,,&
\gamma_{\phi,47} & = \tfrac{1}{96}\,,&
\gamma_{\phi,48} & = -\tfrac{79}{96}\,,\\
\gamma_{\phi,49} & = -\tfrac{7 \zeta_3}{8}+\tfrac{7}{24}+\tfrac{\pi ^4}{240}\,,&
\gamma_{\phi,50} & = -\tfrac{17}{96}\,,&
\gamma_{\phi,51} & = \tfrac{79}{96}-\tfrac{\zeta_3}{2}\,,\\
\gamma_{\phi,52} & = \tfrac{35}{24}-2 \zeta_3\,,&
\gamma_{\phi,53} & = \tfrac{1}{64}\,,&
\gamma_{\phi,54} & = \tfrac{35}{96}-\zeta_3\,,\\
\gamma_{\phi,55} & = -\tfrac{9 \zeta_3}{4}+\tfrac{35}{24}-\tfrac{\pi ^4}{120}\,,&
\gamma_{\phi,56} & = 3 \zeta_3-\tfrac{19}{8}\,,&
\gamma_{\phi,57} & = \tfrac{79}{192}-\tfrac{\zeta_3}{4}\,,\\
\gamma_{\phi,58} & = \tfrac{\zeta_3}{4}-\tfrac{3}{64}\,,&
\gamma_{\phi,59} & = \tfrac{\zeta_3}{2}-\tfrac{3}{32}\,,&
\gamma_{\phi,60} & = \tfrac{5}{64}\,,\\
\gamma_{\phi,61} & = \tfrac{1}{6}\,,&
\gamma_{\phi,62} & = \tfrac{1}{3}\,,&
\gamma_{\phi,63} & = -\tfrac{1}{6}\,,\\
\gamma_{\phi,64} & = -\tfrac{7}{16}\,,&
\gamma_{\phi,65} & = \zeta_3+\tfrac{1}{24}\,,&
\gamma_{\phi,66} & = \tfrac{47}{32}\,,\\
\gamma_{\phi,67} & = -\tfrac{7}{96}\,,&
\gamma_{\phi,68} & = \tfrac{31}{24}-4 \zeta_3\,,&
\gamma_{\phi,69} & = -3 \zeta_3+\tfrac{37}{12}-\tfrac{\pi ^4}{30}\,,\\
\gamma_{\phi,70} & = \tfrac{\zeta_3}{4}-\tfrac{17}{192}\,,&
\gamma_{\phi,71} & = -\tfrac{1}{192}\,,&
\gamma_{\phi,72} & = \tfrac{\zeta_3}{2}-\tfrac{17}{96}\,,\\
\gamma_{\phi,73} & = -\tfrac{9 \zeta_3}{4}+\tfrac{73}{48}+\tfrac{\pi ^4}{120}\,,&
\gamma_{\phi,74} & = -\tfrac{167}{96}\,,&
\gamma_{\phi,75} & = \tfrac{7}{6}-2 \zeta_3\,,\\
\gamma_{\phi,76} & = \tfrac{47}{24}-\zeta_3\,,&
\gamma_{\phi,77} & = -2 \zeta_3+\tfrac{13}{3}-\tfrac{\pi ^4}{30}\,,&
\gamma_{\phi,78} & = \tfrac{23}{3}-10 \zeta_5\,,\\
\gamma_{\phi,79} & = -\tfrac{5 \zeta_3}{4}+\tfrac{7}{12}+\tfrac{\pi ^4}{120}\,,&
\gamma_{\phi,80} & = \tfrac{29}{48}-\zeta_3\,,&
\gamma_{\phi,81} & = \tfrac{1}{16}\,,\\
\gamma_{\phi,82} & = \tfrac{15}{32}\,,&
\gamma_{\phi,83} & = -\tfrac{31}{96}\,,&
\gamma_{\phi,84} & = \tfrac{5}{24}-2 \zeta_3\,,\\
\gamma_{\phi,85} & = -\tfrac{7}{3}\,,&
\gamma_{\phi,86} & = \tfrac{29}{8}-2 \zeta_3\,,&
\gamma_{\phi,87} & = -\tfrac{\zeta_3}{2}+\tfrac{5}{3}-\tfrac{\pi ^4}{60}\,,\\
\gamma_{\phi,88} & = -\zeta_3-\tfrac{13}{6}\,.& \yestag{}
\end{align*}

\section{Fermion Anomalous Dimension}\label{app:gammaF}
The fermion field anomalous dimensions at four loops can be parametrised as 
\begin{equation}\label{eq:gamF-terms}
  \gamma_\psi^{(4)} = \gamma_{\psi, y^2 \lambda^3}^{(4)} + \gamma_{\psi, y^4 \lambda^2}^{(4)} + \gamma_{\psi, y^6 \lambda}^{(4)} +  \gamma_{\psi, y^2 y^6}^{(4)} + \gamma_{\psi, y^4 y^4}^{(4)} + \gamma_{\psi, y^6 y^2}^{(4)} + \gamma_{\psi, y^8}^{(4)}
\end{equation}
where each term consists of the tensor contractions
\begin{align*}
    \gamma_{\psi, y^2 \lambda^3}^{(4)} &= 
    \gamma_{\psi,1}\ y^{a b}\,\lambda^{acde}\,\lambda^{cdfg}\,\lambda^{efgb}
    +    \gamma_{\psi,2}\ y^{a b}\,\lambda^{acde}\,\lambda^{cdfg}\,\lambda^{efgb}\,,\yestag{}\\[.5em]
    \gamma_{\psi, y^4 \lambda^2}^{(4)} &= 
    \gamma_{\psi,3}\ y^{a b}\,\lambda^{acde}\,\lambda^{bcdf}\,\mathrm{tr} \left(y^{e f}\right)
    +    \gamma_{\psi,4}\ y^{a b}\,\lambda^{abcd}\,\lambda^{cdef}\,\mathrm{tr} \left(y^{e f}\right)
    \\&\phantom{=}
    +    \gamma_{\psi,5}\ \mathcal{S}_2\, y^{a b}\,\lambda^{acde}\,\lambda^{cdef}\,\mathrm{tr} \left(y^{f b}\right)
    +    \gamma_{\psi,6}\ y^{a b c d}\,\lambda^{abef}\,\lambda^{cdef}
    \\&\phantom{=}
    +    \gamma_{\psi,7}\ y^{a b c d}\,\lambda^{acef}\,\lambda^{bdef}
    +    \gamma_{\psi,8}\ y^{a b c d}\,\lambda^{adef}\,\lambda^{bcef}
      \\&\phantom{=}
    +    \gamma_{\psi,9}\ y^{a c c b}\,\lambda^{adef}\,\lambda^{bdef}
    +    \gamma_{\psi,10}\ y^{c a b c}\,\lambda^{adef}\,\lambda^{bdef}
      \\&\phantom{=}
    +    \gamma_{\psi,11}\ \mathcal{S}_2\, y^{a c b c}\,\lambda^{adef}\,\lambda^{bdef} \,,\yestag{}\\[.5em]
    \gamma_{\psi, y^6 \lambda}^{(4)} &= 
    \gamma_{\psi,12}\  y^{a b}\,\lambda^{abcd}\,\mathrm{tr} \left(y^{c e}\right)\,\mathrm{tr} \left(y^{e d}\right)
    +    \gamma_{\psi,13}\ \mathcal{S}_2\, y^{a b}\,\mathrm{tr} \left(y^{a c}\right)\,\lambda^{bcde}\,\mathrm{tr} \left(y^{d e}\right)
    \\&\phantom{=}
    +    \gamma_{\psi,14}\ y^{a b}\,\lambda^{abcd}\,\mathrm{tr} \left(y^{c d e e}\right)
    +    \gamma_{\psi,15}\ y^{a b}\,\lambda^{abcd}\,\mathrm{tr} \left(y^{c e d e}\right)
    \\&\phantom{=}
    +    \gamma_{\psi,16}\ y^{a b}\,\lambda^{acde}\,\mathrm{tr} \left(y^{b c d e}\right)
    +    \gamma_{\psi,17}\ y^{a c c b}\,\lambda^{abde}\,\mathrm{tr} \left(y^{d e}\right)
    \\&\phantom{=}
    +    \gamma_{\psi,18}\ y^{c a b c}\,\lambda^{abde}\,\mathrm{tr} \left(y^{d e}\right)
    +    \gamma_{\psi,19}\ \mathcal{S}_2\, y^{a c b c}\,\lambda^{abde}\,\mathrm{tr} \left(y^{d e}\right)
    \\&\phantom{=}
    +    \gamma_{\psi,20}\ \mathcal{S}_2\, y^{a b c d}\,\lambda^{bcde}\,\mathrm{tr} \left(y^{a e}\right)
    +    \gamma_{\psi,21}\ \mathcal{S}_2\, y^{a b c d}\,\lambda^{acde}\,\mathrm{tr} \left(y^{b e}\right)
    \\&\phantom{=}
    +    \gamma_{\psi,22}\ y^{a b c d e a}\,\lambda^{bcde}
    +    \gamma_{\psi,23}\ \mathcal{S}_2\, y^{a b c d a e}\,\lambda^{bcde}
    +    \gamma_{\psi,24}\ \mathcal{S}_2\, y^{a b c a d e}\,\lambda^{bcde}
    \\&\phantom{=}
    +    \gamma_{\psi,25}\ \mathcal{S}_2\, y^{a b a c d e}\,\lambda^{bcde}
    +    \gamma_{\psi,26}\ y^{b a c d a e}\,\lambda^{bcde}
    +    \gamma_{\psi,27}\ \mathcal{S}_2\, y^{b a c a d e}\,\lambda^{bcde}
    \\&\phantom{=}
    +    \gamma_{\psi,28}\ \mathcal{S}_2\, y^{b a a c d e}\,\lambda^{bcde}
    +    \gamma_{\psi,29}\ y^{b c a a d e}\,\lambda^{bcde} \,,\yestag{}\\[.5em]
    \gamma_{\psi, y^2 y^6}^{(4)} &= 
    \gamma_{\psi,30}\  y^{a b}\,\mathrm{tr} \left(y^{a c}\right)\,\mathrm{tr} \left(y^{c d}\right)\,\mathrm{tr} \left(y^{d b}\right)
    +    \gamma_{\psi,31}\ y^{a b}\,\mathrm{tr} \left(y^{a b c d}\right)\,\mathrm{tr} \left(y^{c d}\right)
    \\&\phantom{=}
    +    \gamma_{\psi,32}\ y^{a b}\,\mathrm{tr} \left(y^{a c b d}\right)\,\mathrm{tr} \left(y^{c d}\right)
    +    \gamma_{\psi,33}\ \mathcal{S}_2\, y^{a b}\,\mathrm{tr} \left(y^{a c d d}\right)\,\mathrm{tr} \left(y^{c b}\right)
    \\&\phantom{=}
    +    \gamma_{\psi,34}\ \mathcal{S}_2\, y^{a b}\,\mathrm{tr} \left(y^{a d c d}\right)\,\mathrm{tr} \left(y^{c b}\right)
    +    \gamma_{\psi,35}\ y^{a b}\,\mathrm{tr} \left(y^{a b c c d d}\right)
    \\&\phantom{=}
    +    \gamma_{\psi,36}\ y^{a b}\,\mathrm{tr} \left(y^{a b c d d c}\right)
    +    \gamma_{\psi,37}\ y^{a b}\,\mathrm{tr} \left(y^{a b c d c d}\right)
    \\&\phantom{=}
    +    \gamma_{\psi,38}\ y^{a b}\,\mathrm{tr} \left(y^{a c b d c d}\right)
    +    \gamma_{\psi,39}\ \mathcal{S}_2\, y^{a b}\,\mathrm{tr} \left(y^{a c b c d d}\right)
    \\&\phantom{=}
    +    \gamma_{\psi,40}\ y^{a b}\,\mathrm{tr} \left(y^{a c c b d d}\right)
    +    \gamma_{\psi,41}\ y^{a b}\,\mathrm{tr} \left(y^{a c d b d c}\right)
    \\&\phantom{=}
    +    \gamma_{\psi,42}\ y^{a b}\,\mathrm{tr} \left(y^{a c d b c d}\right)\,,\yestag{}\\[.5em]
    \gamma_{\psi, y^4 y^4}^{(4)} &= 
    \gamma_{\psi,43}\  y^{a b c d}\,\mathrm{tr} \left(y^{a d}\right)\,\mathrm{tr} \left(y^{b c}\right)
    +    \gamma_{\psi,44}\ y^{a b c d}\,\mathrm{tr} \left(y^{a c}\right)\,\mathrm{tr} \left(y^{b d}\right)
    \\&\phantom{=}
    +    \gamma_{\psi,45}\ y^{a b c a}\,\mathrm{tr} \left(y^{b d}\right)\,\mathrm{tr} \left(y^{d c}\right)
    +    \gamma_{\psi,46}\ \mathcal{S}_2\, y^{a b a c}\,\mathrm{tr} \left(y^{b d}\right)\,\mathrm{tr} \left(y^{d c}\right)
    \\&\phantom{=}
    +    \gamma_{\psi,47}\ y^{b a a c}\,\mathrm{tr} \left(y^{b d}\right)\,\mathrm{tr} \left(y^{d c}\right)
    +    \gamma_{\psi,48}\ y^{a b c a}\,\mathrm{tr} \left(y^{b c d d}\right)
    \\&\phantom{=}
    +    \gamma_{\psi,49}\ y^{a b c a}\,\mathrm{tr} \left(y^{b d c d}\right)
    +    \gamma_{\psi,50}\ \mathcal{S}_2\, y^{a b a c}\,\mathrm{tr} \left(y^{b c d d}\right)
    \\&\phantom{=}
    +    \gamma_{\psi,51}\ \mathcal{S}_2\, y^{a b a c}\,\mathrm{tr} \left(y^{b d c d}\right)
    +    \gamma_{\psi,52}\ y^{b a a c}\,\mathrm{tr} \left(y^{b c d d}\right)
    \\&\phantom{=}
    +    \gamma_{\psi,53}\ y^{b a a c}\,\mathrm{tr} \left(y^{b d c d}\right)
    +    \gamma_{\psi,54}\ y^{a b c d}\,\mathrm{tr} \left(y^{a b c d}\right)
    \\&\phantom{=}
    +    \gamma_{\psi,55}\ \mathcal{S}_2\, y^{a b c d}\,\mathrm{tr} \left(y^{a b d c}\right)
    +    \gamma_{\psi,56}\ \mathcal{S}_2\, y^{a b c d}\,\mathrm{tr} \left(y^{a c b d}\right)\,,\yestag{}\\[.5em]
    \gamma_{\psi, y^6 y^2}^{(4)} &= 
    \gamma_{\psi,57}\  y^{a c d d c b}\,\mathrm{tr} \left(y^{a b}\right)
    +    \gamma_{\psi,58}\ y^{a c d c d b}\,\mathrm{tr} \left(y^{a b}\right)
    \\&\phantom{=}
    +    \gamma_{\psi,59}\ y^{a c c d d b}\,\mathrm{tr} \left(y^{a b}\right)
    +    \gamma_{\psi,60}\ \mathcal{S}_2\, y^{a c d d b c}\,\mathrm{tr} \left(y^{a b}\right)
    \\&\phantom{=}
    +    \gamma_{\psi,61}\ \mathcal{S}_2\, y^{a d c d b c}\,\mathrm{tr} \left(y^{a b}\right)
    +    \gamma_{\psi,62}\ \mathcal{S}_2\, y^{a d d c b c}\,\mathrm{tr} \left(y^{a b}\right)
    \\&\phantom{=}
    +    \gamma_{\psi,63}\ \mathcal{S}_2\, y^{a c d b c d}\,\mathrm{tr} \left(y^{a b}\right)
    +    \gamma_{\psi,64}\ \mathcal{S}_2\, y^{a d c b c d}\,\mathrm{tr} \left(y^{a b}\right)
    \\&\phantom{=}
    +    \gamma_{\psi,65}\ \mathcal{S}_2\, y^{a c b d d c}\,\mathrm{tr} \left(y^{a b}\right)
    +    \gamma_{\psi,66}\ \mathcal{S}_2\, y^{a c b d c d}\,\mathrm{tr} \left(y^{a b}\right)
    \\&\phantom{=}
    +    \gamma_{\psi,67}\ y^{c a d d b c}\,\mathrm{tr} \left(y^{a b}\right)
    +    \gamma_{\psi,68}\ y^{c a c d b d}\,\mathrm{tr} \left(y^{a b}\right)
    \\&\phantom{=}
    +    \gamma_{\psi,69}\ y^{c a d c b d}\,\mathrm{tr} \left(y^{a b}\right)
    +    \gamma_{\psi,70}\ \mathcal{S}_2\, y^{c a d b d c}\,\mathrm{tr} \left(y^{a b}\right)
    \\&\phantom{=}
    +    \gamma_{\psi,71}\ \mathcal{S}_2\, y^{c a d b c d}\,\mathrm{tr} \left(y^{a b}\right)
    +    \gamma_{\psi,72}\ \mathcal{S}_2\, y^{c a b d d c}\,\mathrm{tr} \left(y^{a b}\right)
    \\&\phantom{=}
    +    \gamma_{\psi,73}\ \mathcal{S}_2\, y^{c a b d c d}\,\mathrm{tr} \left(y^{a b}\right)
    +    \gamma_{\psi,74}\ y^{c d a b d c}\,\mathrm{tr} \left(y^{a b}\right)
    \\&\phantom{=}
    +    \gamma_{\psi,75}\ y^{c d a b c d}\,\mathrm{tr} \left(y^{a b}\right)\,,\yestag{}\\[.5em]
    \gamma_{\psi, y^8}^{(4)} &= 
    \gamma_{\psi,76}\  y^{a b a c b d c d}
    +    \gamma_{\psi,77}\ \mathcal{S}_2\,  y^{a b a c b d d c}
    +    \gamma_{\psi,78}\ y^{a b a c c d b d}
    \\&\phantom{=}
    +    \gamma_{\psi,79}\ \mathcal{S}_2\, y^{a b a c c d d b}
    +    \gamma_{\psi,80}\ \mathcal{S}_2\, y^{a b a c d b c d}
    +    \gamma_{\psi,81}\ \mathcal{S}_2\, y^{a b a c d b d c}
    \\&\phantom{=}
    +    \gamma_{\psi,82}\ \mathcal{S}_2\, y^{a b a c d c b d}
    +    \gamma_{\psi,83}\ \mathcal{S}_2\, y^{a b a c d c d b}
    +    \gamma_{\psi,84}\ \mathcal{S}_2\, y^{a b a c d d b c}
    \\&\phantom{=}
    +    \gamma_{\psi,85}\ \mathcal{S}_2\, y^{a b a c d d c b}
    +    \gamma_{\psi,86}\ y^{a b b c a d d c}
    +    \gamma_{\psi,87}\ y^{a b b c c d d a}
    \\&\phantom{=}
    +    \gamma_{\psi,88}\ \mathcal{S}_2\, y^{a b b c d a c d}
    +    \gamma_{\psi,89}\ \mathcal{S}_2\, y^{a b b c d a d c}
    +    \gamma_{\psi,90}\ \mathcal{S}_2\, y^{a b b c d c a d}
    \\&\phantom{=}    
    +    \gamma_{\psi,91}\ \mathcal{S}_2\, y^{a b b c d c d a}
    +    \gamma_{\psi,92}\ \mathcal{S}_2\, y^{a b b c d d a c}
    +    \gamma_{\psi,93}\ \mathcal{S}_2\, y^{a b b c d d c a}
    \\&\phantom{=}
    +    \gamma_{\psi,94}\ y^{a b c a d b c d}
    +    \gamma_{\psi,95}\ \mathcal{S}_2\, y^{a b c a d b d c}
    +    \gamma_{\psi,96}\ y^{a b c a d c b d}
    \\&\phantom{=}
    +    \gamma_{\psi,97}\ \mathcal{S}_2\, y^{a b c a d c d b}
    +    \gamma_{\psi,98}\ \mathcal{S}_2\, y^{a b c a d d b c}
    +    \gamma_{\psi,99}\ \mathcal{S}_2\, y^{a b c a d d c b}
    \\&\phantom{=}
    +    \gamma_{\psi,100}\ y^{a b c b d a d c}
    +    \gamma_{\psi,101}\ y^{a b c b d c d a}
    +    \gamma_{\psi,102}\ \mathcal{S}_2\, y^{a b c b d d a c}
    \\&\phantom{=}
    +    \gamma_{\psi,103}\ \mathcal{S}_2\, y^{a b c b d d c a}
    +    \gamma_{\psi,104}\ y^{a b c c d d a b}
    +    \gamma_{\psi,105}\ y^{a b c c d d b a}
    \\&\phantom{=}
    +    \gamma_{\psi,106}\ y^{a b c d a b c d}
    +    \gamma_{\psi,107}\ \mathcal{S}_2\, y^{a b c d a b d c}
    +    \gamma_{\psi,108}\ y^{a b c d a c b d}
    \\&\phantom{=}
    +    \gamma_{\psi,109}\ \mathcal{S}_2\, y^{a b c d a c d b}
    +    \gamma_{\psi,110}\ \mathcal{S}_2\, y^{a b c d a d b c}
    +    \gamma_{\psi,111}\ \mathcal{S}_2\, y^{a b c d a d c b}
    \\&\phantom{=}
    +    \gamma_{\psi,112}\ y^{a b c d b a d c}
    +    \gamma_{\psi,113}\ y^{a b c d b c d a}
    +    \gamma_{\psi,114}\ \mathcal{S}_2\, y^{a b c d b d a c}
    \\&\phantom{=}
    +    \gamma_{\psi,115}\ \mathcal{S}_2\, y^{a b c d b d c a}
    +    \gamma_{\psi,116}\ y^{a b c d c d a b}
    +    \gamma_{\psi,117}\ y^{a b c d c d b a}
    \\&\phantom{=}
    +    \gamma_{\psi,118}\ y^{a b c d d a b c}
    +    \gamma_{\psi,119}\ \mathcal{S}_2\, y^{a b c d d a c b}
    +    \gamma_{\psi,120}\ y^{a b c d d b c a}
    \\&\phantom{=}
    +    \gamma_{\psi,121}\ y^{a b c d d c a b}
    +    \gamma_{\psi,122}\  y^{a b c d d c b a}\,.\yestag{}
\end{align*}
Explicit $\overline{\text{MS}}$ results for the coefficients read
\begin{align*}
\gamma_{\psi,1} & = \tfrac{19}{96}\,,&
\gamma_{\psi,2} & = 0\,,&
\gamma_{\psi,3} & = \tfrac{65}{192}\,,\\
\gamma_{\psi,4} & = 0\,,&
\gamma_{\psi,5} & = -\tfrac{3}{128}\,,&
\gamma_{\psi,6} & = -\tfrac{13}{24}\,,\\
\gamma_{\psi,7} & = \tfrac{3 \zeta_3}{4}-\tfrac{11}{8}\,,&
\gamma_{\psi,8} & = -\tfrac{7}{24}\,,&
\gamma_{\psi,9} & = -\tfrac{7}{128}\,,\\
\gamma_{\psi,10} & = \tfrac{41}{384}\,,&
\gamma_{\psi,11} & = -\tfrac{5}{144}\,,&
\gamma_{\psi,12} & = 0\,,\\
\gamma_{\psi,13} & = 0\,,&
\gamma_{\psi,14} & = 0\,,&
\gamma_{\psi,15} & = 0\,,\\
\gamma_{\psi,16} & = -\tfrac{73}{96}\,,&
\gamma_{\psi,17} & = 0\,,&
\gamma_{\psi,18} & = 0\,,\\
\gamma_{\psi,19} & = 0\,,&
\gamma_{\psi,20} & = -\tfrac{7}{32}\,,&
\gamma_{\psi,21} & = -\tfrac{5}{6}\,,\\
\gamma_{\psi,22} & = -\tfrac{5}{48}\,,&
\gamma_{\psi,23} & = \tfrac{\zeta_3}{2}-\tfrac{7}{4}\,,&
\gamma_{\psi,24} & = \tfrac{\zeta_3}{2}-\tfrac{5}{3}\,,\\
\gamma_{\psi,25} & = -\tfrac{1}{2}\,,&
\gamma_{\psi,26} & = \zeta_3-\tfrac{13}{6}\,,&
\gamma_{\psi,27} & = \tfrac{\zeta_3}{2}-\tfrac{7}{12}\,,\\
\gamma_{\psi,28} & = -\tfrac{1}{32}\,,&
\gamma_{\psi,29} & = -\tfrac{7}{24}\,,&
\gamma_{\psi,30} & = \tfrac{\zeta_3}{8}-\tfrac{3}{128}\,,\\
\gamma_{\psi,31} & = -\tfrac{105}{64}\,,&
\gamma_{\psi,32} & = -\tfrac{\zeta_3}{2}-\tfrac{95}{192}\,,&
\gamma_{\psi,33} & = \tfrac{7}{32}\,,\\
\gamma_{\psi,34} & = \tfrac{\zeta_3}{4}+\tfrac{3}{32}\,,&
\gamma_{\psi,35} & = \tfrac{1}{64}-\tfrac{\zeta_3}{4}\,,&
\gamma_{\psi,36} & = -\tfrac{5}{8}\,,\\
\gamma_{\psi,37} & = \tfrac{1}{32}-\tfrac{\zeta_3}{2}\,,&
\gamma_{\psi,38} & = \tfrac{67}{48}-2 \zeta_3\,,&
\gamma_{\psi,39} & = -\tfrac{149}{192}\,,\\
\gamma_{\psi,40} & = -\tfrac{89}{384}\,,&
\gamma_{\psi,41} & = \tfrac{3 \zeta_3}{2}-\tfrac{65}{24}\,,&
\gamma_{\psi,42} & = -\tfrac{9 \zeta_3}{8}+\tfrac{35}{48}-\tfrac{\pi ^4}{240}\,,\\
\gamma_{\psi,43} & = \tfrac{1}{96}\,,&
\gamma_{\psi,44} & = -\tfrac{11}{64}\,,&
\gamma_{\psi,45} & = \tfrac{7}{128}-\tfrac{\zeta_3}{8}\,,\\
\gamma_{\psi,46} & = \tfrac{7}{96}\,,&
\gamma_{\psi,47} & = \tfrac{\zeta_3}{8}-\tfrac{19}{384}\,,&
\gamma_{\psi,48} & = -1\,,\\
\gamma_{\psi,49} & = -\tfrac{\zeta_3}{4}-\tfrac{17}{32}\,,&
\gamma_{\psi,50} & = \tfrac{13}{32}\,,&
\gamma_{\psi,51} & = \tfrac{17}{48}\,,\\
\gamma_{\psi,52} & = \tfrac{97}{192}\,,&
\gamma_{\psi,53} & = \tfrac{\zeta_3}{4}+\tfrac{11}{48}\,,&
\gamma_{\psi,54} & = -2 \zeta_3\,,\\
\gamma_{\psi,55} & = -\tfrac{7 \zeta_3}{4}+\tfrac{23}{12}-\tfrac{\pi ^4}{120}\,,&
\gamma_{\psi,56} & = -\tfrac{5 \zeta_3}{4}+\tfrac{5}{3}-\tfrac{\pi ^4}{120}\,,&
\gamma_{\psi,57} & = -\tfrac{7}{384}\,,\\
\gamma_{\psi,58} & = \tfrac{\zeta_3}{4}-\tfrac{29}{192}\,,&
\gamma_{\psi,59} & = \tfrac{\zeta_3}{8}-\tfrac{29}{384}\,,&
\gamma_{\psi,60} & = \tfrac{13}{128}\,,\\
\gamma_{\psi,61} & = \tfrac{5}{192}\,,&
\gamma_{\psi,62} & = 0\,,&
\gamma_{\psi,63} & = -\tfrac{9 \zeta_3}{8}+\tfrac{7}{32}+\tfrac{\pi ^4}{240}\,,\\
\gamma_{\psi,64} & = \tfrac{115}{192}\,,&
\gamma_{\psi,65} & = 0\,,&
\gamma_{\psi,66} & = -\tfrac{37}{64}\,,\\
\gamma_{\psi,67} & = \tfrac{13}{128}\,,&
\gamma_{\psi,68} & = \tfrac{15}{32}-\tfrac{\zeta_3}{2}\,,&
\gamma_{\psi,69} & = -\tfrac{5 \zeta_3}{8}+\tfrac{7}{24}+\tfrac{\pi ^4}{240}\,,\\
\gamma_{\psi,70} & = \tfrac{121}{384}\,,&
\gamma_{\psi,71} & = -\tfrac{37}{48}\,,&
\gamma_{\psi,72} & = \tfrac{59}{384}\,,\\
\gamma_{\psi,73} & = -\tfrac{13}{64}\,,&
\gamma_{\psi,74} & = -\tfrac{37}{192}\,,&
\gamma_{\psi,75} & = -\tfrac{143}{192}\,,\\
\gamma_{\psi,76} & = 0\,,&
\gamma_{\psi,77} & = -\tfrac{11}{64}\,,&
\gamma_{\psi,78} & = \tfrac{17}{32}-\tfrac{\zeta_3}{2}\,,\\
\gamma_{\psi,79} & = -\tfrac{7}{96}\,,&
\gamma_{\psi,80} & = \tfrac{1}{2}\,,&
\gamma_{\psi,81} & = \tfrac{3}{8}-\tfrac{\zeta_3}{2}\,,\\
\gamma_{\psi,82} & = \tfrac{11}{24}-\tfrac{\zeta_3}{2}\,,&
\gamma_{\psi,83} & = -\tfrac{7}{48}\,,&
\gamma_{\psi,84} & = \tfrac{1}{12}\,,\\
\gamma_{\psi,85} & = -\tfrac{19}{192}\,,&
\gamma_{\psi,86} & = \tfrac{11}{64}\,,&
\gamma_{\psi,87} & = \tfrac{\zeta_3}{8}-\tfrac{13}{128}\,,\\
\gamma_{\psi,88} & = -\tfrac{3 \zeta_3}{8}+\tfrac{1}{32}+\tfrac{\pi ^4}{240}\,,&
\gamma_{\psi,89} & = \tfrac{133}{192}\,,&
\gamma_{\psi,90} & = -\tfrac{13}{192}\,,\\
\gamma_{\psi,91} & = \tfrac{\zeta_3}{4}-\tfrac{13}{64}\,,&
\gamma_{\psi,92} & = \tfrac{19}{128}\,,&
\gamma_{\psi,93} & = \tfrac{1}{32}\,,\\
\gamma_{\psi,94} & = \tfrac{5}{3}-\tfrac{\pi ^4}{60}\,,&
\gamma_{\psi,95} & = \tfrac{7}{12}\,,&
\gamma_{\psi,96} & = \tfrac{23}{6}-5 \zeta_5\,,\\
\gamma_{\psi,97} & = \tfrac{\zeta_3}{2}-\tfrac{7}{8}\,,&
\gamma_{\psi,98} & = -\tfrac{7 \zeta_3}{8}+\tfrac{5}{6}+\tfrac{\pi ^4}{240}\,,&
\gamma_{\psi,99} & = \tfrac{\zeta_3}{2}-\tfrac{37}{48}\,,\\
\gamma_{\psi,100} & = \tfrac{31}{12}-2 \zeta_3\,,&
\gamma_{\psi,101} & = \tfrac{3 \zeta_3}{2}-\tfrac{55}{48}\,,&
\gamma_{\psi,102} & = \tfrac{55}{48}-\zeta_3\,,\\
\gamma_{\psi,103} & = \tfrac{7}{128}\,,&
\gamma_{\psi,104} & = \tfrac{1}{2}-\tfrac{\zeta_3}{2}\,,&
\gamma_{\psi,105} & = \tfrac{19}{128}-\tfrac{\zeta_3}{8}\,,\\
\gamma_{\psi,106} & = -\tfrac{5}{3}\,,&
\gamma_{\psi,107} & = \tfrac{\zeta_3}{2}+\tfrac{5}{3}-\tfrac{\pi ^4}{60}\,,&
\gamma_{\psi,108} & = \tfrac{13}{6}-\tfrac{\pi ^4}{60}\,,\\
\gamma_{\psi,109} & = -\tfrac{\zeta_3}{2}+\tfrac{7}{4}-\tfrac{\pi ^4}{60}\,,&
\gamma_{\psi,110} & = \tfrac{11}{4}-2 \zeta_3\,,&
\gamma_{\psi,111} & = \tfrac{\zeta_3}{2}-\tfrac{11}{24}\,,\\
\gamma_{\psi,112} & = \tfrac{10}{3}-5 \zeta_5\,,&
\gamma_{\psi,113} & = -\tfrac{3 \zeta_3}{8}+\tfrac{5}{48}-\tfrac{\pi ^4}{240}\,,&
\gamma_{\psi,114} & = \tfrac{31}{24}-2 \zeta_3\,,\\
\gamma_{\psi,115} & = \tfrac{11}{16}-\zeta_3\,,&
\gamma_{\psi,116} & = 1-\zeta_3\,,&
\gamma_{\psi,117} & = \tfrac{19}{64}-\tfrac{\zeta_3}{4}\,,\\
\gamma_{\psi,118} & = -\tfrac{11 \zeta_3}{8}+\tfrac{11}{8}+\tfrac{\pi ^4}{240}\,,&
\gamma_{\psi,119} & = -\tfrac{11}{48}\,,&
\gamma_{\psi,120} & = \tfrac{61}{192}-\tfrac{\zeta_3}{2}\,,\\
\gamma_{\psi,121} & = -\tfrac{11}{64}\,,&
\gamma_{\psi,122} & = -\tfrac{5}{128}\,.&\tag{\stepcounter{equation}\theequation}
\end{align*}

\section{Yukawa Vertex}\label{app:betaY}
The Yukawa $\beta$-function is composed of leg and vertex corrections~\eq{beta-yuk}, with the former consisting of terms \eq{gamS-terms} and \eq{gamF-terms}. Similarly, the four-loop vertex correction is composed of the terms 
\begin{equation}
  \hat{\beta}_y^{(4)} = \beta_{y, y^3 \lambda^3}^{(4)} + \beta_{y, y^{3,2} \lambda^2}^{(4)} + \beta_{y, y^5 \lambda^2}^{(4)} + \beta_{y, y^{3,4} \lambda}^{(4)} + \beta_{y, y^{5,2} \lambda}^{(4)} + \beta_{y, y^7 \lambda}^{(4)} + \beta_{y, y^{3,6}}^{(4)} + \beta_{y, y^{5,4}}^{(4)} +  \beta_{y, y^{7,2}}^{(4)} + \beta_{y, y^{9}}^{(4)}
\end{equation}
which contain the contractions 
 \begin{align*}
    \beta_{y, y^3 \lambda^3}^{(4)\,a} &= 
    \beta_{y,1}\ y^{b a c}\,\lambda^{bcde}\,\lambda^{dfgh}\,\lambda^{efgh}
    + \beta_{y,2}\ y^{b a c}\,\lambda^{bdef}\,\lambda^{cdgh}\,\lambda^{efgh}
    \\&\phantom{=}
    + \beta_{y,3}\ y^{b c d}\,\lambda^{abde}\,\lambda^{efgh}\,\lambda^{cfgh}
    + \beta_{y,4}\ \mathcal{S}_{2}\,y^{b d c}\,\lambda^{abde}\,\lambda^{efgh}\,\lambda^{cfgh}
    \\&\phantom{=}
    + \beta_{y,5}\ y^{b c d}\,\lambda^{bdef}\,\lambda^{efgh}\,\lambda^{acgh}
    + \beta_{y,6}\ \mathcal{S}_{2}\,y^{c b d}\,\lambda^{bdef}\,\lambda^{efgh}\,\lambda^{acgh}
    \\&\phantom{=}
    + \beta_{y,7}\ y^{b c d}\,\lambda^{befg}\,\lambda^{defh}\,\lambda^{acgh}
    + \beta_{y,8}\ \mathcal{S}_{2}\,y^{c b d}\,\lambda^{befg}\,\lambda^{defh}\,\lambda^{acgh}
    \\&\phantom{=}
    + \beta_{y,9}\ y^{b c d}\,\lambda^{bdef}\,\lambda^{cegh}\,\lambda^{afgh}
    + \beta_{y,10}\ \mathcal{S}_{2}\,y^{c b d}\,\lambda^{bdef}\,\lambda^{cegh}\,\lambda^{afgh}\,,\yestag{}\\[.5em]
    \beta_{y, y^{3,2} \lambda^2}^{(4)\,a} &= 
    \beta_{y,11}\ \mathcal{S}_{2}\,y^{b a c}\,\lambda^{bdef}\,\lambda^{defg}\,\mathrm{tr} \left(y^{g c}\right)
    + \beta_{y,12}\ y^{b a c}\,\lambda^{bcde}\,\lambda^{defg}\,\mathrm{tr} \left(y^{f g}\right)
    \\&\phantom{=}
    + \beta_{y,13}\ y^{b a c}\,\lambda^{bdef}\,\lambda^{cdeg}\,\mathrm{tr} \left(y^{f g}\right)
    + \beta_{y,14}\ \mathcal{S}_{2}\,y^{b c d}\,\lambda^{abef}\,\lambda^{cefg}\,\mathrm{tr} \left(y^{g d}\right)
    \\&\phantom{=}
    + \beta_{y,15}\ \mathcal{S}_{2}\,y^{b d c}\,\lambda^{abef}\,\lambda^{cefg}\,\mathrm{tr} \left(y^{g d}\right)
    + \beta_{y,16}\ \mathcal{S}_{2}\,y^{c b d}\,\lambda^{abef}\,\lambda^{cefg}\,\mathrm{tr} \left(y^{g d}\right)
    \\&\phantom{=}
    + \beta_{y,17}\ y^{b c d}\,\lambda^{abde}\,\lambda^{cefg}\,\mathrm{tr} \left(y^{f g}\right)
    + \beta_{y,18}\ \mathcal{S}_{2}\,y^{c b d}\,\lambda^{abde}\,\lambda^{cefg}\,\mathrm{tr} \left(y^{f g}\right)
    \\&\phantom{=}
    + \beta_{y,19}\ y^{b c d}\,\lambda^{acef}\,\lambda^{bdeg}\,\mathrm{tr} \left(y^{f g}\right)
    + \beta_{y,20}\ \mathcal{S}_{2}\,y^{c b d}\,\lambda^{acef}\,\lambda^{bdeg}\,\mathrm{tr} \left(y^{f g}\right)
    \\&\phantom{=}
    + \beta_{y,21}\ y^{b c d}\,\mathrm{tr} \left(y^{c e}\right)\,\lambda^{aefg}\,\lambda^{bdfg}
    + \beta_{y,22}\ \mathcal{S}_{2}\,y^{b d c}\,\mathrm{tr} \left(y^{c e}\right)\,\lambda^{aefg}\,\lambda^{bdfg}\,,\yestag{}\\[.5em]
    \beta_{y, y^5 \lambda^2}^{(4)\,a} &= 
    \beta_{y,23}\ \mathcal{S}_{2}\,y^{b a c d b}\,\lambda^{cefg}\,\lambda^{defg}
    + \beta_{y,24}\ \mathcal{S}_{2}\,y^{c a b b d}\,\lambda^{cefg}\,\lambda^{defg}
    \\&\phantom{=}
    + \beta_{y,25}\ \mathcal{S}_{2}\,y^{b a c b d}\,\lambda^{cefg}\,\lambda^{defg}
    + \beta_{y,26}\ \mathcal{S}_{2}\,y^{c a b d b}\,\lambda^{cefg}\,\lambda^{defg}
    \\&\phantom{=}
    + \beta_{y,27}\ y^{b c a d b}\,\lambda^{cefg}\,\lambda^{defg}
    + \beta_{y,28}\ y^{c b a b d}\,\lambda^{cefg}\,\lambda^{defg}
    \\&\phantom{=}
    + \beta_{y,29}\ \mathcal{S}_{2}\,y^{c b a d b}\,\lambda^{cefg}\,\lambda^{defg}
    + \beta_{y,30}\ \mathcal{S}_{2}\,y^{b a c d e}\,\lambda^{bcfg}\,\lambda^{defg}
    \\&\phantom{=}
    + \beta_{y,31}\ \mathcal{S}_{2}\,y^{b a d c e}\,\lambda^{bcfg}\,\lambda^{defg}
    + \beta_{y,32}\ \mathcal{S}_{2}\,y^{b a d e c}\,\lambda^{bcfg}\,\lambda^{defg}
    \\&\phantom{=}
    + \beta_{y,33}\ y^{b c a d e}\,\lambda^{bcfg}\,\lambda^{defg}
    + \beta_{y,34}\ y^{b d a e c}\,\lambda^{bcfg}\,\lambda^{defg}
    \\&\phantom{=}
    + \beta_{y,35}\ y^{b d a c e}\,\lambda^{bcfg}\,\lambda^{defg}
    + \beta_{y,36}\ \mathcal{S}_{2}\,y^{e b c d e}\,\lambda^{abfg}\,\lambda^{cdfg}
    \\&\phantom{=}
    + \beta_{y,37}\ y^{e c b d e}\,\lambda^{abfg}\,\lambda^{cdfg}
    + \beta_{y,38}\ \mathcal{S}_{2}\,y^{e b c e d}\,\lambda^{abfg}\,\lambda^{cdfg}
    \\&\phantom{=}
    + \beta_{y,39}\ \mathcal{S}_{2}\,y^{e c b e d}\,\lambda^{abfg}\,\lambda^{cdfg}
    + \beta_{y,40}\ \mathcal{S}_{2}\,y^{e c d e b}\,\lambda^{abfg}\,\lambda^{cdfg}
    \\&\phantom{=}
    + \beta_{y,41}\ \mathcal{S}_{2}\,y^{e b e c d}\,\lambda^{abfg}\,\lambda^{cdfg}
    + \beta_{y,42}\ \mathcal{S}_{2}\,y^{e c e b d}\,\lambda^{abfg}\,\lambda^{cdfg}
    \\&\phantom{=}
    + \beta_{y,43}\ \mathcal{S}_{2}\,y^{e c e d b}\,\lambda^{abfg}\,\lambda^{cdfg}
    + \beta_{y,44}\ \mathcal{S}_{2}\,y^{b e c e d}\,\lambda^{abfg}\,\lambda^{cdfg}
    \\&\phantom{=}
    + \beta_{y,45}\ y^{c e b e d}\,\lambda^{abfg}\,\lambda^{cdfg}
    + \beta_{y,46}\ \mathcal{S}_{2}\,y^{b e e c d}\,\lambda^{abfg}\,\lambda^{cdfg}
    \\&\phantom{=}
    + \beta_{y,47}\ \mathcal{S}_{2}\,y^{c e e b d}\,\lambda^{abfg}\,\lambda^{cdfg}
    + \beta_{y,48}\ \mathcal{S}_{2}\,y^{c e e d b}\,\lambda^{abfg}\,\lambda^{cdfg}
    \\&\phantom{=}
    + \beta_{y,49}\ y^{b d e f c}\,\lambda^{abcg}\,\lambda^{defg}
    + \beta_{y,50}\ \mathcal{S}_{2}\,y^{b d e c f}\,\lambda^{abcg}\,\lambda^{defg}
    \\&\phantom{=}
    + \beta_{y,51}\ \mathcal{S}_{2}\,y^{b d c e f}\,\lambda^{abcg}\,\lambda^{defg}
    + \beta_{y,52}\ \mathcal{S}_{2}\,y^{b c d e f}\,\lambda^{abcg}\,\lambda^{defg}
    \\&\phantom{=}
    + \beta_{y,53}\ y^{d b e c f}\,\lambda^{abcg}\,\lambda^{defg}
    + \beta_{y,54}\ \mathcal{S}_{2}\,y^{d b c e f}\,\lambda^{abcg}\,\lambda^{defg}\,,\yestag{}\\[.5em]
    \beta_{y, y^{3,4} \lambda}^{(4)\,a} &= 
    \beta_{y,55}\ y^{b c d}\,\mathrm{tr} \left(y^{b e}\right)\,\mathrm{tr} \left(y^{d f}\right)\,\lambda^{acef}
    + \beta_{y,56}\ \mathcal{S}_{2}\,y^{c b d}\,\mathrm{tr} \left(y^{b e}\right)\,\mathrm{tr} \left(y^{d f}\right)\,\lambda^{acef}
    \\&\phantom{=}
    + \beta_{y,57}\ y^{b c d}\,\mathrm{tr} \left(y^{c e}\right)\,\mathrm{tr} \left(y^{e f}\right)\,\lambda^{abdf}
    + \beta_{y,58}\ \mathcal{S}_{2}\,y^{c b d}\,\mathrm{tr} \left(y^{c e}\right)\,\mathrm{tr} \left(y^{e f}\right)\,\lambda^{abdf}
    \\&\phantom{=}
     + \beta_{y,59}\ \mathcal{S}_{2}\,y^{b a c}\,\lambda^{bdef}\,\mathrm{tr} \left(y^{c d e f}\right)
    + \beta_{y,60}\ y^{b c d}\,\lambda^{bdef}\,\mathrm{tr} \left(y^{a e c f}\right)
    \\&\phantom{=}
    + \beta_{y,61}\ \mathcal{S}_{2}\,y^{c b d}\,\lambda^{bdef}\,\mathrm{tr} \left(y^{a e c f}\right)
    + \beta_{y,62}\ y^{b c d}\,\lambda^{bdef}\,\mathrm{tr} \left(y^{a c e f}\right)
    \\&\phantom{=}
    + \beta_{y,63}\ \mathcal{S}_{2}\,y^{c b d}\,\lambda^{bdef}\,\mathrm{tr} \left(y^{a c e f}\right)
    + \beta_{y,64}\ y^{b c d}\,\lambda^{acef}\,\mathrm{tr} \left(y^{b e d f}\right)
    \\&\phantom{=}
    + \beta_{y,65}\ \mathcal{S}_{2}\,y^{c b d}\,\lambda^{acef}\,\mathrm{tr} \left(y^{b e d f}\right)
    + \beta_{y,66}\ y^{b c d}\,\lambda^{acef}\,\mathrm{tr} \left(y^{b d e f}\right)
    \\&\phantom{=}
    + \beta_{y,67}\ \mathcal{S}_{2}\,y^{c b d}\,\lambda^{acef}\,\mathrm{tr} \left(y^{b d e f}\right)
    + \beta_{y,68}\ y^{b c d}\,\lambda^{abde}\,\mathrm{tr} \left(y^{c f e f}\right)
    \\&\phantom{=}
    + \beta_{y,69}\ \mathcal{S}_{2}\,y^{c b d}\,\lambda^{abde}\,\mathrm{tr} \left(y^{c f e f}\right)
    + \beta_{y,70}\ y^{b c d}\,\lambda^{abde}\,\mathrm{tr} \left(y^{c e f f}\right)
    \\&\phantom{=}
    + \beta_{y,71}\ \mathcal{S}_{2}\,y^{c b d}\,\lambda^{abde}\,\mathrm{tr} \left(y^{c e f f}\right)\,,\yestag{}\\[.5em]
    \beta_{y, y^{5,2} \lambda}^{(4)\,a} &= 
    \beta_{y,72}\ y^{e b c d f}\,\lambda^{abcd}\,\mathrm{tr} \left(y^{e f}\right)
    + \beta_{y,73}\ \mathcal{S}_{2}\,y^{e b c f d}\,\lambda^{abcd}\,\mathrm{tr} \left(y^{e f}\right)
    \\&\phantom{=}
    + \beta_{y,74}\ \mathcal{S}_{2}\,y^{e b f c d}\,\lambda^{abcd}\,\mathrm{tr} \left(y^{e f}\right)
    + \beta_{y,75}\ y^{b e c f d}\,\lambda^{abcd}\,\mathrm{tr} \left(y^{e f}\right)
    \\&\phantom{=}
    + \beta_{y,76}\ \mathcal{S}_{2}\,y^{b e f c d}\,\lambda^{abcd}\,\mathrm{tr} \left(y^{e f}\right)
    + \beta_{y,77}\ y^{e b d c e}\,\lambda^{abcf}\,\mathrm{tr} \left(y^{f d}\right)
    \\&\phantom{=}
    + \beta_{y,78}\ \mathcal{S}_{2}\,y^{e d b c e}\,\lambda^{abcf}\,\mathrm{tr} \left(y^{f d}\right)
    + \beta_{y,79}\ \mathcal{S}_{2}\,y^{e b c e d}\,\lambda^{abcf}\,\mathrm{tr} \left(y^{f d}\right)
    \\&\phantom{=}
    + \beta_{y,80}\ \mathcal{S}_{2}\,y^{e b d e c}\,\lambda^{abcf}\,\mathrm{tr} \left(y^{f d}\right)
    + \beta_{y,81}\ \mathcal{S}_{2}\,y^{e d b e c}\,\lambda^{abcf}\,\mathrm{tr} \left(y^{f d}\right)
    \\&\phantom{=}
    + \beta_{y,82}\ \mathcal{S}_{2}\,y^{e d e b c}\,\lambda^{abcf}\,\mathrm{tr} \left(y^{f d}\right)
    + \beta_{y,83}\ \mathcal{S}_{2}\,y^{e b e d c}\,\lambda^{abcf}\,\mathrm{tr} \left(y^{f d}\right)
    \\&\phantom{=}
    + \beta_{y,84}\ \mathcal{S}_{2}\,y^{e b e c d}\,\lambda^{abcf}\,\mathrm{tr} \left(y^{f d}\right)
    + \beta_{y,85}\ y^{b e d e c}\,\lambda^{abcf}\,\mathrm{tr} \left(y^{f d}\right)
    \\&\phantom{=}
    + \beta_{y,86}\ \mathcal{S}_{2}\,y^{b e c e d}\,\lambda^{abcf}\,\mathrm{tr} \left(y^{f d}\right)
    + \beta_{y,87}\ \mathcal{S}_{2}\,y^{b e e c d}\,\lambda^{abcf}\,\mathrm{tr} \left(y^{f d}\right)
    \\&\phantom{=}
    + \beta_{y,88}\ \mathcal{S}_{2}\,y^{b e e d c}\,\lambda^{abcf}\,\mathrm{tr} \left(y^{f d}\right)
    + \beta_{y,89}\ \mathcal{S}_{2}\,y^{d e e b c}\,\lambda^{abcf}\,\mathrm{tr} \left(y^{f d}\right)
    \\&\phantom{=}
    + \beta_{y,90}\ \mathcal{S}_{2}\,y^{b a c d e}\,\lambda^{bcdf}\,\mathrm{tr} \left(y^{f e}\right)
    + \beta_{y,91}\ \mathcal{S}_{2}\,y^{b a c e d}\,\lambda^{bcdf}\,\mathrm{tr} \left(y^{f e}\right)
    \\&\phantom{=}
    + \beta_{y,92}\ \mathcal{S}_{2}\,y^{b a e c d}\,\lambda^{bcdf}\,\mathrm{tr} \left(y^{f e}\right)
    + \beta_{y,93}\ \mathcal{S}_{2}\,y^{e a b c d}\,\lambda^{bcdf}\,\mathrm{tr} \left(y^{f e}\right)
    \\&\phantom{=}
    + \beta_{y,94}\ \mathcal{S}_{2}\,y^{e b a c d}\,\lambda^{bcdf}\,\mathrm{tr} \left(y^{f e}\right)
    + \beta_{y,95}\ \mathcal{S}_{2}\,y^{b e a c d}\,\lambda^{bcdf}\,\mathrm{tr} \left(y^{f e}\right)\,,\yestag{}\\[.5em]
    \beta_{y, y^7 \lambda}^{(4)\,a} &= 
    \beta_{y,96}\ \mathcal{S}_{2}\,y^{f a b c d e f}\,\lambda^{bcde}
    + \beta_{y,97}\ \mathcal{S}_{2}\,y^{f a b c d f e}\,\lambda^{bcde}
    + \beta_{y,98}\ \mathcal{S}_{2}\,y^{f a b c f d e}\,\lambda^{bcde}
    \\&\phantom{=}
    + \beta_{y,99}\ \mathcal{S}_{2}\,y^{f a b f c d e}\,\lambda^{bcde}
    + \beta_{y,100}\ \mathcal{S}_{2}\,y^{b a f c d e f}\,\lambda^{bcde}
    + \beta_{y,101}\ \mathcal{S}_{2}\,y^{b a f c d f e}\,\lambda^{bcde}
    \\&\phantom{=}
    + \beta_{y,102}\ \mathcal{S}_{2}\,y^{b a f c f d e}\,\lambda^{bcde}
    + \beta_{y,103}\ \mathcal{S}_{2}\,y^{b a f f c d e}\,\lambda^{bcde}
    + \beta_{y,104}\ \mathcal{S}_{2}\,y^{b a c f d e f}\,\lambda^{bcde}
    \\&\phantom{=}
    + \beta_{y,105}\ \mathcal{S}_{2}\,y^{b a c f d f e}\,\lambda^{bcde}
    + \beta_{y,106}\ \mathcal{S}_{2}\,y^{b a c f f d e}\,\lambda^{bcde}
    + \beta_{y,107}\ \mathcal{S}_{2}\,y^{b a c d f e f}\,\lambda^{bcde}
    \\&\phantom{=}
    + \beta_{y,108}\ \mathcal{S}_{2}\,y^{b a c d f f e}\,\lambda^{bcde}
    + \beta_{y,109}\ \mathcal{S}_{2}\,y^{f b a c d e f}\,\lambda^{bcde}
    + \beta_{y,110}\ \mathcal{S}_{2}\,y^{f b a c d f e}\,\lambda^{bcde}
    \\&\phantom{=}
    + \beta_{y,111}\ \mathcal{S}_{2}\,y^{f b a c f d e}\,\lambda^{bcde}
    + \beta_{y,112}\ \mathcal{S}_{2}\,y^{f b a f c d e}\,\lambda^{bcde}
    + \beta_{y,113}\ \mathcal{S}_{2}\,y^{b f a c d e f}\,\lambda^{bcde}
    \\&\phantom{=}
    + \beta_{y,114}\ \mathcal{S}_{2}\,y^{b f a c d f e}\,\lambda^{bcde}
    + \beta_{y,115}\ \mathcal{S}_{2}\,y^{b f a c f d e}\,\lambda^{bcde}
    + \beta_{y,116}\ \mathcal{S}_{2}\,y^{b f a f c d e}\,\lambda^{bcde}
    \\&\phantom{=}
    + \beta_{y,117}\ \mathcal{S}_{2}\,y^{b c a f d e f}\,\lambda^{bcde}
    + \beta_{y,118}\ \mathcal{S}_{2}\,y^{b c a f d f e}\,\lambda^{bcde}
    + \beta_{y,119}\ \mathcal{S}_{2}\,y^{b c a f f d e}\,\lambda^{bcde}
    \\&\phantom{=}
    + \beta_{y,120}\ \mathcal{S}_{2}\,y^{b c a d f e f}\,\lambda^{bcde}
    + \beta_{y,121}\ \mathcal{S}_{2}\,y^{b c a d f f e}\,\lambda^{bcde}
    + \beta_{y,122}\ \mathcal{S}_{2}\,y^{b f f a c d e}\,\lambda^{bcde}
    \\&\phantom{=}
    + \beta_{y,123}\ \mathcal{S}_{2}\,y^{f b f a c d e}\,\lambda^{bcde}
    + \beta_{y,124}\ y^{f b c a d e f}\,\lambda^{bcde}
    + \beta_{y,125}\ \mathcal{S}_{2}\,y^{f b c a d f e}\,\lambda^{bcde}
    \\&\phantom{=}
    + \beta_{y,126}\ \mathcal{S}_{2}\,y^{f b c a f d e}\,\lambda^{bcde}
    + \beta_{y,127}\ y^{b f c a d f e}\,\lambda^{bcde}
    + \beta_{y,128}\ \mathcal{S}_{2}\,y^{b f c a f d e}\,\lambda^{bcde}
    \\&\phantom{=}
    + \beta_{y,129}\ y^{b c f a f d e}\,\lambda^{bcde}
    + \beta_{y,130}\ \mathcal{S}_{2}\,y^{b c e d f e f}\,\lambda^{abcd}
    + \beta_{y,131}\ \mathcal{S}_{2}\,y^{b c e d f f e}\,\lambda^{abcd}
    \\&\phantom{=}
    + \beta_{y,132}\ \mathcal{S}_{2}\,y^{b c e f d e f}\,\lambda^{abcd}
    + \beta_{y,133}\ \mathcal{S}_{2}\,y^{b c e f d f e}\,\lambda^{abcd}
    + \beta_{y,134}\ \mathcal{S}_{2}\,y^{b c e e f d f}\,\lambda^{abcd}
    \\&\phantom{=}
    + \beta_{y,135}\ \mathcal{S}_{2}\,y^{b c e f e d f}\,\lambda^{abcd}
    + \beta_{y,136}\ \mathcal{S}_{2}\,y^{b c f e e d f}\,\lambda^{abcd}
    + \beta_{y,137}\ \mathcal{S}_{2}\,y^{b c e e f f d}\,\lambda^{abcd}
    \\&\phantom{=}
    + \beta_{y,138}\ \mathcal{S}_{2}\,y^{b c e f e f d}\,\lambda^{abcd}
    + \beta_{y,139}\ \mathcal{S}_{2}\,y^{b c e f f e d}\,\lambda^{abcd}
    + \beta_{y,140}\ \mathcal{S}_{2}\,y^{b e c d f f e}\,\lambda^{abcd}
    \\&\phantom{=}
    + \beta_{y,141}\ \mathcal{S}_{2}\,y^{b e c d f e f}\,\lambda^{abcd}
    + \beta_{y,142}\ \mathcal{S}_{2}\,y^{b e c f d f e}\,\lambda^{abcd}
    + \beta_{y,143}\ \mathcal{S}_{2}\,y^{b e c f d e f}\,\lambda^{abcd}
    \\&\phantom{=}
    + \beta_{y,144}\ \mathcal{S}_{2}\,y^{b e c f f d e}\,\lambda^{abcd}
    + \beta_{y,145}\ \mathcal{S}_{2}\,y^{b e c f e d f}\,\lambda^{abcd}
    + \beta_{y,146}\ \mathcal{S}_{2}\,y^{b e c e f d f}\,\lambda^{abcd}
    \\&\phantom{=}
    + \beta_{y,147}\ \mathcal{S}_{2}\,y^{b e c e f f d}\,\lambda^{abcd}
    + \beta_{y,148}\ \mathcal{S}_{2}\,y^{b e c f e f d}\,\lambda^{abcd}
    + \beta_{y,149}\ \mathcal{S}_{2}\,y^{b e c f f e d}\,\lambda^{abcd}
    \\&\phantom{=}
    + \beta_{y,150}\ \mathcal{S}_{2}\,y^{b e f c d e f}\,\lambda^{abcd}
    + \beta_{y,151}\ \mathcal{S}_{2}\,y^{b e f c d f e}\,\lambda^{abcd}
    + \beta_{y,152}\ \mathcal{S}_{2}\,y^{b e f c e d f}\,\lambda^{abcd}
    \\&\phantom{=}
    + \beta_{y,153}\ \mathcal{S}_{2}\,y^{b e f c f d e}\,\lambda^{abcd}
    + \beta_{y,154}\ \mathcal{S}_{2}\,y^{b e e c f d f}\,\lambda^{abcd}
    + \beta_{y,155}\ y^{b e e c f f d}\,\lambda^{abcd}
    \\&\phantom{=}
    + \beta_{y,156}\ y^{b e f c f e d}\,\lambda^{abcd}
    + \beta_{y,157}\ y^{b e f c e f d}\,\lambda^{abcd}
    + \beta_{y,158}\ \mathcal{S}_{2}\,y^{e b c d f f e}\,\lambda^{abcd}
    \\&\phantom{=}
    + \beta_{y,159}\ \mathcal{S}_{2}\,y^{e b c d f e f}\,\lambda^{abcd}
    + \beta_{y,160}\ \mathcal{S}_{2}\,y^{e b c f d f e}\,\lambda^{abcd}
    + \beta_{y,161}\ \mathcal{S}_{2}\,y^{e b c f d e f}\,\lambda^{abcd}
    \\&\phantom{=}
    + \beta_{y,162}\ \mathcal{S}_{2}\,y^{e b c e f d f}\,\lambda^{abcd}
    + \beta_{y,163}\ \mathcal{S}_{2}\,y^{e b c f e d f}\,\lambda^{abcd}
    + \beta_{y,164}\ \mathcal{S}_{2}\,y^{e b c f f d e}\,\lambda^{abcd}
    \\&\phantom{=}
    + \beta_{y,165}\ \mathcal{S}_{2}\,y^{e b c f f e d}\,\lambda^{abcd}
    + \beta_{y,166}\ \mathcal{S}_{2}\,y^{e b c f e f d}\,\lambda^{abcd}
    + \beta_{y,167}\ \mathcal{S}_{2}\,y^{e b c e f f d}\,\lambda^{abcd}
    \\&\phantom{=}
    + \beta_{y,168}\ y^{e b f c f d e}\,\lambda^{abcd}
    + \beta_{y,169}\ y^{e b e c f d f}\,\lambda^{abcd}
    + \beta_{y,170}\ y^{e b f c e d f}\,\lambda^{abcd}
    \\&\phantom{=}
    + \beta_{y,171}\ \mathcal{S}_{2}\,y^{e b f c d f e}\,\lambda^{abcd}
    + \beta_{y,172}\ \mathcal{S}_{2}\,y^{e b f c d e f}\,\lambda^{abcd}
    + \beta_{y,173}\ y^{e f b c d f e}\,\lambda^{abcd}
    \\&\phantom{=}
    + \beta_{y,174}\ y^{e f b c d e f}\,\lambda^{abcd} \,,\yestag{}\\[.5em]
    \beta_{y, y^{3,6}}^{(4)\,a} &= \beta_{y,175}\  y^{b a c}\,\mathrm{tr} \left(y^{b d}\right)\,\mathrm{tr} \left(y^{d e}\right)\,\mathrm{tr} \left(y^{e c}\right)
    \\&\phantom{=}
    + \beta_{y,176}\ y^{b a c}\,\mathrm{tr} \left(y^{b c d e}\right)\,\mathrm{tr} \left(y^{d e}\right)
    + \beta_{y,177}\ y^{b a c}\,\mathrm{tr} \left(y^{b d c e}\right)\,\mathrm{tr} \left(y^{d e}\right)
    \\&\phantom{=}
    + \beta_{y,178}\ \mathcal{S}_{2}\,y^{b a c}\,\mathrm{tr} \left(y^{b d d e}\right)\,\mathrm{tr} \left(y^{c e}\right)
    + \beta_{y,179}\ \mathcal{S}_{2}\,y^{b a c}\,\mathrm{tr} \left(y^{b d e d}\right)\,\mathrm{tr} \left(y^{c e}\right)
    \\&\phantom{=}
    + \beta_{y,180}\ y^{b c d}\,\mathrm{tr} \left(y^{c e}\right)\,\mathrm{tr} \left(y^{a b e d}\right)
    + \beta_{y,181}\ \mathcal{S}_{2}\,y^{c b d}\,\mathrm{tr} \left(y^{c e}\right)\,\mathrm{tr} \left(y^{a b e d}\right)
    \\&\phantom{=}
    + \beta_{y,182}\ \mathcal{S}_{2}\,y^{b c d}\,\mathrm{tr} \left(y^{c e}\right)\,\mathrm{tr} \left(y^{a e b d}\right)
    + \beta_{y,183}\ \mathcal{S}_{2}\,y^{c b d}\,\mathrm{tr} \left(y^{c e}\right)\,\mathrm{tr} \left(y^{a e b d}\right)
    \\&\phantom{=}
    + \beta_{y,184}\ \mathcal{S}_{2}\,y^{c d b}\,\mathrm{tr} \left(y^{c e}\right)\,\mathrm{tr} \left(y^{a e b d}\right)\\&\phantom{=}
    + \beta_{y,185}\ y^{b a c}\,\mathrm{tr} \left(y^{b c d d e e}\right)
    + \beta_{y,186}\ y^{b a c}\,\mathrm{tr} \left(y^{b c d e e d}\right)
    + \beta_{y,187}\ y^{b a c}\,\mathrm{tr} \left(y^{b c d e d e}\right)
    \\&\phantom{=}
    + \beta_{y,188}\ y^{b a c}\,\mathrm{tr} \left(y^{b d c e d e}\right)
    + \beta_{y,189}\ \mathcal{S}_{2}\,y^{b a c}\,\mathrm{tr} \left(y^{b d c d e e}\right)
    + \beta_{y,190}\ y^{b a c}\,\mathrm{tr} \left(y^{b d d c e e}\right)
    \\&\phantom{=}
    + \beta_{y,191}\ y^{b a c}\,\mathrm{tr} \left(y^{b d e c d e}\right)
    + \beta_{y,192}\ y^{b a c}\,\mathrm{tr} \left(y^{b d e c e d}\right)
    + \beta_{y,193}\ y^{b c d}\,\mathrm{tr} \left(y^{a e b c d e}\right)
    \\&\phantom{=}
    + \beta_{y,194}\ \mathcal{S}_{2}\,y^{c b d}\,\mathrm{tr} \left(y^{a e b c d e}\right)
    + \beta_{y,195}\ \mathcal{S}_{2}\,y^{b c d}\,\mathrm{tr} \left(y^{a e b c e d}\right)
    + \beta_{y,196}\ \mathcal{S}_{2}\,y^{b d c}\,\mathrm{tr} \left(y^{a e b c e d}\right)
    \\&\phantom{=}
    + \beta_{y,197}\ \mathcal{S}_{2}\,y^{c b d}\,\mathrm{tr} \left(y^{a e b c e d}\right)
    + \beta_{y,198}\ \mathcal{S}_{2}\,y^{b c d}\,\mathrm{tr} \left(y^{a e b e c d}\right)
    + \beta_{y,199}\ \mathcal{S}_{2}\,y^{b d c}\,\mathrm{tr} \left(y^{a e b e c d}\right)
    \\&\phantom{=}
    + \beta_{y,200}\ \mathcal{S}_{2}\,y^{c b d}\,\mathrm{tr} \left(y^{a e b e c d}\right)
    + \beta_{y,201}\ \mathcal{S}_{2}\,y^{b c d}\,\mathrm{tr} \left(y^{a e e b c d}\right)
    + \beta_{y,202}\ \mathcal{S}_{2}\,y^{b d c}\,\mathrm{tr} \left(y^{a e e b c d}\right)
    \\&\phantom{=}
    + \beta_{y,203}\ \mathcal{S}_{2}\,y^{c b d}\,\mathrm{tr} \left(y^{a e e b c d}\right)
    + \beta_{y,204}\ y^{b c d}\,\mathrm{tr} \left(y^{a b e c e d}\right)
    + \beta_{y,205}\ \mathcal{S}_{2}\,y^{b d c}\,\mathrm{tr} \left(y^{a b e c e d}\right)
    \\&\phantom{=}
    + \beta_{y,206}\ \mathcal{S}_{2}\,y^{b c d}\,\mathrm{tr} \left(y^{a b e e c d}\right)
    + \beta_{y,207}\ \mathcal{S}_{2}\,y^{b d c}\,\mathrm{tr} \left(y^{a b e e c d}\right)
    + \beta_{y,208}\ \mathcal{S}_{2}\,y^{c b d}\,\mathrm{tr} \left(y^{a b e e c d}\right)\,,\yestag{}\\[.5em]
    \beta_{y, y^{5,4}}^{(4)\,a} &= 
    \beta_{y,209}\ \mathcal{S}_{2}\,y^{b a d e c}\,\mathrm{tr} \left(y^{b c}\right)\,\mathrm{tr} \left(y^{d e}\right)
    + \beta_{y,210}\ \mathcal{S}_{2}\,y^{b a d c e}\,\mathrm{tr} \left(y^{b c}\right)\,\mathrm{tr} \left(y^{d e}\right)
    \\&\phantom{=}
    + \beta_{y,211}\ y^{b d a e c}\,\mathrm{tr} \left(y^{b c}\right)\,\mathrm{tr} \left(y^{d e}\right)
    + \beta_{y,212}\ y^{d b a e c}\,\mathrm{tr} \left(y^{b c}\right)\,\mathrm{tr} \left(y^{d e}\right)
    \\&\phantom{=}
    + \beta_{y,213}\ \mathcal{S}_{2}\,y^{b a d d c}\,\mathrm{tr} \left(y^{b e}\right)\,\mathrm{tr} \left(y^{e c}\right)
    + \beta_{y,214}\ \mathcal{S}_{2}\,y^{b a d c d}\,\mathrm{tr} \left(y^{b e}\right)\,\mathrm{tr} \left(y^{e c}\right)
    \\&\phantom{=}
    + \beta_{y,215}\ \mathcal{S}_{2}\,y^{d a b c d}\,\mathrm{tr} \left(y^{b e}\right)\,\mathrm{tr} \left(y^{e c}\right)
    + \beta_{y,216}\ \mathcal{S}_{2}\,y^{d a b d c}\,\mathrm{tr} \left(y^{b e}\right)\,\mathrm{tr} \left(y^{e c}\right)
    \\&\phantom{=}
    + \beta_{y,217}\ y^{b d a d c}\,\mathrm{tr} \left(y^{b e}\right)\,\mathrm{tr} \left(y^{e c}\right)
    + \beta_{y,218}\ y^{d b a c d}\,\mathrm{tr} \left(y^{b e}\right)\,\mathrm{tr} \left(y^{e c}\right)
    \\&\phantom{=}
    + \beta_{y,219}\ \mathcal{S}_{2}\,y^{d b a d c}\,\mathrm{tr} \left(y^{b e}\right)\,\mathrm{tr} \left(y^{e c}\right)\\&\phantom{=}
    + \beta_{y,220}\ y^{e b c d e}\,\mathrm{tr} \left(y^{a b c d}\right)
    + \beta_{y,221}\ \mathcal{S}_{2}\,y^{e b d c e}\,\mathrm{tr} \left(y^{a b c d}\right)
    + \beta_{y,222}\ \mathcal{S}_{2}\,y^{e b c e d}\,\mathrm{tr} \left(y^{a b c d}\right)
    \\&\phantom{=}
    + \beta_{y,223}\ \mathcal{S}_{2}\,y^{e c b e d}\,\mathrm{tr} \left(y^{a b c d}\right)
    + \beta_{y,224}\ \mathcal{S}_{2}\,y^{e b d e c}\,\mathrm{tr} \left(y^{a b c d}\right)
    + \beta_{y,225}\ \mathcal{S}_{2}\,y^{e b e d c}\,\mathrm{tr} \left(y^{a b c d}\right)
    \\&\phantom{=}
    + \beta_{y,226}\ \mathcal{S}_{2}\,y^{e b e c d}\,\mathrm{tr} \left(y^{a b c d}\right)
    + \beta_{y,227}\ \mathcal{S}_{2}\,y^{e c e b d}\,\mathrm{tr} \left(y^{a b c d}\right)
    + \beta_{y,228}\ y^{b e c e d}\,\mathrm{tr} \left(y^{a b c d}\right)
    \\&\phantom{=}
    + \beta_{y,229}\ \mathcal{S}_{2}\,y^{c e b e d}\,\mathrm{tr} \left(y^{a b c d}\right)
    + \beta_{y,230}\ \mathcal{S}_{2}\,y^{b e e d c}\,\mathrm{tr} \left(y^{a b c d}\right)
    + \beta_{y,231}\ \mathcal{S}_{2}\,y^{b e e c d}\,\mathrm{tr} \left(y^{a b c d}\right)
    \\&\phantom{=}
    + \beta_{y,232}\ \mathcal{S}_{2}\,y^{c e e b d}\,\mathrm{tr} \left(y^{a b c d}\right)
    + \beta_{y,233}\ \mathcal{S}_{2}\,y^{b a c d e}\,\mathrm{tr} \left(y^{b c d e}\right)
    + \beta_{y,234}\ \mathcal{S}_{2}\,y^{b a c e d}\,\mathrm{tr} \left(y^{b c d e}\right)
    \\&\phantom{=}
    + \beta_{y,235}\ \mathcal{S}_{2}\,y^{b a d c e}\,\mathrm{tr} \left(y^{b c d e}\right)
    + \beta_{y,236}\ y^{b c a d e}\,\mathrm{tr} \left(y^{b c d e}\right)
    + \beta_{y,237}\ \mathcal{S}_{2}\,y^{b c a e d}\,\mathrm{tr} \left(y^{b c d e}\right)
    \\&\phantom{=}
    + \beta_{y,238}\ y^{b d a c e}\,\mathrm{tr} \left(y^{b c d e}\right)
    + \beta_{y,239}\ \mathcal{S}_{2}\,y^{b a c d b}\,\mathrm{tr} \left(y^{c d e e}\right)
    + \beta_{y,240}\ \mathcal{S}_{2}\,y^{b a c b d}\,\mathrm{tr} \left(y^{c d e e}\right)
    \\&\phantom{=}
    + \beta_{y,241}\ \mathcal{S}_{2}\,y^{c a b b d}\,\mathrm{tr} \left(y^{c d e e}\right)
    + \beta_{y,242}\ \mathcal{S}_{2}\,y^{c a b d b}\,\mathrm{tr} \left(y^{c d e e}\right)
    + \beta_{y,243}\ y^{b c a d b}\,\mathrm{tr} \left(y^{c d e e}\right)
    \\&\phantom{=}
    + \beta_{y,244}\ y^{c b a b d}\,\mathrm{tr} \left(y^{c d e e}\right)
    + \beta_{y,245}\ \mathcal{S}_{2}\,y^{c b a d b}\,\mathrm{tr} \left(y^{c d e e}\right)
    + \beta_{y,246}\ \mathcal{S}_{2}\,y^{b a c d b}\,\mathrm{tr} \left(y^{c e d e}\right)
    \\&\phantom{=}
    + \beta_{y,247}\ \mathcal{S}_{2}\,y^{b a c b d}\,\mathrm{tr} \left(y^{c e d e}\right)
    + \beta_{y,248}\ \mathcal{S}_{2}\,y^{c a b b d}\,\mathrm{tr} \left(y^{c e d e}\right)
    + \beta_{y,249}\ \mathcal{S}_{2}\,y^{c a b d b}\,\mathrm{tr} \left(y^{c e d e}\right)
    \\&\phantom{=}
    + \beta_{y,250}\ y^{b c a d b}\,\mathrm{tr} \left(y^{c e d e}\right)
    + \beta_{y,251}\ y^{c b a b d}\,\mathrm{tr} \left(y^{c e d e}\right)
    + \beta_{y,252}\ \mathcal{S}_{2}\,y^{c b a d b}\,\mathrm{tr} \left(y^{c e d e}\right)\,,\yestag{}\\[.5em]
    \beta_{y, y^{7,2}}^{(4)\,a} &= 
    \beta_{y,253}\ \mathcal{S}_{2}\,y^{b a d d e e c}\,\mathrm{tr} \left(y^{b c}\right)
    + \beta_{y,254}\ \mathcal{S}_{2}\,y^{b a d e d e c}\,\mathrm{tr} \left(y^{b c}\right)
    + \beta_{y,255}\ \mathcal{S}_{2}\,y^{b a d e e d c}\,\mathrm{tr} \left(y^{b c}\right)
    \\&\phantom{=}
    + \beta_{y,256}\ \mathcal{S}_{2}\,y^{b a d d e c e}\,\mathrm{tr} \left(y^{b c}\right)
    + \beta_{y,257}\ \mathcal{S}_{2}\,y^{b a d e d c e}\,\mathrm{tr} \left(y^{b c}\right)
    + \beta_{y,258}\ \mathcal{S}_{2}\,y^{b a d e e c d}\,\mathrm{tr} \left(y^{b c}\right)
    \\&\phantom{=}
    + \beta_{y,259}\ \mathcal{S}_{2}\,y^{b a d e c d e}\,\mathrm{tr} \left(y^{b c}\right)
    + \beta_{y,260}\ \mathcal{S}_{2}\,y^{b a d e c e d}\,\mathrm{tr} \left(y^{b c}\right)
    + \beta_{y,261}\ \mathcal{S}_{2}\,y^{b a d c e e d}\,\mathrm{tr} \left(y^{b c}\right)
    \\&\phantom{=}
    + \beta_{y,262}\ \mathcal{S}_{2}\,y^{b a d c e d e}\,\mathrm{tr} \left(y^{b c}\right)
    + \beta_{y,263}\ \mathcal{S}_{2}\,y^{d a b d e e c}\,\mathrm{tr} \left(y^{b c}\right)
    + \beta_{y,264}\ \mathcal{S}_{2}\,y^{d a b e d e c}\,\mathrm{tr} \left(y^{b c}\right)
    \\&\phantom{=}
    + \beta_{y,265}\ \mathcal{S}_{2}\,y^{d a b e e d c}\,\mathrm{tr} \left(y^{b c}\right)
    + \beta_{y,266}\ \mathcal{S}_{2}\,y^{d a b e e c d}\,\mathrm{tr} \left(y^{b c}\right)
    + \beta_{y,267}\ \mathcal{S}_{2}\,y^{d a b e d c e}\,\mathrm{tr} \left(y^{b c}\right)
    \\&\phantom{=}
    + \beta_{y,268}\ \mathcal{S}_{2}\,y^{d a b d e c e}\,\mathrm{tr} \left(y^{b c}\right)
    + \beta_{y,269}\ \mathcal{S}_{2}\,y^{d a b e c e d}\,\mathrm{tr} \left(y^{b c}\right)
    + \beta_{y,270}\ \mathcal{S}_{2}\,y^{d a b e c d e}\,\mathrm{tr} \left(y^{b c}\right)
    \\&\phantom{=}
    + \beta_{y,271}\ \mathcal{S}_{2}\,y^{d a b c e e d}\,\mathrm{tr} \left(y^{b c}\right)
    + \beta_{y,272}\ \mathcal{S}_{2}\,y^{d a b c e d e}\,\mathrm{tr} \left(y^{b c}\right)
    + \beta_{y,273}\ \mathcal{S}_{2}\,y^{d a e b e d c}\,\mathrm{tr} \left(y^{b c}\right)
    \\&\phantom{=}
    + \beta_{y,274}\ \mathcal{S}_{2}\,y^{d a e b d e c}\,\mathrm{tr} \left(y^{b c}\right)
    + \beta_{y,275}\ \mathcal{S}_{2}\,y^{d a e b e c d}\,\mathrm{tr} \left(y^{b c}\right)
    + \beta_{y,276}\ \mathcal{S}_{2}\,y^{d a e b d c e}\,\mathrm{tr} \left(y^{b c}\right)
    \\&\phantom{=}
    + \beta_{y,277}\ \mathcal{S}_{2}\,y^{d a e b c d e}\,\mathrm{tr} \left(y^{b c}\right)
    + \beta_{y,278}\ \mathcal{S}_{2}\,y^{d a e b c e d}\,\mathrm{tr} \left(y^{b c}\right)
    + \beta_{y,279}\ \mathcal{S}_{2}\,y^{d a e d b e c}\,\mathrm{tr} \left(y^{b c}\right)
    \\&\phantom{=}
    + \beta_{y,280}\ \mathcal{S}_{2}\,y^{d a e e b d c}\,\mathrm{tr} \left(y^{b c}\right)
    + \beta_{y,281}\ \mathcal{S}_{2}\,y^{d a e e b c d}\,\mathrm{tr} \left(y^{b c}\right)
    + \beta_{y,282}\ \mathcal{S}_{2}\,y^{d a e d b c e}\,\mathrm{tr} \left(y^{b c}\right)
    \\&\phantom{=}
    + \beta_{y,283}\ \mathcal{S}_{2}\,y^{b d a e e d c}\,\mathrm{tr} \left(y^{b c}\right)
    + \beta_{y,284}\ \mathcal{S}_{2}\,y^{b d a e d e c}\,\mathrm{tr} \left(y^{b c}\right)
    + \beta_{y,285}\ \mathcal{S}_{2}\,y^{b d a d e e c}\,\mathrm{tr} \left(y^{b c}\right)
    \\&\phantom{=}
    + \beta_{y,286}\ \mathcal{S}_{2}\,y^{b d a e e c d}\,\mathrm{tr} \left(y^{b c}\right)
    + \beta_{y,287}\ \mathcal{S}_{2}\,y^{b d a e d c e}\,\mathrm{tr} \left(y^{b c}\right)
    + \beta_{y,288}\ \mathcal{S}_{2}\,y^{b d a d e c e}\,\mathrm{tr} \left(y^{b c}\right)
    \\&\phantom{=}
    + \beta_{y,289}\ \mathcal{S}_{2}\,y^{b d a e c e d}\,\mathrm{tr} \left(y^{b c}\right)
    + \beta_{y,290}\ \mathcal{S}_{2}\,y^{b d a e c d e}\,\mathrm{tr} \left(y^{b c}\right)
    + \beta_{y,291}\ \mathcal{S}_{2}\,y^{b d a c e e d}\,\mathrm{tr} \left(y^{b c}\right)
    \\&\phantom{=}
    + \beta_{y,292}\ \mathcal{S}_{2}\,y^{b d a c e d e}\,\mathrm{tr} \left(y^{b c}\right)
    + \beta_{y,293}\ \mathcal{S}_{2}\,y^{d b a e e d c}\,\mathrm{tr} \left(y^{b c}\right)
    + \beta_{y,294}\ \mathcal{S}_{2}\,y^{d b a e d e c}\,\mathrm{tr} \left(y^{b c}\right)
    \\&\phantom{=}
    + \beta_{y,295}\ \mathcal{S}_{2}\,y^{d b a d e e c}\,\mathrm{tr} \left(y^{b c}\right)
    + \beta_{y,296}\ \mathcal{S}_{2}\,y^{d b a e e c d}\,\mathrm{tr} \left(y^{b c}\right)
    + \beta_{y,297}\ \mathcal{S}_{2}\,y^{d b a e d c e}\,\mathrm{tr} \left(y^{b c}\right)
    \\&\phantom{=}
    + \beta_{y,298}\ \mathcal{S}_{2}\,y^{d b a d e c e}\,\mathrm{tr} \left(y^{b c}\right)
    + \beta_{y,299}\ \mathcal{S}_{2}\,y^{d b a e c e d}\,\mathrm{tr} \left(y^{b c}\right)
    + \beta_{y,300}\ \mathcal{S}_{2}\,y^{d b a e c d e}\,\mathrm{tr} \left(y^{b c}\right)
    \\&\phantom{=}
    + \beta_{y,301}\ \mathcal{S}_{2}\,y^{d b a c e e d}\,\mathrm{tr} \left(y^{b c}\right)
    + \beta_{y,302}\ \mathcal{S}_{2}\,y^{d b a c e d e}\,\mathrm{tr} \left(y^{b c}\right)
    + \beta_{y,303}\ \mathcal{S}_{2}\,y^{d e a b d e c}\,\mathrm{tr} \left(y^{b c}\right)
    \\&\phantom{=}
    + \beta_{y,304}\ \mathcal{S}_{2}\,y^{d e a b e d c}\,\mathrm{tr} \left(y^{b c}\right)
    + \beta_{y,305}\ \mathcal{S}_{2}\,y^{d e a b d c e}\,\mathrm{tr} \left(y^{b c}\right)
    + \beta_{y,306}\ \mathcal{S}_{2}\,y^{d e a b e c d}\,\mathrm{tr} \left(y^{b c}\right)
    \\&\phantom{=}
    + \beta_{y,307}\ \mathcal{S}_{2}\,y^{d e a b c d e}\,\mathrm{tr} \left(y^{b c}\right)
    + \beta_{y,308}\ \mathcal{S}_{2}\,y^{d e a b c e d}\,\mathrm{tr} \left(y^{b c}\right)
    + \beta_{y,309}\ \mathcal{S}_{2}\,y^{d e a d b e c}\,\mathrm{tr} \left(y^{b c}\right)
    \\&\phantom{=}
    + \beta_{y,310}\ \mathcal{S}_{2}\,y^{d e a e b d c}\,\mathrm{tr} \left(y^{b c}\right)
    + \beta_{y,311}\ \mathcal{S}_{2}\,y^{d e a e b c d}\,\mathrm{tr} \left(y^{b c}\right)
    + \beta_{y,312}\ \mathcal{S}_{2}\,y^{d e a d b c e}\,\mathrm{tr} \left(y^{b c}\right)
    \\&\phantom{=}
    + \beta_{y,313}\ y^{b d d a e e c}\,\mathrm{tr} \left(y^{b c}\right)
    + \beta_{y,314}\ y^{b d e a e d c}\,\mathrm{tr} \left(y^{b c}\right)
    + \beta_{y,315}\ y^{b d e a d e c}\,\mathrm{tr} \left(y^{b c}\right)
    \\&\phantom{=}
    + \beta_{y,316}\ \mathcal{S}_{2}\,y^{b d d a e c e}\,\mathrm{tr} \left(y^{b c}\right)
    + \beta_{y,317}\ \mathcal{S}_{2}\,y^{b d e a d c e}\,\mathrm{tr} \left(y^{b c}\right)
    + \beta_{y,318}\ \mathcal{S}_{2}\,y^{b d e a e c d}\,\mathrm{tr} \left(y^{b c}\right)
    \\&\phantom{=}
    + \beta_{y,319}\ \mathcal{S}_{2}\,y^{b d e a c d e}\,\mathrm{tr} \left(y^{b c}\right)
    + \beta_{y,320}\ \mathcal{S}_{2}\,y^{b d e a c e d}\,\mathrm{tr} \left(y^{b c}\right)
    + \beta_{y,321}\ y^{d b d a e c e}\,\mathrm{tr} \left(y^{b c}\right)
    \\&\phantom{=}
    + \beta_{y,322}\ y^{d b e a d c e}\,\mathrm{tr} \left(y^{b c}\right)
    + \beta_{y,323}\ y^{d b e a e c d}\,\mathrm{tr} \left(y^{b c}\right)
    + \beta_{y,324}\ \mathcal{S}_{2}\,y^{d b e a c e d}\,\mathrm{tr} \left(y^{b c}\right)
    \\&\phantom{=}
    + \beta_{y,325}\ \mathcal{S}_{2}\,y^{d b e a c d e}\,\mathrm{tr} \left(y^{b c}\right)
    + \beta_{y,326}\ y^{d e b a c d e}\,\mathrm{tr} \left(y^{b c}\right)
    + \beta_{y,327}\ y^{d e b a c e d}\,\mathrm{tr} \left(y^{b c}\right)
    \\&\phantom{=}
    + \beta_{y,328}\ \mathcal{S}_{2}\,y^{d b c a e e d}\,\mathrm{tr} \left(y^{b c}\right)
    + \beta_{y,329}\ \mathcal{S}_{2}\,y^{d b c a e d e}\,\mathrm{tr} \left(y^{b c}\right)
    + \beta_{y,330}\ \mathcal{S}_{2}\,y^{b d c a e e d}\,\mathrm{tr} \left(y^{b c}\right)
    \\&\phantom{=}
    + \beta_{y,331}\ \mathcal{S}_{2}\,y^{b d c a e d e}\,\mathrm{tr} \left(y^{b c}\right)\,,\yestag{}\\[.5em]
    \beta_{y, y^9}^{(4)\,a} &= 
    \beta_{y,332}\ \mathcal{S}_{2}\,y^{b a c b d c e d e}
    + \beta_{y,333}\ \mathcal{S}_{2}\,y^{b a c b d c e e d}
    + \beta_{y,334}\ \mathcal{S}_{2}\,y^{b a c b d d e c e}
    + \beta_{y,335}\ \mathcal{S}_{2}\,y^{b a c b d d e e c}
    \\&\phantom{=}
    + \beta_{y,336}\ \mathcal{S}_{2}\,y^{b a c b d e c d e}
    + \beta_{y,337}\ \mathcal{S}_{2}\,y^{b a c b d e c e d}
    + \beta_{y,338}\ \mathcal{S}_{2}\,y^{b a c b d e d c e}
    + \beta_{y,339}\ \mathcal{S}_{2}\,y^{b a c b d e d e c}
    \\&\phantom{=}
    + \beta_{y,340}\ \mathcal{S}_{2}\,y^{b a c b d e e c d}
    + \beta_{y,341}\ \mathcal{S}_{2}\,y^{b a c b d e e d c}
    + \beta_{y,342}\ \mathcal{S}_{2}\,y^{b a c c d b e d e}
    + \beta_{y,343}\ \mathcal{S}_{2}\,y^{b a c c d b e e d}
    \\&\phantom{=}
    + \beta_{y,344}\ \mathcal{S}_{2}\,y^{b a c c d d e b e}
    + \beta_{y,345}\ \mathcal{S}_{2}\,y^{b a c c d d e e b}
    + \beta_{y,346}\ \mathcal{S}_{2}\,y^{b a c c d e b d e}
    + \beta_{y,347}\ \mathcal{S}_{2}\,y^{b a c c d e b e d}
    \\&\phantom{=}
    + \beta_{y,348}\ \mathcal{S}_{2}\,y^{b a c c d e d b e}
    + \beta_{y,349}\ \mathcal{S}_{2}\,y^{b a c c d e d e b}
    + \beta_{y,350}\ \mathcal{S}_{2}\,y^{b a c c d e e b d}
    + \beta_{y,351}\ \mathcal{S}_{2}\,y^{b a c c d e e d b}
    \\&\phantom{=}
    + \beta_{y,352}\ \mathcal{S}_{2}\,y^{b a c d b c e d e}
    + \beta_{y,353}\ \mathcal{S}_{2}\,y^{b a c d b c e e d}
    + \beta_{y,354}\ \mathcal{S}_{2}\,y^{b a c d b d e c e}
    + \beta_{y,355}\ \mathcal{S}_{2}\,y^{b a c d b d e e c}
    \\&\phantom{=}
    + \beta_{y,356}\ \mathcal{S}_{2}\,y^{b a c d b e c d e}
    + \beta_{y,357}\ \mathcal{S}_{2}\,y^{b a c d b e c e d}
    + \beta_{y,358}\ \mathcal{S}_{2}\,y^{b a c d b e d c e}
    + \beta_{y,359}\ \mathcal{S}_{2}\,y^{b a c d b e d e c}
    \\&\phantom{=}
    + \beta_{y,360}\ \mathcal{S}_{2}\,y^{b a c d b e e c d}
    + \beta_{y,361}\ \mathcal{S}_{2}\,y^{b a c d b e e d c}
    + \beta_{y,362}\ \mathcal{S}_{2}\,y^{b a c d c b e d e}
    + \beta_{y,363}\ \mathcal{S}_{2}\,y^{b a c d c b e e d}
    \\&\phantom{=}
    + \beta_{y,364}\ \mathcal{S}_{2}\,y^{b a c d c d e b e}
    + \beta_{y,365}\ \mathcal{S}_{2}\,y^{b a c d c d e e b}
    + \beta_{y,366}\ \mathcal{S}_{2}\,y^{b a c d c e b d e}
    + \beta_{y,367}\ \mathcal{S}_{2}\,y^{b a c d c e b e d}
    \\&\phantom{=}
    + \beta_{y,368}\ \mathcal{S}_{2}\,y^{b a c d c e d b e}
    + \beta_{y,369}\ \mathcal{S}_{2}\,y^{b a c d c e d e b}
    + \beta_{y,370}\ \mathcal{S}_{2}\,y^{b a c d c e e b d}
    + \beta_{y,371}\ \mathcal{S}_{2}\,y^{b a c d c e e d b}
    \\&\phantom{=}
    + \beta_{y,372}\ \mathcal{S}_{2}\,y^{b a c d d b e c e}
    + \beta_{y,373}\ \mathcal{S}_{2}\,y^{b a c d d b e e c}
    + \beta_{y,374}\ \mathcal{S}_{2}\,y^{b a c d d c e b e}
    + \beta_{y,375}\ \mathcal{S}_{2}\,y^{b a c d d c e e b}
    \\&\phantom{=}
    + \beta_{y,376}\ \mathcal{S}_{2}\,y^{b a c d d e b c e}
    + \beta_{y,377}\ \mathcal{S}_{2}\,y^{b a c d d e b e c}
    + \beta_{y,378}\ \mathcal{S}_{2}\,y^{b a c d d e c b e}
    + \beta_{y,379}\ \mathcal{S}_{2}\,y^{b a c d d e c e b}
    \\&\phantom{=}
    + \beta_{y,380}\ \mathcal{S}_{2}\,y^{b a c d d e e b c}
    + \beta_{y,381}\ \mathcal{S}_{2}\,y^{b a c d d e e c b}
    + \beta_{y,382}\ \mathcal{S}_{2}\,y^{b a c d e b c d e}
    + \beta_{y,383}\ \mathcal{S}_{2}\,y^{b a c d e b c e d}
    \\&\phantom{=}
    + \beta_{y,384}\ \mathcal{S}_{2}\,y^{b a c d e b d c e}
    + \beta_{y,385}\ \mathcal{S}_{2}\,y^{b a c d e b d e c}
    + \beta_{y,386}\ \mathcal{S}_{2}\,y^{b a c d e b e c d}
    + \beta_{y,387}\ \mathcal{S}_{2}\,y^{b a c d e b e d c}
    \\&\phantom{=}
    + \beta_{y,388}\ \mathcal{S}_{2}\,y^{b a c d e c b d e}
    + \beta_{y,389}\ \mathcal{S}_{2}\,y^{b a c d e c b e d}
    + \beta_{y,390}\ \mathcal{S}_{2}\,y^{b a c d e c d b e}
    + \beta_{y,391}\ \mathcal{S}_{2}\,y^{b a c d e c d e b}
    \\&\phantom{=}
    + \beta_{y,392}\ \mathcal{S}_{2}\,y^{b a c d e c e b d}
    + \beta_{y,393}\ \mathcal{S}_{2}\,y^{b a c d e c e d b}
    + \beta_{y,394}\ \mathcal{S}_{2}\,y^{b a c d e d b c e}
    + \beta_{y,395}\ \mathcal{S}_{2}\,y^{b a c d e d b e c}
    \\&\phantom{=}
    + \beta_{y,396}\ \mathcal{S}_{2}\,y^{b a c d e d c b e}
    + \beta_{y,397}\ \mathcal{S}_{2}\,y^{b a c d e d c e b}
    + \beta_{y,398}\ \mathcal{S}_{2}\,y^{b a c d e d e b c}
    + \beta_{y,399}\ \mathcal{S}_{2}\,y^{b a c d e d e c b}
    \\&\phantom{=}
    + \beta_{y,400}\ \mathcal{S}_{2}\,y^{b a c d e e b c d}
    + \beta_{y,401}\ \mathcal{S}_{2}\,y^{b a c d e e b d c}
    + \beta_{y,402}\ \mathcal{S}_{2}\,y^{b a c d e e c b d}
    + \beta_{y,403}\ \mathcal{S}_{2}\,y^{b a c d e e c d b}
    \\&\phantom{=}
    + \beta_{y,404}\ \mathcal{S}_{2}\,y^{b a c d e e d b c}
    + \beta_{y,405}\ \mathcal{S}_{2}\,y^{b a c d e e d c b}
    + \beta_{y,406}\ \mathcal{S}_{2}\,y^{b c a b d c e d e}
    + \beta_{y,407}\ \mathcal{S}_{2}\,y^{b c a b d c e e d}
    \\&\phantom{=}
    + \beta_{y,408}\ \mathcal{S}_{2}\,y^{b c a b d d e c e}
    + \beta_{y,409}\ \mathcal{S}_{2}\,y^{b c a b d d e e c}
    + \beta_{y,410}\ \mathcal{S}_{2}\,y^{b c a b d e c d e}
    + \beta_{y,411}\ \mathcal{S}_{2}\,y^{b c a b d e c e d}
    \\&\phantom{=}
    + \beta_{y,412}\ \mathcal{S}_{2}\,y^{b c a b d e d c e}
    + \beta_{y,413}\ \mathcal{S}_{2}\,y^{b c a b d e d e c}
    + \beta_{y,414}\ \mathcal{S}_{2}\,y^{b c a b d e e c d}
    + \beta_{y,415}\ \mathcal{S}_{2}\,y^{b c a b d e e d c}
    \\&\phantom{=}
    + \beta_{y,416}\ \mathcal{S}_{2}\,y^{b c a c d b e d e}
    + \beta_{y,417}\ \mathcal{S}_{2}\,y^{b c a c d b e e d}
    + \beta_{y,418}\ \mathcal{S}_{2}\,y^{b c a c d d e b e}
    + \beta_{y,419}\ \mathcal{S}_{2}\,y^{b c a c d d e e b}
    \\&\phantom{=}
    + \beta_{y,420}\ \mathcal{S}_{2}\,y^{b c a c d e b d e}
    + \beta_{y,421}\ \mathcal{S}_{2}\,y^{b c a c d e b e d}
    + \beta_{y,422}\ \mathcal{S}_{2}\,y^{b c a c d e d b e}
    + \beta_{y,423}\ \mathcal{S}_{2}\,y^{b c a c d e d e b}
    \\&\phantom{=}
    + \beta_{y,424}\ \mathcal{S}_{2}\,y^{b c a c d e e b d}
    + \beta_{y,425}\ \mathcal{S}_{2}\,y^{b c a c d e e d b}
    + \beta_{y,426}\ \mathcal{S}_{2}\,y^{b c a d b c e d e}
    + \beta_{y,427}\ \mathcal{S}_{2}\,y^{b c a d b c e e d}
    \\&\phantom{=}
    + \beta_{y,428}\ \mathcal{S}_{2}\,y^{b c a d b d e c e}
    + \beta_{y,429}\ \mathcal{S}_{2}\,y^{b c a d b d e e c}
    + \beta_{y,430}\ \mathcal{S}_{2}\,y^{b c a d b e c d e}
    + \beta_{y,431}\ \mathcal{S}_{2}\,y^{b c a d b e c e d}
    \\&\phantom{=}
    + \beta_{y,432}\ \mathcal{S}_{2}\,y^{b c a d b e d c e}
    + \beta_{y,433}\ \mathcal{S}_{2}\,y^{b c a d b e d e c}
    + \beta_{y,434}\ \mathcal{S}_{2}\,y^{b c a d b e e c d}
    + \beta_{y,435}\ \mathcal{S}_{2}\,y^{b c a d b e e d c}
    \\&\phantom{=}
    + \beta_{y,436}\ \mathcal{S}_{2}\,y^{b c a d c b e d e}
    + \beta_{y,437}\ \mathcal{S}_{2}\,y^{b c a d c b e e d}
    + \beta_{y,438}\ \mathcal{S}_{2}\,y^{b c a d c d e b e}
    + \beta_{y,439}\ \mathcal{S}_{2}\,y^{b c a d c d e e b}
    \\&\phantom{=}
    + \beta_{y,440}\ \mathcal{S}_{2}\,y^{b c a d c e b d e}
    + \beta_{y,441}\ \mathcal{S}_{2}\,y^{b c a d c e b e d}
    + \beta_{y,442}\ \mathcal{S}_{2}\,y^{b c a d c e d b e}
    + \beta_{y,443}\ \mathcal{S}_{2}\,y^{b c a d c e d e b}
    \\&\phantom{=}
    + \beta_{y,444}\ \mathcal{S}_{2}\,y^{b c a d c e e b d}
    + \beta_{y,445}\ \mathcal{S}_{2}\,y^{b c a d c e e d b}
    + \beta_{y,446}\ \mathcal{S}_{2}\,y^{b c a d d b e c e}
    + \beta_{y,447}\ \mathcal{S}_{2}\,y^{b c a d d b e e c}
    \\&\phantom{=}
    + \beta_{y,448}\ \mathcal{S}_{2}\,y^{b c a d d c e b e}
    + \beta_{y,449}\ \mathcal{S}_{2}\,y^{b c a d d c e e b}
    + \beta_{y,450}\ \mathcal{S}_{2}\,y^{b c a d d e b c e}
    + \beta_{y,451}\ \mathcal{S}_{2}\,y^{b c a d d e b e c}
    \\&\phantom{=}
    + \beta_{y,452}\ \mathcal{S}_{2}\,y^{b c a d d e c b e}
    + \beta_{y,453}\ \mathcal{S}_{2}\,y^{b c a d d e c e b}
    + \beta_{y,454}\ \mathcal{S}_{2}\,y^{b c a d d e e b c}
    + \beta_{y,455}\ \mathcal{S}_{2}\,y^{b c a d d e e c b}
    \\&\phantom{=}
    + \beta_{y,456}\ \mathcal{S}_{2}\,y^{b c a d e b c d e}
    + \beta_{y,457}\ \mathcal{S}_{2}\,y^{b c a d e b c e d}
    + \beta_{y,458}\ \mathcal{S}_{2}\,y^{b c a d e b d c e}
    + \beta_{y,459}\ \mathcal{S}_{2}\,y^{b c a d e b d e c}
    \\&\phantom{=}
    + \beta_{y,460}\ \mathcal{S}_{2}\,y^{b c a d e b e c d}
    + \beta_{y,461}\ \mathcal{S}_{2}\,y^{b c a d e b e d c}
    + \beta_{y,462}\ \mathcal{S}_{2}\,y^{b c a d e c b d e}
    + \beta_{y,463}\ \mathcal{S}_{2}\,y^{b c a d e c b e d}
    \\&\phantom{=}
    + \beta_{y,464}\ \mathcal{S}_{2}\,y^{b c a d e c d b e}
    + \beta_{y,465}\ \mathcal{S}_{2}\,y^{b c a d e c d e b}
    + \beta_{y,466}\ \mathcal{S}_{2}\,y^{b c a d e c e b d}
    + \beta_{y,467}\ \mathcal{S}_{2}\,y^{b c a d e c e d b}
    \\&\phantom{=}
    + \beta_{y,468}\ \mathcal{S}_{2}\,y^{b c a d e d b c e}
    + \beta_{y,469}\ \mathcal{S}_{2}\,y^{b c a d e d b e c}
    + \beta_{y,470}\ \mathcal{S}_{2}\,y^{b c a d e d c b e}
    + \beta_{y,471}\ \mathcal{S}_{2}\,y^{b c a d e d c e b}
    \\&\phantom{=}
    + \beta_{y,472}\ \mathcal{S}_{2}\,y^{b c a d e d e b c}
    + \beta_{y,473}\ \mathcal{S}_{2}\,y^{b c a d e d e c b}
    + \beta_{y,474}\ \mathcal{S}_{2}\,y^{b c a d e e b c d}
    + \beta_{y,475}\ \mathcal{S}_{2}\,y^{b c a d e e b d c}
    \\&\phantom{=}
    + \beta_{y,476}\ \mathcal{S}_{2}\,y^{b c a d e e c b d}
    + \beta_{y,477}\ \mathcal{S}_{2}\,y^{b c a d e e c d b}
    + \beta_{y,478}\ \mathcal{S}_{2}\,y^{b c a d e e d b c}
    + \beta_{y,479}\ \mathcal{S}_{2}\,y^{b c a d e e d c b}
    \\&\phantom{=}
    + \beta_{y,480}\ \mathcal{S}_{2}\,y^{b c b a d c e d e}
    + \beta_{y,481}\ \mathcal{S}_{2}\,y^{b c b a d c e e d}
    + \beta_{y,482}\ \mathcal{S}_{2}\,y^{b c b a d d e c e}
    + \beta_{y,483}\ \mathcal{S}_{2}\,y^{b c b a d d e e c}
    \\&\phantom{=}
    + \beta_{y,484}\ \mathcal{S}_{2}\,y^{b c b a d e c d e}
    + \beta_{y,485}\ \mathcal{S}_{2}\,y^{b c b a d e c e d}
    + \beta_{y,486}\ \mathcal{S}_{2}\,y^{b c b a d e d c e}
    + \beta_{y,487}\ \mathcal{S}_{2}\,y^{b c b a d e d e c}
    \\&\phantom{=}
    + \beta_{y,488}\ \mathcal{S}_{2}\,y^{b c b a d e e c d}
    + \beta_{y,489}\ \mathcal{S}_{2}\,y^{b c b a d e e d c}
    + \beta_{y,490}\ \mathcal{S}_{2}\,y^{b c c a d b e d e}
    + \beta_{y,491}\ \mathcal{S}_{2}\,y^{b c c a d b e e d}
    \\&\phantom{=}
    + \beta_{y,492}\ \mathcal{S}_{2}\,y^{b c c a d d e b e}
    + \beta_{y,493}\ \mathcal{S}_{2}\,y^{b c c a d d e e b}
    + \beta_{y,494}\ \mathcal{S}_{2}\,y^{b c c a d e b d e}
    + \beta_{y,495}\ \mathcal{S}_{2}\,y^{b c c a d e b e d}
    \\&\phantom{=}
    + \beta_{y,496}\ \mathcal{S}_{2}\,y^{b c c a d e d b e}
    + \beta_{y,497}\ \mathcal{S}_{2}\,y^{b c c a d e d e b}
    + \beta_{y,498}\ \mathcal{S}_{2}\,y^{b c c a d e e b d}
    + \beta_{y,499}\ \mathcal{S}_{2}\,y^{b c c a d e e d b}
    \\&\phantom{=}
    + \beta_{y,500}\ \mathcal{S}_{2}\,y^{b c d a b c e d e}
    + \beta_{y,501}\ \mathcal{S}_{2}\,y^{b c d a b c e e d}
    + \beta_{y,502}\ \mathcal{S}_{2}\,y^{b c d a b d e c e}
    + \beta_{y,503}\ \mathcal{S}_{2}\,y^{b c d a b d e e c}
    \\&\phantom{=}
    + \beta_{y,504}\ \mathcal{S}_{2}\,y^{b c d a b e c d e}
    + \beta_{y,505}\ \mathcal{S}_{2}\,y^{b c d a b e c e d}
    + \beta_{y,506}\ \mathcal{S}_{2}\,y^{b c d a b e d c e}
    + \beta_{y,507}\ \mathcal{S}_{2}\,y^{b c d a b e d e c}
    \\&\phantom{=}
    + \beta_{y,508}\ \mathcal{S}_{2}\,y^{b c d a b e e c d}
    + \beta_{y,509}\ \mathcal{S}_{2}\,y^{b c d a b e e d c}
    + \beta_{y,510}\ \mathcal{S}_{2}\,y^{b c d a c b e d e}
    + \beta_{y,511}\ \mathcal{S}_{2}\,y^{b c d a c b e e d}
    \\&\phantom{=}
    + \beta_{y,512}\ \mathcal{S}_{2}\,y^{b c d a c d e b e}
    + \beta_{y,513}\ \mathcal{S}_{2}\,y^{b c d a c d e e b}
    + \beta_{y,514}\ \mathcal{S}_{2}\,y^{b c d a c e b d e}
    + \beta_{y,515}\ \mathcal{S}_{2}\,y^{b c d a c e b e d}
    \\&\phantom{=}
    + \beta_{y,516}\ \mathcal{S}_{2}\,y^{b c d a c e d b e}
    + \beta_{y,517}\ \mathcal{S}_{2}\,y^{b c d a c e d e b}
    + \beta_{y,518}\ \mathcal{S}_{2}\,y^{b c d a c e e b d}
    + \beta_{y,519}\ \mathcal{S}_{2}\,y^{b c d a c e e d b}
    \\&\phantom{=}
    + \beta_{y,520}\ \mathcal{S}_{2}\,y^{b c d a d b e c e}
    + \beta_{y,521}\ \mathcal{S}_{2}\,y^{b c d a d b e e c}
    + \beta_{y,522}\ \mathcal{S}_{2}\,y^{b c d a d c e b e}
    + \beta_{y,523}\ \mathcal{S}_{2}\,y^{b c d a d c e e b}
    \\&\phantom{=}
    + \beta_{y,524}\ \mathcal{S}_{2}\,y^{b c d a d e b c e}
    + \beta_{y,525}\ \mathcal{S}_{2}\,y^{b c d a d e b e c}
    + \beta_{y,526}\ \mathcal{S}_{2}\,y^{b c d a d e c b e}
    + \beta_{y,527}\ \mathcal{S}_{2}\,y^{b c d a d e c e b}
    \\&\phantom{=}
    + \beta_{y,528}\ \mathcal{S}_{2}\,y^{b c d a d e e b c}
    + \beta_{y,529}\ \mathcal{S}_{2}\,y^{b c d a d e e c b}
    + \beta_{y,530}\ \mathcal{S}_{2}\,y^{b c d a e b c d e}
    + \beta_{y,531}\ \mathcal{S}_{2}\,y^{b c d a e b c e d}
    \\&\phantom{=}
    + \beta_{y,532}\ \mathcal{S}_{2}\,y^{b c d a e b d c e}
    + \beta_{y,533}\ \mathcal{S}_{2}\,y^{b c d a e b d e c}
    + \beta_{y,534}\ \mathcal{S}_{2}\,y^{b c d a e b e c d}
    + \beta_{y,535}\ \mathcal{S}_{2}\,y^{b c d a e b e d c}
    \\&\phantom{=}
    + \beta_{y,536}\ \mathcal{S}_{2}\,y^{b c d a e c b d e}
    + \beta_{y,537}\ \mathcal{S}_{2}\,y^{b c d a e c b e d}
    + \beta_{y,538}\ \mathcal{S}_{2}\,y^{b c d a e c d b e}
    + \beta_{y,539}\ \mathcal{S}_{2}\,y^{b c d a e c d e b}
    \\&\phantom{=}
    + \beta_{y,540}\ \mathcal{S}_{2}\,y^{b c d a e c e b d}
    + \beta_{y,541}\ \mathcal{S}_{2}\,y^{b c d a e c e d b}
    + \beta_{y,542}\ \mathcal{S}_{2}\,y^{b c d a e d b c e}
    + \beta_{y,543}\ \mathcal{S}_{2}\,y^{b c d a e d b e c}
    \\&\phantom{=}
    + \beta_{y,544}\ \mathcal{S}_{2}\,y^{b c d a e d c b e}
    + \beta_{y,545}\ \mathcal{S}_{2}\,y^{b c d a e d c e b}
    + \beta_{y,546}\ \mathcal{S}_{2}\,y^{b c d a e d e b c}
    + \beta_{y,547}\ \mathcal{S}_{2}\,y^{b c d a e d e c b}
    \\&\phantom{=}
    + \beta_{y,548}\ \mathcal{S}_{2}\,y^{b c d a e e b c d}
    + \beta_{y,549}\ \mathcal{S}_{2}\,y^{b c d a e e b d c}
    + \beta_{y,550}\ \mathcal{S}_{2}\,y^{b c d a e e c b d}
    + \beta_{y,551}\ \mathcal{S}_{2}\,y^{b c d a e e c d b}
    \\&\phantom{=}
    + \beta_{y,552}\ \mathcal{S}_{2}\,y^{b c d a e e d b c}
    + \beta_{y,553}\ \mathcal{S}_{2}\,y^{b c d a e e d c b}
    + \beta_{y,554}\ y^{b c b d a c e d e}
    + \beta_{y,555}\ y^{b c b d a d e c e}
    \\&\phantom{=}
    + \beta_{y,556}\ y^{b c c d a b e e d}
    + \beta_{y,557}\ y^{b c c d a d e e b}
    + \beta_{y,558}\ y^{b c d b a e c d e}
    + \beta_{y,559}\ y^{b c d b a e d c e}
    \\&\phantom{=}
    + \beta_{y,560}\ y^{b c d c a e b e d}
    + \beta_{y,561}\ y^{b c d c a e d e b}
    + \beta_{y,562}\ y^{b c d d a e e b c}
    + \beta_{y,563}\ y^{b c d d a e e c b}
    \\&\phantom{=}
    + \beta_{y,564}\ y^{b c d e a b c d e}
    + \beta_{y,565}\ y^{b c d e a b d c e}
    + \beta_{y,566}\ y^{b c d e a c b e d}
    + \beta_{y,567}\ y^{b c d e a c d e b}
    \\&\phantom{=}
    + \beta_{y,568}\ y^{b c d e a d e b c}
    + \beta_{y,569}\ y^{b c d e a d e c b}
    + \beta_{y,570}\ y^{b c d e a e b c d}
    + \beta_{y,571}\ y^{b c d e a e c d b}
    \\&\phantom{=}
    + \beta_{y,572}\ y^{b c d e a e d b c}
    + \beta_{y,573}\ y^{b c d e a e d c b}
    + \beta_{y,574}\ \mathcal{S}_{2}\,y^{b c b d a c e e d}
    + \beta_{y,575}\ \mathcal{S}_{2}\,y^{b c b d a d e e c}
    \\&\phantom{=}
    + \beta_{y,576}\ \mathcal{S}_{2}\,y^{b c b d a e c d e}
    + \beta_{y,577}\ \mathcal{S}_{2}\,y^{b c b d a e c e d}
    + \beta_{y,578}\ \mathcal{S}_{2}\,y^{b c b d a e d c e}
    + \beta_{y,579}\ \mathcal{S}_{2}\,y^{b c b d a e d e c}
    \\&\phantom{=}
    + \beta_{y,580}\ \mathcal{S}_{2}\,y^{b c b d a e e c d}
    + \beta_{y,581}\ \mathcal{S}_{2}\,y^{b c b d a e e d c}
    + \beta_{y,582}\ \mathcal{S}_{2}\,y^{b c c d a e b d e}
    + \beta_{y,583}\ \mathcal{S}_{2}\,y^{b c c d a e b e d}
    \\&\phantom{=}
    + \beta_{y,584}\ \mathcal{S}_{2}\,y^{b c c d a e d b e}
    + \beta_{y,585}\ \mathcal{S}_{2}\,y^{b c c d a e d e b}
    + \beta_{y,586}\ \mathcal{S}_{2}\,y^{b c c d a e e b d}
    + \beta_{y,587}\ \mathcal{S}_{2}\,y^{b c c d a e e d b}
    \\&\phantom{=}
    + \beta_{y,588}\ \mathcal{S}_{2}\,y^{b c d b a e c e d}
    + \beta_{y,589}\ \mathcal{S}_{2}\,y^{b c d b a e d e c}
    + \beta_{y,590}\ \mathcal{S}_{2}\,y^{b c d b a e e c d}
    + \beta_{y,591}\ \mathcal{S}_{2}\,y^{b c d b a e e d c}
    \\&\phantom{=}
    + \beta_{y,592}\ \mathcal{S}_{2}\,y^{b c d c a e e b d}
    + \beta_{y,593}\ \mathcal{S}_{2}\,y^{b c d c a e e d b}
    + \beta_{y,594}\ \mathcal{S}_{2}\,y^{b c d e a b c e d}
    + \beta_{y,595}\ \mathcal{S}_{2}\,y^{b c d e a b d e c}
    \\&\phantom{=}
    + \beta_{y,596}\ \mathcal{S}_{2}\,y^{b c d e a b e c d}
    + \beta_{y,597}\ \mathcal{S}_{2}\,y^{b c d e a b e d c}
    + \beta_{y,598}\ \mathcal{S}_{2}\,y^{b c d e a c e b d}
    + \beta_{y,599}\ \mathcal{S}_{2}\,y^{b c d e a c e d b}
    \\&\phantom{=}
    + \beta_{y,600}\ \mathcal{S}_{2}\,y^{b c d e a e b d c}\,. \tag{\stepcounter{equation}\theequation}
\end{align*}
The coefficients $\beta_{y,i}$ in the $\overline{\text{MS}}$ scheme read
\begin{align*}
  \beta_{y,1} & = 0\,,&
\beta_{y,2} & = \tfrac{7}{12}\,,&
\beta_{y,3} & = \tfrac{11}{24}\,,\\
\beta_{y,4} & = \tfrac{7}{24}\,,&
\beta_{y,5} & = -\tfrac{1}{6}\,,&
\beta_{y,6} & = \tfrac{1}{6}\,,\\
\beta_{y,7} & = 3 \zeta_3-\tfrac{19}{6}\,,&
\beta_{y,8} & = -\tfrac{5}{6}\,,&
\beta_{y,9} & = -\tfrac{19}{6}\,,\\
\beta_{y,10} & = -\tfrac{11}{6}\,,&
\beta_{y,11} & = -\tfrac{37}{288}\,,&
\beta_{y,12} & = 0\,,\\
\beta_{y,13} & = \tfrac{9}{8}\,,&
\beta_{y,14} & = \tfrac{2}{3}\,,&
\beta_{y,15} & = \tfrac{1}{2}\,,\\
\beta_{y,16} & = \tfrac{1}{2}\,,&
\beta_{y,17} & = 0\,,&
\beta_{y,18} & = 0\,,\\
\beta_{y,19} & = -\tfrac{19}{6}\,,&
\beta_{y,20} & = -\tfrac{11}{6}\,,&
\beta_{y,21} & = -\tfrac{1}{6}\,,\\
\beta_{y,22} & = \tfrac{1}{6}\,,&
\beta_{y,23} & = \tfrac{53}{96}\,,&
\beta_{y,24} & = -\tfrac{37}{288}\,,\\
\beta_{y,25} & = \tfrac{79}{144}\,,&
\beta_{y,26} & = -\tfrac{79}{144}\,,&
\beta_{y,27} & = \tfrac{121}{144}\,,\\
\beta_{y,28} & = -\tfrac{37}{144}\,,&
\beta_{y,29} & = -\tfrac{7}{24}\,,&
\beta_{y,30} & = -1\,,\\
\beta_{y,31} & = 3 \zeta_3-\tfrac{11}{3}\,,&
\beta_{y,32} & = -1\,,&
\beta_{y,33} & = -\tfrac{1}{3}\,,\\
\beta_{y,34} & = -\tfrac{7}{3}\,,&
\beta_{y,35} & = 3 \zeta_3-\tfrac{7}{3}\,,&
\beta_{y,36} & = -\zeta_3-\tfrac{7}{3}\,,\\
\beta_{y,37} & = 2 \zeta_3-6\,,&
\beta_{y,38} & = -\zeta_3-\tfrac{4}{3}\,,&
\beta_{y,39} & = 2 \zeta_3-\tfrac{4}{3}\,,\\
\beta_{y,40} & = \tfrac{2}{3}-\zeta_3\,,&
\beta_{y,41} & = \tfrac{2}{3}\,,&
\beta_{y,42} & = \tfrac{7}{3}\,,\\
\beta_{y,43} & = \tfrac{5}{3}\,,&
\beta_{y,44} & = 1-\zeta_3\,,&
\beta_{y,45} & = 2 \zeta_3-\tfrac{1}{3}\,,\\
\beta_{y,46} & = \tfrac{1}{6}\,,&
\beta_{y,47} & = \tfrac{1}{2}\,,&
\beta_{y,48} & = \tfrac{2}{3}\,,\\
\beta_{y,49} & = -6\,,&
\beta_{y,50} & = -6 \zeta_3\,,&
\beta_{y,51} & = -6 \zeta_3\,,\\
\beta_{y,52} & = -2\,,&
\beta_{y,53} & = -12 \zeta_3\,,&
\beta_{y,54} & = -6 \zeta_3\,,\\
\beta_{y,55} & = \tfrac{2}{3}\,,&
\beta_{y,56} & = \tfrac{1}{2}\,,&
\beta_{y,57} & = -\tfrac{1}{6}\,,\\
\beta_{y,58} & = \tfrac{1}{6}\,,&
\beta_{y,59} & = -2\,,&
\beta_{y,60} & = \tfrac{19}{3}-6 \zeta_3\,,\\
\beta_{y,61} & = \tfrac{11}{3}-6 \zeta_3\,,&
\beta_{y,62} & = \tfrac{2}{3}\,,&
\beta_{y,63} & = -\tfrac{2}{3}\,,\\
\beta_{y,64} & = \tfrac{19}{3}-12 \zeta_3\,,&
\beta_{y,65} & = \tfrac{5}{3}-6 \zeta_3\,,&
\beta_{y,66} & = \tfrac{2}{3}-12 \zeta_3\,,\\
\beta_{y,67} & = -\tfrac{14}{3}\,,&
\beta_{y,68} & = -\tfrac{37}{12}\,,&
\beta_{y,69} & = -\tfrac{17}{12}\,,\\
\beta_{y,70} & = -\tfrac{109}{24}\,,&
\beta_{y,71} & = -\tfrac{65}{24}\,,&
\beta_{y,72} & = \tfrac{11}{6}-\zeta_3\,,\\
\beta_{y,73} & = -\zeta_3-\tfrac{1}{3}\,,&
\beta_{y,74} & = -\tfrac{3}{2}\,,&
\beta_{y,75} & = -\zeta_3-\tfrac{19}{6}\,,\\
\beta_{y,76} & = -\tfrac{65}{48}\,,&
\beta_{y,77} & = 2 \zeta_3-6\,,&
\beta_{y,78} & = -\zeta_3-\tfrac{7}{3}\,,\\
\beta_{y,79} & = \tfrac{2}{3}-\zeta_3\,,&
\beta_{y,80} & = 2 \zeta_3-\tfrac{4}{3}\,,&
\beta_{y,81} & = -\zeta_3-\tfrac{4}{3}\,,\\
\beta_{y,82} & = \tfrac{2}{3}\,,&
\beta_{y,83} & = \tfrac{7}{3}\,,&
\beta_{y,84} & = \tfrac{5}{3}\,,\\
\beta_{y,85} & = 2 \zeta_3-\tfrac{1}{3}\,,&
\beta_{y,86} & = 1-\zeta_3\,,&
\beta_{y,87} & = \tfrac{2}{3}\,,\\
\beta_{y,88} & = \tfrac{1}{2}\,,&
\beta_{y,89} & = \tfrac{1}{6}\,,&
\beta_{y,90} & = -2\,,\\
\beta_{y,91} & = -\tfrac{11}{3}\,,&
\beta_{y,92} & = -1\,,&
\beta_{y,93} & = \tfrac{7}{6}\,,\\
\beta_{y,94} & = -\tfrac{1}{3}\,,&
\beta_{y,95} & = -\tfrac{4}{3}\,,&
\beta_{y,96} & = -\tfrac{5}{3}\,,\\
\beta_{y,97} & = -\tfrac{1}{3}\,,&
\beta_{y,98} & = -\tfrac{16}{3}\,,&
\beta_{y,99} & = -\tfrac{13}{3}\,,\\
\beta_{y,100} & = 2 \zeta_3-\tfrac{17}{3}\,,&
\beta_{y,101} & = 2 \zeta_3-\tfrac{16}{3}\,,&
\beta_{y,102} & = \tfrac{10}{3}\,,\\
\beta_{y,103} & = \tfrac{7}{6}\,,&
\beta_{y,104} & = 2 \zeta_3-\tfrac{16}{3}\,,&
\beta_{y,105} & = 2 \zeta_3-\tfrac{22}{3}\,,\\
\beta_{y,106} & = -1\,,&
\beta_{y,107} & = -\tfrac{2}{3}\,,&
\beta_{y,108} & = -2\,,\\
\beta_{y,109} & = -\tfrac{13}{3}\,,&
\beta_{y,110} & = 0\,,&
\beta_{y,111} & = -6 \zeta_3\,,\\
\beta_{y,112} & = 2\,,&
\beta_{y,113} & = 6-6 \zeta_3\,,&
\beta_{y,114} & = -6 \zeta_3\,,\\
\beta_{y,115} & = 0\,,&
\beta_{y,116} & = \tfrac{7}{3}\,,&
\beta_{y,117} & = -2\,,\\
\beta_{y,118} & = -\tfrac{8}{3}\,,&
\beta_{y,119} & = -\tfrac{1}{3}\,,&
\beta_{y,120} & = \tfrac{4}{3}\,,\\
\beta_{y,121} & = -\tfrac{1}{3}\,,&
\beta_{y,122} & = \tfrac{7}{6}\,,&
\beta_{y,123} & = \tfrac{13}{3}\,,\\
\beta_{y,124} & = -\tfrac{22}{3}\,,&
\beta_{y,125} & = -6 \zeta_3\,,&
\beta_{y,126} & = -6 \zeta_3\,,\\
\beta_{y,127} & = -12 \zeta_3\,,&
\beta_{y,128} & = -6 \zeta_3\,,&
\beta_{y,129} & = -\tfrac{2}{3}\,,\\
\beta_{y,130} & = -3\,,&
\beta_{y,131} & = -\tfrac{3}{2}\,,&
\beta_{y,132} & = 2-6 \zeta_3\,,\\
\beta_{y,133} & = -1\,,&
\beta_{y,134} & = \tfrac{2}{3}\,,&
\beta_{y,135} & = 1\,,\\
\beta_{y,136} & = -\tfrac{1}{2}\,,&
\beta_{y,137} & = \tfrac{1}{6}\,,&
\beta_{y,138} & = \tfrac{1}{3}\,,\\
\beta_{y,139} & = -\tfrac{23}{48}\,,&
\beta_{y,140} & = -\zeta_3-\tfrac{1}{3}\,,&
\beta_{y,141} & = \tfrac{4}{3}-2 \zeta_3\,,\\
\beta_{y,142} & = -2 \zeta_3-\tfrac{14}{3}\,,&
\beta_{y,143} & = -6 \zeta_3\,,&
\beta_{y,144} & = 2 \zeta_3-\tfrac{7}{3}\,,\\
\beta_{y,145} & = 0\,,&
\beta_{y,146} & = \tfrac{14}{3}-2 \zeta_3\,,&
\beta_{y,147} & = 1-\zeta_3\,,\\
\beta_{y,148} & = -2 \zeta_3-\tfrac{1}{3}\,,&
\beta_{y,149} & = 2 \zeta_3-\tfrac{19}{6}\,,&
\beta_{y,150} & = -6 \zeta_3\,,\\
\beta_{y,151} & = -2\,,&
\beta_{y,152} & = -6 \zeta_3\,,&
\beta_{y,153} & = 10 \zeta_3-\tfrac{14}{3}\,,\\
\beta_{y,154} & = \tfrac{5}{3}\,,&
\beta_{y,155} & = \tfrac{2}{3}\,,&
\beta_{y,156} & = 10 \zeta_3-\tfrac{19}{3}\,,\\
\beta_{y,157} & = 6-6 \zeta_3\,,&
\beta_{y,158} & = \tfrac{11}{6}-\zeta_3\,,&
\beta_{y,159} & = \tfrac{23}{3}-2 \zeta_3\,,\\
\beta_{y,160} & = \tfrac{8}{3}-2 \zeta_3\,,&
\beta_{y,161} & = 0\,,&
\beta_{y,162} & = \tfrac{10}{3}-2 \zeta_3\,,\\
\beta_{y,163} & = -6 \zeta_3\,,&
\beta_{y,164} & = 2 \zeta_3-\tfrac{7}{3}\,,&
\beta_{y,165} & = 2 \zeta_3-\tfrac{4}{3}\,,\\
\beta_{y,166} & = -2 \zeta_3-\tfrac{8}{3}\,,&
\beta_{y,167} & = \tfrac{2}{3}-\zeta_3\,,&
\beta_{y,168} & = 12 \zeta_3-12\,,\\
\beta_{y,169} & = 2\,,&
\beta_{y,170} & = -12 \zeta_3\,,&
\beta_{y,171} & = 2 \zeta_3-\tfrac{22}{3}\,,\\
\beta_{y,172} & = -6 \zeta_3\,,&
\beta_{y,173} & = 2 \zeta_3-\tfrac{35}{3}\,,&
\beta_{y,174} & = 6\,,\\
\beta_{y,175} & = \tfrac{\zeta_3}{2}-\tfrac{1}{4}\,,&
\beta_{y,176} & = -\tfrac{169}{32}\,,&
\beta_{y,177} & = -2 \zeta_3-\tfrac{5}{4}\,,\\
\beta_{y,178} & = \tfrac{101}{96}\,,&
\beta_{y,179} & = \zeta_3+\tfrac{13}{48}\,,&
\beta_{y,180} & = -\tfrac{13}{3}\,,\\
\beta_{y,181} & = -\zeta_3-\tfrac{7}{3}+\tfrac{\pi ^4}{60}\,,&
\beta_{y,182} & = -5 \zeta_3-\tfrac{2}{3}+\tfrac{\pi ^4}{60}\,,&
\beta_{y,183} & = -2\,,\\
\beta_{y,184} & = -3 \zeta_3-\tfrac{4}{3}+\tfrac{\pi ^4}{60}\,,&
\beta_{y,185} & = \tfrac{1}{4}-\zeta_3\,,&
\beta_{y,186} & = -\tfrac{63}{32}\,,\\
\beta_{y,187} & = \tfrac{1}{2}-2 \zeta_3\,,&
\beta_{y,188} & = 5-8 \zeta_3\,,&
\beta_{y,189} & = -\tfrac{25}{12}\,,\\
\beta_{y,190} & = -\tfrac{17}{24}\,,&
\beta_{y,191} & = -3 \zeta_3+\tfrac{5}{3}-\tfrac{\pi ^4}{60}\,,&
\beta_{y,192} & = 6 \zeta_3-\tfrac{26}{3}\,,\\
\beta_{y,193} & = \tfrac{20}{3}\,,&
\beta_{y,194} & = -8 \zeta_3+10+\tfrac{\pi ^4}{30}\,,&
\beta_{y,195} & = 4 \zeta_3-20 \zeta_5+10\,,\\
\beta_{y,196} & = 6 \zeta_3-20 \zeta_5+12\,,&
\beta_{y,197} & = 10-6 \zeta_3\,,&
\beta_{y,198} & = \tfrac{8}{3}-8 \zeta_3\,,\\
\beta_{y,199} & = -12 \zeta_3+10+\tfrac{\pi ^4}{30}\,,&
\beta_{y,200} & = -8 \zeta_3+8+\tfrac{\pi ^4}{30}\,,&
\beta_{y,201} & = \tfrac{10}{3}\,,\\
\beta_{y,202} & = -4 \zeta_3+5+\tfrac{\pi ^4}{60}\,,&
\beta_{y,203} & = -4 \zeta_3+5+\tfrac{\pi ^4}{60}\,,&
\beta_{y,204} & = -\tfrac{4}{3}\,,\\
\beta_{y,205} & = -8 \zeta_3+8+\tfrac{\pi ^4}{30}\,,&
\beta_{y,206} & = \tfrac{4}{3}\,,&
\beta_{y,207} & = -5 \zeta_3+5+\tfrac{\pi ^4}{60}\,,\\
\beta_{y,208} & = -3 \zeta_3+4+\tfrac{\pi ^4}{60}\,,&
\beta_{y,209} & = \tfrac{101}{192}\,,&
\beta_{y,210} & = 0\,,\\
\beta_{y,211} & = \tfrac{2}{3}\,,&
\beta_{y,212} & = -\tfrac{2}{3}\,,&
\beta_{y,213} & = \tfrac{\zeta_3}{2}-\tfrac{1}{4}\,,\\
\beta_{y,214} & = \zeta_3-\tfrac{2}{3}\,,&
\beta_{y,215} & = \tfrac{1}{8}-\tfrac{\zeta_3}{2}\,,&
\beta_{y,216} & = \tfrac{2}{3}-\zeta_3\,,\\
\beta_{y,217} & = \zeta_3-\tfrac{1}{2}\,,&
\beta_{y,218} & = \tfrac{5}{6}-\zeta_3\,,&
\beta_{y,219} & = -\tfrac{1}{6}\,,\\
\beta_{y,220} & = 2-4 \zeta_3\,,&
\beta_{y,221} & = -12 \zeta_3+\tfrac{50}{3}-\tfrac{\pi ^4}{30}\,,&
\beta_{y,222} & = 8 \zeta_3-20 \zeta_5+\tfrac{10}{3}\,,\\
\beta_{y,223} & = \tfrac{10}{3}-2 \zeta_3\,,&
\beta_{y,224} & = 10 \zeta_3-20 \zeta_5+\tfrac{16}{3}\,,&
\beta_{y,225} & = \tfrac{\pi ^4}{30}-6\,,\\
\beta_{y,226} & = -\tfrac{10}{3}\,,&
\beta_{y,227} & = 4 \zeta_3-8+\tfrac{\pi ^4}{30}\,,&
\beta_{y,228} & = -4 \zeta_3-\tfrac{26}{3}\,,\\
\beta_{y,229} & = -8 \zeta_3-\tfrac{4}{3}+\tfrac{\pi ^4}{30}\,,&
\beta_{y,230} & = -3 \zeta_3-\tfrac{4}{3}+\tfrac{\pi ^4}{60}\,,&
\beta_{y,231} & = -2\,,\\
\beta_{y,232} & = -\zeta_3-\tfrac{7}{3}+\tfrac{\pi ^4}{60}\,,&
\beta_{y,233} & = -8 \zeta_3-\tfrac{4}{3}\,,&
\beta_{y,234} & = -10 \zeta_3+\tfrac{28}{3}-\tfrac{\pi ^4}{30}\,,\\
\beta_{y,235} & = -10 \zeta_3+\tfrac{28}{3}-\tfrac{\pi ^4}{30}\,,&
\beta_{y,236} & = 4 \zeta_3-20 \zeta_5+\tfrac{22}{3}\,,&
\beta_{y,237} & = \tfrac{11}{3}-3 \zeta_3\,,\\
\beta_{y,238} & = 6 \zeta_3-20 \zeta_5+\tfrac{28}{3}\,,&
\beta_{y,239} & = -\tfrac{169}{32}\,,&
\beta_{y,240} & = -\tfrac{77}{16}\,,\\
\beta_{y,241} & = \tfrac{101}{96}\,,&
\beta_{y,242} & = \tfrac{77}{16}\,,&
\beta_{y,243} & = -\tfrac{361}{48}\,,\\
\beta_{y,244} & = \tfrac{101}{48}\,,&
\beta_{y,245} & = \tfrac{65}{24}\,,&
\beta_{y,246} & = -\zeta_3-\tfrac{49}{16}\,,\\
\beta_{y,247} & = -2 \zeta_3-\tfrac{47}{24}\,,&
\beta_{y,248} & = \zeta_3+\tfrac{13}{48}\,,&
\beta_{y,249} & = 2 \zeta_3+\tfrac{47}{24}\,,\\
\beta_{y,250} & = -2 \zeta_3-\tfrac{27}{8}\,,&
\beta_{y,251} & = 2 \zeta_3+\tfrac{13}{24}\,,&
\beta_{y,252} & = \tfrac{17}{12}\,,\\
\beta_{y,253} & = \tfrac{\zeta_3}{2}-\tfrac{1}{4}\,,&
\beta_{y,254} & = \zeta_3-\tfrac{1}{2}\,,&
\beta_{y,255} & = \tfrac{9}{64}\,,\\
\beta_{y,256} & = \zeta_3-\tfrac{2}{3}\,,&
\beta_{y,257} & = \zeta_3-\tfrac{5}{6}\,,&
\beta_{y,258} & = \tfrac{1}{6}\,,\\
\beta_{y,259} & = -3 \zeta_3-\tfrac{7}{6}+\tfrac{\pi ^4}{60}\,,&
\beta_{y,260} & = \zeta_3+\tfrac{1}{3}\,,&
\beta_{y,261} & = 0\,,\\
\beta_{y,262} & = -\tfrac{3}{2}\,,&
\beta_{y,263} & = \tfrac{2}{3}-\zeta_3\,,&
\beta_{y,264} & = \tfrac{2}{3}-\zeta_3\,,\\
\beta_{y,265} & = \tfrac{1}{3}\,,&
\beta_{y,266} & = \tfrac{5}{12}\,,&
\beta_{y,267} & = -4 \zeta_3+1+\tfrac{\pi ^4}{60}\,,\\
\beta_{y,268} & = 2-2 \zeta_3\,,&
\beta_{y,269} & = \tfrac{29}{24}\,,&
\beta_{y,270} & = 2 \zeta_3-\tfrac{5}{6}\,,\\
\beta_{y,271} & = \tfrac{101}{192}\,,&
\beta_{y,272} & = \tfrac{77}{32}\,,&
\beta_{y,273} & = \tfrac{8}{3}-\zeta_3\,,\\
\beta_{y,274} & = -3 \zeta_3+2+\tfrac{\pi ^4}{60}\,,&
\beta_{y,275} & = \tfrac{29}{24}\,,&
\beta_{y,276} & = -2 \zeta_3-\tfrac{11}{6}\,,\\
\beta_{y,277} & = -\tfrac{143}{96}\,,&
\beta_{y,278} & = -\tfrac{215}{192}\,,&
\beta_{y,279} & = -\tfrac{3}{2}\,,\\
\beta_{y,280} & = 0\,,&
\beta_{y,281} & = \tfrac{101}{192}\,,&
\beta_{y,282} & = -\tfrac{77}{32}\,,\\
\beta_{y,283} & = \tfrac{2}{3}\,,&
\beta_{y,284} & = \zeta_3-\tfrac{1}{2}\,,&
\beta_{y,285} & = \zeta_3-\tfrac{1}{2}\,,\\
\beta_{y,286} & = -\tfrac{1}{2}\,,&
\beta_{y,287} & = \zeta_3+\tfrac{2}{3}\,,&
\beta_{y,288} & = 2 \zeta_3-\tfrac{4}{3}\,,\\
\beta_{y,289} & = -2 \zeta_3-1+\tfrac{\pi ^4}{60}\,,&
\beta_{y,290} & = -3 \zeta_3+\tfrac{1}{3}+\tfrac{\pi ^4}{60}\,,&
\beta_{y,291} & = -\tfrac{2}{3}\,,\\
\beta_{y,292} & = -\tfrac{5}{3}\,,&
\beta_{y,293} & = -\tfrac{1}{2}\,,&
\beta_{y,294} & = -3 \zeta_3-1+\tfrac{\pi ^4}{60}\,,\\
\beta_{y,295} & = -\tfrac{1}{6}\,,&
\beta_{y,296} & = \tfrac{5}{6}\,,&
\beta_{y,297} & = -5 \zeta_3+\tfrac{4}{3}+\tfrac{\pi ^4}{60}\,,\\
\beta_{y,298} & = -\tfrac{2}{3}\,,&
\beta_{y,299} & = 4-2 \zeta_3\,,&
\beta_{y,300} & = \tfrac{11}{3}-2 \zeta_3\,,\\
\beta_{y,301} & = \tfrac{2}{3}\,,&
\beta_{y,302} & = \tfrac{19}{6}\,,&
\beta_{y,303} & = -3 \zeta_3+\tfrac{1}{3}+\tfrac{\pi ^4}{60}\,,\\
\beta_{y,304} & = \zeta_3-\tfrac{1}{3}\,,&
\beta_{y,305} & = -5 \zeta_3+\tfrac{19}{6}+\tfrac{\pi ^4}{60}\,,&
\beta_{y,306} & = \zeta_3-\tfrac{8}{3}\,,\\
\beta_{y,307} & = \tfrac{109}{48}\,,&
\beta_{y,308} & = -\tfrac{361}{96}\,,&
\beta_{y,309} & = \tfrac{3}{2}\,,\\
\beta_{y,310} & = 0\,,&
\beta_{y,311} & = \tfrac{101}{96}\,,&
\beta_{y,312} & = \tfrac{65}{48}\,,\\
\beta_{y,313} & = \tfrac{\zeta_3}{2}-\tfrac{1}{4}\,,&
\beta_{y,314} & = \tfrac{4}{3}\,,&
\beta_{y,315} & = -2 \zeta_3-\tfrac{11}{6}+\tfrac{\pi ^4}{60}\,,\\
\beta_{y,316} & = \zeta_3-\tfrac{2}{3}\,,&
\beta_{y,317} & = -4 \zeta_3+\tfrac{1}{3}+\tfrac{\pi ^4}{60}\,,&
\beta_{y,318} & = -1\,,\\
\beta_{y,319} & = -\tfrac{1}{3}\,,&
\beta_{y,320} & = -2 \zeta_3-\tfrac{2}{3}+\tfrac{\pi ^4}{60}\,,&
\beta_{y,321} & = 2 \zeta_3-2\,,\\
\beta_{y,322} & = -\zeta_3-\tfrac{7}{3}\,,&
\beta_{y,323} & = \tfrac{5}{3}\,,&
\beta_{y,324} & = -4 \zeta_3+\tfrac{7}{3}+\tfrac{\pi ^4}{60}\,,\\
\beta_{y,325} & = -5 \zeta_3+\tfrac{7}{3}+\tfrac{\pi ^4}{60}\,,&
\beta_{y,326} & = \tfrac{19}{6}\,,&
\beta_{y,327} & = -5\,,\\
\beta_{y,328} & = \tfrac{101}{192}\,,&
\beta_{y,329} & = \tfrac{77}{32}\,,&
\beta_{y,330} & = 0\,,\\
\beta_{y,331} & = \tfrac{3}{2}\,,&
\beta_{y,332} & = 0\,,&
\beta_{y,333} & = -\tfrac{3}{2}\,,\\
\beta_{y,334} & = 2-2 \zeta_3\,,&
\beta_{y,335} & = \tfrac{2}{3}-\zeta_3\,,&
\beta_{y,336} & = \tfrac{13}{3}-\tfrac{\pi ^4}{30}\,,\\
\beta_{y,337} & = -2 \zeta_3-\tfrac{5}{3}\,,&
\beta_{y,338} & = \tfrac{8}{3}-2 \zeta_3\,,&
\beta_{y,339} & = \tfrac{4}{3}-2 \zeta_3\,,\\
\beta_{y,340} & = -\tfrac{5}{6}\,,&
\beta_{y,341} & = -\tfrac{73}{96}\,,&
\beta_{y,342} & = -\tfrac{3}{2}\,,\\
\beta_{y,343} & = 0\,,&
\beta_{y,344} & = \zeta_3-\tfrac{2}{3}\,,&
\beta_{y,345} & = \tfrac{\zeta_3}{2}-\tfrac{1}{4}\,,\\
\beta_{y,346} & = -3 \zeta_3-\tfrac{7}{6}+\tfrac{\pi ^4}{60}\,,&
\beta_{y,347} & = \zeta_3+\tfrac{1}{3}\,,&
\beta_{y,348} & = \zeta_3-\tfrac{5}{6}\,,\\
\beta_{y,349} & = \zeta_3-\tfrac{1}{2}\,,&
\beta_{y,350} & = \tfrac{1}{6}\,,&
\beta_{y,351} & = \tfrac{9}{64}\,,\\
\beta_{y,352} & = \tfrac{2}{3}+\tfrac{\pi ^4}{30}\,,&
\beta_{y,353} & = -3 \zeta_3+2+\tfrac{\pi ^4}{60}\,,&
\beta_{y,354} & = \tfrac{2}{3}-2 \zeta_3\,,\\
\beta_{y,355} & = \tfrac{2}{3}-\zeta_3\,,&
\beta_{y,356} & = \tfrac{16}{3}-\tfrac{\pi ^4}{30}\,,&
\beta_{y,357} & = \tfrac{22}{3}-6 \zeta_3\,,\\
\beta_{y,358} & = 2 \zeta_3-20 \zeta_5+\tfrac{32}{3}\,,&
\beta_{y,359} & = 2-8 \zeta_3\,,&
\beta_{y,360} & = -5 \zeta_3+\tfrac{11}{3}+\tfrac{\pi ^4}{60}\,,\\
\beta_{y,361} & = \zeta_3-\tfrac{11}{6}\,,&
\beta_{y,362} & = \tfrac{4}{3}-2 \zeta_3\,,&
\beta_{y,363} & = \tfrac{8}{3}-\zeta_3\,,\\
\beta_{y,364} & = 2 \zeta_3-\tfrac{4}{3}\,,&
\beta_{y,365} & = \zeta_3-\tfrac{1}{2}\,,&
\beta_{y,366} & = \tfrac{\pi ^4}{30}-\tfrac{10}{3}\,,\\
\beta_{y,367} & = 6-8 \zeta_3\,,&
\beta_{y,368} & = 10 \zeta_3-\tfrac{20}{3}\,,&
\beta_{y,369} & = 6 \zeta_3-\tfrac{14}{3}\,,\\
\beta_{y,370} & = 3-2 \zeta_3\,,&
\beta_{y,371} & = \tfrac{1}{8}\,,&
\beta_{y,372} & = \tfrac{1}{3}\,,\\
\beta_{y,373} & = \tfrac{1}{3}\,,&
\beta_{y,374} & = \tfrac{73}{96}\,,&
\beta_{y,375} & = \tfrac{9}{64}\,,\\
\beta_{y,376} & = -2 \zeta_3+1+\tfrac{\pi ^4}{60}\,,&
\beta_{y,377} & = 2-2 \zeta_3\,,&
\beta_{y,378} & = \zeta_3-\tfrac{5}{6}\,,\\
\beta_{y,379} & = \tfrac{1}{8}\,,&
\beta_{y,380} & = 1-\zeta_3\,,&
\beta_{y,381} & = \tfrac{17}{24}-\tfrac{\zeta_3}{2}\,,\\
\beta_{y,382} & = -\tfrac{16}{3}\,,&
\beta_{y,383} & = 2 \zeta_3+\tfrac{16}{3}-\tfrac{\pi ^4}{30}\,,&
\beta_{y,384} & = 2 \zeta_3+\tfrac{16}{3}-\tfrac{\pi ^4}{30}\,,\\
\beta_{y,385} & = -2 \zeta_3+\tfrac{17}{3}-\tfrac{\pi ^4}{30}\,,&
\beta_{y,386} & = -6 \zeta_3+\tfrac{22}{3}+\tfrac{\pi ^4}{30}\,,&
\beta_{y,387} & = 10 \zeta_3-\tfrac{11}{3}\,,\\
\beta_{y,388} & = \tfrac{16}{3}-\tfrac{\pi ^4}{30}\,,&
\beta_{y,389} & = 2 \zeta_3-20 \zeta_5+\tfrac{32}{3}\,,&
\beta_{y,390} & = \tfrac{1}{3}-\tfrac{\pi ^4}{30}\,,\\
\beta_{y,391} & = -3 \zeta_3+\tfrac{5}{3}-\tfrac{\pi ^4}{60}\,,&
\beta_{y,392} & = 4-4 \zeta_3\,,&
\beta_{y,393} & = \tfrac{17}{6}-4 \zeta_3\,,\\
\beta_{y,394} & = \tfrac{22}{3}-2 \zeta_3\,,&
\beta_{y,395} & = 2-4 \zeta_3\,,&
\beta_{y,396} & = \tfrac{11}{3}-8 \zeta_3\,,\\
\beta_{y,397} & = \tfrac{17}{6}-4 \zeta_3\,,&
\beta_{y,398} & = 2-2 \zeta_3\,,&
\beta_{y,399} & = \tfrac{17}{12}-\zeta_3\,,\\
\beta_{y,400} & = -4 \zeta_3+\tfrac{11}{3}+\tfrac{\pi ^4}{60}\,,&
\beta_{y,401} & = 2 \zeta_3-\tfrac{11}{6}\,,&
\beta_{y,402} & = \tfrac{11}{6}-2 \zeta_3\,,\\
\beta_{y,403} & = 1-2 \zeta_3\,,&
\beta_{y,404} & = -\tfrac{11}{32}\,,&
\beta_{y,405} & = -\tfrac{49}{192}\,,\\
\beta_{y,406} & = 3\,,&
\beta_{y,407} & = \tfrac{3}{2}\,,&
\beta_{y,408} & = -\tfrac{2}{3}\,,\\
\beta_{y,409} & = -\tfrac{1}{6}\,,&
\beta_{y,410} & = 6 \zeta_3-2\,,&
\beta_{y,411} & = 1\,,\\
\beta_{y,412} & = -1\,,&
\beta_{y,413} & = -\tfrac{1}{3}\,,&
\beta_{y,414} & = \tfrac{1}{2}\,,\\
\beta_{y,415} & = \tfrac{23}{48}\,,&
\beta_{y,416} & = -3\,,&
\beta_{y,417} & = 0\,,\\
\beta_{y,418} & = 2 \zeta_3-\tfrac{4}{3}\,,&
\beta_{y,419} & = \zeta_3-\tfrac{1}{2}\,,&
\beta_{y,420} & = -6 \zeta_3-\tfrac{7}{3}+\tfrac{\pi ^4}{30}\,,\\
\beta_{y,421} & = 2 \zeta_3+\tfrac{2}{3}\,,&
\beta_{y,422} & = 2 \zeta_3-\tfrac{5}{3}\,,&
\beta_{y,423} & = 2 \zeta_3-1\,,\\
\beta_{y,424} & = \tfrac{1}{3}\,,&
\beta_{y,425} & = \tfrac{9}{32}\,,&
\beta_{y,426} & = \tfrac{\pi ^4}{30}-\tfrac{4}{3}\,,\\
\beta_{y,427} & = -3 \zeta_3+\tfrac{1}{3}+\tfrac{\pi ^4}{60}\,,&
\beta_{y,428} & = \tfrac{\pi ^4}{30}-\tfrac{14}{3}\,,&
\beta_{y,429} & = -3 \zeta_3-1+\tfrac{\pi ^4}{60}\,,\\
\beta_{y,430} & = 18 \zeta_3-20 \zeta_5\,,&
\beta_{y,431} & = -6 \zeta_3+\tfrac{14}{3}+\tfrac{\pi ^4}{30}\,,&
\beta_{y,432} & = 12 \zeta_3-20 \zeta_5\,,\\
\beta_{y,433} & = -6 \zeta_3+\tfrac{1}{3}+\tfrac{\pi ^4}{30}\,,&
\beta_{y,434} & = -4 \zeta_3+\tfrac{7}{3}+\tfrac{\pi ^4}{60}\,,&
\beta_{y,435} & = -4 \zeta_3+\tfrac{19}{6}+\tfrac{\pi ^4}{60}\,,\\
\beta_{y,436} & = 2 \zeta_3-\tfrac{14}{3}\,,&
\beta_{y,437} & = \zeta_3-\tfrac{1}{3}\,,&
\beta_{y,438} & = 2 \zeta_3-\tfrac{4}{3}\,,\\
\beta_{y,439} & = \zeta_3-\tfrac{1}{2}\,,&
\beta_{y,440} & = 12 \zeta_3-20 \zeta_5\,,&
\beta_{y,441} & = 6 \zeta_3-20 \zeta_5+\tfrac{22}{3}\,,\\
\beta_{y,442} & = 6 \zeta_3\,,&
\beta_{y,443} & = 10 \zeta_3-\tfrac{25}{3}\,,&
\beta_{y,444} & = \tfrac{11}{3}-2 \zeta_3\,,\\
\beta_{y,445} & = 2 \zeta_3-\tfrac{8}{3}\,,&
\beta_{y,446} & = -\tfrac{7}{3}\,,&
\beta_{y,447} & = -\tfrac{1}{2}\,,\\
\beta_{y,448} & = \tfrac{19}{6}\,,&
\beta_{y,449} & = \tfrac{2}{3}\,,&
\beta_{y,450} & = -4 \zeta_3+\tfrac{1}{3}+\tfrac{\pi ^4}{60}\,,\\
\beta_{y,451} & = -4 \zeta_3+\tfrac{1}{3}+\tfrac{\pi ^4}{60}\,,&
\beta_{y,452} & = \zeta_3+\tfrac{5}{3}\,,&
\beta_{y,453} & = 4-\zeta_3\,,\\
\beta_{y,454} & = \tfrac{1}{6}\,,&
\beta_{y,455} & = \tfrac{5}{6}-\zeta_3\,,&
\beta_{y,456} & = 16 \zeta_3-20 \zeta_5\,,\\
\beta_{y,457} & = 6 \zeta_3\,,&
\beta_{y,458} & = 18 \zeta_3-20 \zeta_5\,,&
\beta_{y,459} & = 12 \zeta_3-6\,,\\
\beta_{y,460} & = \tfrac{14}{3}+\tfrac{\pi ^4}{30}\,,&
\beta_{y,461} & = -4 \zeta_3+\tfrac{19}{3}+\tfrac{\pi ^4}{30}\,,&
\beta_{y,462} & = -4 \zeta_3\,,\\
\beta_{y,463} & = -8 \zeta_3\,,&
\beta_{y,464} & = 12 \zeta_3-20 \zeta_5\,,&
\beta_{y,465} & = -6 \zeta_3+\tfrac{13}{3}-\tfrac{\pi ^4}{30}\,,\\
\beta_{y,466} & = \tfrac{28}{3}-10 \zeta_3\,,&
\beta_{y,467} & = 2-4 \zeta_3\,,&
\beta_{y,468} & = -2 \zeta_3+\tfrac{8}{3}+\tfrac{\pi ^4}{30}\,,\\
\beta_{y,469} & = -2 \zeta_3+\tfrac{1}{3}+\tfrac{\pi ^4}{30}\,,&
\beta_{y,470} & = 6 \zeta_3-20 \zeta_5+\tfrac{16}{3}\,,&
\beta_{y,471} & = \tfrac{19}{3}-8 \zeta_3\,,\\
\beta_{y,472} & = \tfrac{1}{3}\,,&
\beta_{y,473} & = \tfrac{5}{3}-2 \zeta_3\,,&
\beta_{y,474} & = -2 \zeta_3+\tfrac{4}{3}+\tfrac{\pi ^4}{60}\,,\\
\beta_{y,475} & = -3 \zeta_3+\tfrac{19}{6}+\tfrac{\pi ^4}{60}\,,&
\beta_{y,476} & = \tfrac{8}{3}-2 \zeta_3\,,&
\beta_{y,477} & = 1-2 \zeta_3\,,\\
\beta_{y,478} & = \tfrac{43}{48}\,,&
\beta_{y,479} & = -\tfrac{119}{96}\,,&
\beta_{y,480} & = 0\,,\\
\beta_{y,481} & = \tfrac{3}{2}\,,&
\beta_{y,482} & = 2 \zeta_3-2\,,&
\beta_{y,483} & = \zeta_3-\tfrac{2}{3}\,,\\
\beta_{y,484} & = \tfrac{\pi ^4}{30}-\tfrac{13}{3}\,,&
\beta_{y,485} & = 2 \zeta_3+\tfrac{5}{3}\,,&
\beta_{y,486} & = 2 \zeta_3-\tfrac{8}{3}\,,\\
\beta_{y,487} & = 2 \zeta_3-\tfrac{4}{3}\,,&
\beta_{y,488} & = \tfrac{5}{6}\,,&
\beta_{y,489} & = \tfrac{73}{96}\,,\\
\beta_{y,490} & = -\tfrac{3}{2}\,,&
\beta_{y,491} & = 0\,,&
\beta_{y,492} & = \zeta_3-\tfrac{2}{3}\,,\\
\beta_{y,493} & = \tfrac{\zeta_3}{2}-\tfrac{1}{4}\,,&
\beta_{y,494} & = -3 \zeta_3-\tfrac{7}{6}+\tfrac{\pi ^4}{60}\,,&
\beta_{y,495} & = \zeta_3+\tfrac{1}{3}\,,\\
\beta_{y,496} & = \zeta_3-\tfrac{5}{6}\,,&
\beta_{y,497} & = \zeta_3-\tfrac{1}{2}\,,&
\beta_{y,498} & = \tfrac{1}{6}\,,\\
\beta_{y,499} & = \tfrac{9}{64}\,,&
\beta_{y,500} & = \tfrac{4}{3}\,,&
\beta_{y,501} & = -\tfrac{1}{3}\,,\\
\beta_{y,502} & = -2 \zeta_3-\tfrac{4}{3}+\tfrac{\pi ^4}{30}\,,&
\beta_{y,503} & = -4 \zeta_3+\tfrac{1}{3}+\tfrac{\pi ^4}{60}\,,&
\beta_{y,504} & = 16 \zeta_3-20 \zeta_5\,,\\
\beta_{y,505} & = -\tfrac{14}{3}\,,&
\beta_{y,506} & = -4 \zeta_3\,,&
\beta_{y,507} & = -8 \zeta_3+\tfrac{14}{3}+\tfrac{\pi ^4}{30}\,,\\
\beta_{y,508} & = \zeta_3-\tfrac{4}{3}\,,&
\beta_{y,509} & = -3 \zeta_3+\tfrac{4}{3}+\tfrac{\pi ^4}{60}\,,&
\beta_{y,510} & = 2 \zeta_3-\tfrac{10}{3}+\tfrac{\pi ^4}{30}\,,\\
\beta_{y,511} & = -2 \zeta_3-\tfrac{2}{3}+\tfrac{\pi ^4}{60}\,,&
\beta_{y,512} & = 2 \zeta_3-\tfrac{23}{3}+\tfrac{\pi ^4}{30}\,,&
\beta_{y,513} & = -2 \zeta_3-\tfrac{11}{6}+\tfrac{\pi ^4}{60}\,,\\
\beta_{y,514} & = 18 \zeta_3-20 \zeta_5\,,&
\beta_{y,515} & = -4 \zeta_3+\tfrac{8}{3}+\tfrac{\pi ^4}{30}\,,&
\beta_{y,516} & = 12 \zeta_3-20 \zeta_5\,,\\
\beta_{y,517} & = -4 \zeta_3-\tfrac{8}{3}+\tfrac{\pi ^4}{30}\,,&
\beta_{y,518} & = -3 \zeta_3+\tfrac{4}{3}+\tfrac{\pi ^4}{60}\,,&
\beta_{y,519} & = -3 \zeta_3+\tfrac{7}{3}+\tfrac{\pi ^4}{60}\,,\\
\beta_{y,520} & = -\tfrac{14}{3}\,,&
\beta_{y,521} & = -1\,,&
\beta_{y,522} & = \tfrac{19}{3}\,,\\
\beta_{y,523} & = \tfrac{4}{3}\,,&
\beta_{y,524} & = -8 \zeta_3+\tfrac{2}{3}+\tfrac{\pi ^4}{30}\,,&
\beta_{y,525} & = -8 \zeta_3+\tfrac{2}{3}+\tfrac{\pi ^4}{30}\,,\\
\beta_{y,526} & = 2 \zeta_3+\tfrac{10}{3}\,,&
\beta_{y,527} & = 8-2 \zeta_3\,,&
\beta_{y,528} & = \tfrac{1}{3}\,,\\
\beta_{y,529} & = \tfrac{5}{3}-2 \zeta_3\,,&
\beta_{y,530} & = -2 \zeta_3\,,&
\beta_{y,531} & = 16 \zeta_3-20 \zeta_5\,,\\
\beta_{y,532} & = -8 \zeta_3\,,&
\beta_{y,533} & = 18 \zeta_3-20 \zeta_5\,,&
\beta_{y,534} & = 4 \zeta_3-\tfrac{8}{3}\,,\\
\beta_{y,535} & = 2 \zeta_3+\tfrac{8}{3}+\tfrac{\pi ^4}{30}\,,&
\beta_{y,536} & = -8 \zeta_3\,,&
\beta_{y,537} & = -4 \zeta_3\,,\\
\beta_{y,538} & = -4 \zeta_3\,,&
\beta_{y,539} & = -6 \zeta_3+\tfrac{22}{3}-\tfrac{\pi ^4}{30}\,,&
\beta_{y,540} & = -2 \zeta_3+\tfrac{14}{3}+\tfrac{\pi ^4}{30}\,,\\
\beta_{y,541} & = 12-8 \zeta_3\,,&
\beta_{y,542} & = -4 \zeta_3\,,&
\beta_{y,543} & = 12 \zeta_3-20 \zeta_5\,,\\
\beta_{y,544} & = -8 \zeta_3\,,&
\beta_{y,545} & = 2 \zeta_3-20 \zeta_5+\tfrac{44}{3}\,,&
\beta_{y,546} & = \tfrac{1}{3}\,,\\
\beta_{y,547} & = \tfrac{5}{3}-2 \zeta_3\,,&
\beta_{y,548} & = 2 \zeta_3-\tfrac{1}{3}\,,&
\beta_{y,549} & = -4 \zeta_3+\tfrac{4}{3}+\tfrac{\pi ^4}{60}\,,\\
\beta_{y,550} & = -4 \zeta_3+\tfrac{4}{3}+\tfrac{\pi ^4}{60}\,,&
\beta_{y,551} & = -5 \zeta_3+6+\tfrac{\pi ^4}{60}\,,&
\beta_{y,552} & = \tfrac{19}{6}\,,\\
\beta_{y,553} & = -5\,,&
\beta_{y,554} & = -2\,,&
\beta_{y,555} & = 4 \zeta_3-4\,,\\
\beta_{y,556} & = -\tfrac{2}{3}\,,&
\beta_{y,557} & = \zeta_3-\tfrac{1}{2}\,,&
\beta_{y,558} & = 6 \zeta_3\,,\\
\beta_{y,559} & = -8 \zeta_3\,,&
\beta_{y,560} & = -4 \zeta_3+\tfrac{2}{3}+\tfrac{\pi ^4}{15}\,,&
\beta_{y,561} & = \tfrac{16}{3}-8 \zeta_3\,,\\
\beta_{y,562} & = \tfrac{1}{6}\,,&
\beta_{y,563} & = \tfrac{5}{6}-\zeta_3\,,&
\beta_{y,564} & = 6 \zeta_3+2\,,\\
\beta_{y,565} & = -2 \zeta_3\,,&
\beta_{y,566} & = -8 \zeta_3\,,&
\beta_{y,567} & = -\tfrac{22}{3}\,,\\
\beta_{y,568} & = 6 \zeta_3-6\,,&
\beta_{y,569} & = -8 \zeta_3+\tfrac{35}{3}-\tfrac{\pi ^4}{30}\,,&
\beta_{y,570} & = 4 \zeta_3-\tfrac{2}{3}\,,\\
\beta_{y,571} & = -10 \zeta_3+12+\tfrac{\pi ^4}{30}\,,&
\beta_{y,572} & = \tfrac{19}{3}\,,&
\beta_{y,573} & = -10\,,\\
\beta_{y,574} & = -\tfrac{5}{3}\,,&
\beta_{y,575} & = 2 \zeta_3-\tfrac{4}{3}\,,&
\beta_{y,576} & = \tfrac{\pi ^4}{30}-\tfrac{4}{3}\,,\\
\beta_{y,577} & = 2 \zeta_3-\tfrac{14}{3}+\tfrac{\pi ^4}{30}\,,&
\beta_{y,578} & = 2 \zeta_3-\tfrac{8}{3}\,,&
\beta_{y,579} & = 2 \zeta_3-\tfrac{4}{3}\,,\\
\beta_{y,580} & = -\tfrac{7}{3}\,,&
\beta_{y,581} & = \tfrac{19}{6}\,,&
\beta_{y,582} & = -3 \zeta_3+\tfrac{1}{3}+\tfrac{\pi ^4}{60}\,,\\
\beta_{y,583} & = -2 \zeta_3-1+\tfrac{\pi ^4}{60}\,,&
\beta_{y,584} & = \zeta_3+\tfrac{2}{3}\,,&
\beta_{y,585} & = \zeta_3-\tfrac{1}{2}\,,\\
\beta_{y,586} & = -\tfrac{1}{2}\,,&
\beta_{y,587} & = \tfrac{2}{3}\,,&
\beta_{y,588} & = -2 \zeta_3+\tfrac{8}{3}+\tfrac{\pi ^4}{30}\,,\\
\beta_{y,589} & = 6 \zeta_3-20 \zeta_5+\tfrac{22}{3}\,,&
\beta_{y,590} & = -4 \zeta_3+\tfrac{1}{3}+\tfrac{\pi ^4}{60}\,,&
\beta_{y,591} & = \zeta_3+\tfrac{5}{3}\,,\\
\beta_{y,592} & = -4 \zeta_3+\tfrac{1}{3}+\tfrac{\pi ^4}{60}\,,&
\beta_{y,593} & = 4-\zeta_3\,,&
\beta_{y,594} & = -2 \zeta_3\,,\\
\beta_{y,595} & = 16 \zeta_3-20 \zeta_5\,,&
\beta_{y,596} & = 16 \zeta_3-20 \zeta_5\,,&
\beta_{y,597} & = 6 \zeta_3\,,\\
\beta_{y,598} & = 18 \zeta_3-20 \zeta_5\,,&
\beta_{y,599} & = -4 \zeta_3+\tfrac{22}{3}-\tfrac{\pi ^4}{30}\,,&
\beta_{y,600} & = -8 \zeta_3+\tfrac{8}{3}+\tfrac{\pi ^4}{30}\,. \tag{\stepcounter{equation}\theequation}
\end{align*}

\section{Quartic Vertex}\label{app:betaQ}
The four-loop quartic $\beta$-function~\eq{beta-lam} consists of scalar leg corrections detailed in App.~\ref{app:gammaS} as well as vertex terms 
\begin{align*}
  \hat{\beta}_\lambda^{(4)} &= \beta_{\lambda, \lambda^5}^{(4)} 
  + \beta_{\lambda, \lambda^4 y^2}^{(4)} + 
  + \beta_{\lambda, \lambda^3 y^2 y^2}^{(4)}
  + \beta_{\lambda, \lambda^3 y^4}^{(4)}
  + \beta_{\lambda, \lambda^2 y^2 y^2 y^2}^{(4)}
  + \beta_{\lambda, \lambda^2 y^4 y^2}^{(4)}
  + \beta_{\lambda, \lambda^2 y^6}^{(4)} 
  + \beta_{\lambda, \lambda y^4 y^2 y^2 }^{(4)}\\
  &\phantom{=}
  + \beta_{\lambda, \lambda y^4 y^4 }^{(4)}
  + \beta_{\lambda, \lambda y^6 y^2 }^{(4)}
  + \beta_{\lambda, \lambda y^8 }^{(4)}
  + \beta_{\lambda, y^6 y^2 y^2}^{(4)}
  + \beta_{\lambda, y^6 y^4}^{(4)}
  + \beta_{\lambda, y^4 y^4 y^2}^{(4)}
  + \beta_{\lambda, y^{10}}^{(4)}\,. \yestag{}
\end{align*}
The first term does not contain any Yukawa interactions and was known prior to this work~\cite{Kazakov:1979ik,Jack:1990eb,Jack:2018oec}. 
Overall, they read
\begin{align*}
    \beta_{\lambda, \lambda^5}^{(4)\,abcd} &= 
    \beta_{\lambda,1}\ \mathcal{S}_{3}\,\lambda^{abef}\,\lambda^{cdgh}\,\lambda^{fhij}\,\lambda^{eikl}\,\lambda^{gjkl}
    + \beta_{\lambda,2}\ \mathcal{S}_{3}\,\lambda^{abef}\,\lambda^{cdgh}\,\lambda^{fhij}\,\lambda^{egkl}\,\lambda^{ijkl}
    \\&\phantom{=}
    + \beta_{\lambda,3}\ \mathcal{S}_{3}\,\lambda^{abef}\,\lambda^{cdge}\,\lambda^{fhij}\,\lambda^{ghkl}\,\lambda^{ijkl}
    + \beta_{\lambda,4}\ \mathcal{S}_{6}\,\lambda^{abef}\,\lambda^{eghi}\,\lambda^{fghj}\,\lambda^{cikl}\,\lambda^{djkl}
    \\&\phantom{=}
    + \beta_{\lambda,5}\ \mathcal{S}_{6}\,\lambda^{abef}\,\lambda^{cegh}\,\lambda^{dfgi}\,\lambda^{hjkl}\,\lambda^{ijkl}
    + \beta_{\lambda,6}\ \mathcal{S}_{6}\,\lambda^{abef}\,\lambda^{cegh}\,\lambda^{dfij}\,\lambda^{gikl}\,\lambda^{hjkl}
    \\&\phantom{=}
    + \beta_{\lambda,7}\ \mathcal{S}_{6}\,\lambda^{abef}\,\lambda^{cegh}\,\lambda^{dfij}\,\lambda^{ghkl}\,\lambda^{ijkl}
    + \beta_{\lambda,8}\ \mathcal{S}_{6}\,\lambda^{abef}\,\lambda^{eghi}\,\lambda^{chij}\,\lambda^{fgkl}\,\lambda^{djkl}
    \\&\phantom{=}
    + \beta_{\lambda,9}\ \mathcal{S}_{6}\,\lambda^{abef}\,\lambda^{eghi}\,\lambda^{chij}\,\lambda^{fjkl}\,\lambda^{dgkl}
    + \beta_{\lambda,10}\ \mathcal{S}_{12}\,\lambda^{abef}\,\lambda^{cegh}\,\lambda^{dghi}\,\lambda^{fjkl}\,\lambda^{ijkl}
    \\&\phantom{=}
    + \beta_{\lambda,11}\ \mathcal{S}_{12}\,\lambda^{abef}\,\lambda^{cegh}\,\lambda^{djkl}\,\lambda^{ghij}\,\lambda^{fikl}
    + \beta_{\lambda,12}\ \mathcal{S}_{12}\,\lambda^{abef}\,\lambda^{cegh}\,\lambda^{dgkl}\,\lambda^{fhij}\,\lambda^{ijkl}
    \\&\phantom{=}
    + \beta_{\lambda,13}\ \mathcal{S}_{12}\,\lambda^{abef}\,\lambda^{cegh}\,\lambda^{dgkl}\,\lambda^{fijl}\,\lambda^{hijk}
    + \beta_{\lambda,14}\ \mathcal{S}_{12}\,\lambda^{abef}\,\lambda^{cegh}\,\lambda^{djkl}\,\lambda^{fgij}\,\lambda^{hikl}
    \\&\phantom{=}
    + \beta_{\lambda,15}\ \mathcal{S}_{12}\,\lambda^{aefg}\,\lambda^{befh}\,\lambda^{cgij}\,\lambda^{dhkl}\,\lambda^{ijkl}
    + \beta_{\lambda,16}\ \mathcal{S}_{12}\,\lambda^{aefh}\,\lambda^{begi}\,\lambda^{cfgj}\,\lambda^{djkl}\,\lambda^{hikl}
    \\&\phantom{=}
    + \beta_{\lambda,17}\ \mathcal{S}_{6}\,\lambda^{abef}\,\lambda^{eghi}\,\lambda^{fgjk}\,\lambda^{chjl}\,\lambda^{dikl}
    + \beta_{\lambda,18}\ \mathcal{S}_{24}\,\lambda^{aefg}\,\lambda^{befh}\,\lambda^{chij}\,\lambda^{djkl}\,\lambda^{gikl}
    \\&\phantom{=}
    + \beta_{\lambda,19}\ \mathcal{S}_{3}\,\lambda^{aefi}\,\lambda^{bfgj}\,\lambda^{cghk}\,\lambda^{dhel}\,\lambda^{ijkl}
    + \beta_{\lambda,20}\ \mathcal{S}_{3}\,\lambda^{abef}\,\lambda^{efgh}\,\lambda^{ghij}\,\lambda^{ijkl}\,\lambda^{klcd}
    \\&\phantom{=}
    + \beta_{\lambda,21}\ \mathcal{S}_{3}\,\lambda^{aefg}\,\lambda^{befh}\,\lambda^{ghij}\,\lambda^{cikl}\,\lambda^{djkl}
    + \beta_{\lambda,22}\ \mathcal{S}_{6}\,\lambda^{abef}\,\lambda^{efgh}\,\lambda^{ghij}\,\lambda^{cikl}\,\lambda^{djkl}
    \\&\phantom{=}
    + \beta_{\lambda,23}\ \mathcal{S}_{6}\,\lambda^{abef}\,\lambda^{efgh}\,\lambda^{cgij}\,\lambda^{djkl}\,\lambda^{ijkl}
    + \beta_{\lambda,24}\ \mathcal{S}_{6}\,\lambda^{abef}\,\lambda^{efgh}\,\lambda^{gijk}\,\lambda^{ijkl}\,\lambda^{cdhl}
    \\&\phantom{=}
    + \beta_{\lambda,25}\ \mathcal{S}_{6}\,\lambda^{abef}\,\lambda^{efgh}\,\lambda^{gijk}\,\lambda^{hijl}\,\lambda^{cdkl}
    + \beta_{\lambda,26}\ \mathcal{S}_{12}\,\lambda^{abef}\,\lambda^{efgh}\,\lambda^{gijk}\,\lambda^{cijl}\,\lambda^{dhkl}\,,\yestag{}\\[.5em]
    \beta_{\lambda, \lambda^4 y^2}^{(4)\,abcd} &= 
    \beta_{\lambda,27}\ \mathcal{S}_{3}\,\lambda^{abef}\,\lambda^{eghi}\,\lambda^{fgjk}\,\lambda^{cdhj}\,\mathrm{tr} \left(y^{i k}\right)
    + \beta_{\lambda,28}\ \mathcal{S}_{3}\,\lambda^{abef}\,\lambda^{efgh}\,\lambda^{gijk}\,\lambda^{cdjk}\,\mathrm{tr} \left(y^{h i}\right)
    \\&\phantom{=}
    + \beta_{\lambda,29}\ \mathcal{S}_{3}\,\lambda^{abef}\,\lambda^{cdeg}\,\lambda^{fhij}\,\lambda^{ghik}\,\mathrm{tr} \left(y^{j k}\right)
    + \beta_{\lambda,30}\ \mathcal{S}_{6}\,\lambda^{abef}\,\lambda^{eghi}\,\lambda^{fghj}\,\lambda^{cdik}\,\mathrm{tr} \left(y^{j k}\right)
    \\&\phantom{=}
    + \beta_{\lambda,31}\ \mathcal{S}_{6}\,\lambda^{abef}\,\lambda^{efgh}\,\lambda^{ghij}\,\lambda^{cdik}\,\mathrm{tr} \left(y^{j k}\right)
    + \beta_{\lambda,32}\ \mathcal{S}_{6}\,\lambda^{abef}\,\lambda^{fghi}\,\lambda^{ghij}\,\lambda^{cdek}\,\mathrm{tr} \left(y^{j k}\right)
    \\&\phantom{=}
    + \beta_{\lambda,33}\ \mathcal{S}_{3}\,\lambda^{abef}\,\lambda^{fghi}\,\lambda^{ghij}\,\lambda^{cdjk}\,\mathrm{tr} \left(y^{e k}\right)
    + \beta_{\lambda,34}\ \mathcal{S}_{12}\,\lambda^{abef}\,\lambda^{efgh}\,\lambda^{chij}\,\lambda^{dijk}\,\mathrm{tr} \left(y^{g k}\right)
    \\&\phantom{=}
    + \beta_{\lambda,35}\ \mathcal{S}_{6}\,\lambda^{abef}\,\lambda^{efgh}\,\lambda^{chij}\,\lambda^{dgik}\,\mathrm{tr} \left(y^{j k}\right)
    + \beta_{\lambda,36}\ \mathcal{S}_{6}\,\lambda^{abef}\,\lambda^{eghi}\,\lambda^{chjk}\,\lambda^{dijk}\,\mathrm{tr} \left(y^{f g}\right)
    \\&\phantom{=}
    + \beta_{\lambda,37}\ \mathcal{S}_{12}\,\lambda^{abef}\,\lambda^{cegh}\,\lambda^{dgij}\,\lambda^{fhik}\,\mathrm{tr} \left(y^{j k}\right)
    + \beta_{\lambda,38}\ \mathcal{S}_{12}\,\lambda^{abef}\,\lambda^{cegh}\,\lambda^{dgij}\,\lambda^{fkij}\,\mathrm{tr} \left(y^{h k}\right)
    \\&\phantom{=}
    + \beta_{\lambda,39}\ \mathcal{S}_{12}\,\lambda^{abef}\,\lambda^{cegh}\,\lambda^{dgij}\,\lambda^{khij}\,\mathrm{tr} \left(y^{f k}\right)
    + \beta_{\lambda,40}\ \mathcal{S}_{12}\,\lambda^{abef}\,\lambda^{cegh}\,\lambda^{dkij}\,\lambda^{fhij}\,\mathrm{tr} \left(y^{g k}\right)
    \\&\phantom{=}
    + \beta_{\lambda,41}\ \mathcal{S}_{12}\,\lambda^{abef}\,\lambda^{ckgh}\,\lambda^{dgij}\,\lambda^{fhij}\,\mathrm{tr} \left(y^{e k}\right)
    + \beta_{\lambda,42}\ \mathcal{S}_{12}\,\lambda^{abef}\,\lambda^{cegh}\,\lambda^{dkij}\,\lambda^{ghij}\,\mathrm{tr} \left(y^{f k}\right)
    \\&\phantom{=}
    + \beta_{\lambda,43}\ \mathcal{S}_{12}\,\lambda^{abef}\,\lambda^{cegh}\,\lambda^{dfij}\,\lambda^{ghik}\,\mathrm{tr} \left(y^{j k}\right)
    + \beta_{\lambda,44}\ \mathcal{S}_{6}\,\lambda^{aefg}\,\lambda^{behi}\,\lambda^{cfhj}\,\lambda^{dgik}\,\mathrm{tr} \left(y^{j k}\right)
    \\&\phantom{=}
    + \beta_{\lambda,45}\ \mathcal{S}_{12}\,\lambda^{aefg}\,\lambda^{behi}\,\lambda^{chij}\,\lambda^{dfgk}\,\mathrm{tr} \left(y^{j k}\right)
    + \beta_{\lambda,46}\ \mathcal{S}_{12}\,\lambda^{aefg}\,\lambda^{bfgh}\,\lambda^{chij}\,\lambda^{deik}\,\mathrm{tr} \left(y^{j k}\right)\,,\yestag{}\\[.5em]
    \beta_{\lambda, \lambda^3 y^2 y^2}^{(4)\,abcd} &= 
    \beta_{\lambda,47}\ \mathcal{S}_{3}\,\lambda^{abef}\,\mathrm{tr} \left(y^{e g}\right)\,\lambda^{fghi}\,\mathrm{tr} \left(y^{h j}\right)\,\lambda^{cdij}
    \\&\phantom{=}
    + \beta_{\lambda,48}\ \mathcal{S}_{6}\,\lambda^{abef}\,\mathrm{tr} \left(y^{e g}\right)\,\mathrm{tr} \left(y^{f h}\right)\,\lambda^{ghij}\,\lambda^{cdij}
    \\&\phantom{=}
    + \beta_{\lambda,49}\ \mathcal{S}_{6}\,\lambda^{abef}\,\mathrm{tr} \left(y^{e g}\right)\,\mathrm{tr} \left(y^{g h}\right)\,\lambda^{fhij}\,\lambda^{cdij}
    \\&\phantom{=}
    + \beta_{\lambda,50}\ \mathcal{S}_{6}\,\lambda^{abef}\,\lambda^{cegi}\,\lambda^{dfhj}\,\mathrm{tr} \left(y^{g h}\right)\,\mathrm{tr} \left(y^{i j}\right)
    \\&\phantom{=}
    + \beta_{\lambda,51}\ \mathcal{S}_{12}\,\lambda^{abef}\,\lambda^{cghi}\,\lambda^{dehj}\,\mathrm{tr} \left(y^{f g}\right)\,\mathrm{tr} \left(y^{i j}\right)
    \\&\phantom{=}
    + \beta_{\lambda,52}\ \mathcal{S}_{6}\,\lambda^{abef}\,\mathrm{tr} \left(y^{e g}\right)\,\mathrm{tr} \left(y^{f h}\right)\,\lambda^{cgij}\,\lambda^{dhij}
    \\&\phantom{=}
    + \beta_{\lambda,53}\ \mathcal{S}_{6}\,\lambda^{abef}\,\mathrm{tr} \left(y^{h i}\right)\,\mathrm{tr} \left(y^{i j}\right)\,\lambda^{cegh}\,\lambda^{dfgj}
    \\&\phantom{=}
    + \beta_{\lambda,54}\ \mathcal{S}_{12}\,\lambda^{abef}\,\mathrm{tr} \left(y^{f g}\right)\,\mathrm{tr} \left(y^{g h}\right)\,\lambda^{chij}\,\lambda^{deij}\,,\yestag{}\\[.5em]
    \beta_{\lambda, \lambda^3 y^4}^{(4)\,abcd} &= 
    \beta_{\lambda,55}\ \mathcal{S}_{6}\,\lambda^{abef}\,\lambda^{cdfg}\,\lambda^{ghij}\,\mathrm{tr} \left(y^{e h i j}\right)
    + \beta_{\lambda,56}\ \mathcal{S}_{3}\,\lambda^{abei}\,\lambda^{fgij}\,\lambda^{cdhj}\,\mathrm{tr} \left(y^{e f h g}\right)
    \\&\phantom{=}
    + \beta_{\lambda,57}\ \mathcal{S}_{3}\,\lambda^{abei}\,\lambda^{fgij}\,\lambda^{cdhj}\,\mathrm{tr} \left(y^{e f g h}\right)
    + \beta_{\lambda,58}\ \mathcal{S}_{6}\,\lambda^{abef}\,\lambda^{ghij}\,\lambda^{cdij}\,\mathrm{tr} \left(y^{e g f h}\right)
    \\&\phantom{=}
    + \beta_{\lambda,59}\ \mathcal{S}_{6}\,\lambda^{abef}\,\lambda^{ghij}\,\lambda^{cdij}\,\mathrm{tr} \left(y^{e f g h}\right)
    + \beta_{\lambda,60}\ \mathcal{S}_{6}\,\lambda^{abef}\,\lambda^{efgh}\,\lambda^{cdgi}\,\mathrm{tr} \left(y^{h j i j}\right)
    \\&\phantom{=}
    + \beta_{\lambda,61}\ \mathcal{S}_{6}\,\lambda^{abef}\,\lambda^{efgh}\,\lambda^{cdgi}\,\mathrm{tr} \left(y^{h i j j}\right)
    + \beta_{\lambda,62}\ \mathcal{S}_{6}\,\lambda^{abef}\,\lambda^{cegh}\,\lambda^{dfij}\,\mathrm{tr} \left(y^{g i h j}\right)
    \\&\phantom{=}
    + \beta_{\lambda,63}\ \mathcal{S}_{6}\,\lambda^{abef}\,\lambda^{cegh}\,\lambda^{dfij}\,\mathrm{tr} \left(y^{g h i j}\right)
    + \beta_{\lambda,64}\ \mathcal{S}_{12}\,\lambda^{abef}\,\lambda^{cegh}\,\lambda^{dgij}\,\mathrm{tr} \left(y^{f h i j}\right)
    \\&\phantom{=}
    + \beta_{\lambda,65}\ \mathcal{S}_{12}\,\lambda^{abef}\,\lambda^{cegh}\,\lambda^{dgij}\,\mathrm{tr} \left(y^{f i h j}\right)
    + \beta_{\lambda,66}\ \mathcal{S}_{6}\,\lambda^{abef}\,\lambda^{cghi}\,\lambda^{dghj}\,\mathrm{tr} \left(y^{e i f j}\right)
    \\&\phantom{=}
    + \beta_{\lambda,67}\ \mathcal{S}_{6}\,\lambda^{abef}\,\lambda^{cghi}\,\lambda^{dghj}\,\mathrm{tr} \left(y^{e f i j}\right)
    + \beta_{\lambda,68}\ \mathcal{S}_{6}\,\lambda^{abef}\,\lambda^{cegh}\,\lambda^{dfgi}\,\mathrm{tr} \left(y^{h j i j}\right)
    \\&\phantom{=}
    + \beta_{\lambda,69}\ \mathcal{S}_{12}\,\lambda^{abef}\,\lambda^{cegh}\,\lambda^{dghi}\,\mathrm{tr} \left(y^{f j i j}\right)
    + \beta_{\lambda,70}\ \mathcal{S}_{6}\,\lambda^{abef}\,\lambda^{cegh}\,\lambda^{dfgi}\,\mathrm{tr} \left(y^{h i j j}\right)
    \\&\phantom{=}
    + \beta_{\lambda,71}\ \mathcal{S}_{12}\,\lambda^{abef}\,\lambda^{cegh}\,\lambda^{dghi}\,\mathrm{tr} \left(y^{f i j j}\right)
    + \beta_{\lambda,72}\ \mathcal{S}_{12}\,\lambda^{abef}\,\lambda^{cegh}\,\lambda^{fgij}\,\mathrm{tr} \left(y^{d i h j}\right)
    \\&\phantom{=}
    + \beta_{\lambda,73}\ \mathcal{S}_{12}\,\lambda^{abef}\,\lambda^{cegh}\,\lambda^{fgij}\,\mathrm{tr} \left(y^{d h i j}\right)
    + \beta_{\lambda,74}\ \mathcal{S}_{12}\,\lambda^{abef}\,\lambda^{cegh}\,\lambda^{ghij}\,\mathrm{tr} \left(y^{d i f j}\right)
    \\&\phantom{=}
    + \beta_{\lambda,75}\ \mathcal{S}_{12}\,\lambda^{abef}\,\lambda^{cegh}\,\lambda^{ghij}\,\mathrm{tr} \left(y^{d f i j}\right)
    + \beta_{\lambda,76}\ \mathcal{S}_{12}\,\lambda^{abef}\,\lambda^{efgh}\,\lambda^{cgij}\,\mathrm{tr} \left(y^{d h i j}\right)
    \\&\phantom{=}
    + \beta_{\lambda,77}\ \mathcal{S}_{12}\,\lambda^{abef}\,\lambda^{efgh}\,\lambda^{cgij}\,\mathrm{tr} \left(y^{d i h j}\right)
    + \beta_{\lambda,78}\ \mathcal{S}_{12}\,\lambda^{abef}\,\lambda^{eghi}\,\lambda^{cghj}\,\mathrm{tr} \left(y^{d j f i}\right)
    \\&\phantom{=}
    + \beta_{\lambda,79}\ \mathcal{S}_{12}\,\lambda^{abef}\,\lambda^{eghi}\,\lambda^{cghj}\,\mathrm{tr} \left(y^{d f j i}\right)
    + \beta_{\lambda,80}\ \mathcal{S}_{12}\,\lambda^{abef}\,\lambda^{eghi}\,\lambda^{cghj}\,\mathrm{tr} \left(y^{d f i j}\right)
    \\&\phantom{=}
    + \beta_{\lambda,81}\ \mathcal{S}_{12}\,\lambda^{aefg}\,\lambda^{cehi}\,\lambda^{bghj}\,\mathrm{tr} \left(y^{d f i j}\right)
    + \beta_{\lambda,82}\ \mathcal{S}_{24}\,\lambda^{aefg}\,\lambda^{befh}\,\lambda^{chij}\,\mathrm{tr} \left(y^{d g i j}\right)
    \\&\phantom{=}
    + \beta_{\lambda,83}\ \mathcal{S}_{24}\,\lambda^{aefg}\,\lambda^{befh}\,\lambda^{chij}\,\mathrm{tr} \left(y^{d i g j}\right)
    + \beta_{\lambda,84}\ \mathcal{S}_{6}\,\lambda^{abef}\,\lambda^{eghi}\,\lambda^{fghj}\,\mathrm{tr} \left(y^{c i d j}\right)
    \\&\phantom{=}
    + \beta_{\lambda,85}\ \mathcal{S}_{6}\,\lambda^{abef}\,\lambda^{efgh}\,\lambda^{ghij}\,\mathrm{tr} \left(y^{c i d j}\right)
    + \beta_{\lambda,86}\ \mathcal{S}_{6}\,\lambda^{abef}\,\lambda^{eghi}\,\lambda^{ghij}\,\mathrm{tr} \left(y^{c f d j}\right)
    \\&\phantom{=}
    + \beta_{\lambda,87}\ \mathcal{S}_{12}\,\lambda^{abef}\,\lambda^{eghi}\,\lambda^{ghij}\,\mathrm{tr} \left(y^{c d f j}\right)
    + \beta_{\lambda,88}\ \mathcal{S}_{6}\,\lambda^{abef}\,\lambda^{eghi}\,\lambda^{fghj}\,\mathrm{tr} \left(y^{c d i j}\right)
    \\&\phantom{=}
    + \beta_{\lambda,89}\ \mathcal{S}_{6}\,\lambda^{abef}\,\lambda^{efgh}\,\lambda^{ghij}\,\mathrm{tr} \left(y^{c d i j}\right)
    + \beta_{\lambda,90}\ \mathcal{S}_{6}\,\lambda^{aefg}\,\lambda^{befh}\,\lambda^{ghij}\,\mathrm{tr} \left(y^{c i d j}\right)
    \\&\phantom{=}
    + \beta_{\lambda,91}\ \mathcal{S}_{6}\,\lambda^{aefg}\,\lambda^{befh}\,\lambda^{ghij}\,\mathrm{tr} \left(y^{c d i j}\right)
    + \beta_{\lambda,92}\ \mathcal{S}_{12}\,\lambda^{cefg}\,\lambda^{dehi}\,\lambda^{fghj}\,\mathrm{tr} \left(y^{a i b j}\right)
    \\&\phantom{=}
    + \beta_{\lambda,93}\ \mathcal{S}_{24}\,\lambda^{cefg}\,\lambda^{dehi}\,\lambda^{fghj}\,\mathrm{tr} \left(y^{a b i j}\right)
    + \beta_{\lambda,94}\ \mathcal{S}_{6}\,\lambda^{aefg}\,\lambda^{efhi}\,\lambda^{bhij}\,\mathrm{tr} \left(y^{c g d j}\right)
    \\&\phantom{=}
    + \beta_{\lambda,95}\ \mathcal{S}_{12}\,\lambda^{aefg}\,\lambda^{efhi}\,\lambda^{bhij}\,\mathrm{tr} \left(y^{c d g j}\right)\,,\yestag{}\\[.5em]
    \beta_{\lambda, \lambda^2 y^2 y^2 y^2}^{(4)\,abcd} &= 
    \beta_{\lambda,96}\ \mathcal{S}_{3}\,\lambda^{abef}\,\lambda^{cdei}\,\mathrm{tr} \left(y^{f g}\right)\,\mathrm{tr} \left(y^{g h}\right)\,\mathrm{tr} \left(y^{h i}\right)
    \\&\phantom{=}
    + \beta_{\lambda,97}\ \mathcal{S}_{3}\,\lambda^{abef}\,\lambda^{cdgi}\,\mathrm{tr} \left(y^{f g}\right)\,\mathrm{tr} \left(y^{e h}\right)\,\mathrm{tr} \left(y^{h i}\right)\,,\yestag{}\\[.5em]
    \beta_{\lambda, \lambda^2 y^4 y^2}^{(4)\,abcd} &= 
    \beta_{\lambda,98}\ \mathcal{S}_{3}\,\lambda^{abef}\,\lambda^{cdgh}\,\mathrm{tr} \left(y^{e g}\right)\,\mathrm{tr} \left(y^{f h i i}\right)
    + \beta_{\lambda,99}\ \mathcal{S}_{3}\,\lambda^{abef}\,\lambda^{cdgh}\,\mathrm{tr} \left(y^{e g}\right)\,\mathrm{tr} \left(y^{f i h i}\right)
    \\&\phantom{=}
    + \beta_{\lambda,100}\ \mathcal{S}_{6}\,\lambda^{abef}\,\lambda^{cdeg}\,\mathrm{tr} \left(y^{f h}\right)\,\mathrm{tr} \left(y^{g h i i}\right)
    + \beta_{\lambda,101}\ \mathcal{S}_{6}\,\lambda^{abef}\,\lambda^{cdeg}\,\mathrm{tr} \left(y^{f h}\right)\,\mathrm{tr} \left(y^{g i h i}\right)
    \\&\phantom{=}
    + \beta_{\lambda,102}\ \mathcal{S}_{3}\,\lambda^{abef}\,\lambda^{cdeg}\,\mathrm{tr} \left(y^{f g h i}\right)\,\mathrm{tr} \left(y^{h i}\right)
    + \beta_{\lambda,103}\ \mathcal{S}_{3}\,\lambda^{abef}\,\lambda^{cdeg}\,\mathrm{tr} \left(y^{f h g i}\right)\,\mathrm{tr} \left(y^{h i}\right)
    \\&\phantom{=}
    + \beta_{\lambda,104}\ \mathcal{S}_{6}\,\lambda^{abef}\,\lambda^{cdgh}\,\mathrm{tr} \left(y^{e f g i}\right)\,\mathrm{tr} \left(y^{h i}\right)
    + \beta_{\lambda,105}\ \mathcal{S}_{6}\,\lambda^{abef}\,\lambda^{cdgh}\,\mathrm{tr} \left(y^{e g f i}\right)\,\mathrm{tr} \left(y^{h i}\right)
    \\&\phantom{=}
    + \beta_{\lambda,106}\ \mathcal{S}_{12}\,\lambda^{abef}\,\lambda^{cfgh}\,\mathrm{tr} \left(y^{d g i h}\right)\,\mathrm{tr} \left(y^{e i}\right)
    + \beta_{\lambda,107}\ \mathcal{S}_{12}\,\lambda^{abef}\,\lambda^{cigh}\,\mathrm{tr} \left(y^{d g e h}\right)\,\mathrm{tr} \left(y^{f i}\right)
    \\&\phantom{=}
    + \beta_{\lambda,108}\ \mathcal{S}_{12}\,\lambda^{abef}\,\lambda^{cfgh}\,\mathrm{tr} \left(y^{d i e h}\right)\,\mathrm{tr} \left(y^{g i}\right)
    + \beta_{\lambda,109}\ \mathcal{S}_{12}\,\lambda^{abef}\,\lambda^{cfgh}\,\mathrm{tr} \left(y^{d g i e}\right)\,\mathrm{tr} \left(y^{h i}\right)
    \\&\phantom{=}
    + \beta_{\lambda,110}\ \mathcal{S}_{12}\,\lambda^{abef}\,\lambda^{cfgh}\,\mathrm{tr} \left(y^{d g h i}\right)\,\mathrm{tr} \left(y^{e i}\right)
    + \beta_{\lambda,111}\ \mathcal{S}_{12}\,\lambda^{abef}\,\lambda^{cigh}\,\mathrm{tr} \left(y^{d g h e}\right)\,\mathrm{tr} \left(y^{f i}\right)
    \\&\phantom{=}
    + \beta_{\lambda,112}\ \mathcal{S}_{12}\,\lambda^{abef}\,\lambda^{cfgh}\,\mathrm{tr} \left(y^{d i h e}\right)\,\mathrm{tr} \left(y^{g i}\right)
    + \beta_{\lambda,113}\ \mathcal{S}_{12}\,\lambda^{aefg}\,\lambda^{beih}\,\mathrm{tr} \left(y^{c d g h}\right)\,\mathrm{tr} \left(y^{f i}\right)
    \\&\phantom{=}
    + \beta_{\lambda,114}\ \mathcal{S}_{6}\,\lambda^{aefg}\,\lambda^{beih}\,\mathrm{tr} \left(y^{c g d h}\right)\,\mathrm{tr} \left(y^{f i}\right)
    + \beta_{\lambda,115}\ \mathcal{S}_{24}\,\lambda^{aefg}\,\lambda^{befh}\,\mathrm{tr} \left(y^{c d g i}\right)\,\mathrm{tr} \left(y^{h i}\right)
    \\&\phantom{=}
    + \beta_{\lambda,116}\ \mathcal{S}_{12}\,\lambda^{cefg}\,\lambda^{defh}\,\mathrm{tr} \left(y^{a g b i}\right)\,\mathrm{tr} \left(y^{h i}\right)
    + \beta_{\lambda,117}\ \mathcal{S}_{6}\,\lambda^{abef}\,\lambda^{ifgh}\,\mathrm{tr} \left(y^{c d g h}\right)\,\mathrm{tr} \left(y^{e i}\right)
    \\&\phantom{=}
    + \beta_{\lambda,118}\ \mathcal{S}_{6}\,\lambda^{abef}\,\lambda^{ifgh}\,\mathrm{tr} \left(y^{c g d h}\right)\,\mathrm{tr} \left(y^{e i}\right)
    + \beta_{\lambda,119}\ \mathcal{S}_{12}\,\lambda^{abef}\,\lambda^{efgi}\,\mathrm{tr} \left(y^{c d g h}\right)\,\mathrm{tr} \left(y^{h i}\right)
    \\&\phantom{=}
    + \beta_{\lambda,120}\ \mathcal{S}_{6}\,\lambda^{abef}\,\lambda^{efgi}\,\mathrm{tr} \left(y^{c g d h}\right)\,\mathrm{tr} \left(y^{h i}\right)\,,\yestag{}\\[.5em]
    \beta_{\lambda, \lambda^2 y^6}^{(4)\,abcd} &=     
    + \beta_{\lambda,121}\ \mathcal{S}_{3}\,\lambda^{abef}\,\lambda^{cdeg}\,\mathrm{tr} \left(y^{f g h h i i}\right)
    + \beta_{\lambda,122}\ \mathcal{S}_{3}\,\lambda^{abef}\,\lambda^{cdeg}\,\mathrm{tr} \left(y^{f g h i h i}\right)
    \\&\phantom{=}
    + \beta_{\lambda,123}\ \mathcal{S}_{3}\,\lambda^{abef}\,\lambda^{cdeg}\,\mathrm{tr} \left(y^{f g h i i h}\right)
    + \beta_{\lambda,124}\ \mathcal{S}_{6}\,\lambda^{abef}\,\lambda^{cdeg}\,\mathrm{tr} \left(y^{f h g h i i}\right)
    \\&\phantom{=}
    + \beta_{\lambda,125}\ \mathcal{S}_{3}\,\lambda^{abef}\,\lambda^{cdeg}\,\mathrm{tr} \left(y^{f h g i h i}\right)
    + \beta_{\lambda,126}\ \mathcal{S}_{3}\,\lambda^{abef}\,\lambda^{cdeg}\,\mathrm{tr} \left(y^{f h h g i i}\right)
    \\&\phantom{=}
    + \beta_{\lambda,127}\ \mathcal{S}_{3}\,\lambda^{abef}\,\lambda^{cdeg}\,\mathrm{tr} \left(y^{f h i g h i}\right)
    + \beta_{\lambda,128}\ \mathcal{S}_{3}\,\lambda^{abef}\,\lambda^{cdeg}\,\mathrm{tr} \left(y^{f h i g i h}\right)
    \\&\phantom{=}
    + \beta_{\lambda,129}\ \mathcal{S}_{3}\,\lambda^{abef}\,\lambda^{cdgh}\,\mathrm{tr} \left(y^{i i e f g h}\right)
    + \beta_{\lambda,130}\ \mathcal{S}_{3}\,\lambda^{abef}\,\lambda^{cdgh}\,\mathrm{tr} \left(y^{i i e g f h}\right)
    \\&\phantom{=}
    + \beta_{\lambda,131}\ \mathcal{S}_{6}\,\lambda^{abef}\,\lambda^{cdgh}\,\mathrm{tr} \left(y^{i i e g h f}\right)
    + \beta_{\lambda,132}\ \mathcal{S}_{6}\,\lambda^{abef}\,\lambda^{cdgh}\,\mathrm{tr} \left(y^{i e i f g h}\right)
    \\&\phantom{=}
    + \beta_{\lambda,133}\ \mathcal{S}_{6}\,\lambda^{abef}\,\lambda^{cdgh}\,\mathrm{tr} \left(y^{i e i g f h}\right)
    + \beta_{\lambda,134}\ \mathcal{S}_{3}\,\lambda^{abef}\,\lambda^{cdgh}\,\mathrm{tr} \left(y^{i e f i g h}\right)
    \\&\phantom{=}
    + \beta_{\lambda,135}\ \mathcal{S}_{3}\,\lambda^{abef}\,\lambda^{cdgh}\,\mathrm{tr} \left(y^{i e g i f h}\right)
    + \beta_{\lambda,136}\ \mathcal{S}_{3}\,\lambda^{abef}\,\lambda^{cdgh}\,\mathrm{tr} \left(y^{i e g i h f}\right)
    \\&\phantom{=}
    + \beta_{\lambda,137}\ \mathcal{S}_{12}\,\lambda^{abef}\,\lambda^{cghi}\,\mathrm{tr} \left(y^{d e f g h i}\right)
    + \beta_{\lambda,138}\ \mathcal{S}_{12}\,\lambda^{abef}\,\lambda^{cghi}\,\mathrm{tr} \left(y^{d e g f h i}\right)
    \\&\phantom{=}
    + \beta_{\lambda,139}\ \mathcal{S}_{12}\,\lambda^{abef}\,\lambda^{cghi}\,\mathrm{tr} \left(y^{d e g h f i}\right)
    + \beta_{\lambda,140}\ \mathcal{S}_{12}\,\lambda^{abef}\,\lambda^{cghi}\,\mathrm{tr} \left(y^{d e g h i f}\right)
    \\&\phantom{=}
    + \beta_{\lambda,141}\ \mathcal{S}_{12}\,\lambda^{abef}\,\lambda^{cghi}\,\mathrm{tr} \left(y^{d g e h f i}\right)
    + \beta_{\lambda,142}\ \mathcal{S}_{12}\,\lambda^{abef}\,\lambda^{cghi}\,\mathrm{tr} \left(y^{d g e f h i}\right)
    \\&\phantom{=}
    + \beta_{\lambda,143}\ \mathcal{S}_{12}\,\lambda^{abef}\,\lambda^{cegh}\,\mathrm{tr} \left(y^{d f g h i i}\right)
    + \beta_{\lambda,144}\ \mathcal{S}_{12}\,\lambda^{abef}\,\lambda^{cegh}\,\mathrm{tr} \left(y^{d f g i h i}\right)
    \\&\phantom{=}
    + \beta_{\lambda,145}\ \mathcal{S}_{12}\,\lambda^{abef}\,\lambda^{cegh}\,\mathrm{tr} \left(y^{d f i g h i}\right)
    + \beta_{\lambda,146}\ \mathcal{S}_{12}\,\lambda^{abef}\,\lambda^{cegh}\,\mathrm{tr} \left(y^{d f g i i h}\right)
    \\&\phantom{=}
    + \beta_{\lambda,147}\ \mathcal{S}_{12}\,\lambda^{abef}\,\lambda^{cegh}\,\mathrm{tr} \left(y^{d f i i g h}\right)
    + \beta_{\lambda,148}\ \mathcal{S}_{12}\,\lambda^{abef}\,\lambda^{cegh}\,\mathrm{tr} \left(y^{d f i g i h}\right)
    \\&\phantom{=}
    + \beta_{\lambda,149}\ \mathcal{S}_{12}\,\lambda^{abef}\,\lambda^{cegh}\,\mathrm{tr} \left(y^{d i f g h i}\right)
    + \beta_{\lambda,150}\ \mathcal{S}_{12}\,\lambda^{abef}\,\lambda^{cegh}\,\mathrm{tr} \left(y^{d i f g i h}\right)
    \\&\phantom{=}
    + \beta_{\lambda,151}\ \mathcal{S}_{12}\,\lambda^{abef}\,\lambda^{cegh}\,\mathrm{tr} \left(y^{d i f i g h}\right)
    + \beta_{\lambda,152}\ \mathcal{S}_{12}\,\lambda^{abef}\,\lambda^{cegh}\,\mathrm{tr} \left(y^{d g f h i i}\right)
    \\&\phantom{=}
    + \beta_{\lambda,153}\ \mathcal{S}_{12}\,\lambda^{abef}\,\lambda^{cegh}\,\mathrm{tr} \left(y^{d g f i h i}\right)
    + \beta_{\lambda,154}\ \mathcal{S}_{12}\,\lambda^{abef}\,\lambda^{cegh}\,\mathrm{tr} \left(y^{d g f i i h}\right)
    \\&\phantom{=}
    + \beta_{\lambda,155}\ \mathcal{S}_{12}\,\lambda^{abef}\,\lambda^{cegh}\,\mathrm{tr} \left(y^{d g h f i i}\right)
    + \beta_{\lambda,156}\ \mathcal{S}_{12}\,\lambda^{abef}\,\lambda^{cegh}\,\mathrm{tr} \left(y^{d g i f h i}\right)
    \\&\phantom{=}
    + \beta_{\lambda,157}\ \mathcal{S}_{12}\,\lambda^{abef}\,\lambda^{cegh}\,\mathrm{tr} \left(y^{d g i f i h}\right)
    + \beta_{\lambda,158}\ \mathcal{S}_{12}\,\lambda^{abef}\,\lambda^{cegh}\,\mathrm{tr} \left(y^{d i g f h i}\right)
    \\&\phantom{=}
    + \beta_{\lambda,159}\ \mathcal{S}_{24}\,\lambda^{aefg}\,\lambda^{befh}\,\mathrm{tr} \left(y^{c d g h i i}\right)
    + \beta_{\lambda,160}\ \mathcal{S}_{24}\,\lambda^{aefg}\,\lambda^{befh}\,\mathrm{tr} \left(y^{c d g i h i}\right)
    \\&\phantom{=}
    + \beta_{\lambda,161}\ \mathcal{S}_{12}\,\lambda^{aefg}\,\lambda^{befh}\,\mathrm{tr} \left(y^{c d g i i h}\right)
    + \beta_{\lambda,162}\ \mathcal{S}_{12}\,\lambda^{aefg}\,\lambda^{befh}\,\mathrm{tr} \left(y^{c d i g h i}\right)
    \\&\phantom{=}
    + \beta_{\lambda,163}\ \mathcal{S}_{24}\,\lambda^{aefg}\,\lambda^{befh}\,\mathrm{tr} \left(y^{c i d g h i}\right)
    + \beta_{\lambda,164}\ \mathcal{S}_{12}\,\lambda^{aefg}\,\lambda^{befh}\,\mathrm{tr} \left(y^{c i d g i h}\right)
    \\&\phantom{=}
    + \beta_{\lambda,165}\ \mathcal{S}_{24}\,\lambda^{aefg}\,\lambda^{befh}\,\mathrm{tr} \left(y^{c g d h i i}\right)
    + \beta_{\lambda,166}\ \mathcal{S}_{12}\,\lambda^{cefg}\,\lambda^{defh}\,\mathrm{tr} \left(y^{a g b i h i}\right)
    \\&\phantom{=}
    + \beta_{\lambda,167}\ \mathcal{S}_{12}\,\lambda^{aefg}\,\lambda^{befh}\,\mathrm{tr} \left(y^{c g h d i i}\right)
    + \beta_{\lambda,168}\ \mathcal{S}_{12}\,\lambda^{aefg}\,\lambda^{befh}\,\mathrm{tr} \left(y^{c g i d h i}\right)
    \\&\phantom{=}
    + \beta_{\lambda,169}\ \mathcal{S}_{12}\,\lambda^{aefg}\,\lambda^{befh}\,\mathrm{tr} \left(y^{c g i d i h}\right)
    + \beta_{\lambda,170}\ \mathcal{S}_{12}\,\lambda^{aefg}\,\lambda^{behi}\,\mathrm{tr} \left(y^{c d f g h i}\right)
    \\&\phantom{=}
    + \beta_{\lambda,171}\ \mathcal{S}_{12}\,\lambda^{aefg}\,\lambda^{behi}\,\mathrm{tr} \left(y^{c d f h g i}\right)
    + \beta_{\lambda,172}\ \mathcal{S}_{12}\,\lambda^{cefg}\,\lambda^{dehi}\,\mathrm{tr} \left(y^{a b f h i g}\right)
    \\&\phantom{=}
    + \beta_{\lambda,173}\ \mathcal{S}_{24}\,\lambda^{aefg}\,\lambda^{behi}\,\mathrm{tr} \left(y^{c f d h i g}\right)
    + \beta_{\lambda,174}\ \mathcal{S}_{12}\,\lambda^{cefg}\,\lambda^{dehi}\,\mathrm{tr} \left(y^{a f b h g i}\right)
    \\&\phantom{=}
    + \beta_{\lambda,175}\ \mathcal{S}_{6}\,\lambda^{aefg}\,\lambda^{behi}\,\mathrm{tr} \left(y^{c f g d h i}\right)
    + \beta_{\lambda,176}\ \mathcal{S}_{6}\,\lambda^{aefg}\,\lambda^{behi}\,\mathrm{tr} \left(y^{c f h d g i}\right)
    \\&\phantom{=}
    + \beta_{\lambda,177}\ \mathcal{S}_{12}\,\lambda^{aefg}\,\lambda^{behi}\,\mathrm{tr} \left(y^{c f h d i g}\right)
    + \beta_{\lambda,178}\ \mathcal{S}_{12}\,\lambda^{abef}\,\lambda^{efgh}\,\mathrm{tr} \left(y^{c d g h i i}\right)
    \\&\phantom{=}
    + \beta_{\lambda,179}\ \mathcal{S}_{12}\,\lambda^{abef}\,\lambda^{efgh}\,\mathrm{tr} \left(y^{c d g i h i}\right)
    + \beta_{\lambda,180}\ \mathcal{S}_{6}\,\lambda^{abef}\,\lambda^{efgh}\,\mathrm{tr} \left(y^{c d g i i h}\right)
    \\&\phantom{=}
    + \beta_{\lambda,181}\ \mathcal{S}_{6}\,\lambda^{abef}\,\lambda^{efgh}\,\mathrm{tr} \left(y^{c d i g h i}\right)
    + \beta_{\lambda,182}\ \mathcal{S}_{12}\,\lambda^{abef}\,\lambda^{efgh}\,\mathrm{tr} \left(y^{c i d g h i}\right)
    \\&\phantom{=}
    + \beta_{\lambda,183}\ \mathcal{S}_{6}\,\lambda^{abef}\,\lambda^{efgh}\,\mathrm{tr} \left(y^{c i d g i h}\right)
    + \beta_{\lambda,184}\ \mathcal{S}_{12}\,\lambda^{abef}\,\lambda^{efgh}\,\mathrm{tr} \left(y^{c g d h i i}\right)
    \\&\phantom{=}
    + \beta_{\lambda,185}\ \mathcal{S}_{6}\,\lambda^{abef}\,\lambda^{efgh}\,\mathrm{tr} \left(y^{c g d i h i}\right)
    + \beta_{\lambda,186}\ \mathcal{S}_{6}\,\lambda^{abef}\,\lambda^{efgh}\,\mathrm{tr} \left(y^{c g h d i i}\right)
    \\&\phantom{=}
    + \beta_{\lambda,187}\ \mathcal{S}_{6}\,\lambda^{abef}\,\lambda^{efgh}\,\mathrm{tr} \left(y^{c g i d h i}\right)
    + \beta_{\lambda,188}\ \mathcal{S}_{12}\,\lambda^{abef}\,\lambda^{efgh}\,\mathrm{tr} \left(y^{c g i d i h}\right)
    \\&\phantom{=}
    + \beta_{\lambda,189}\ \mathcal{S}_{12}\,\lambda^{abef}\,\lambda^{eghi}\,\mathrm{tr} \left(y^{c d f g h i}\right)
    + \beta_{\lambda,190}\ \mathcal{S}_{12}\,\lambda^{abef}\,\lambda^{eghi}\,\mathrm{tr} \left(y^{c d g f h i}\right)
    \\&\phantom{=}
    + \beta_{\lambda,191}\ \mathcal{S}_{6}\,\lambda^{abef}\,\lambda^{eghi}\,\mathrm{tr} \left(y^{c f d g h i}\right)
    + \beta_{\lambda,192}\ \mathcal{S}_{12}\,\lambda^{abef}\,\lambda^{eghi}\,\mathrm{tr} \left(y^{c g d f h i}\right)
    \\&\phantom{=}
    + \beta_{\lambda,193}\ \mathcal{S}_{6}\,\lambda^{abef}\,\lambda^{eghi}\,\mathrm{tr} \left(y^{c g d h f i}\right)
    + \beta_{\lambda,194}\ \mathcal{S}_{12}\,\lambda^{abef}\,\lambda^{eghi}\,\mathrm{tr} \left(y^{c f g d h i}\right)
    \\&\phantom{=}
    + \beta_{\lambda,195}\ \mathcal{S}_{24}\,\lambda^{cefg}\,\lambda^{efhi}\,\mathrm{tr} \left(y^{g h i a d b}\right)
    + \beta_{\lambda,196}\ \mathcal{S}_{24}\,\lambda^{cefg}\,\lambda^{efhi}\,\mathrm{tr} \left(y^{g h a i d b}\right)
    \\&\phantom{=}
    + \beta_{\lambda,197}\ \mathcal{S}_{24}\,\lambda^{cefg}\,\lambda^{efhi}\,\mathrm{tr} \left(y^{g h a d i b}\right)
    + \beta_{\lambda,198}\ \mathcal{S}_{12}\,\lambda^{cefg}\,\lambda^{efhi}\,\mathrm{tr} \left(y^{g h a d b i}\right)
    \\&\phantom{=}
    + \beta_{\lambda,199}\ \mathcal{S}_{12}\,\lambda^{cefg}\,\lambda^{efhi}\,\mathrm{tr} \left(y^{g a h d i b}\right)
    + \beta_{\lambda,200}\ \mathcal{S}_{24}\,\lambda^{cefg}\,\lambda^{efhi}\,\mathrm{tr} \left(y^{g a h i d b}\right)
    \\&\phantom{=}
    + \beta_{\lambda,201}\ \mathcal{S}_{12}\,\lambda^{efgh}\,\lambda^{efgi}\,\mathrm{tr} \left(y^{h i a c d b}\right)
    + \beta_{\lambda,202}\ \mathcal{S}_{12}\,\lambda^{efgh}\,\lambda^{efgi}\,\mathrm{tr} \left(y^{h c i a d b}\right)
    \\&\phantom{=}
    + \beta_{\lambda,203}\ \mathcal{S}_{6}\,\lambda^{efgh}\,\lambda^{efgi}\,\mathrm{tr} \left(y^{h a b i c d}\right)\,,\yestag{}\\[.5em]
    \beta_{\lambda, \lambda y^4 y^2 y^2 }^{(4)\,abcd} &= 
    \beta_{\lambda,204}\ \mathcal{S}_{6}\,\lambda^{abef}\,\mathrm{tr} \left(y^{e g}\right)\,\mathrm{tr} \left(y^{f h}\right)\,\mathrm{tr} \left(y^{c d g h}\right)
    \\&\phantom{=}
    + \beta_{\lambda,205}\ \mathcal{S}_{6}\,\lambda^{abef}\,\mathrm{tr} \left(y^{e g}\right)\,\mathrm{tr} \left(y^{f h}\right)\,\mathrm{tr} \left(y^{c g d h}\right)
    \\&\phantom{=}
    + \beta_{\lambda,206}\ \mathcal{S}_{12}\,\lambda^{abef}\,\mathrm{tr} \left(y^{e g}\right)\,\mathrm{tr} \left(y^{g h}\right)\,\mathrm{tr} \left(y^{c d f h}\right)
    \\&\phantom{=}
    + \beta_{\lambda,207}\ \mathcal{S}_{6}\,\lambda^{abef}\,\mathrm{tr} \left(y^{e g}\right)\,\mathrm{tr} \left(y^{g h}\right)\,\mathrm{tr} \left(y^{c f d h}\right)\,,\yestag{}\\[.5em]
    \beta_{\lambda, \lambda y^4 y^4 }^{(4)\,abcd} &= 
    \beta_{\lambda,208}\ \mathcal{S}_{6}\,\lambda^{abef}\,\mathrm{tr} \left(y^{e f g h}\right)\,\mathrm{tr} \left(y^{c d g h}\right)
    + \beta_{\lambda,209}\ \mathcal{S}_{6}\,\lambda^{abef}\,\mathrm{tr} \left(y^{e g f h}\right)\,\mathrm{tr} \left(y^{c d g h}\right)
    \\&\phantom{=}
    + \beta_{\lambda,210}\ \mathcal{S}_{6}\,\lambda^{abef}\,\mathrm{tr} \left(y^{e f g h}\right)\,\mathrm{tr} \left(y^{c g d h}\right)
    + \beta_{\lambda,211}\ \mathcal{S}_{6}\,\lambda^{abef}\,\mathrm{tr} \left(y^{e g f h}\right)\,\mathrm{tr} \left(y^{c g d h}\right)
    \\&\phantom{=}
    + \beta_{\lambda,212}\ \mathcal{S}_{12}\,\lambda^{abef}\,\mathrm{tr} \left(y^{e g g h}\right)\,\mathrm{tr} \left(y^{c d f h}\right)
    + \beta_{\lambda,213}\ \mathcal{S}_{12}\,\lambda^{abef}\,\mathrm{tr} \left(y^{e g h g}\right)\,\mathrm{tr} \left(y^{c d f h}\right)
    \\&\phantom{=}
    + \beta_{\lambda,214}\ \mathcal{S}_{6}\,\lambda^{abef}\,\mathrm{tr} \left(y^{e g g h}\right)\,\mathrm{tr} \left(y^{c f d h}\right)
    + \beta_{\lambda,215}\ \mathcal{S}_{6}\,\lambda^{abef}\,\mathrm{tr} \left(y^{e g h g}\right)\,\mathrm{tr} \left(y^{c f d h}\right)
    \\&\phantom{=}
    + \beta_{\lambda,216}\ \mathcal{S}_{6}\,\lambda^{abef}\,\mathrm{tr} \left(y^{c e g h}\right)\,\mathrm{tr} \left(y^{d f g h}\right)
    + \beta_{\lambda,217}\ \mathcal{S}_{6}\,\lambda^{abef}\,\mathrm{tr} \left(y^{c e g h}\right)\,\mathrm{tr} \left(y^{d f h g}\right)
    \\&\phantom{=}
    + \beta_{\lambda,218}\ \mathcal{S}_{6}\,\lambda^{abef}\,\mathrm{tr} \left(y^{c g e h}\right)\,\mathrm{tr} \left(y^{d g f h}\right)
    + \beta_{\lambda,219}\ \mathcal{S}_{12}\,\lambda^{abef}\,\mathrm{tr} \left(y^{c e g h}\right)\,\mathrm{tr} \left(y^{d g f h}\right)
    \\&\phantom{=}
    + \beta_{\lambda,220}\ \mathcal{S}_{24}\,\lambda^{cefg}\,\mathrm{tr} \left(y^{a b e h}\right)\,\mathrm{tr} \left(y^{d f g h}\right)
    + \beta_{\lambda,221}\ \mathcal{S}_{12}\,\lambda^{cefg}\,\mathrm{tr} \left(y^{a e b h}\right)\,\mathrm{tr} \left(y^{d f g h}\right)
    \\&\phantom{=}
    + \beta_{\lambda,222}\ \mathcal{S}_{24}\,\lambda^{cefg}\,\mathrm{tr} \left(y^{a b e h}\right)\,\mathrm{tr} \left(y^{d f h g}\right)
    + \beta_{\lambda,223}\ \mathcal{S}_{12}\,\lambda^{cefg}\,\mathrm{tr} \left(y^{a e b h}\right)\,\mathrm{tr} \left(y^{d f h g}\right)
    \\&\phantom{=}
    + \beta_{\lambda,224}\ \mathcal{S}_{3}\,\mathrm{tr} \left(y^{a b e f}\right)\,\lambda^{efgh}\,\mathrm{tr} \left(y^{c d g h}\right)
    + \beta_{\lambda,225}\ \mathcal{S}_{6}\,\mathrm{tr} \left(y^{a b e f}\right)\,\lambda^{efgh}\,\mathrm{tr} \left(y^{c g d h}\right)
    \\&\phantom{=}
    + \beta_{\lambda,226}\ \mathcal{S}_{3}\,\mathrm{tr} \left(y^{a e b f}\right)\,\lambda^{efgh}\,\mathrm{tr} \left(y^{c g d h}\right)\,,\yestag{}\\[.5em]
    \beta_{\lambda, \lambda y^6 y^2 }^{(4)\,abcd} &= 
    \beta_{\lambda,227}\ \mathcal{S}_{12}\,\lambda^{abef}\,\mathrm{tr} \left(y^{g h}\right)\,\mathrm{tr} \left(y^{c d e f g h}\right)
    + \beta_{\lambda,228}\ \mathcal{S}_{6}\,\lambda^{abef}\,\mathrm{tr} \left(y^{g h}\right)\,\mathrm{tr} \left(y^{c d e g h f}\right)
    \\&\phantom{=}
    + \beta_{\lambda,229}\ \mathcal{S}_{6}\,\lambda^{abef}\,\mathrm{tr} \left(y^{g h}\right)\,\mathrm{tr} \left(y^{c d g e f h}\right)
    + \beta_{\lambda,230}\ \mathcal{S}_{12}\,\lambda^{abef}\,\mathrm{tr} \left(y^{g h}\right)\,\mathrm{tr} \left(y^{c d e g f h}\right)
    \\&\phantom{=}
    + \beta_{\lambda,231}\ \mathcal{S}_{6}\,\lambda^{abef}\,\mathrm{tr} \left(y^{g h}\right)\,\mathrm{tr} \left(y^{c e d g f h}\right)
    + \beta_{\lambda,232}\ \mathcal{S}_{6}\,\lambda^{abef}\,\mathrm{tr} \left(y^{g h}\right)\,\mathrm{tr} \left(y^{c g d e h f}\right)
    \\&\phantom{=}
    + \beta_{\lambda,233}\ \mathcal{S}_{12}\,\lambda^{abef}\,\mathrm{tr} \left(y^{g h}\right)\,\mathrm{tr} \left(y^{c e d f g h}\right)
    + \beta_{\lambda,234}\ \mathcal{S}_{12}\,\lambda^{abef}\,\mathrm{tr} \left(y^{g h}\right)\,\mathrm{tr} \left(y^{c g d h e f}\right)
    \\&\phantom{=}
    + \beta_{\lambda,235}\ \mathcal{S}_{6}\,\lambda^{abef}\,\mathrm{tr} \left(y^{g h}\right)\,\mathrm{tr} \left(y^{c e f d g h}\right)
    + \beta_{\lambda,236}\ \mathcal{S}_{6}\,\lambda^{abef}\,\mathrm{tr} \left(y^{g h}\right)\,\mathrm{tr} \left(y^{c e g d f h}\right)
    \\&\phantom{=}
    + \beta_{\lambda,237}\ \mathcal{S}_{12}\,\lambda^{abef}\,\mathrm{tr} \left(y^{g h}\right)\,\mathrm{tr} \left(y^{c e g d h f}\right)
    + \beta_{\lambda,238}\ \mathcal{S}_{12}\,\lambda^{abef}\,\mathrm{tr} \left(y^{e g}\right)\,\mathrm{tr} \left(y^{c d f g h h}\right)
    \\&\phantom{=}
    + \beta_{\lambda,239}\ \mathcal{S}_{12}\,\lambda^{abef}\,\mathrm{tr} \left(y^{e g}\right)\,\mathrm{tr} \left(y^{c d g f h h}\right)
    + \beta_{\lambda,240}\ \mathcal{S}_{12}\,\lambda^{abef}\,\mathrm{tr} \left(y^{e g}\right)\,\mathrm{tr} \left(y^{c d h f g h}\right)
    \\&\phantom{=}
    + \beta_{\lambda,241}\ \mathcal{S}_{12}\,\lambda^{abef}\,\mathrm{tr} \left(y^{e g}\right)\,\mathrm{tr} \left(y^{c d f h h g}\right)
    + \beta_{\lambda,242}\ \mathcal{S}_{12}\,\lambda^{abef}\,\mathrm{tr} \left(y^{e g}\right)\,\mathrm{tr} \left(y^{c d h f h g}\right)
    \\&\phantom{=}
    + \beta_{\lambda,243}\ \mathcal{S}_{12}\,\lambda^{abef}\,\mathrm{tr} \left(y^{e g}\right)\,\mathrm{tr} \left(y^{c d h g h f}\right)
    + \beta_{\lambda,244}\ \mathcal{S}_{6}\,\lambda^{abef}\,\mathrm{tr} \left(y^{e g}\right)\,\mathrm{tr} \left(y^{c f d h g h}\right)
    \\&\phantom{=}
    + \beta_{\lambda,245}\ \mathcal{S}_{6}\,\lambda^{abef}\,\mathrm{tr} \left(y^{e g}\right)\,\mathrm{tr} \left(y^{c g d h f h}\right)
    + \beta_{\lambda,246}\ \mathcal{S}_{12}\,\lambda^{abef}\,\mathrm{tr} \left(y^{e g}\right)\,\mathrm{tr} \left(y^{c h d f h g}\right)
    \\&\phantom{=}
    + \beta_{\lambda,247}\ \mathcal{S}_{12}\,\lambda^{abef}\,\mathrm{tr} \left(y^{e g}\right)\,\mathrm{tr} \left(y^{c f d g h h}\right)
    + \beta_{\lambda,248}\ \mathcal{S}_{12}\,\lambda^{abef}\,\mathrm{tr} \left(y^{e g}\right)\,\mathrm{tr} \left(y^{c g d f h h}\right)
    \\&\phantom{=}
    + \beta_{\lambda,249}\ \mathcal{S}_{12}\,\lambda^{abef}\,\mathrm{tr} \left(y^{e g}\right)\,\mathrm{tr} \left(y^{c h d f g h}\right)
    + \beta_{\lambda,250}\ \mathcal{S}_{12}\,\lambda^{abef}\,\mathrm{tr} \left(y^{e g}\right)\,\mathrm{tr} \left(y^{c h h d f g}\right)
    \\&\phantom{=}
    + \beta_{\lambda,251}\ \mathcal{S}_{12}\,\lambda^{abef}\,\mathrm{tr} \left(y^{e g}\right)\,\mathrm{tr} \left(y^{c h f d g h}\right)
    + \beta_{\lambda,252}\ \mathcal{S}_{12}\,\lambda^{abef}\,\mathrm{tr} \left(y^{e g}\right)\,\mathrm{tr} \left(y^{c f h d g h}\right)
    \\&\phantom{=}
    + \beta_{\lambda,253}\ \mathcal{S}_{12}\,\lambda^{cefg}\,\mathrm{tr} \left(y^{e h}\right)\,\mathrm{tr} \left(y^{h f a d b g}\right)
    + \beta_{\lambda,254}\ \mathcal{S}_{24}\,\lambda^{cefg}\,\mathrm{tr} \left(y^{e h}\right)\,\mathrm{tr} \left(y^{h f a d g b}\right)
    \\&\phantom{=}
    + \beta_{\lambda,255}\ \mathcal{S}_{24}\,\lambda^{cefg}\,\mathrm{tr} \left(y^{e h}\right)\,\mathrm{tr} \left(y^{h f a g d b}\right)
    + \beta_{\lambda,256}\ \mathcal{S}_{24}\,\lambda^{cefg}\,\mathrm{tr} \left(y^{e h}\right)\,\mathrm{tr} \left(y^{h f g a d b}\right)
    \\&\phantom{=}
    + \beta_{\lambda,257}\ \mathcal{S}_{12}\,\lambda^{cefg}\,\mathrm{tr} \left(y^{e h}\right)\,\mathrm{tr} \left(y^{h a f d g b}\right)
    + \beta_{\lambda,258}\ \mathcal{S}_{24}\,\lambda^{cefg}\,\mathrm{tr} \left(y^{e h}\right)\,\mathrm{tr} \left(y^{h a f g d b}\right)\,,\yestag{}\\[.5em]
    \beta_{\lambda, \lambda y^8 }^{(4)\,abcd} &= 
    \beta_{\lambda,259}\ \mathcal{S}_{12}\,\lambda^{abef}\,\mathrm{tr} \left(y^{c d e f g g h h}\right)
    + \beta_{\lambda,260}\ \mathcal{S}_{12}\,\lambda^{abef}\,\mathrm{tr} \left(y^{c d e f g h h g}\right)
    \\&\phantom{=}
    + \beta_{\lambda,261}\ \mathcal{S}_{12}\,\lambda^{abef}\,\mathrm{tr} \left(y^{c d e f g h g h}\right)
    + \beta_{\lambda,262}\ \mathcal{S}_{12}\,\lambda^{abef}\,\mathrm{tr} \left(y^{c d e g f h g h}\right)
    \\&\phantom{=}
    + \beta_{\lambda,263}\ \mathcal{S}_{12}\,\lambda^{abef}\,\mathrm{tr} \left(y^{c d e g f g h h}\right)
    + \beta_{\lambda,264}\ \mathcal{S}_{12}\,\lambda^{abef}\,\mathrm{tr} \left(y^{c d e g f h h g}\right)
    \\&\phantom{=}
    + \beta_{\lambda,265}\ \mathcal{S}_{12}\,\lambda^{abef}\,\mathrm{tr} \left(y^{c d e g g f h h}\right)
    + \beta_{\lambda,266}\ \mathcal{S}_{12}\,\lambda^{abef}\,\mathrm{tr} \left(y^{c d e g h f g h}\right)
    \\&\phantom{=}
    + \beta_{\lambda,267}\ \mathcal{S}_{12}\,\lambda^{abef}\,\mathrm{tr} \left(y^{c d e g h f h g}\right)
    + \beta_{\lambda,268}\ \mathcal{S}_{12}\,\lambda^{abef}\,\mathrm{tr} \left(y^{c d e g g h f h}\right)
    \\&\phantom{=}
    + \beta_{\lambda,269}\ \mathcal{S}_{12}\,\lambda^{abef}\,\mathrm{tr} \left(y^{c d e g h g f h}\right)
    + \beta_{\lambda,270}\ \mathcal{S}_{12}\,\lambda^{abef}\,\mathrm{tr} \left(y^{c d e h g g f h}\right)
    \\&\phantom{=}
    + \beta_{\lambda,271}\ \mathcal{S}_{6}\,\lambda^{abef}\,\mathrm{tr} \left(y^{c d e g g h h f}\right)
    + \beta_{\lambda,272}\ \mathcal{S}_{6}\,\lambda^{abef}\,\mathrm{tr} \left(y^{c d e g h h g f}\right)
    \\&\phantom{=}
    + \beta_{\lambda,273}\ \mathcal{S}_{6}\,\lambda^{abef}\,\mathrm{tr} \left(y^{c d e g h g h f}\right)
    + \beta_{\lambda,274}\ \mathcal{S}_{12}\,\lambda^{abef}\,\mathrm{tr} \left(y^{c d g e f g h h}\right)
    \\&\phantom{=}
    + \beta_{\lambda,275}\ \mathcal{S}_{12}\,\lambda^{abef}\,\mathrm{tr} \left(y^{c d g e f h g h}\right)
    + \beta_{\lambda,276}\ \mathcal{S}_{12}\,\lambda^{abef}\,\mathrm{tr} \left(y^{c d g e f h h g}\right)
    \\&\phantom{=}
    + \beta_{\lambda,277}\ \mathcal{S}_{12}\,\lambda^{abef}\,\mathrm{tr} \left(y^{c d g e g f h h}\right)
    + \beta_{\lambda,278}\ \mathcal{S}_{12}\,\lambda^{abef}\,\mathrm{tr} \left(y^{c d g e h f g h}\right)
    \\&\phantom{=}
    + \beta_{\lambda,279}\ \mathcal{S}_{12}\,\lambda^{abef}\,\mathrm{tr} \left(y^{c d g e h f h g}\right)
    + \beta_{\lambda,280}\ \mathcal{S}_{6}\,\lambda^{abef}\,\mathrm{tr} \left(y^{c d g e g h f h}\right)
    \\&\phantom{=}
    + \beta_{\lambda,281}\ \mathcal{S}_{6}\,\lambda^{abef}\,\mathrm{tr} \left(y^{c d g e h g f h}\right)
    + \beta_{\lambda,282}\ \mathcal{S}_{6}\,\lambda^{abef}\,\mathrm{tr} \left(y^{c d g e h h f g}\right)
    \\&\phantom{=}
    + \beta_{\lambda,283}\ \mathcal{S}_{6}\,\lambda^{abef}\,\mathrm{tr} \left(y^{c d g g e f h h}\right)
    + \beta_{\lambda,284}\ \mathcal{S}_{6}\,\lambda^{abef}\,\mathrm{tr} \left(y^{c d g h e f g h}\right)
    \\&\phantom{=}
    + \beta_{\lambda,285}\ \mathcal{S}_{6}\,\lambda^{abef}\,\mathrm{tr} \left(y^{c d g h e f h g}\right)
    + \beta_{\lambda,286}\ \mathcal{S}_{12}\,\lambda^{abef}\,\mathrm{tr} \left(y^{c e d f g g h h}\right)
    \\&\phantom{=}
    + \beta_{\lambda,287}\ \mathcal{S}_{12}\,\lambda^{abef}\,\mathrm{tr} \left(y^{c e d f g h g h}\right)
    + \beta_{\lambda,288}\ \mathcal{S}_{12}\,\lambda^{abef}\,\mathrm{tr} \left(y^{c e d f g h h g}\right)
    \\&\phantom{=}
    + \beta_{\lambda,289}\ \mathcal{S}_{12}\,\lambda^{abef}\,\mathrm{tr} \left(y^{c e d g f g h h}\right)
    + \beta_{\lambda,290}\ \mathcal{S}_{12}\,\lambda^{abef}\,\mathrm{tr} \left(y^{c e d g f h g h}\right)
    \\&\phantom{=}
    + \beta_{\lambda,291}\ \mathcal{S}_{12}\,\lambda^{abef}\,\mathrm{tr} \left(y^{c e d g f h h g}\right)
    + \beta_{\lambda,292}\ \mathcal{S}_{6}\,\lambda^{abef}\,\mathrm{tr} \left(y^{c e d g g f h h}\right)
    \\&\phantom{=}
    + \beta_{\lambda,293}\ \mathcal{S}_{6}\,\lambda^{abef}\,\mathrm{tr} \left(y^{c e d g h f g h}\right)
    + \beta_{\lambda,294}\ \mathcal{S}_{6}\,\lambda^{abef}\,\mathrm{tr} \left(y^{c e d g h f h g}\right)
    \\&\phantom{=}
    + \beta_{\lambda,295}\ \mathcal{S}_{12}\,\lambda^{abef}\,\mathrm{tr} \left(y^{c g d e f g h h}\right)
    + \beta_{\lambda,296}\ \mathcal{S}_{12}\,\lambda^{abef}\,\mathrm{tr} \left(y^{c g d e f h g h}\right)
    \\&\phantom{=}
    + \beta_{\lambda,297}\ \mathcal{S}_{12}\,\lambda^{abef}\,\mathrm{tr} \left(y^{c g d e f h h g}\right)
    + \beta_{\lambda,298}\ \mathcal{S}_{12}\,\lambda^{abef}\,\mathrm{tr} \left(y^{c g d e g f h h}\right)
    \\&\phantom{=}
    + \beta_{\lambda,299}\ \mathcal{S}_{12}\,\lambda^{abef}\,\mathrm{tr} \left(y^{c g d e h f g h}\right)
    + \beta_{\lambda,300}\ \mathcal{S}_{12}\,\lambda^{abef}\,\mathrm{tr} \left(y^{c g d e h f h g}\right)
    \\&\phantom{=}
    + \beta_{\lambda,301}\ \mathcal{S}_{12}\,\lambda^{abef}\,\mathrm{tr} \left(y^{c g d e g h f h}\right)
    + \beta_{\lambda,302}\ \mathcal{S}_{12}\,\lambda^{abef}\,\mathrm{tr} \left(y^{c g d e h g f h}\right)
    \\&\phantom{=}
    + \beta_{\lambda,303}\ \mathcal{S}_{12}\,\lambda^{abef}\,\mathrm{tr} \left(y^{c g d e h h f g}\right)
    + \beta_{\lambda,304}\ \mathcal{S}_{12}\,\lambda^{abef}\,\mathrm{tr} \left(y^{c g d e g h h f}\right)
    \\&\phantom{=}
    + \beta_{\lambda,305}\ \mathcal{S}_{6}\,\lambda^{abef}\,\mathrm{tr} \left(y^{c g d e h g h f}\right)
    + \beta_{\lambda,306}\ \mathcal{S}_{12}\,\lambda^{abef}\,\mathrm{tr} \left(y^{c g d g e f h h}\right)
    \\&\phantom{=}
    + \beta_{\lambda,307}\ \mathcal{S}_{12}\,\lambda^{abef}\,\mathrm{tr} \left(y^{c g d h e f g h}\right)
    + \beta_{\lambda,308}\ \mathcal{S}_{12}\,\lambda^{abef}\,\mathrm{tr} \left(y^{c g d h e f h g}\right)
    \\&\phantom{=}
    + \beta_{\lambda,309}\ \mathcal{S}_{12}\,\lambda^{abef}\,\mathrm{tr} \left(y^{c g d g e h f h}\right)
    + \beta_{\lambda,310}\ \mathcal{S}_{6}\,\lambda^{abef}\,\mathrm{tr} \left(y^{c g d h e g f h}\right)
    \\&\phantom{=}
    + \beta_{\lambda,311}\ \mathcal{S}_{6}\,\lambda^{abef}\,\mathrm{tr} \left(y^{c e f d g g h h}\right)
    + \beta_{\lambda,312}\ \mathcal{S}_{6}\,\lambda^{abef}\,\mathrm{tr} \left(y^{c e f d g h g h}\right)
    \\&\phantom{=}
    + \beta_{\lambda,313}\ \mathcal{S}_{6}\,\lambda^{abef}\,\mathrm{tr} \left(y^{c e f d g h h g}\right)
    + \beta_{\lambda,314}\ \mathcal{S}_{12}\,\lambda^{abef}\,\mathrm{tr} \left(y^{c e g d f g h h}\right)
    \\&\phantom{=}
    + \beta_{\lambda,315}\ \mathcal{S}_{12}\,\lambda^{abef}\,\mathrm{tr} \left(y^{c e g d f h g h}\right)
    + \beta_{\lambda,316}\ \mathcal{S}_{12}\,\lambda^{abef}\,\mathrm{tr} \left(y^{c e g d f h h g}\right)
    \\&\phantom{=}
    + \beta_{\lambda,317}\ \mathcal{S}_{12}\,\lambda^{abef}\,\mathrm{tr} \left(y^{c e g d g f h h}\right)
    + \beta_{\lambda,318}\ \mathcal{S}_{12}\,\lambda^{abef}\,\mathrm{tr} \left(y^{c e g d h f g h}\right)
    \\&\phantom{=}
    + \beta_{\lambda,319}\ \mathcal{S}_{12}\,\lambda^{abef}\,\mathrm{tr} \left(y^{c e g d h f h g}\right)
    + \beta_{\lambda,320}\ \mathcal{S}_{12}\,\lambda^{abef}\,\mathrm{tr} \left(y^{c e g d g h f h}\right)
    \\&\phantom{=}
    + \beta_{\lambda,321}\ \mathcal{S}_{12}\,\lambda^{abef}\,\mathrm{tr} \left(y^{c e g d h g f h}\right)
    + \beta_{\lambda,322}\ \mathcal{S}_{12}\,\lambda^{abef}\,\mathrm{tr} \left(y^{c e g d h h f g}\right)
    \\&\phantom{=}
    + \beta_{\lambda,323}\ \mathcal{S}_{12}\,\lambda^{abef}\,\mathrm{tr} \left(y^{c e g d g h h f}\right)
    + \beta_{\lambda,324}\ \mathcal{S}_{12}\,\lambda^{abef}\,\mathrm{tr} \left(y^{c e g d h g h f}\right)
    \\&\phantom{=}
    + \beta_{\lambda,325}\ \mathcal{S}_{12}\,\lambda^{abef}\,\mathrm{tr} \left(y^{c e g d h h g f}\right)
    + \beta_{\lambda,326}\ \mathcal{S}_{12}\,\lambda^{abef}\,\mathrm{tr} \left(y^{c g g d e f h h}\right)
    \\&\phantom{=}
    + \beta_{\lambda,327}\ \mathcal{S}_{12}\,\lambda^{abef}\,\mathrm{tr} \left(y^{c g g d e h f h}\right)
    + \beta_{\lambda,328}\ \mathcal{S}_{6}\,\lambda^{abef}\,\mathrm{tr} \left(y^{c g g d e h h f}\right)
    \\&\phantom{=}
    + \beta_{\lambda,329}\ \mathcal{S}_{6}\,\lambda^{abef}\,\mathrm{tr} \left(y^{c g g d h e f h}\right)
    + \beta_{\lambda,330}\ \mathcal{S}_{12}\,\lambda^{abef}\,\mathrm{tr} \left(y^{c g h d e f g h}\right)
    \\&\phantom{=}
    + \beta_{\lambda,331}\ \mathcal{S}_{12}\,\lambda^{abef}\,\mathrm{tr} \left(y^{c g h d e f h g}\right)
    + \beta_{\lambda,332}\ \mathcal{S}_{12}\,\lambda^{abef}\,\mathrm{tr} \left(y^{c g h d e g f h}\right)
    \\&\phantom{=}
    + \beta_{\lambda,333}\ \mathcal{S}_{12}\,\lambda^{abef}\,\mathrm{tr} \left(y^{c g h d e h f g}\right)
    + \beta_{\lambda,334}\ \mathcal{S}_{6}\,\lambda^{abef}\,\mathrm{tr} \left(y^{c g h d e g h f}\right)
    \\&\phantom{=}
    + \beta_{\lambda,335}\ \mathcal{S}_{6}\,\lambda^{abef}\,\mathrm{tr} \left(y^{c g h d e h g f}\right)
    + \beta_{\lambda,336}\ \mathcal{S}_{6}\,\lambda^{abef}\,\mathrm{tr} \left(y^{c g h d g e f h}\right)
    \\&\phantom{=}
    + \beta_{\lambda,337}\ \mathcal{S}_{6}\,\lambda^{abef}\,\mathrm{tr} \left(y^{c g h d h e f g}\right)
    + \beta_{\lambda,338}\ \mathcal{S}_{12}\,\lambda^{abef}\,\mathrm{tr} \left(y^{c e f g d g h h}\right)
    \\&\phantom{=}
    + \beta_{\lambda,339}\ \mathcal{S}_{12}\,\lambda^{abef}\,\mathrm{tr} \left(y^{c e f g d h g h}\right)
    + \beta_{\lambda,340}\ \mathcal{S}_{12}\,\lambda^{abef}\,\mathrm{tr} \left(y^{c e f g d h h g}\right)
    \\&\phantom{=}
    + \beta_{\lambda,341}\ \mathcal{S}_{12}\,\lambda^{abef}\,\mathrm{tr} \left(y^{c e g f d g h h}\right)
    + \beta_{\lambda,342}\ \mathcal{S}_{6}\,\lambda^{abef}\,\mathrm{tr} \left(y^{c e g f d h g h}\right)
    \\&\phantom{=}
    + \beta_{\lambda,343}\ \mathcal{S}_{6}\,\lambda^{abef}\,\mathrm{tr} \left(y^{c e g g d f h h}\right)
    + \beta_{\lambda,344}\ \mathcal{S}_{6}\,\lambda^{abef}\,\mathrm{tr} \left(y^{c e g h d f g h}\right)
    \\&\phantom{=}
    + \beta_{\lambda,345}\ \mathcal{S}_{6}\,\lambda^{abef}\,\mathrm{tr} \left(y^{c e g h d f h g}\right)
    + \beta_{\lambda,346}\ \mathcal{S}_{12}\,\lambda^{abef}\,\mathrm{tr} \left(y^{c e g g d h f h}\right)
    \\&\phantom{=}
    + \beta_{\lambda,347}\ \mathcal{S}_{12}\,\lambda^{abef}\,\mathrm{tr} \left(y^{c e g h d g f h}\right)
    + \beta_{\lambda,348}\ \mathcal{S}_{12}\,\lambda^{abef}\,\mathrm{tr} \left(y^{c e g h d h f g}\right)
    \\&\phantom{=}
    + \beta_{\lambda,349}\ \mathcal{S}_{12}\,\lambda^{abef}\,\mathrm{tr} \left(y^{c e g g d h h f}\right)
    + \beta_{\lambda,350}\ \mathcal{S}_{12}\,\lambda^{abef}\,\mathrm{tr} \left(y^{c e g h d g h f}\right)
    \\&\phantom{=}
    + \beta_{\lambda,351}\ \mathcal{S}_{12}\,\lambda^{abef}\,\mathrm{tr} \left(y^{c e g h d h g f}\right)
    + \beta_{\lambda,352}\ \mathcal{S}_{6}\,\lambda^{abef}\,\mathrm{tr} \left(y^{c g e g d h f h}\right)
    \\&\phantom{=}
    + \beta_{\lambda,353}\ \mathcal{S}_{6}\,\lambda^{abef}\,\mathrm{tr} \left(y^{c g e h d g f h}\right)
    + \beta_{\lambda,354}\ \mathcal{S}_{6}\,\lambda^{abef}\,\mathrm{tr} \left(y^{c g e h d h f g}\right)
    \\&\phantom{=}
    + \beta_{\lambda,355}\ \mathcal{S}_{24}\,\lambda^{cefg}\,\mathrm{tr} \left(y^{a d b h h e f g}\right)
    + \beta_{\lambda,356}\ \mathcal{S}_{24}\,\lambda^{cefg}\,\mathrm{tr} \left(y^{a d b h e h f g}\right)
    \\&\phantom{=}
    + \beta_{\lambda,357}\ \mathcal{S}_{24}\,\lambda^{cefg}\,\mathrm{tr} \left(y^{a d b h e f h g}\right)
    + \beta_{\lambda,358}\ \mathcal{S}_{12}\,\lambda^{cefg}\,\mathrm{tr} \left(y^{a d b h e f g h}\right)
    \\&\phantom{=}
    + \beta_{\lambda,359}\ \mathcal{S}_{24}\,\lambda^{cefg}\,\mathrm{tr} \left(y^{a d b e h h f g}\right)
    + \beta_{\lambda,360}\ \mathcal{S}_{12}\,\lambda^{cefg}\,\mathrm{tr} \left(y^{a d b e h f h g}\right)
    \\&\phantom{=}
    + \beta_{\lambda,361}\ \mathcal{S}_{24}\,\lambda^{cefg}\,\mathrm{tr} \left(y^{a d h b h e f g}\right)
    + \beta_{\lambda,362}\ \mathcal{S}_{24}\,\lambda^{cefg}\,\mathrm{tr} \left(y^{a d h b e h f g}\right)
    \\&\phantom{=}
    + \beta_{\lambda,363}\ \mathcal{S}_{24}\,\lambda^{cefg}\,\mathrm{tr} \left(y^{a d h b e f h g}\right)
    + \beta_{\lambda,364}\ \mathcal{S}_{24}\,\lambda^{cefg}\,\mathrm{tr} \left(y^{a d h b e f g h}\right)
    \\&\phantom{=}
    + \beta_{\lambda,365}\ \mathcal{S}_{24}\,\lambda^{cefg}\,\mathrm{tr} \left(y^{a d e b h h f g}\right)
    + \beta_{\lambda,366}\ \mathcal{S}_{24}\,\lambda^{cefg}\,\mathrm{tr} \left(y^{a d e b h f h g}\right)
    \\&\phantom{=}
    + \beta_{\lambda,367}\ \mathcal{S}_{24}\,\lambda^{cefg}\,\mathrm{tr} \left(y^{a d e b h f g h}\right)
    + \beta_{\lambda,368}\ \mathcal{S}_{24}\,\lambda^{cefg}\,\mathrm{tr} \left(y^{a d e b f h h g}\right)
    \\&\phantom{=}
    + \beta_{\lambda,369}\ \mathcal{S}_{24}\,\lambda^{cefg}\,\mathrm{tr} \left(y^{a d e b f h g h}\right)
    + \beta_{\lambda,370}\ \mathcal{S}_{24}\,\lambda^{cefg}\,\mathrm{tr} \left(y^{a d e b f g h h}\right)
    \\&\phantom{=}
    + \beta_{\lambda,371}\ \mathcal{S}_{24}\,\lambda^{cefg}\,\mathrm{tr} \left(y^{a d h h b e f g}\right)
    + \beta_{\lambda,372}\ \mathcal{S}_{24}\,\lambda^{cefg}\,\mathrm{tr} \left(y^{a d e h b h f g}\right)
    \\&\phantom{=}
    + \beta_{\lambda,373}\ \mathcal{S}_{24}\,\lambda^{cefg}\,\mathrm{tr} \left(y^{a d e h b f h g}\right)
    + \beta_{\lambda,374}\ \mathcal{S}_{24}\,\lambda^{cefg}\,\mathrm{tr} \left(y^{a d e h b f g h}\right)
    \\&\phantom{=}
    + \beta_{\lambda,375}\ \mathcal{S}_{24}\,\lambda^{cefg}\,\mathrm{tr} \left(y^{a d h e b h f g}\right)
    + \beta_{\lambda,376}\ \mathcal{S}_{24}\,\lambda^{cefg}\,\mathrm{tr} \left(y^{a d h e b f h g}\right)
    \\&\phantom{=}
    + \beta_{\lambda,377}\ \mathcal{S}_{24}\,\lambda^{cefg}\,\mathrm{tr} \left(y^{a d h e b f g h}\right)
    + \beta_{\lambda,378}\ \mathcal{S}_{24}\,\lambda^{cefg}\,\mathrm{tr} \left(y^{a d e f b g h h}\right)
    \\&\phantom{=}
    + \beta_{\lambda,379}\ \mathcal{S}_{24}\,\lambda^{cefg}\,\mathrm{tr} \left(y^{a d e f b h g h}\right)
    + \beta_{\lambda,380}\ \mathcal{S}_{24}\,\lambda^{cefg}\,\mathrm{tr} \left(y^{a d e f b h h g}\right)
    \\&\phantom{=}
    + \beta_{\lambda,381}\ \mathcal{S}_{24}\,\lambda^{cefg}\,\mathrm{tr} \left(y^{a e d f b g h h}\right)
    + \beta_{\lambda,382}\ \mathcal{S}_{12}\,\lambda^{cefg}\,\mathrm{tr} \left(y^{a e d f b h g h}\right)
    \\&\phantom{=}
    + \beta_{\lambda,383}\ \mathcal{S}_{12}\,\lambda^{cefg}\,\mathrm{tr} \left(y^{a h d h b e f g}\right)
    + \beta_{\lambda,384}\ \mathcal{S}_{24}\,\lambda^{cefg}\,\mathrm{tr} \left(y^{a e d h b f g h}\right)
    \\&\phantom{=}
    + \beta_{\lambda,385}\ \mathcal{S}_{24}\,\lambda^{cefg}\,\mathrm{tr} \left(y^{a e d h b f h g}\right)
    + \beta_{\lambda,386}\ \mathcal{S}_{24}\,\lambda^{cefg}\,\mathrm{tr} \left(y^{a e d h b h f g}\right)
    \\&\phantom{=}
    + \beta_{\lambda,387}\ \mathcal{S}_{24}\,\lambda^{cefg}\,\mathrm{tr} \left(y^{a e b f g d h h}\right)
    + \beta_{\lambda,388}\ \mathcal{S}_{24}\,\lambda^{cefg}\,\mathrm{tr} \left(y^{a e b f h d g h}\right)
    \\&\phantom{=}
    + \beta_{\lambda,389}\ \mathcal{S}_{12}\,\lambda^{cefg}\,\mathrm{tr} \left(y^{a e b h f d g h}\right)
    + \beta_{\lambda,390}\ \mathcal{S}_{12}\,\lambda^{cefg}\,\mathrm{tr} \left(y^{a e b f h d h g}\right)
    \\&\phantom{=}
    + \beta_{\lambda,391}\ \mathcal{S}_{24}\,\lambda^{cefg}\,\mathrm{tr} \left(y^{a h b h e d f g}\right)
    + \beta_{\lambda,392}\ \mathcal{S}_{24}\,\lambda^{cefg}\,\mathrm{tr} \left(y^{a h b e h d f g}\right)
    \\&\phantom{=}
    + \beta_{\lambda,393}\ \mathcal{S}_{12}\,\lambda^{efgh}\,\mathrm{tr} \left(y^{a c d b e f g h}\right)
    + \beta_{\lambda,394}\ \mathcal{S}_{24}\,\lambda^{efgh}\,\mathrm{tr} \left(y^{a c b e d f g h}\right)
    \\&\phantom{=}
    + \beta_{\lambda,395}\ \mathcal{S}_{12}\,\lambda^{efgh}\,\mathrm{tr} \left(y^{a c b e f d g h}\right)
    + \beta_{\lambda,396}\ \mathcal{S}_{12}\,\lambda^{efgh}\,\mathrm{tr} \left(y^{a c e d b f g h}\right)
    \\&\phantom{=}
    + \beta_{\lambda,397}\ \mathcal{S}_{6}\,\lambda^{efgh}\,\mathrm{tr} \left(y^{a b e f c d g h}\right)
    + \beta_{\lambda,398}\ \mathcal{S}_{12}\,\lambda^{efgh}\,\mathrm{tr} \left(y^{a b e d f g c h}\right)
    \\&\phantom{=}
    + \beta_{\lambda,399}\ \mathcal{S}_{24}\,\lambda^{efgh}\,\mathrm{tr} \left(y^{a b e c f d g h}\right)
    + \beta_{\lambda,400}\ \mathcal{S}_{3}\,\lambda^{efgh}\,\mathrm{tr} \left(y^{a e c f b g d h}\right)\,,\yestag{}\\[.5em]
    \beta_{\lambda, y^6 y^2 y^2}^{(4)\,abcd} &= 
    \beta_{\lambda,401}\ \mathcal{S}_{12}\,\mathrm{tr} \left(y^{c a b d e f}\right)\,\mathrm{tr} \left(y^{e g}\right)\,\mathrm{tr} \left(y^{g f}\right)
    \\&\phantom{=}
    + \beta_{\lambda,402}\ \mathcal{S}_{24}\,\mathrm{tr} \left(y^{c a b e d f}\right)\,\mathrm{tr} \left(y^{e g}\right)\,\mathrm{tr} \left(y^{g f}\right)
    \\&\phantom{=}
    + \beta_{\lambda,403}\ \mathcal{S}_{6}\,\mathrm{tr} \left(y^{a b e c d f}\right)\,\mathrm{tr} \left(y^{e g}\right)\,\mathrm{tr} \left(y^{g f}\right)\,,\yestag{}\\[.5em]
    \beta_{\lambda, y^6 y^4}^{(4)\,abcd} &= 
    \beta_{\lambda,404}\ \mathcal{S}_{12}\,\mathrm{tr} \left(y^{c a b d e f}\right)\,\mathrm{tr} \left(y^{e g g f}\right)
    + \beta_{\lambda,405}\ \mathcal{S}_{12}\,\mathrm{tr} \left(y^{c a b d e f}\right)\,\mathrm{tr} \left(y^{e g f g}\right)
    \\&\phantom{=}
    + \beta_{\lambda,406}\ \mathcal{S}_{12}\,\mathrm{tr} \left(y^{a c b e d f}\right)\,\mathrm{tr} \left(y^{e g g f}\right)
    + \beta_{\lambda,407}\ \mathcal{S}_{12}\,\mathrm{tr} \left(y^{a c b e d f}\right)\,\mathrm{tr} \left(y^{e g f g}\right)
    \\&\phantom{=}
    + \beta_{\lambda,408}\ \mathcal{S}_{6}\,\mathrm{tr} \left(y^{a b e c d f}\right)\,\mathrm{tr} \left(y^{e g g f}\right)
    + \beta_{\lambda,409}\ \mathcal{S}_{6}\,\mathrm{tr} \left(y^{a b e c d f}\right)\,\mathrm{tr} \left(y^{e g f g}\right)
    \\&\phantom{=}
    + \beta_{\lambda,410}\ \mathcal{S}_{24}\,\mathrm{tr} \left(y^{c d e f g g}\right)\,\mathrm{tr} \left(y^{a b e f}\right)
    + \beta_{\lambda,411}\ \mathcal{S}_{12}\,\mathrm{tr} \left(y^{c d e f g g}\right)\,\mathrm{tr} \left(y^{a e b f}\right)
    \\&\phantom{=}
    + \beta_{\lambda,412}\ \mathcal{S}_{24}\,\mathrm{tr} \left(y^{c d e g f g}\right)\,\mathrm{tr} \left(y^{a b e f}\right)
    + \beta_{\lambda,413}\ \mathcal{S}_{12}\,\mathrm{tr} \left(y^{c d e g f g}\right)\,\mathrm{tr} \left(y^{a e b f}\right)
    \\&\phantom{=}
    + \beta_{\lambda,414}\ \mathcal{S}_{12}\,\mathrm{tr} \left(y^{c d e g g f}\right)\,\mathrm{tr} \left(y^{a b e f}\right)
    + \beta_{\lambda,415}\ \mathcal{S}_{6}\,\mathrm{tr} \left(y^{c d e g g f}\right)\,\mathrm{tr} \left(y^{a e b f}\right)
    \\&\phantom{=}
    + \beta_{\lambda,416}\ \mathcal{S}_{12}\,\mathrm{tr} \left(y^{c d g e f g}\right)\,\mathrm{tr} \left(y^{a b e f}\right)
    + \beta_{\lambda,417}\ \mathcal{S}_{6}\,\mathrm{tr} \left(y^{c d g e f g}\right)\,\mathrm{tr} \left(y^{a e b f}\right)
    \\&\phantom{=}
    + \beta_{\lambda,418}\ \mathcal{S}_{24}\,\mathrm{tr} \left(y^{c e d f g g}\right)\,\mathrm{tr} \left(y^{a b e f}\right)
    + \beta_{\lambda,419}\ \mathcal{S}_{12}\,\mathrm{tr} \left(y^{c e d f g g}\right)\,\mathrm{tr} \left(y^{a e b f}\right)
    \\&\phantom{=}
    + \beta_{\lambda,420}\ \mathcal{S}_{12}\,\mathrm{tr} \left(y^{a e b g f g}\right)\,\mathrm{tr} \left(y^{c d e f}\right)
    + \beta_{\lambda,421}\ \mathcal{S}_{6}\,\mathrm{tr} \left(y^{c e d g f g}\right)\,\mathrm{tr} \left(y^{a e b f}\right)
    \\&\phantom{=}
    + \beta_{\lambda,422}\ \mathcal{S}_{24}\,\mathrm{tr} \left(y^{c g d g e f}\right)\,\mathrm{tr} \left(y^{a b e f}\right)
    + \beta_{\lambda,423}\ \mathcal{S}_{12}\,\mathrm{tr} \left(y^{c g d g e f}\right)\,\mathrm{tr} \left(y^{a e b f}\right)
    \\&\phantom{=}
    + \beta_{\lambda,424}\ \mathcal{S}_{12}\,\mathrm{tr} \left(y^{c g d e g f}\right)\,\mathrm{tr} \left(y^{a b e f}\right)
    + \beta_{\lambda,425}\ \mathcal{S}_{6}\,\mathrm{tr} \left(y^{c g d e g f}\right)\,\mathrm{tr} \left(y^{a e b f}\right)
    \\&\phantom{=}
    + \beta_{\lambda,426}\ \mathcal{S}_{12}\,\mathrm{tr} \left(y^{c e f d g g}\right)\,\mathrm{tr} \left(y^{a b e f}\right)
    + \beta_{\lambda,427}\ \mathcal{S}_{6}\,\mathrm{tr} \left(y^{c e f d g g}\right)\,\mathrm{tr} \left(y^{a e b f}\right)
    \\&\phantom{=}
    + \beta_{\lambda,428}\ \mathcal{S}_{12}\,\mathrm{tr} \left(y^{c e g d f g}\right)\,\mathrm{tr} \left(y^{a b e f}\right)
    + \beta_{\lambda,429}\ \mathcal{S}_{6}\,\mathrm{tr} \left(y^{c e g d f g}\right)\,\mathrm{tr} \left(y^{a e b f}\right)
    \\&\phantom{=}
    + \beta_{\lambda,430}\ \mathcal{S}_{12}\,\mathrm{tr} \left(y^{c e g d g f}\right)\,\mathrm{tr} \left(y^{a b e f}\right)
    + \beta_{\lambda,431}\ \mathcal{S}_{12}\,\mathrm{tr} \left(y^{c e g d g f}\right)\,\mathrm{tr} \left(y^{a e b f}\right)
    \\&\phantom{=}
    + \beta_{\lambda,432}\ \mathcal{S}_{24}\,\mathrm{tr} \left(y^{a c b e f g}\right)\,\mathrm{tr} \left(y^{d e g f}\right)
    + \beta_{\lambda,433}\ \mathcal{S}_{12}\,\mathrm{tr} \left(y^{a c b e g f}\right)\,\mathrm{tr} \left(y^{d e g f}\right)
    \\&\phantom{=}
    + \beta_{\lambda,434}\ \mathcal{S}_{24}\,\mathrm{tr} \left(y^{a b e c f g}\right)\,\mathrm{tr} \left(y^{d e g f}\right)
    + \beta_{\lambda,435}\ \mathcal{S}_{24}\,\mathrm{tr} \left(y^{a b e c g f}\right)\,\mathrm{tr} \left(y^{d e g f}\right)
    \\&\phantom{=}
    + \beta_{\lambda,436}\ \mathcal{S}_{24}\,\mathrm{tr} \left(y^{a b g c e f}\right)\,\mathrm{tr} \left(y^{d e g f}\right)
    + \beta_{\lambda,437}\ \mathcal{S}_{12}\,\mathrm{tr} \left(y^{a g b e c f}\right)\,\mathrm{tr} \left(y^{d e g f}\right)
    \\&\phantom{=}
    + \beta_{\lambda,438}\ \mathcal{S}_{24}\,\mathrm{tr} \left(y^{c a b d e f g g}\right)\,\mathrm{tr} \left(y^{e f}\right)
    + \beta_{\lambda,439}\ \mathcal{S}_{12}\,\mathrm{tr} \left(y^{c a b d e g g f}\right)\,\mathrm{tr} \left(y^{e f}\right)
    \\&\phantom{=}
    + \beta_{\lambda,440}\ \mathcal{S}_{12}\,\mathrm{tr} \left(y^{c a b d g e f g}\right)\,\mathrm{tr} \left(y^{e f}\right)
    + \beta_{\lambda,441}\ \mathcal{S}_{24}\,\mathrm{tr} \left(y^{c a b d e g f g}\right)\,\mathrm{tr} \left(y^{e f}\right)
    \\&\phantom{=}
    + \beta_{\lambda,442}\ \mathcal{S}_{24}\,\mathrm{tr} \left(y^{c a b e d f g g}\right)\,\mathrm{tr} \left(y^{e f}\right)
    + \beta_{\lambda,443}\ \mathcal{S}_{24}\,\mathrm{tr} \left(y^{c a b e d g f g}\right)\,\mathrm{tr} \left(y^{e f}\right)
    \\&\phantom{=}
    + \beta_{\lambda,444}\ \mathcal{S}_{24}\,\mathrm{tr} \left(y^{c a b e d g g f}\right)\,\mathrm{tr} \left(y^{e f}\right)
    + \beta_{\lambda,445}\ \mathcal{S}_{24}\,\mathrm{tr} \left(y^{c a b g d g e f}\right)\,\mathrm{tr} \left(y^{e f}\right)
    \\&\phantom{=}
    + \beta_{\lambda,446}\ \mathcal{S}_{24}\,\mathrm{tr} \left(y^{c a b g d e g f}\right)\,\mathrm{tr} \left(y^{e f}\right)
    + \beta_{\lambda,447}\ \mathcal{S}_{24}\,\mathrm{tr} \left(y^{c a b g d e f g}\right)\,\mathrm{tr} \left(y^{e f}\right)
    \\&\phantom{=}
    + \beta_{\lambda,448}\ \mathcal{S}_{24}\,\mathrm{tr} \left(y^{a c b e f d g g}\right)\,\mathrm{tr} \left(y^{e f}\right)
    + \beta_{\lambda,449}\ \mathcal{S}_{12}\,\mathrm{tr} \left(y^{a c b e g d g f}\right)\,\mathrm{tr} \left(y^{e f}\right)
    \\&\phantom{=}
    + \beta_{\lambda,450}\ \mathcal{S}_{12}\,\mathrm{tr} \left(y^{a c b g e d f g}\right)\,\mathrm{tr} \left(y^{e f}\right)
    + \beta_{\lambda,451}\ \mathcal{S}_{24}\,\mathrm{tr} \left(y^{a c b e g d f g}\right)\,\mathrm{tr} \left(y^{e f}\right)
    \\&\phantom{=}
    + \beta_{\lambda,452}\ \mathcal{S}_{24}\,\mathrm{tr} \left(y^{a b e c f d g g}\right)\,\mathrm{tr} \left(y^{e f}\right)
    + \beta_{\lambda,453}\ \mathcal{S}_{24}\,\mathrm{tr} \left(y^{a b e c g d f g}\right)\,\mathrm{tr} \left(y^{e f}\right)
    \\&\phantom{=}
    + \beta_{\lambda,454}\ \mathcal{S}_{24}\,\mathrm{tr} \left(y^{a b g c e d f g}\right)\,\mathrm{tr} \left(y^{e f}\right)
    + \beta_{\lambda,455}\ \mathcal{S}_{24}\,\mathrm{tr} \left(y^{a b e c g d g f}\right)\,\mathrm{tr} \left(y^{e f}\right)
    \\&\phantom{=}
    + \beta_{\lambda,456}\ \mathcal{S}_{24}\,\mathrm{tr} \left(y^{a b g c e d g f}\right)\,\mathrm{tr} \left(y^{e f}\right)
    + \beta_{\lambda,457}\ \mathcal{S}_{24}\,\mathrm{tr} \left(y^{a b g c g d e f}\right)\,\mathrm{tr} \left(y^{e f}\right)
    \\&\phantom{=}
    + \beta_{\lambda,458}\ \mathcal{S}_{12}\,\mathrm{tr} \left(y^{a c e f d b g g}\right)\,\mathrm{tr} \left(y^{e f}\right)
    + \beta_{\lambda,459}\ \mathcal{S}_{12}\,\mathrm{tr} \left(y^{a c e g b d f g}\right)\,\mathrm{tr} \left(y^{e f}\right)
    \\&\phantom{=}
    + \beta_{\lambda,460}\ \mathcal{S}_{12}\,\mathrm{tr} \left(y^{a b e g c d g f}\right)\,\mathrm{tr} \left(y^{e f}\right)
    + \beta_{\lambda,461}\ \mathcal{S}_{12}\,\mathrm{tr} \left(y^{a e c f b g d g}\right)\,\mathrm{tr} \left(y^{e f}\right)
    \\&\phantom{=}
    + \beta_{\lambda,462}\ \mathcal{S}_{6}\,\mathrm{tr} \left(y^{a e b g c f d g}\right)\,\mathrm{tr} \left(y^{e f}\right)
    + \beta_{\lambda,463}\ \mathcal{S}_{24}\,\mathrm{tr} \left(y^{a b e c f g d g}\right)\,\mathrm{tr} \left(y^{e f}\right)
    \\&\phantom{=}
    + \beta_{\lambda,464}\ \mathcal{S}_{24}\,\mathrm{tr} \left(y^{a b e c g f d g}\right)\,\mathrm{tr} \left(y^{e f}\right)
    + \beta_{\lambda,465}\ \mathcal{S}_{12}\,\mathrm{tr} \left(y^{a b g c e f d g}\right)\,\mathrm{tr} \left(y^{e f}\right)
    \\&\phantom{=}
    + \beta_{\lambda,466}\ \mathcal{S}_{12}\,\mathrm{tr} \left(y^{a b e c g g d f}\right)\,\mathrm{tr} \left(y^{e f}\right)
    + \beta_{\lambda,467}\ \mathcal{S}_{24}\,\mathrm{tr} \left(y^{a c e d b f g g}\right)\,\mathrm{tr} \left(y^{e f}\right)
    \\&\phantom{=}
    + \beta_{\lambda,468}\ \mathcal{S}_{12}\,\mathrm{tr} \left(y^{a c e d b g f g}\right)\,\mathrm{tr} \left(y^{e f}\right)
    + \beta_{\lambda,469}\ \mathcal{S}_{24}\,\mathrm{tr} \left(y^{a c g d b g e f}\right)\,\mathrm{tr} \left(y^{e f}\right)
    \\&\phantom{=}
    + \beta_{\lambda,470}\ \mathcal{S}_{12}\,\mathrm{tr} \left(y^{a c g d b e g f}\right)\,\mathrm{tr} \left(y^{e f}\right)\,,\yestag{}\\[.5em]
    \beta_{\lambda, y^4 y^4 y^2}^{(4)\,abcd} &= 
    \beta_{\lambda,471}\ \mathcal{S}_{12}\,\mathrm{tr} \left(y^{a c e g}\right)\,\mathrm{tr} \left(y^{b d f g}\right)\,\mathrm{tr} \left(y^{e f}\right)
    \\&\phantom{=}
    + \beta_{\lambda,472}\ \mathcal{S}_{12}\,\mathrm{tr} \left(y^{a e b g}\right)\,\mathrm{tr} \left(y^{c d f g}\right)\,\mathrm{tr} \left(y^{e f}\right)
    \\&\phantom{=}
    + \beta_{\lambda,473}\ \mathcal{S}_{6}\,\mathrm{tr} \left(y^{a e b g}\right)\,\mathrm{tr} \left(y^{c f d g}\right)\,\mathrm{tr} \left(y^{e f}\right)\,,\yestag{}\\[.5em]
    \beta_{\lambda, y^{10}}^{(4)\,abcd} &= 
    \beta_{\lambda,474}\ \mathcal{S}_{12}\,\mathrm{tr} \left(y^{c a b d e e f f g g}\right)
    + \beta_{\lambda,475}\ \mathcal{S}_{24}\,\mathrm{tr} \left(y^{c a b d e e f g f g}\right)
    + \beta_{\lambda,476}\ \mathcal{S}_{24}\,\mathrm{tr} \left(y^{c a b d e e f g g f}\right)
    \\&\phantom{=}
    + \beta_{\lambda,477}\ \mathcal{S}_{12}\,\mathrm{tr} \left(y^{c a b d e f e g f g}\right)
    + \beta_{\lambda,478}\ \mathcal{S}_{24}\,\mathrm{tr} \left(y^{c a b d e f e g g f}\right)
    + \beta_{\lambda,479}\ \mathcal{S}_{12}\,\mathrm{tr} \left(y^{c a b d e f f g g e}\right)
    \\&\phantom{=}
    + \beta_{\lambda,480}\ \mathcal{S}_{12}\,\mathrm{tr} \left(y^{c a b d e f g e f g}\right)
    + \beta_{\lambda,481}\ \mathcal{S}_{24}\,\mathrm{tr} \left(y^{c a b d e f g e g f}\right)
    + \beta_{\lambda,482}\ \mathcal{S}_{12}\,\mathrm{tr} \left(y^{c a b d e f g f g e}\right)
    \\&\phantom{=}
    + \beta_{\lambda,483}\ \mathcal{S}_{12}\,\mathrm{tr} \left(y^{c a b d e f g g e f}\right)
    + \beta_{\lambda,484}\ \mathcal{S}_{12}\,\mathrm{tr} \left(y^{c a b d e f g g f e}\right)
    + \beta_{\lambda,485}\ \mathcal{S}_{24}\,\mathrm{tr} \left(y^{a b c e d e f f g g}\right)
    \\&\phantom{=}
    + \beta_{\lambda,486}\ \mathcal{S}_{24}\,\mathrm{tr} \left(y^{a b c e d e f g f g}\right)
    + \beta_{\lambda,487}\ \mathcal{S}_{24}\,\mathrm{tr} \left(y^{a b c e d e f g g f}\right)
    + \beta_{\lambda,488}\ \mathcal{S}_{24}\,\mathrm{tr} \left(y^{a b c e d f e f g g}\right)
    \\&\phantom{=}
    + \beta_{\lambda,489}\ \mathcal{S}_{24}\,\mathrm{tr} \left(y^{a b c e d f e g f g}\right)
    + \beta_{\lambda,490}\ \mathcal{S}_{24}\,\mathrm{tr} \left(y^{a b c e d f e g g f}\right)
    + \beta_{\lambda,491}\ \mathcal{S}_{24}\,\mathrm{tr} \left(y^{a b c e d f f e g g}\right)
    \\&\phantom{=}
    + \beta_{\lambda,492}\ \mathcal{S}_{24}\,\mathrm{tr} \left(y^{a b c e d f f g e g}\right)
    + \beta_{\lambda,493}\ \mathcal{S}_{24}\,\mathrm{tr} \left(y^{a b c e d f f g g e}\right)
    + \beta_{\lambda,494}\ \mathcal{S}_{24}\,\mathrm{tr} \left(y^{a b c e d f g e f g}\right)
    \\&\phantom{=}
    + \beta_{\lambda,495}\ \mathcal{S}_{24}\,\mathrm{tr} \left(y^{a b c e d f g e g f}\right)
    + \beta_{\lambda,496}\ \mathcal{S}_{24}\,\mathrm{tr} \left(y^{a b c e d f g f e g}\right)
    + \beta_{\lambda,497}\ \mathcal{S}_{24}\,\mathrm{tr} \left(y^{a b c e d f g f g e}\right)
    \\&\phantom{=}
    + \beta_{\lambda,498}\ \mathcal{S}_{24}\,\mathrm{tr} \left(y^{a b c e d f g g e f}\right)
    + \beta_{\lambda,499}\ \mathcal{S}_{24}\,\mathrm{tr} \left(y^{a b c e d f g g f e}\right)
    + \beta_{\lambda,500}\ \mathcal{S}_{24}\,\mathrm{tr} \left(y^{a b c e e d f f g g}\right)
    \\&\phantom{=}
    + \beta_{\lambda,501}\ \mathcal{S}_{24}\,\mathrm{tr} \left(y^{a b c e e d f g f g}\right)
    + \beta_{\lambda,502}\ \mathcal{S}_{24}\,\mathrm{tr} \left(y^{a b c e e d f g g f}\right)
    + \beta_{\lambda,503}\ \mathcal{S}_{24}\,\mathrm{tr} \left(y^{a b c e f d e f g g}\right)
    \\&\phantom{=}
    + \beta_{\lambda,504}\ \mathcal{S}_{24}\,\mathrm{tr} \left(y^{a b c e f d f e g g}\right)
    + \beta_{\lambda,505}\ \mathcal{S}_{24}\,\mathrm{tr} \left(y^{a b c e f d e g f g}\right)
    + \beta_{\lambda,506}\ \mathcal{S}_{24}\,\mathrm{tr} \left(y^{a b c e f d e g g f}\right)
    \\&\phantom{=}
    + \beta_{\lambda,507}\ \mathcal{S}_{24}\,\mathrm{tr} \left(y^{a b c e f d f g e g}\right)
    + \beta_{\lambda,508}\ \mathcal{S}_{24}\,\mathrm{tr} \left(y^{a b c e f d f g g e}\right)
    + \beta_{\lambda,509}\ \mathcal{S}_{24}\,\mathrm{tr} \left(y^{a b c e f d g e f g}\right)
    \\&\phantom{=}
    + \beta_{\lambda,510}\ \mathcal{S}_{24}\,\mathrm{tr} \left(y^{a b c e f d g e g f}\right)
    + \beta_{\lambda,511}\ \mathcal{S}_{24}\,\mathrm{tr} \left(y^{a b c e f d g f e g}\right)
    + \beta_{\lambda,512}\ \mathcal{S}_{24}\,\mathrm{tr} \left(y^{a b c e f d g f g e}\right)
    \\&\phantom{=}
    + \beta_{\lambda,513}\ \mathcal{S}_{24}\,\mathrm{tr} \left(y^{a b c e f d g g e f}\right)
    + \beta_{\lambda,514}\ \mathcal{S}_{24}\,\mathrm{tr} \left(y^{a b c e f d g g f e}\right)
    + \beta_{\lambda,515}\ \mathcal{S}_{12}\,\mathrm{tr} \left(y^{a c b e e f d f g g}\right)
    \\&\phantom{=}
    + \beta_{\lambda,516}\ \mathcal{S}_{24}\,\mathrm{tr} \left(y^{a c b e e f d g f g}\right)
    + \beta_{\lambda,517}\ \mathcal{S}_{24}\,\mathrm{tr} \left(y^{a c b e e f d g g f}\right)
    + \beta_{\lambda,518}\ \mathcal{S}_{12}\,\mathrm{tr} \left(y^{a c b e f e d g f g}\right)
    \\&\phantom{=}
    + \beta_{\lambda,519}\ \mathcal{S}_{24}\,\mathrm{tr} \left(y^{a c b e f e d g g f}\right)
    + \beta_{\lambda,520}\ \mathcal{S}_{12}\,\mathrm{tr} \left(y^{a c b e f f d g g e}\right)
    + \beta_{\lambda,521}\ \mathcal{S}_{12}\,\mathrm{tr} \left(y^{a c b e f g d e f g}\right)
    \\&\phantom{=}
    + \beta_{\lambda,522}\ \mathcal{S}_{24}\,\mathrm{tr} \left(y^{a c b e f g d e g f}\right)
    + \beta_{\lambda,523}\ \mathcal{S}_{12}\,\mathrm{tr} \left(y^{a c b e f g d f g e}\right)
    + \beta_{\lambda,524}\ \mathcal{S}_{12}\,\mathrm{tr} \left(y^{a c b e f g d g e f}\right)
    \\&\phantom{=}
    + \beta_{\lambda,525}\ \mathcal{S}_{12}\,\mathrm{tr} \left(y^{a c b e f g d g f e}\right)
    + \beta_{\lambda,526}\ \mathcal{S}_{24}\,\mathrm{tr} \left(y^{c a e b d e f f g g}\right)
    + \beta_{\lambda,527}\ \mathcal{S}_{24}\,\mathrm{tr} \left(y^{c a e b d e f g f g}\right)
    \\&\phantom{=}
    + \beta_{\lambda,528}\ \mathcal{S}_{24}\,\mathrm{tr} \left(y^{c a e b d e f g g f}\right)
    + \beta_{\lambda,529}\ \mathcal{S}_{24}\,\mathrm{tr} \left(y^{c a e b d f e f g g}\right)
    + \beta_{\lambda,530}\ \mathcal{S}_{24}\,\mathrm{tr} \left(y^{c a e b d f e g f g}\right)
    \\&\phantom{=}
    + \beta_{\lambda,531}\ \mathcal{S}_{24}\,\mathrm{tr} \left(y^{c a e b d f e g g f}\right)
    + \beta_{\lambda,532}\ \mathcal{S}_{12}\,\mathrm{tr} \left(y^{c a e b d f f e g g}\right)
    + \beta_{\lambda,533}\ \mathcal{S}_{12}\,\mathrm{tr} \left(y^{c a e b d f g e f g}\right)
    \\&\phantom{=}
    + \beta_{\lambda,534}\ \mathcal{S}_{12}\,\mathrm{tr} \left(y^{c a e b d f g e g f}\right)
    + \beta_{\lambda,535}\ \mathcal{S}_{24}\,\mathrm{tr} \left(y^{a b e c e d f f g g}\right)
    + \beta_{\lambda,536}\ \mathcal{S}_{24}\,\mathrm{tr} \left(y^{a b e c e d f g f g}\right)
    \\&\phantom{=}
    + \beta_{\lambda,537}\ \mathcal{S}_{24}\,\mathrm{tr} \left(y^{a b e c e d f g g f}\right)
    + \beta_{\lambda,538}\ \mathcal{S}_{24}\,\mathrm{tr} \left(y^{a b e c f d e f g g}\right)
    + \beta_{\lambda,539}\ \mathcal{S}_{24}\,\mathrm{tr} \left(y^{a b e c f d e g f g}\right)
    \\&\phantom{=}
    + \beta_{\lambda,540}\ \mathcal{S}_{24}\,\mathrm{tr} \left(y^{a b e c f d e g g f}\right)
    + \beta_{\lambda,541}\ \mathcal{S}_{24}\,\mathrm{tr} \left(y^{a b e c f d f e g g}\right)
    + \beta_{\lambda,542}\ \mathcal{S}_{24}\,\mathrm{tr} \left(y^{a b e c f d f g e g}\right)
    \\&\phantom{=}
    + \beta_{\lambda,543}\ \mathcal{S}_{24}\,\mathrm{tr} \left(y^{a b e c f d f g g e}\right)
    + \beta_{\lambda,544}\ \mathcal{S}_{24}\,\mathrm{tr} \left(y^{a b e c f d g e f g}\right)
    + \beta_{\lambda,545}\ \mathcal{S}_{24}\,\mathrm{tr} \left(y^{a b e c f d g e g f}\right)
    \\&\phantom{=}
    + \beta_{\lambda,546}\ \mathcal{S}_{24}\,\mathrm{tr} \left(y^{a b e c f d g f e g}\right)
    + \beta_{\lambda,547}\ \mathcal{S}_{24}\,\mathrm{tr} \left(y^{a b e c f d g f g e}\right)
    + \beta_{\lambda,548}\ \mathcal{S}_{24}\,\mathrm{tr} \left(y^{a b e c f d g g e f}\right)
    \\&\phantom{=}
    + \beta_{\lambda,549}\ \mathcal{S}_{24}\,\mathrm{tr} \left(y^{a b e c f d g g f e}\right)
    + \beta_{\lambda,550}\ \mathcal{S}_{24}\,\mathrm{tr} \left(y^{a b e c e f d f g g}\right)
    + \beta_{\lambda,551}\ \mathcal{S}_{24}\,\mathrm{tr} \left(y^{a b e c e f d g f g}\right)
    \\&\phantom{=}
    + \beta_{\lambda,552}\ \mathcal{S}_{24}\,\mathrm{tr} \left(y^{a b e c e f d g g f}\right)
    + \beta_{\lambda,553}\ \mathcal{S}_{24}\,\mathrm{tr} \left(y^{a b e c f e d f g g}\right)
    + \beta_{\lambda,554}\ \mathcal{S}_{24}\,\mathrm{tr} \left(y^{a b e c f e d g f g}\right)
    \\&\phantom{=}
    + \beta_{\lambda,555}\ \mathcal{S}_{24}\,\mathrm{tr} \left(y^{a b e c f e d g g f}\right)
    + \beta_{\lambda,556}\ \mathcal{S}_{24}\,\mathrm{tr} \left(y^{a b e c f f d e g g}\right)
    + \beta_{\lambda,557}\ \mathcal{S}_{24}\,\mathrm{tr} \left(y^{a b e c f f d g e g}\right)
    \\&\phantom{=}
    + \beta_{\lambda,558}\ \mathcal{S}_{24}\,\mathrm{tr} \left(y^{a b e c f f d g g e}\right)
    + \beta_{\lambda,559}\ \mathcal{S}_{24}\,\mathrm{tr} \left(y^{a b e c f g d e f g}\right)
    + \beta_{\lambda,560}\ \mathcal{S}_{24}\,\mathrm{tr} \left(y^{a b e c f g d e g f}\right)
    \\&\phantom{=}
    + \beta_{\lambda,561}\ \mathcal{S}_{24}\,\mathrm{tr} \left(y^{a b e c f g d f e g}\right)
    + \beta_{\lambda,562}\ \mathcal{S}_{24}\,\mathrm{tr} \left(y^{a b e c f g d f g e}\right)
    + \beta_{\lambda,563}\ \mathcal{S}_{24}\,\mathrm{tr} \left(y^{a b e c f g d g e f}\right)
    \\&\phantom{=}
    + \beta_{\lambda,564}\ \mathcal{S}_{24}\,\mathrm{tr} \left(y^{a b e c f g d g f e}\right)
    + \beta_{\lambda,565}\ \mathcal{S}_{24}\,\mathrm{tr} \left(y^{a b e c e f f d g g}\right)
    + \beta_{\lambda,566}\ \mathcal{S}_{24}\,\mathrm{tr} \left(y^{a b e c e f g d f g}\right)
    \\&\phantom{=}
    + \beta_{\lambda,567}\ \mathcal{S}_{24}\,\mathrm{tr} \left(y^{a b e c e f g d g f}\right)
    + \beta_{\lambda,568}\ \mathcal{S}_{24}\,\mathrm{tr} \left(y^{a b e c f e f d g g}\right)
    + \beta_{\lambda,569}\ \mathcal{S}_{24}\,\mathrm{tr} \left(y^{a b e c f e g d f g}\right)
    \\&\phantom{=}
    + \beta_{\lambda,570}\ \mathcal{S}_{24}\,\mathrm{tr} \left(y^{a b e c f e g d g f}\right)
    + \beta_{\lambda,571}\ \mathcal{S}_{24}\,\mathrm{tr} \left(y^{a b e c f f e d g g}\right)
    + \beta_{\lambda,572}\ \mathcal{S}_{24}\,\mathrm{tr} \left(y^{a b e c f f g d e g}\right)
    \\&\phantom{=}
    + \beta_{\lambda,573}\ \mathcal{S}_{24}\,\mathrm{tr} \left(y^{a b e c f f g d g e}\right)
    + \beta_{\lambda,574}\ \mathcal{S}_{24}\,\mathrm{tr} \left(y^{a b e c f g e d f g}\right)
    + \beta_{\lambda,575}\ \mathcal{S}_{24}\,\mathrm{tr} \left(y^{a b e c f g e d g f}\right)
    \\&\phantom{=}
    + \beta_{\lambda,576}\ \mathcal{S}_{24}\,\mathrm{tr} \left(y^{a b e c f g f d e g}\right)
    + \beta_{\lambda,577}\ \mathcal{S}_{24}\,\mathrm{tr} \left(y^{a b e c f g f d g e}\right)
    + \beta_{\lambda,578}\ \mathcal{S}_{24}\,\mathrm{tr} \left(y^{a b e c f g g d e f}\right)
    \\&\phantom{=}
    + \beta_{\lambda,579}\ \mathcal{S}_{24}\,\mathrm{tr} \left(y^{a b e c f g g d f e}\right)
    + \beta_{\lambda,580}\ \mathcal{S}_{12}\,\mathrm{tr} \left(y^{a b e c e f f g d g}\right)
    + \beta_{\lambda,581}\ \mathcal{S}_{24}\,\mathrm{tr} \left(y^{a b e c e f g f d g}\right)
    \\&\phantom{=}
    + \beta_{\lambda,582}\ \mathcal{S}_{24}\,\mathrm{tr} \left(y^{a b e c e f g g d f}\right)
    + \beta_{\lambda,583}\ \mathcal{S}_{12}\,\mathrm{tr} \left(y^{a b e c f e g f d g}\right)
    + \beta_{\lambda,584}\ \mathcal{S}_{24}\,\mathrm{tr} \left(y^{a b e c f e g g d f}\right)
    \\&\phantom{=}
    + \beta_{\lambda,585}\ \mathcal{S}_{12}\,\mathrm{tr} \left(y^{a b e c f f g g d e}\right)
    + \beta_{\lambda,586}\ \mathcal{S}_{12}\,\mathrm{tr} \left(y^{a b e c f g e f d g}\right)
    + \beta_{\lambda,587}\ \mathcal{S}_{24}\,\mathrm{tr} \left(y^{a b e c f g e g d f}\right)
    \\&\phantom{=}
    + \beta_{\lambda,588}\ \mathcal{S}_{12}\,\mathrm{tr} \left(y^{a b e c f g f g d e}\right)
    + \beta_{\lambda,589}\ \mathcal{S}_{12}\,\mathrm{tr} \left(y^{a b e c f g g e d f}\right)
    + \beta_{\lambda,590}\ \mathcal{S}_{12}\,\mathrm{tr} \left(y^{a b e c f g g f d e}\right)
    \\&\phantom{=}
    + \beta_{\lambda,591}\ \mathcal{S}_{12}\,\mathrm{tr} \left(y^{c a e e b d f f g g}\right)
    + \beta_{\lambda,592}\ \mathcal{S}_{12}\,\mathrm{tr} \left(y^{c a e e b d f g f g}\right)
    + \beta_{\lambda,593}\ \mathcal{S}_{12}\,\mathrm{tr} \left(y^{c a e e b d f g g f}\right)
    \\&\phantom{=}
    + \beta_{\lambda,594}\ \mathcal{S}_{24}\,\mathrm{tr} \left(y^{c a e f b d e f g g}\right)
    + \beta_{\lambda,595}\ \mathcal{S}_{24}\,\mathrm{tr} \left(y^{c a e f b d f e g g}\right)
    + \beta_{\lambda,596}\ \mathcal{S}_{24}\,\mathrm{tr} \left(y^{c a e f b d e g f g}\right)
    \\&\phantom{=}
    + \beta_{\lambda,597}\ \mathcal{S}_{12}\,\mathrm{tr} \left(y^{c a e f b d e g g f}\right)
    + \beta_{\lambda,598}\ \mathcal{S}_{24}\,\mathrm{tr} \left(y^{c a e f b d f g e g}\right)
    + \beta_{\lambda,599}\ \mathcal{S}_{12}\,\mathrm{tr} \left(y^{c a e f b d f g g e}\right)
    \\&\phantom{=}
    + \beta_{\lambda,600}\ \mathcal{S}_{12}\,\mathrm{tr} \left(y^{c a e f b d g e f g}\right)
    + \beta_{\lambda,601}\ \mathcal{S}_{12}\,\mathrm{tr} \left(y^{c a e f b d g f e g}\right)
    + \beta_{\lambda,602}\ \mathcal{S}_{24}\,\mathrm{tr} \left(y^{a b e e c f d f g g}\right)
    \\&\phantom{=}
    + \beta_{\lambda,603}\ \mathcal{S}_{24}\,\mathrm{tr} \left(y^{a b e e c f d g f g}\right)
    + \beta_{\lambda,604}\ \mathcal{S}_{24}\,\mathrm{tr} \left(y^{a b e e c f d g g f}\right)
    + \beta_{\lambda,605}\ \mathcal{S}_{24}\,\mathrm{tr} \left(y^{a b e f c e d f g g}\right)
    \\&\phantom{=}
    + \beta_{\lambda,606}\ \mathcal{S}_{24}\,\mathrm{tr} \left(y^{a b e f c f d e g g}\right)
    + \beta_{\lambda,607}\ \mathcal{S}_{24}\,\mathrm{tr} \left(y^{a b e f c e d g f g}\right)
    + \beta_{\lambda,608}\ \mathcal{S}_{24}\,\mathrm{tr} \left(y^{a b e f c e d g g f}\right)
    \\&\phantom{=}
    + \beta_{\lambda,609}\ \mathcal{S}_{24}\,\mathrm{tr} \left(y^{a b e f c f d g e g}\right)
    + \beta_{\lambda,610}\ \mathcal{S}_{24}\,\mathrm{tr} \left(y^{a b e f c f d g g e}\right)
    + \beta_{\lambda,611}\ \mathcal{S}_{24}\,\mathrm{tr} \left(y^{a b e f c g d e f g}\right)
    \\&\phantom{=}
    + \beta_{\lambda,612}\ \mathcal{S}_{24}\,\mathrm{tr} \left(y^{a b e f c g d e g f}\right)
    + \beta_{\lambda,613}\ \mathcal{S}_{24}\,\mathrm{tr} \left(y^{a b e f c g d f e g}\right)
    + \beta_{\lambda,614}\ \mathcal{S}_{24}\,\mathrm{tr} \left(y^{a b e f c g d f g e}\right)
    \\&\phantom{=}
    + \beta_{\lambda,615}\ \mathcal{S}_{24}\,\mathrm{tr} \left(y^{a b e f c g d g e f}\right)
    + \beta_{\lambda,616}\ \mathcal{S}_{24}\,\mathrm{tr} \left(y^{a b e f c g d g f e}\right)
    + \beta_{\lambda,617}\ \mathcal{S}_{12}\,\mathrm{tr} \left(y^{a b e e c f f d g g}\right)
    \\&\phantom{=}
    + \beta_{\lambda,618}\ \mathcal{S}_{24}\,\mathrm{tr} \left(y^{a b e f c e f d g g}\right)
    + \beta_{\lambda,619}\ \mathcal{S}_{24}\,\mathrm{tr} \left(y^{a b e f c f e d g g}\right)
    + \beta_{\lambda,620}\ \mathcal{S}_{12}\,\mathrm{tr} \left(y^{a b e f c e g d f g}\right)
    \\&\phantom{=}
    + \beta_{\lambda,621}\ \mathcal{S}_{24}\,\mathrm{tr} \left(y^{a b e f c f g d e g}\right)
    + \beta_{\lambda,622}\ \mathcal{S}_{12}\,\mathrm{tr} \left(y^{a b e f c f g d g e}\right)
    + \beta_{\lambda,623}\ \mathcal{S}_{12}\,\mathrm{tr} \left(y^{a b e f c g e d f g}\right)
    \\&\phantom{=}
    + \beta_{\lambda,624}\ \mathcal{S}_{24}\,\mathrm{tr} \left(y^{a b e f c g f d e g}\right)
    + \beta_{\lambda,625}\ \mathcal{S}_{12}\,\mathrm{tr} \left(y^{a b e f c g f d g e}\right)
    + \beta_{\lambda,626}\ \mathcal{S}_{12}\,\mathrm{tr} \left(y^{a b e f c g g d e f}\right)
    \\&\phantom{=}
    + \beta_{\lambda,627}\ \mathcal{S}_{12}\,\mathrm{tr} \left(y^{a b e f c g g d f e}\right)
    + \beta_{\lambda,628}\ \mathcal{S}_{12}\,\mathrm{tr} \left(y^{a b e e f c d f g g}\right)
    + \beta_{\lambda,629}\ \mathcal{S}_{24}\,\mathrm{tr} \left(y^{a b e f e c d f g g}\right)
    \\&\phantom{=}
    + \beta_{\lambda,630}\ \mathcal{S}_{12}\,\mathrm{tr} \left(y^{a c e f f b d e g g}\right)
    + \beta_{\lambda,631}\ \mathcal{S}_{6}\,\mathrm{tr} \left(y^{a b e f e c d g f g}\right)
    + \beta_{\lambda,632}\ \mathcal{S}_{6}\,\mathrm{tr} \left(y^{a c e f g b d e f g}\right)
    \\&\phantom{=}
    + \beta_{\lambda,633}\ \mathcal{S}_{12}\,\mathrm{tr} \left(y^{a b e f g c d f e g}\right)
    + \beta_{\lambda,634}\ \mathcal{S}_{12}\,\mathrm{tr} \left(y^{a b e f g c d f g e}\right)
    + \beta_{\lambda,635}\ \mathcal{S}_{6}\,\mathrm{tr} \left(y^{a b e f g c d g f e}\right)
    \\&\phantom{=}
    + \beta_{\lambda,636}\ \mathcal{S}_{24}\,\mathrm{tr} \left(y^{c e a e b f d f g g}\right)
    + \beta_{\lambda,637}\ \mathcal{S}_{24}\,\mathrm{tr} \left(y^{c e a e b f d g f g}\right)
    + \beta_{\lambda,638}\ \mathcal{S}_{24}\,\mathrm{tr} \left(y^{c e a e b f d g g f}\right)
    \\&\phantom{=}
    + \beta_{\lambda,639}\ \mathcal{S}_{24}\,\mathrm{tr} \left(y^{c e a f b g d g e f}\right)
    + \beta_{\lambda,640}\ \mathcal{S}_{12}\,\mathrm{tr} \left(y^{c e a f b g d g f e}\right)
    + \beta_{\lambda,641}\ \mathcal{S}_{24}\,\mathrm{tr} \left(y^{c e a f b e d f g g}\right)
    \\&\phantom{=}
    + \beta_{\lambda,642}\ \mathcal{S}_{12}\,\mathrm{tr} \left(y^{c e a f b e d g f g}\right)
    + \beta_{\lambda,643}\ \mathcal{S}_{24}\,\mathrm{tr} \left(y^{c e a f b g d f e g}\right)
    + \beta_{\lambda,644}\ \mathcal{S}_{12}\,\mathrm{tr} \left(y^{c e a f b g d e f g}\right)
    \\&\phantom{=}
    + \beta_{\lambda,645}\ \mathcal{S}_{12}\,\mathrm{tr} \left(y^{a e c e b f f d g g}\right)
    + \beta_{\lambda,646}\ \mathcal{S}_{12}\,\mathrm{tr} \left(y^{a e c e b f g d f g}\right)
    + \beta_{\lambda,647}\ \mathcal{S}_{12}\,\mathrm{tr} \left(y^{a e c e b f g d g f}\right)
    \\&\phantom{=}
    + \beta_{\lambda,648}\ \mathcal{S}_{12}\,\mathrm{tr} \left(y^{a e c f b f g d g e}\right)
    + \beta_{\lambda,649}\ \mathcal{S}_{24}\,\mathrm{tr} \left(y^{a e c f b f g d e g}\right)
    + \beta_{\lambda,650}\ \mathcal{S}_{24}\,\mathrm{tr} \left(y^{a e c f b f e d g g}\right)
    \\&\phantom{=}
    + \beta_{\lambda,651}\ \mathcal{S}_{24}\,\mathrm{tr} \left(y^{a e c f b e f d g g}\right)
    + \beta_{\lambda,652}\ \mathcal{S}_{24}\,\mathrm{tr} \left(y^{a e c f b e g d f g}\right)
    + \beta_{\lambda,653}\ \mathcal{S}_{12}\,\mathrm{tr} \left(y^{a e c f b e g d g f}\right)
    \\&\phantom{=}
    + \beta_{\lambda,654}\ \mathcal{S}_{12}\,\mathrm{tr} \left(y^{a e c f b g e d f g}\right)
    + \beta_{\lambda,655}\ \mathcal{S}_{12}\,\mathrm{tr} \left(y^{a e c f b g f d e g}\right)
    + \beta_{\lambda,656}\ \mathcal{S}_{12}\,\mathrm{tr} \left(y^{c e a e f b f d g g}\right)
    \\&\phantom{=}
    + \beta_{\lambda,657}\ \mathcal{S}_{24}\,\mathrm{tr} \left(y^{c e a e f b g d f g}\right)
    + \beta_{\lambda,658}\ \mathcal{S}_{12}\,\mathrm{tr} \left(y^{d e a e f b g c g f}\right)
    + \beta_{\lambda,659}\ \mathcal{S}_{12}\,\mathrm{tr} \left(y^{c e a f e b f d g g}\right)
    \\&\phantom{=}
    + \beta_{\lambda,660}\ \mathcal{S}_{12}\,\mathrm{tr} \left(y^{a e b f e c g d f g}\right)
    + \beta_{\lambda,661}\ \mathcal{S}_{6}\,\mathrm{tr} \left(y^{c e a f f b e d g g}\right)
    + \beta_{\lambda,662}\ \mathcal{S}_{6}\,\mathrm{tr} \left(y^{c e a f g b e d f g}\right)
    \\&\phantom{=}
    + \beta_{\lambda,663}\ \mathcal{S}_{6}\,\mathrm{tr} \left(y^{c e a f g b e d g f}\right)\,.\yestag{}
  \end{align*}

  The specific values of the coefficients in the $\overline{\text{MS}}$ are
  \begin{align*}
 \beta_{\lambda,1} & = 2 \zeta_3-\tfrac{11}{3}\,,&
\beta_{\lambda,2} & = 1-\zeta_3\,,&
\beta_{\lambda,3} & = \tfrac{7}{12}\,,\\
\beta_{\lambda,4} & = \tfrac{1}{2}\,,&
\beta_{\lambda,5} & = \tfrac{121}{144}\,,&
\beta_{\lambda,6} & = 1-2 \zeta_3\,,\\
\beta_{\lambda,7} & = \tfrac{\zeta_3}{2}-\tfrac{1}{4}\,,&
\beta_{\lambda,8} & = \tfrac{5}{6}-\zeta_3\,,&
\beta_{\lambda,9} & = \tfrac{5}{6}\,,\\
\beta_{\lambda,10} & = -\tfrac{37}{288}\,,&
\beta_{\lambda,11} & = \tfrac{2}{3}\,,&
\beta_{\lambda,12} & = \tfrac{5}{6}-\zeta_3\,,\\
\beta_{\lambda,13} & = 4 \zeta_3-5\,,&
\beta_{\lambda,14} & = -5\,,&
\beta_{\lambda,15} & = \tfrac{\zeta_3}{2}-\tfrac{1}{4}\,,\\
\beta_{\lambda,16} & = \tfrac{\pi ^4}{30}-6 \zeta_3\,,&
\beta_{\lambda,17} & = -6 \zeta_3-\tfrac{\pi ^4}{30}\,,&
\beta_{\lambda,18} & = \tfrac{2}{3}\,,\\
\beta_{\lambda,19} & = -40 \zeta_5\,,&
\beta_{\lambda,20} & = 0\,,&
\beta_{\lambda,21} & = 0\,,\\
\beta_{\lambda,22} & = 0\,,&
\beta_{\lambda,23} & = 0\,,&
\beta_{\lambda,24} & = 0\,,\\
\beta_{\lambda,25} & = 0\,,&
\beta_{\lambda,26} & = 0\,,&
\beta_{\lambda,27} & = -\tfrac{11}{6}\,,\\
\beta_{\lambda,28} & = 0\,,&
\beta_{\lambda,29} & = \tfrac{9}{8}\,,&
\beta_{\lambda,30} & = \tfrac{1}{2}\,,\\
\beta_{\lambda,31} & = 0\,,&
\beta_{\lambda,32} & = -\tfrac{37}{288}\,,&
\beta_{\lambda,33} & = -\tfrac{37}{288}\,,\\
\beta_{\lambda,34} & = 0\,,&
\beta_{\lambda,35} & = 0\,,&
\beta_{\lambda,36} & = 0\,,\\
\beta_{\lambda,37} & = -5\,,&
\beta_{\lambda,38} & = \tfrac{5}{6}\,,&
\beta_{\lambda,39} & = \tfrac{2}{3}\,,\\
\beta_{\lambda,40} & = \tfrac{5}{6}-\zeta_3\,,&
\beta_{\lambda,41} & = \tfrac{2}{3}\,,&
\beta_{\lambda,42} & = \tfrac{\zeta_3}{2}-\tfrac{1}{4}\,,\\
\beta_{\lambda,43} & = \tfrac{2}{3}\,,&
\beta_{\lambda,44} & = \tfrac{\pi ^4}{30}-6 \zeta_3\,,&
\beta_{\lambda,45} & = \tfrac{\zeta_3}{2}-\tfrac{1}{4}\,,\\
\beta_{\lambda,46} & = \tfrac{2}{3}\,,&
\beta_{\lambda,47} & = 0\,,&
\beta_{\lambda,48} & = 0\,,\\
\beta_{\lambda,49} & = 0\,,&
\beta_{\lambda,50} & = \tfrac{5}{12}\,,&
\beta_{\lambda,51} & = \tfrac{2}{3}\,,\\
\beta_{\lambda,52} & = \tfrac{\zeta_3}{2}-\tfrac{1}{4}\,,&
\beta_{\lambda,53} & = \tfrac{5}{6}-\zeta_3\,,&
\beta_{\lambda,54} & = \tfrac{\zeta_3}{2}-\tfrac{1}{4}\,,\\
\beta_{\lambda,55} & = -2\,,&
\beta_{\lambda,56} & = -4 \zeta_3+\tfrac{22}{3}-\tfrac{\pi ^4}{15}\,,&
\beta_{\lambda,57} & = -8\,,\\
\beta_{\lambda,58} & = 0\,,&
\beta_{\lambda,59} & = 0\,,&
\beta_{\lambda,60} & = 0\,,\\
\beta_{\lambda,61} & = 0\,,&
\beta_{\lambda,62} & = \zeta_3-2-\tfrac{\pi ^4}{60}\,,&
\beta_{\lambda,63} & = 8 \zeta_3-\tfrac{22}{3}\,,\\
\beta_{\lambda,64} & = -12 \zeta_3-\tfrac{10}{3}\,,&
\beta_{\lambda,65} & = -14 \zeta_3+10-\tfrac{\pi ^4}{30}\,,&
\beta_{\lambda,66} & = -3 \zeta_3-1+\tfrac{\pi ^4}{60}\,,\\
\beta_{\lambda,67} & = -3\,,&
\beta_{\lambda,68} & = -2 \zeta_3-\tfrac{27}{8}\,,&
\beta_{\lambda,69} & = \zeta_3+\tfrac{13}{48}\,,\\
\beta_{\lambda,70} & = -\tfrac{361}{48}\,,&
\beta_{\lambda,71} & = \tfrac{101}{96}\,,&
\beta_{\lambda,72} & = -6 \zeta_3+10-\tfrac{\pi ^4}{30}\,,\\
\beta_{\lambda,73} & = 4 \zeta_3-\tfrac{10}{3}\,,&
\beta_{\lambda,74} & = -3 \zeta_3-\tfrac{4}{3}+\tfrac{\pi ^4}{60}\,,&
\beta_{\lambda,75} & = 1-2 \zeta_3\,,\\
\beta_{\lambda,76} & = 0\,,&
\beta_{\lambda,77} & = 0\,,&
\beta_{\lambda,78} & = -4 \zeta_3-\tfrac{10}{3}+\tfrac{\pi ^4}{30}\,,\\
\beta_{\lambda,79} & = 4 \zeta_3-8\,,&
\beta_{\lambda,80} & = 2 \zeta_3-8\,,&
\beta_{\lambda,81} & = 24 \zeta_3-40 \zeta_5-\tfrac{2 \pi ^4}{15}\,,\\
\beta_{\lambda,82} & = 1-2 \zeta_3\,,&
\beta_{\lambda,83} & = -3 \zeta_3-\tfrac{4}{3}+\tfrac{\pi ^4}{60}\,,&
\beta_{\lambda,84} & = -1\,,\\
\beta_{\lambda,85} & = 0\,,&
\beta_{\lambda,86} & = \tfrac{37}{72}\,,&
\beta_{\lambda,87} & = \tfrac{79}{72}\,,\\
\beta_{\lambda,88} & = -\tfrac{7}{3}\,,&
\beta_{\lambda,89} & = 0\,,&
\beta_{\lambda,90} & = 0\,,\\
\beta_{\lambda,91} & = 0\,,&
\beta_{\lambda,92} & = -\tfrac{8}{3}\,,&
\beta_{\lambda,93} & = -\tfrac{19}{3}\,,\\
\beta_{\lambda,94} & = 1-2 \zeta_3\,,&
\beta_{\lambda,95} & = \tfrac{4}{3}-2 \zeta_3\,,&
\beta_{\lambda,96} & = \tfrac{\zeta_3}{2}-\tfrac{1}{4}\,,\\
\beta_{\lambda,97} & = \tfrac{\zeta_3}{2}-\tfrac{1}{4}\,,&
\beta_{\lambda,98} & = \tfrac{101}{96}\,,&
\beta_{\lambda,99} & = \zeta_3+\tfrac{13}{48}\,,\\
\beta_{\lambda,100} & = \tfrac{101}{96}\,,&
\beta_{\lambda,101} & = \zeta_3+\tfrac{13}{48}\,,&
\beta_{\lambda,102} & = -\tfrac{169}{32}\,,\\
\beta_{\lambda,103} & = -2 \zeta_3-\tfrac{5}{4}\,,&
\beta_{\lambda,104} & = -3\,,&
\beta_{\lambda,105} & = -3 \zeta_3-1+\tfrac{\pi ^4}{60}\,,\\
\beta_{\lambda,106} & = -3 \zeta_3-\tfrac{4}{3}+\tfrac{\pi ^4}{60}\,,&
\beta_{\lambda,107} & = -3 \zeta_3-\tfrac{4}{3}+\tfrac{\pi ^4}{60}\,,&
\beta_{\lambda,108} & = -4 \zeta_3-\tfrac{10}{3}+\tfrac{\pi ^4}{30}\,,\\
\beta_{\lambda,109} & = 4 \zeta_3-8\,,&
\beta_{\lambda,110} & = 1-2 \zeta_3\,,&
\beta_{\lambda,111} & = 1-2 \zeta_3\,,\\
\beta_{\lambda,112} & = 2 \zeta_3-8\,,&
\beta_{\lambda,113} & = -\tfrac{19}{3}\,,&
\beta_{\lambda,114} & = -\tfrac{8}{3}\,,\\
\beta_{\lambda,115} & = \tfrac{4}{3}-2 \zeta_3\,,&
\beta_{\lambda,116} & = 1-2 \zeta_3\,,&
\beta_{\lambda,117} & = 0\,,\\
\beta_{\lambda,118} & = 0\,,&
\beta_{\lambda,119} & = 0\,,&
\beta_{\lambda,120} & = 0\,,\\
\beta_{\lambda,121} & = \tfrac{1}{4}-\zeta_3\,,&
\beta_{\lambda,122} & = \tfrac{1}{2}-2 \zeta_3\,,&
\beta_{\lambda,123} & = -\tfrac{63}{32}\,,\\
\beta_{\lambda,124} & = -\tfrac{25}{12}\,,&
\beta_{\lambda,125} & = 5-8 \zeta_3\,,&
\beta_{\lambda,126} & = -\tfrac{17}{24}\,,\\
\beta_{\lambda,127} & = -3 \zeta_3+\tfrac{5}{3}-\tfrac{\pi ^4}{60}\,,&
\beta_{\lambda,128} & = 6 \zeta_3-\tfrac{26}{3}\,,&
\beta_{\lambda,129} & = 4-4 \zeta_3\,,\\
\beta_{\lambda,130} & = -8 \zeta_3+\tfrac{22}{3}+\tfrac{\pi ^4}{30}\,,&
\beta_{\lambda,131} & = \tfrac{1}{3}\,,&
\beta_{\lambda,132} & = \tfrac{4}{3}-4 \zeta_3\,,\\
\beta_{\lambda,133} & = \tfrac{22}{3}-4 \zeta_3\,,&
\beta_{\lambda,134} & = -2\,,&
\beta_{\lambda,135} & = \tfrac{34}{3}-\tfrac{\pi ^4}{15}\,,\\
\beta_{\lambda,136} & = \tfrac{34}{3}-20 \zeta_5\,,&
\beta_{\lambda,137} & = \tfrac{46}{3}-4 \zeta_3\,,&
\beta_{\lambda,138} & = 4 \zeta_3+\tfrac{\pi ^4}{15}\,,\\
\beta_{\lambda,139} & = \tfrac{\pi ^4}{15}-4 \zeta_3\,,&
\beta_{\lambda,140} & = \tfrac{35}{3}-2 \zeta_3\,,&
\beta_{\lambda,141} & = 12 \zeta_3-20 \zeta_5+\tfrac{\pi ^4}{15}\,,\\
\beta_{\lambda,142} & = \tfrac{\pi ^4}{15}\,,&
\beta_{\lambda,143} & = \tfrac{16}{3}-2 \zeta_3\,,&
\beta_{\lambda,144} & = -4\,,\\
\beta_{\lambda,145} & = -2\,,&
\beta_{\lambda,146} & = 4 \zeta_3-2\,,&
\beta_{\lambda,147} & = \tfrac{16}{3}-4 \zeta_3\,,\\
\beta_{\lambda,148} & = 8 \zeta_3-4\,,&
\beta_{\lambda,149} & = \tfrac{32}{3}-4 \zeta_3\,,&
\beta_{\lambda,150} & = 4 \zeta_3-40 \zeta_5+\tfrac{70}{3}\,,\\
\beta_{\lambda,151} & = \tfrac{32}{3}-20 \zeta_3\,,&
\beta_{\lambda,152} & = -6 \zeta_3+10+\tfrac{\pi ^4}{30}\,,&
\beta_{\lambda,153} & = 20-8 \zeta_3\,,\\
\beta_{\lambda,154} & = -10 \zeta_3+10+\tfrac{\pi ^4}{30}\,,&
\beta_{\lambda,155} & = \tfrac{16}{3}-2 \zeta_3\,,&
\beta_{\lambda,156} & = -8 \zeta_3+\tfrac{70}{3}-\tfrac{\pi ^4}{15}\,,\\
\beta_{\lambda,157} & = -14 \zeta_3+10+\tfrac{\pi ^4}{30}\,,&
\beta_{\lambda,158} & = -6 \zeta_3+10+\tfrac{\pi ^4}{30}\,,&
\beta_{\lambda,159} & = 0\,,\\
\beta_{\lambda,160} & = -2 \zeta_3-\tfrac{2}{3}\,,&
\beta_{\lambda,161} & = -\tfrac{1}{3}\,,&
\beta_{\lambda,162} & = -3\,,\\
\beta_{\lambda,163} & = 0\,,&
\beta_{\lambda,164} & = 2 \zeta_3-\tfrac{11}{3}\,,&
\beta_{\lambda,165} & = -\tfrac{4}{3}\,,\\
\beta_{\lambda,166} & = -6 \zeta_3-\tfrac{8}{3}+\tfrac{\pi ^4}{30}\,,&
\beta_{\lambda,167} & = 0\,,&
\beta_{\lambda,168} & = -4 \zeta_3-\tfrac{11}{3}+\tfrac{\pi ^4}{30}\,,\\
\beta_{\lambda,169} & = -\tfrac{8}{3}\,,&
\beta_{\lambda,170} & = 40 \zeta_5-8 \zeta_3\,,&
\beta_{\lambda,171} & = 40 \zeta_5\,,\\
\beta_{\lambda,172} & = 24 \zeta_3\,,&
\beta_{\lambda,173} & = 40 \zeta_5\,,&
\beta_{\lambda,174} & = 40 \zeta_5\,,\\
\beta_{\lambda,175} & = 24 \zeta_3\,,&
\beta_{\lambda,176} & = 20 \zeta_5\,,&
\beta_{\lambda,177} & = 20 \zeta_5\,,\\
\beta_{\lambda,178} & = 0\,,&
\beta_{\lambda,179} & = 0\,,&
\beta_{\lambda,180} & = 0\,,\\
\beta_{\lambda,181} & = 0\,,&
\beta_{\lambda,182} & = 0\,,&
\beta_{\lambda,183} & = 0\,,\\
\beta_{\lambda,184} & = 0\,,&
\beta_{\lambda,185} & = 0\,,&
\beta_{\lambda,186} & = 0\,,\\
\beta_{\lambda,187} & = 0\,,&
\beta_{\lambda,188} & = 0\,,&
\beta_{\lambda,189} & = \tfrac{34}{3}-4 \zeta_3\,,\\
\beta_{\lambda,190} & = \tfrac{\pi ^4}{15}\,,&
\beta_{\lambda,191} & = \tfrac{70}{3}-16 \zeta_3\,,&
\beta_{\lambda,192} & = 4 \zeta_3+\tfrac{\pi ^4}{15}\,,\\
\beta_{\lambda,193} & = 24 \zeta_3-40 \zeta_5+\tfrac{2 \pi ^4}{15}\,,&
\beta_{\lambda,194} & = \tfrac{\pi ^4}{15}-4 \zeta_3\,,&
\beta_{\lambda,195} & = 2 \zeta_3+\tfrac{14}{3}\,,\\
\beta_{\lambda,196} & = 10 \zeta_3-\tfrac{\pi ^4}{30}\,,&
\beta_{\lambda,197} & = 6 \zeta_3-\tfrac{\pi ^4}{30}\,,&
\beta_{\lambda,198} & = 12-4 \zeta_3\,,\\
\beta_{\lambda,199} & = 12 \zeta_3-\tfrac{\pi ^4}{15}\,,&
\beta_{\lambda,200} & = 8 \zeta_3-\tfrac{\pi ^4}{30}\,,&
\beta_{\lambda,201} & = -\tfrac{53}{48}\,,\\
\beta_{\lambda,202} & = -\tfrac{121}{72}\,,&
\beta_{\lambda,203} & = -\tfrac{11}{12}\,,&
\beta_{\lambda,204} & = \tfrac{4}{3}-2 \zeta_3\,,\\
\beta_{\lambda,205} & = \tfrac{1}{2}-\zeta_3\,,&
\beta_{\lambda,206} & = \tfrac{4}{3}-2 \zeta_3\,,&
\beta_{\lambda,207} & = 1-2 \zeta_3\,,\\
\beta_{\lambda,208} & = 0\,,&
\beta_{\lambda,209} & = \tfrac{14}{3}-\tfrac{\pi ^4}{15}\,,&
\beta_{\lambda,210} & = 6\,,\\
\beta_{\lambda,211} & = 6 \zeta_3+2-\tfrac{\pi ^4}{30}\,,&
\beta_{\lambda,212} & = -\tfrac{77}{8}\,,&
\beta_{\lambda,213} & = -4 \zeta_3-\tfrac{47}{12}\,,\\
\beta_{\lambda,214} & = -\tfrac{101}{24}\,,&
\beta_{\lambda,215} & = -4 \zeta_3-\tfrac{13}{12}\,,&
\beta_{\lambda,216} & = 8 \zeta_3-40 \zeta_5+\tfrac{76}{3}\,,\\
\beta_{\lambda,217} & = \tfrac{76}{3}-20 \zeta_3\,,&
\beta_{\lambda,218} & = 20 \zeta_3-20 \zeta_5+\tfrac{20}{3}-\tfrac{2 \pi ^4}{15}\,,&
\beta_{\lambda,219} & = 12 \zeta_3-40 \zeta_5+32-\tfrac{2 \pi ^4}{15}\,,\\
\beta_{\lambda,220} & = 8 \zeta_3-\tfrac{16}{3}\,,&
\beta_{\lambda,221} & = 8 \zeta_3-4\,,&
\beta_{\lambda,222} & = \tfrac{38}{3}-\tfrac{\pi ^4}{15}\,,\\
\beta_{\lambda,223} & = 12 \zeta_3+\tfrac{16}{3}-\tfrac{\pi ^4}{15}\,,&
\beta_{\lambda,224} & = 0\,,&
\beta_{\lambda,225} & = 0\,,\\
\beta_{\lambda,226} & = 0\,,&
\beta_{\lambda,227} & = \tfrac{77}{16}\,,&
\beta_{\lambda,228} & = \tfrac{143}{48}\,,\\
\beta_{\lambda,229} & = \tfrac{14}{3}\,,&
\beta_{\lambda,230} & = 4 \zeta_3+\tfrac{11}{3}\,,&
\beta_{\lambda,231} & = -6 \zeta_3+10+\tfrac{\pi ^4}{30}\,,\\
\beta_{\lambda,232} & = 12-4 \zeta_3\,,&
\beta_{\lambda,233} & = \tfrac{361}{48}\,,&
\beta_{\lambda,234} & = \tfrac{19}{3}\,,\\
\beta_{\lambda,235} & = \tfrac{77}{16}\,,&
\beta_{\lambda,236} & = -10 \zeta_3+12+\tfrac{\pi ^4}{30}\,,&
\beta_{\lambda,237} & = 5\,,\\
\beta_{\lambda,238} & = 0\,,&
\beta_{\lambda,239} & = 0\,,&
\beta_{\lambda,240} & = -3\,,\\
\beta_{\lambda,241} & = -\tfrac{1}{3}\,,&
\beta_{\lambda,242} & = -2 \zeta_3-\tfrac{2}{3}\,,&
\beta_{\lambda,243} & = -2 \zeta_3-\tfrac{2}{3}\,,\\
\beta_{\lambda,244} & = -6 \zeta_3-\tfrac{8}{3}+\tfrac{\pi ^4}{30}\,,&
\beta_{\lambda,245} & = -6 \zeta_3-\tfrac{8}{3}+\tfrac{\pi ^4}{30}\,,&
\beta_{\lambda,246} & = 2 \zeta_3-\tfrac{11}{3}\,,\\
\beta_{\lambda,247} & = -\tfrac{4}{3}\,,&
\beta_{\lambda,248} & = -\tfrac{4}{3}\,,&
\beta_{\lambda,249} & = 0\,,\\
\beta_{\lambda,250} & = 0\,,&
\beta_{\lambda,251} & = -\tfrac{8}{3}\,,&
\beta_{\lambda,252} & = -4 \zeta_3-\tfrac{11}{3}+\tfrac{\pi ^4}{30}\,,\\
\beta_{\lambda,253} & = 12-4 \zeta_3\,,&
\beta_{\lambda,254} & = 6 \zeta_3-\tfrac{\pi ^4}{30}\,,&
\beta_{\lambda,255} & = 10 \zeta_3-\tfrac{\pi ^4}{30}\,,\\
\beta_{\lambda,256} & = 2 \zeta_3+\tfrac{14}{3}\,,&
\beta_{\lambda,257} & = 12 \zeta_3-\tfrac{\pi ^4}{15}\,,&
\beta_{\lambda,258} & = 8 \zeta_3-\tfrac{\pi ^4}{30}\,,\\
\beta_{\lambda,259} & = 2 \zeta_3-\tfrac{4}{3}\,,&
\beta_{\lambda,260} & = \tfrac{73}{48}\,,&
\beta_{\lambda,261} & = 4 \zeta_3-\tfrac{8}{3}\,,\\
\beta_{\lambda,262} & = 16 \zeta_3-4\,,&
\beta_{\lambda,263} & = 2 \zeta_3-\tfrac{4}{3}\,,&
\beta_{\lambda,264} & = \tfrac{11}{3}-2 \zeta_3\,,\\
\beta_{\lambda,265} & = -\tfrac{2}{3}\,,&
\beta_{\lambda,266} & = 4 \zeta_3-\tfrac{34}{3}+\tfrac{\pi ^4}{15}\,,&
\beta_{\lambda,267} & = \tfrac{22}{3}-20 \zeta_3\,,\\
\beta_{\lambda,268} & = 4 \zeta_3-4\,,&
\beta_{\lambda,269} & = 8 \zeta_3-4\,,&
\beta_{\lambda,270} & = \tfrac{11}{3}-4 \zeta_3\,,\\
\beta_{\lambda,271} & = 2 \zeta_3-2\,,&
\beta_{\lambda,272} & = \tfrac{11}{16}\,,&
\beta_{\lambda,273} & = 4 \zeta_3-4\,,\\
\beta_{\lambda,274} & = \tfrac{10}{3}\,,&
\beta_{\lambda,275} & = \tfrac{28}{3}-4 \zeta_3\,,&
\beta_{\lambda,276} & = \tfrac{4}{3}\,,\\
\beta_{\lambda,277} & = 2 \zeta_3-\tfrac{4}{3}\,,&
\beta_{\lambda,278} & = \tfrac{\pi ^4}{15}\,,&
\beta_{\lambda,279} & = -\tfrac{4}{3}\,,\\
\beta_{\lambda,280} & = -8\,,&
\beta_{\lambda,281} & = 0\,,&
\beta_{\lambda,282} & = -\tfrac{2}{3}\,,\\
\beta_{\lambda,283} & = 2 \zeta_3-\tfrac{4}{3}\,,&
\beta_{\lambda,284} & = \tfrac{\pi ^4}{15}\,,&
\beta_{\lambda,285} & = \tfrac{8}{3}\,,\\
\beta_{\lambda,286} & = 2 \zeta_3-\tfrac{5}{3}\,,&
\beta_{\lambda,287} & = 4 \zeta_3-\tfrac{10}{3}\,,&
\beta_{\lambda,288} & = \tfrac{119}{48}\,,\\
\beta_{\lambda,289} & = -4 \zeta_3-\tfrac{10}{3}+\tfrac{\pi ^4}{30}\,,&
\beta_{\lambda,290} & = 4 \zeta_3-\tfrac{10}{3}+\tfrac{\pi ^4}{15}\,,&
\beta_{\lambda,291} & = -10 \zeta_3+10+\tfrac{\pi ^4}{30}\,,\\
\beta_{\lambda,292} & = -\tfrac{5}{3}\,,&
\beta_{\lambda,293} & = 28 \zeta_3-\tfrac{70}{3}+\tfrac{\pi ^4}{15}\,,&
\beta_{\lambda,294} & = -28 \zeta_3+20+\tfrac{\pi ^4}{15}\,,\\
\beta_{\lambda,295} & = \tfrac{19}{3}\,,&
\beta_{\lambda,296} & = 4 \zeta_3-\tfrac{8}{3}\,,&
\beta_{\lambda,297} & = 4 \zeta_3-\tfrac{8}{3}\,,\\
\beta_{\lambda,298} & = 2 \zeta_3+\tfrac{14}{3}\,,&
\beta_{\lambda,299} & = 16 \zeta_3\,,&
\beta_{\lambda,300} & = 4 \zeta_3-\tfrac{8}{3}\,,\\
\beta_{\lambda,301} & = 12 \zeta_3-40 \zeta_5+\tfrac{28}{3}\,,&
\beta_{\lambda,302} & = 20 \zeta_3-40 \zeta_5+\tfrac{\pi ^4}{15}\,,&
\beta_{\lambda,303} & = -\tfrac{4}{3}\,,\\
\beta_{\lambda,304} & = \tfrac{14}{3}-4 \zeta_3\,,&
\beta_{\lambda,305} & = 24-24 \zeta_3\,,&
\beta_{\lambda,306} & = 4 \zeta_3-\tfrac{8}{3}\,,\\
\beta_{\lambda,307} & = \tfrac{\pi ^4}{15}\,,&
\beta_{\lambda,308} & = \tfrac{20}{3}\,,&
\beta_{\lambda,309} & = 4 \zeta_3-\tfrac{8}{3}\,,\\
\beta_{\lambda,310} & = 24 \zeta_3-40 \zeta_5+\tfrac{2 \pi ^4}{15}\,,&
\beta_{\lambda,311} & = 2 \zeta_3-\tfrac{4}{3}\,,&
\beta_{\lambda,312} & = 4 \zeta_3-\tfrac{8}{3}\,,\\
\beta_{\lambda,313} & = \tfrac{73}{48}\,,&
\beta_{\lambda,314} & = -8 \zeta_3+\tfrac{14}{3}+\tfrac{\pi ^4}{30}\,,&
\beta_{\lambda,315} & = -20 \zeta_3+24+\tfrac{\pi ^4}{15}\,,\\
\beta_{\lambda,316} & = -6 \zeta_3+\tfrac{14}{3}+\tfrac{\pi ^4}{30}\,,&
\beta_{\lambda,317} & = -\tfrac{10}{3}\,,&
\beta_{\lambda,318} & = 28 \zeta_3-40 \zeta_5+\tfrac{\pi ^4}{15}\,,\\
\beta_{\lambda,319} & = -8 \zeta_3+\tfrac{28}{3}+\tfrac{\pi ^4}{15}\,,&
\beta_{\lambda,320} & = -8 \zeta_3-\tfrac{20}{3}+\tfrac{\pi ^4}{15}\,,&
\beta_{\lambda,321} & = 32 \zeta_3-40 \zeta_5\,,\\
\beta_{\lambda,322} & = -8 \zeta_3+\tfrac{14}{3}+\tfrac{\pi ^4}{30}\,,&
\beta_{\lambda,323} & = 4 \zeta_3-\tfrac{10}{3}\,,&
\beta_{\lambda,324} & = 4 \zeta_3-\tfrac{10}{3}\,,\\
\beta_{\lambda,325} & = 10\,,&
\beta_{\lambda,326} & = 2 \zeta_3-\tfrac{4}{3}\,,&
\beta_{\lambda,327} & = 2 \zeta_3-\tfrac{4}{3}\,,\\
\beta_{\lambda,328} & = -\tfrac{2}{3}\,,&
\beta_{\lambda,329} & = \tfrac{10}{3}\,,&
\beta_{\lambda,330} & = 4 \zeta_3-\tfrac{46}{3}+\tfrac{\pi ^4}{15}\,,\\
\beta_{\lambda,331} & = \tfrac{38}{3}\,,&
\beta_{\lambda,332} & = 20 \zeta_3-40 \zeta_5+\tfrac{\pi ^4}{15}\,,&
\beta_{\lambda,333} & = 4 \zeta_3+\tfrac{28}{3}\,,\\
\beta_{\lambda,334} & = \tfrac{\pi ^4}{15}-4 \zeta_3\,,&
\beta_{\lambda,335} & = \tfrac{32}{3}-16 \zeta_3\,,&
\beta_{\lambda,336} & = 0\,,\\
\beta_{\lambda,337} & = 8 \zeta_3-\tfrac{16}{3}\,,&
\beta_{\lambda,338} & = 4 \zeta_3-\tfrac{8}{3}\,,&
\beta_{\lambda,339} & = 4 \zeta_3-\tfrac{8}{3}\,,\\
\beta_{\lambda,340} & = \tfrac{19}{3}\,,&
\beta_{\lambda,341} & = 2 \zeta_3+\tfrac{14}{3}\,,&
\beta_{\lambda,342} & = 8 \zeta_3-40 \zeta_5+24\,,\\
\beta_{\lambda,343} & = -\tfrac{5}{3}\,,&
\beta_{\lambda,344} & = 28 \zeta_3-40 \zeta_5+\tfrac{\pi ^4}{15}\,,&
\beta_{\lambda,345} & = \tfrac{32}{3}\,,\\
\beta_{\lambda,346} & = -4 \zeta_3-\tfrac{10}{3}+\tfrac{\pi ^4}{30}\,,&
\beta_{\lambda,347} & = 28 \zeta_3-40 \zeta_5+\tfrac{\pi ^4}{15}\,,&
\beta_{\lambda,348} & = -16 \zeta_3+\tfrac{28}{3}+\tfrac{\pi ^4}{15}\,,\\
\beta_{\lambda,349} & = \zeta_3-\tfrac{5}{6}\,,&
\beta_{\lambda,350} & = 8 \zeta_3-\tfrac{35}{3}+\tfrac{\pi ^4}{30}\,,&
\beta_{\lambda,351} & = 10\,,\\
\beta_{\lambda,352} & = \tfrac{\pi ^4}{15}-\tfrac{10}{3}\,,&
\beta_{\lambda,353} & = 12 \zeta_3-30 \zeta_5+\tfrac{\pi ^4}{15}\,,&
\beta_{\lambda,354} & = -\tfrac{10}{3}\,,\\
\beta_{\lambda,355} & = 2 \zeta_3-\tfrac{11}{3}\,,&
\beta_{\lambda,356} & = 4 \zeta_3-\tfrac{16}{3}\,,&
\beta_{\lambda,357} & = \tfrac{44}{3}-4 \zeta_3\,,\\
\beta_{\lambda,358} & = \tfrac{70}{3}-4 \zeta_3\,,&
\beta_{\lambda,359} & = \tfrac{14}{3}-4 \zeta_3\,,&
\beta_{\lambda,360} & = 24-24 \zeta_3\,,\\
\beta_{\lambda,361} & = 4 \zeta_3-\tfrac{22}{3}\,,&
\beta_{\lambda,362} & = 16 \zeta_3\,,&
\beta_{\lambda,363} & = 40 \zeta_5-8 \zeta_3\,,\\
\beta_{\lambda,364} & = 8\,,&
\beta_{\lambda,365} & = 4 \zeta_3-\tfrac{\pi ^4}{30}\,,&
\beta_{\lambda,366} & = -4 \zeta_3-\tfrac{\pi ^4}{15}\,,\\
\beta_{\lambda,367} & = 12 \zeta_3\,,&
\beta_{\lambda,368} & = 2 \zeta_3-\tfrac{\pi ^4}{30}\,,&
\beta_{\lambda,369} & = 12 \zeta_3-\tfrac{\pi ^4}{15}\,,\\
\beta_{\lambda,370} & = 6 \zeta_3-\tfrac{\pi ^4}{30}\,,&
\beta_{\lambda,371} & = 2 \zeta_3-\tfrac{11}{3}\,,&
\beta_{\lambda,372} & = 8 \zeta_3-\tfrac{\pi ^4}{15}\,,\\
\beta_{\lambda,373} & = 40 \zeta_5-24 \zeta_3\,,&
\beta_{\lambda,374} & = 24 \zeta_3\,,&
\beta_{\lambda,375} & = 8 \zeta_3\,,\\
\beta_{\lambda,376} & = 40 \zeta_5\,,&
\beta_{\lambda,377} & = 12 \zeta_3\,,&
\beta_{\lambda,378} & = 6 \zeta_3-\tfrac{\pi ^4}{30}\,,\\
\beta_{\lambda,379} & = -\tfrac{\pi ^4}{15}\,,&
\beta_{\lambda,380} & = 4 \zeta_3-\tfrac{\pi ^4}{30}\,,&
\beta_{\lambda,381} & = 12 \zeta_3-\tfrac{\pi ^4}{15}\,,\\
\beta_{\lambda,382} & = 24 \zeta_3-40 \zeta_5-\tfrac{2 \pi ^4}{15}\,,&
\beta_{\lambda,383} & = 4 \zeta_3-\tfrac{22}{3}\,,&
\beta_{\lambda,384} & = 40 \zeta_5\,,\\
\beta_{\lambda,385} & = 40 \zeta_5\,,&
\beta_{\lambda,386} & = 12 \zeta_3-\tfrac{\pi ^4}{15}\,,&
\beta_{\lambda,387} & = 6 \zeta_3-\tfrac{\pi ^4}{30}\,,\\
\beta_{\lambda,388} & = 0\,,&
\beta_{\lambda,389} & = 40 \zeta_5\,,&
\beta_{\lambda,390} & = 24 \zeta_3-\tfrac{2 \pi ^4}{15}\,,\\
\beta_{\lambda,391} & = 12 \zeta_3-\tfrac{\pi ^4}{15}\,,&
\beta_{\lambda,392} & = 40 \zeta_5-24 \zeta_3\,,&
\beta_{\lambda,393} & = \tfrac{10}{3}\,,\\
\beta_{\lambda,394} & = \tfrac{26}{3}\,,&
\beta_{\lambda,395} & = \tfrac{44}{3}\,,&
\beta_{\lambda,396} & = 12\,,\\
\beta_{\lambda,397} & = 16 \zeta_3\,,&
\beta_{\lambda,398} & = 16 \zeta_3\,,&
\beta_{\lambda,399} & = 40 \zeta_5-8 \zeta_3\,,\\
\beta_{\lambda,400} & = 80 \zeta_5\,,&
\beta_{\lambda,401} & = \zeta_3-\tfrac{1}{4}\,,&
\beta_{\lambda,402} & = \zeta_3-\tfrac{5}{6}\,,\\
\beta_{\lambda,403} & = \tfrac{1}{3}\,,&
\beta_{\lambda,404} & = \tfrac{169}{16}\,,&
\beta_{\lambda,405} & = 2 \zeta_3+\tfrac{49}{8}\,,\\
\beta_{\lambda,406} & = \tfrac{361}{24}\,,&
\beta_{\lambda,407} & = 4 \zeta_3+\tfrac{27}{4}\,,&
\beta_{\lambda,408} & = \tfrac{109}{12}\,,\\
\beta_{\lambda,409} & = \tfrac{37}{6}\,,&
\beta_{\lambda,410} & = 6\,,&
\beta_{\lambda,411} & = 0\,,\\
\beta_{\lambda,412} & = 8 \zeta_3+\tfrac{20}{3}\,,&
\beta_{\lambda,413} & = 8 \zeta_3+\tfrac{8}{3}\,,&
\beta_{\lambda,414} & = \tfrac{10}{3}\,,\\
\beta_{\lambda,415} & = \tfrac{4}{3}\,,&
\beta_{\lambda,416} & = 12\,,&
\beta_{\lambda,417} & = 12\,,\\
\beta_{\lambda,418} & = \tfrac{38}{3}\,,&
\beta_{\lambda,419} & = \tfrac{16}{3}\,,&
\beta_{\lambda,420} & = \tfrac{76}{3}-\tfrac{2 \pi ^4}{15}\,,\\
\beta_{\lambda,421} & = 24 \zeta_3+\tfrac{32}{3}-\tfrac{2 \pi ^4}{15}\,,&
\beta_{\lambda,422} & = 12\,,&
\beta_{\lambda,423} & = 0\,,\\
\beta_{\lambda,424} & = \tfrac{92}{3}-8 \zeta_3\,,&
\beta_{\lambda,425} & = \tfrac{44}{3}-8 \zeta_3\,,&
\beta_{\lambda,426} & = 6\,,\\
\beta_{\lambda,427} & = 0\,,&
\beta_{\lambda,428} & = -8 \zeta_3+\tfrac{92}{3}-\tfrac{2 \pi ^4}{15}\,,&
\beta_{\lambda,429} & = 16 \zeta_3+\tfrac{44}{3}-\tfrac{2 \pi ^4}{15}\,,\\
\beta_{\lambda,430} & = \tfrac{76}{3}\,,&
\beta_{\lambda,431} & = \tfrac{16}{3}\,,&
\beta_{\lambda,432} & = 24 \zeta_3-\tfrac{100}{3}+\tfrac{\pi ^4}{15}\,,\\
\beta_{\lambda,433} & = 8 \zeta_3-4\,,&
\beta_{\lambda,434} & = -36 \zeta_3+40 \zeta_5+\tfrac{2 \pi ^4}{15}\,,&
\beta_{\lambda,435} & = -28 \zeta_3+40 \zeta_5+\tfrac{2 \pi ^4}{15}\,,\\
\beta_{\lambda,436} & = -32 \zeta_3+40 \zeta_5+\tfrac{2 \pi ^4}{15}\,,&
\beta_{\lambda,437} & = -48 \zeta_3+80 \zeta_5+\tfrac{4 \pi ^4}{15}\,,&
\beta_{\lambda,438} & = -\tfrac{101}{96}\,,\\
\beta_{\lambda,439} & = -\tfrac{5}{6}\,,&
\beta_{\lambda,440} & = \tfrac{215}{96}\,,&
\beta_{\lambda,441} & = -\tfrac{29}{12}\,,\\
\beta_{\lambda,442} & = -\tfrac{4}{3}\,,&
\beta_{\lambda,443} & = 4 \zeta_3-8\,,&
\beta_{\lambda,444} & = -\tfrac{5}{3}\,,\\
\beta_{\lambda,445} & = -\tfrac{101}{48}\,,&
\beta_{\lambda,446} & = \tfrac{16}{3}-2 \zeta_3\,,&
\beta_{\lambda,447} & = \tfrac{361}{48}\,,\\
\beta_{\lambda,448} & = -\tfrac{101}{96}\,,&
\beta_{\lambda,449} & = -\tfrac{10}{3}\,,&
\beta_{\lambda,450} & = 10\,,\\
\beta_{\lambda,451} & = 8 \zeta_3-\tfrac{14}{3}-\tfrac{\pi ^4}{30}\,,&
\beta_{\lambda,452} & = -\tfrac{4}{3}\,,&
\beta_{\lambda,453} & = 4 \zeta_3-\tfrac{\pi ^4}{30}\,,\\
\beta_{\lambda,454} & = \tfrac{11}{3}\,,&
\beta_{\lambda,455} & = -2\,,&
\beta_{\lambda,456} & = 10 \zeta_3-\tfrac{\pi ^4}{30}\,,\\
\beta_{\lambda,457} & = -\tfrac{101}{48}\,,&
\beta_{\lambda,458} & = -\tfrac{101}{96}\,,&
\beta_{\lambda,459} & = 8 \zeta_3-\tfrac{\pi ^4}{30}\,,\\
\beta_{\lambda,460} & = \tfrac{8}{3}\,,&
\beta_{\lambda,461} & = -\tfrac{8}{3}\,,&
\beta_{\lambda,462} & = 12 \zeta_3-\tfrac{\pi ^4}{15}\,,\\
\beta_{\lambda,463} & = -\tfrac{8}{3}\,,&
\beta_{\lambda,464} & = 2 \zeta_3\,,&
\beta_{\lambda,465} & = \tfrac{65}{24}\,,\\
\beta_{\lambda,466} & = -1\,,&
\beta_{\lambda,467} & = -1\,,&
\beta_{\lambda,468} & = \tfrac{2}{3}-4 \zeta_3\,,\\
\beta_{\lambda,469} & = \tfrac{65}{24}\,,&
\beta_{\lambda,470} & = 2 \zeta_3+\tfrac{19}{3}\,,&
\beta_{\lambda,471} & = 8 \zeta_3-8\,,\\
\beta_{\lambda,472} & = 8 \zeta_3-\tfrac{16}{3}\,,&
\beta_{\lambda,473} & = 4 \zeta_3-2\,,&
\beta_{\lambda,474} & = \tfrac{1}{2}-\zeta_3\,,\\
\beta_{\lambda,475} & = 1-2 \zeta_3\,,&
\beta_{\lambda,476} & = -\tfrac{9}{32}\,,&
\beta_{\lambda,477} & = \tfrac{28}{3}-12 \zeta_3\,,\\
\beta_{\lambda,478} & = -\tfrac{1}{4}\,,&
\beta_{\lambda,479} & = \zeta_3-\tfrac{17}{12}\,,&
\beta_{\lambda,480} & = 6 \zeta_3-\tfrac{10}{3}+\tfrac{\pi ^4}{30}\,,\\
\beta_{\lambda,481} & = 8 \zeta_3-\tfrac{17}{3}\,,&
\beta_{\lambda,482} & = 2 \zeta_3-\tfrac{17}{6}\,,&
\beta_{\lambda,483} & = 4 \zeta_3-2\,,\\
\beta_{\lambda,484} & = \tfrac{49}{96}\,,&
\beta_{\lambda,485} & = 1-2 \zeta_3\,,&
\beta_{\lambda,486} & = 2-4 \zeta_3\,,\\
\beta_{\lambda,487} & = -\tfrac{9}{16}\,,&
\beta_{\lambda,488} & = 1-2 \zeta_3\,,&
\beta_{\lambda,489} & = \tfrac{50}{3}-20 \zeta_3\,,\\
\beta_{\lambda,490} & = \tfrac{16}{3}-4 \zeta_3\,,&
\beta_{\lambda,491} & = -\tfrac{4}{3}\,,&
\beta_{\lambda,492} & = 2 \zeta_3-8\,,\\
\beta_{\lambda,493} & = 2 \zeta_3-\tfrac{5}{3}\,,&
\beta_{\lambda,494} & = 12 \zeta_3-\tfrac{26}{3}+\tfrac{\pi ^4}{15}\,,&
\beta_{\lambda,495} & = 8 \zeta_3-4\,,\\
\beta_{\lambda,496} & = 16 \zeta_3-\tfrac{38}{3}\,,&
\beta_{\lambda,497} & = 4 \zeta_3-\tfrac{10}{3}\,,&
\beta_{\lambda,498} & = 4 \zeta_3-2\,,\\
\beta_{\lambda,499} & = \tfrac{119}{48}\,,&
\beta_{\lambda,500} & = \tfrac{1}{2}-\zeta_3\,,&
\beta_{\lambda,501} & = 1-2 \zeta_3\,,\\
\beta_{\lambda,502} & = -\tfrac{9}{32}\,,&
\beta_{\lambda,503} & = 4 \zeta_3+\tfrac{11}{3}-\tfrac{\pi ^4}{30}\,,&
\beta_{\lambda,504} & = -\tfrac{8}{3}\,,\\
\beta_{\lambda,505} & = 8 \zeta_3+\tfrac{16}{3}-\tfrac{\pi ^4}{15}\,,&
\beta_{\lambda,506} & = 6 \zeta_3-\tfrac{14}{3}-\tfrac{\pi ^4}{30}\,,&
\beta_{\lambda,507} & = 4 \zeta_3-16\,,\\
\beta_{\lambda,508} & = 4 \zeta_3-\tfrac{10}{3}\,,&
\beta_{\lambda,509} & = 12 \zeta_3-\tfrac{44}{3}+\tfrac{\pi ^4}{15}\,,&
\beta_{\lambda,510} & = 16 \zeta_3-24\,,\\
\beta_{\lambda,511} & = -4 \zeta_3+40 \zeta_5-\tfrac{88}{3}\,,&
\beta_{\lambda,512} & = 4 \zeta_3-\tfrac{10}{3}\,,&
\beta_{\lambda,513} & = 10 \zeta_3-12-\tfrac{\pi ^4}{30}\,,\\
\beta_{\lambda,514} & = 10\,,&
\beta_{\lambda,515} & = 1-2 \zeta_3\,,&
\beta_{\lambda,516} & = 1-2 \zeta_3\,,\\
\beta_{\lambda,517} & = -\tfrac{4}{3}\,,&
\beta_{\lambda,518} & = 16 \zeta_3-\tfrac{32}{3}\,,&
\beta_{\lambda,519} & = 2 \zeta_3-8\,,\\
\beta_{\lambda,520} & = 2 \zeta_3-\tfrac{5}{3}\,,&
\beta_{\lambda,521} & = \tfrac{44}{3}\,,&
\beta_{\lambda,522} & = 8 \zeta_3-\tfrac{44}{3}+\tfrac{\pi ^4}{15}\,,\\
\beta_{\lambda,523} & = 16 \zeta_3-\tfrac{70}{3}+\tfrac{\pi ^4}{15}\,,&
\beta_{\lambda,524} & = 20 \zeta_3-24-\tfrac{\pi ^4}{15}\,,&
\beta_{\lambda,525} & = 20\,,\\
\beta_{\lambda,526} & = -\tfrac{1}{3}\,,&
\beta_{\lambda,527} & = -\tfrac{2}{3}\,,&
\beta_{\lambda,528} & = \tfrac{23}{24}\,,\\
\beta_{\lambda,529} & = 2 \zeta_3-2\,,&
\beta_{\lambda,530} & = 4 \zeta_3+\tfrac{2}{3}\,,&
\beta_{\lambda,531} & = \tfrac{19}{3}-4 \zeta_3\,,\\
\beta_{\lambda,532} & = -\tfrac{4}{3}\,,&
\beta_{\lambda,533} & = 12 \zeta_3-12\,,&
\beta_{\lambda,534} & = \tfrac{38}{3}-20 \zeta_3\,,\\
\beta_{\lambda,535} & = 1-2 \zeta_3\,,&
\beta_{\lambda,536} & = 2-4 \zeta_3\,,&
\beta_{\lambda,537} & = -\tfrac{9}{16}\,,\\
\beta_{\lambda,538} & = 6 \zeta_3-\tfrac{\pi ^4}{30}\,,&
\beta_{\lambda,539} & = 12 \zeta_3-\tfrac{\pi ^4}{15}\,,&
\beta_{\lambda,540} & = 2 \zeta_3-\tfrac{\pi ^4}{30}\,,\\
\beta_{\lambda,541} & = -\tfrac{8}{3}\,,&
\beta_{\lambda,542} & = 4 \zeta_3-4\,,&
\beta_{\lambda,543} & = -\tfrac{2}{3}\,,\\
\beta_{\lambda,544} & = 8 \zeta_3\,,&
\beta_{\lambda,545} & = -12 \zeta_3-\tfrac{\pi ^4}{15}\,,&
\beta_{\lambda,546} & = 16 \zeta_3\,,\\
\beta_{\lambda,547} & = -\tfrac{2}{3}\,,&
\beta_{\lambda,548} & = 6 \zeta_3-\tfrac{\pi ^4}{30}\,,&
\beta_{\lambda,549} & = \tfrac{11}{3}\,,\\
\beta_{\lambda,550} & = 2-4 \zeta_3\,,&
\beta_{\lambda,551} & = 2-4 \zeta_3\,,&
\beta_{\lambda,552} & = -\tfrac{8}{3}\,,\\
\beta_{\lambda,553} & = 0\,,&
\beta_{\lambda,554} & = 12 \zeta_3\,,&
\beta_{\lambda,555} & = -2 \zeta_3\,,\\
\beta_{\lambda,556} & = -\tfrac{4}{3}\,,&
\beta_{\lambda,557} & = 2 \zeta_3-2\,,&
\beta_{\lambda,558} & = -\tfrac{1}{3}\,,\\
\beta_{\lambda,559} & = 16 \zeta_3\,,&
\beta_{\lambda,560} & = -4 \zeta_3\,,&
\beta_{\lambda,561} & = 8 \zeta_3\,,\\
\beta_{\lambda,562} & = 12 \zeta_3-8\,,&
\beta_{\lambda,563} & = 12 \zeta_3-\tfrac{\pi ^4}{15}\,,&
\beta_{\lambda,564} & = \tfrac{22}{3}\,,\\
\beta_{\lambda,565} & = 1-2 \zeta_3\,,&
\beta_{\lambda,566} & = 8 \zeta_3+\tfrac{22}{3}-\tfrac{\pi ^4}{15}\,,&
\beta_{\lambda,567} & = -\tfrac{16}{3}\,,\\
\beta_{\lambda,568} & = 1-2 \zeta_3\,,&
\beta_{\lambda,569} & = 40 \zeta_5-16 \zeta_3\,,&
\beta_{\lambda,570} & = -4 \zeta_3\,,\\
\beta_{\lambda,571} & = -\tfrac{4}{3}\,,&
\beta_{\lambda,572} & = 8 \zeta_3-\tfrac{\pi ^4}{30}\,,&
\beta_{\lambda,573} & = -\tfrac{2}{3}\,,\\
\beta_{\lambda,574} & = 40 \zeta_5-16 \zeta_3\,,&
\beta_{\lambda,575} & = 40 \zeta_5-20 \zeta_3\,,&
\beta_{\lambda,576} & = 4 \zeta_3-\tfrac{\pi ^4}{15}\,,\\
\beta_{\lambda,577} & = -\tfrac{2}{3}\,,&
\beta_{\lambda,578} & = 6 \zeta_3-\tfrac{\pi ^4}{30}\,,&
\beta_{\lambda,579} & = \tfrac{11}{3}\,,\\
\beta_{\lambda,580} & = 2-4 \zeta_3\,,&
\beta_{\lambda,581} & = 2-4 \zeta_3\,,&
\beta_{\lambda,582} & = -\tfrac{8}{3}\,,\\
\beta_{\lambda,583} & = 40 \zeta_5-40 \zeta_3\,,&
\beta_{\lambda,584} & = -2 \zeta_3\,,&
\beta_{\lambda,585} & = -\tfrac{1}{3}\,,\\
\beta_{\lambda,586} & = 40 \zeta_5-16 \zeta_3\,,&
\beta_{\lambda,587} & = 8 \zeta_3\,,&
\beta_{\lambda,588} & = -\tfrac{2}{3}\,,\\
\beta_{\lambda,589} & = 0\,,&
\beta_{\lambda,590} & = \tfrac{23}{24}\,,&
\beta_{\lambda,591} & = \tfrac{1}{2}-\zeta_3\,,\\
\beta_{\lambda,592} & = 1-2 \zeta_3\,,&
\beta_{\lambda,593} & = -\tfrac{9}{32}\,,&
\beta_{\lambda,594} & = 6 \zeta_3-\tfrac{\pi ^4}{30}\,,\\
\beta_{\lambda,595} & = \tfrac{2}{3}\,,&
\beta_{\lambda,596} & = -\tfrac{\pi ^4}{15}\,,&
\beta_{\lambda,597} & = 4 \zeta_3-\tfrac{\pi ^4}{30}\,,\\
\beta_{\lambda,598} & = \tfrac{16}{3}\,,&
\beta_{\lambda,599} & = \tfrac{2}{3}\,,&
\beta_{\lambda,600} & = -8 \zeta_3\,,\\
\beta_{\lambda,601} & = 40 \zeta_5-8 \zeta_3\,,&
\beta_{\lambda,602} & = 1-2 \zeta_3\,,&
\beta_{\lambda,603} & = 1-2 \zeta_3\,,\\
\beta_{\lambda,604} & = -\tfrac{4}{3}\,,&
\beta_{\lambda,605} & = 6 \zeta_3-\tfrac{\pi ^4}{30}\,,&
\beta_{\lambda,606} & = -\tfrac{8}{3}\,,\\
\beta_{\lambda,607} & = 24 \zeta_3-40 \zeta_5-\tfrac{\pi ^4}{15}\,,&
\beta_{\lambda,608} & = 8 \zeta_3-\tfrac{\pi ^4}{30}\,,&
\beta_{\lambda,609} & = 4 \zeta_3-4\,,\\
\beta_{\lambda,610} & = -\tfrac{2}{3}\,,&
\beta_{\lambda,611} & = 0\,,&
\beta_{\lambda,612} & = 40 \zeta_5-24 \zeta_3\,,\\
\beta_{\lambda,613} & = 40 \zeta_5\,,&
\beta_{\lambda,614} & = 12 \zeta_3\,,&
\beta_{\lambda,615} & = 12 \zeta_3-\tfrac{\pi ^4}{15}\,,\\
\beta_{\lambda,616} & = \tfrac{4}{3}\,,&
\beta_{\lambda,617} & = \tfrac{1}{2}-\zeta_3\,,&
\beta_{\lambda,618} & = 4 \zeta_3+\tfrac{11}{3}-\tfrac{\pi ^4}{30}\,,\\
\beta_{\lambda,619} & = -\tfrac{8}{3}\,,&
\beta_{\lambda,620} & = -8 \zeta_3\,,&
\beta_{\lambda,621} & = 16 \zeta_3-\tfrac{\pi ^4}{15}\,,\\
\beta_{\lambda,622} & = -\tfrac{4}{3}\,,&
\beta_{\lambda,623} & = 20 \zeta_5\,,&
\beta_{\lambda,624} & = 40 \zeta_5-24 \zeta_3\,,\\
\beta_{\lambda,625} & = 0\,,&
\beta_{\lambda,626} & = 6 \zeta_3-\tfrac{\pi ^4}{30}\,,&
\beta_{\lambda,627} & = \tfrac{2}{3}\,,\\
\beta_{\lambda,628} & = -\tfrac{1}{3}\,,&
\beta_{\lambda,629} & = 2 \zeta_3-2\,,&
\beta_{\lambda,630} & = -\tfrac{4}{3}\,,\\
\beta_{\lambda,631} & = 16 \zeta_3-40 \zeta_5+\tfrac{4}{3}\,,&
\beta_{\lambda,632} & = 24 \zeta_3-40 \zeta_5\,,&
\beta_{\lambda,633} & = 24 \zeta_3\,,\\
\beta_{\lambda,634} & = 12 \zeta_3\,,&
\beta_{\lambda,635} & = 4\,,&
\beta_{\lambda,636} & = 2-4 \zeta_3\,,\\
\beta_{\lambda,637} & = 2-4 \zeta_3\,,&
\beta_{\lambda,638} & = -\tfrac{8}{3}\,,&
\beta_{\lambda,639} & = 12 \zeta_3-\tfrac{\pi ^4}{15}\,,\\
\beta_{\lambda,640} & = -\tfrac{16}{3}\,,&
\beta_{\lambda,641} & = 12 \zeta_3-\tfrac{\pi ^4}{15}\,,&
\beta_{\lambda,642} & = 24 \zeta_3-40 \zeta_5-\tfrac{2 \pi ^4}{15}\,,\\
\beta_{\lambda,643} & = 40 \zeta_5\,,&
\beta_{\lambda,644} & = 40 \zeta_5\,,&
\beta_{\lambda,645} & = 1-2 \zeta_3\,,\\
\beta_{\lambda,646} & = 8 \zeta_3+\tfrac{22}{3}-\tfrac{\pi ^4}{15}\,,&
\beta_{\lambda,647} & = -\tfrac{16}{3}\,,&
\beta_{\lambda,648} & = 4-8 \zeta_3\,,\\
\beta_{\lambda,649} & = 0\,,&
\beta_{\lambda,650} & = -\tfrac{8}{3}\,,&
\beta_{\lambda,651} & = 6 \zeta_3-\tfrac{\pi ^4}{30}\,,\\
\beta_{\lambda,652} & = 0\,,&
\beta_{\lambda,653} & = 24 \zeta_3-\tfrac{2 \pi ^4}{15}\,,&
\beta_{\lambda,654} & = 20 \zeta_5\,,\\
\beta_{\lambda,655} & = 24 \zeta_3\,,&
\beta_{\lambda,656} & = 2-4 \zeta_3\,,&
\beta_{\lambda,657} & = 12 \zeta_3-\tfrac{\pi ^4}{15}\,,\\
\beta_{\lambda,658} & = -\tfrac{16}{3}\,,&
\beta_{\lambda,659} & = 0\,,&
\beta_{\lambda,660} & = 24 \zeta_3-40 \zeta_5\,,\\
\beta_{\lambda,661} & = -\tfrac{4}{3}\,,&
\beta_{\lambda,662} & = -20 \zeta_5\,,&
\beta_{\lambda,663} & = 40 \zeta_5\,.   \tag{\stepcounter{equation}\theequation}
  \end{align*}

\section{\texorpdfstring{$\mathcal{N}=1$}{N=1} Supersymmetry Relations}\label{app:SUSY1}

In this appendix, we detail the constraints from $\mathcal{N}=1$ supersymmetry in four spacetime dimensions. In particular, we focus on the generalised Wess-Zumino theory with the superpotential~\eq{suppot}. 
Integrating out auxiliary modes from the superfields, the interaction Lagrangian is given by \eq{susy-lag} in terms of Weyl fermions $\psi_A$ and complex scalars $\phi_A$. Evidently, supersymmetry enforces that scalar quartic interactions are determined by the super-Yukawas. Moreover, the theory retains a holomorphic structure, where each Yukawa $Y^{ABC}$ couples only to the left-chiral or scalar fields $\psi_A$ and $\phi_A$ but not their conjugates $\psi^A = \psi_A^{*}$ or $\phi^A = \phi_A^{*}$. Thus, each tensor $Y^{ABC}$ can only be contracted with its conjugate $Y_{ABC} = (Y^{ABC})^*$, which outright eliminates many tensor contractions in App.~\ref{app:gammaS}-\ref{app:betaQ}. With supersymmetry manifest, both fermions and scalars exhibit the anomalous dimensions of their associated superfields~\cite{Ferreira:1996az,Ferreira:1997bi,Gracey:2021yvb}. Following the notation of~\cite{Gracey:2021yvb}, this anomalous dimension is given by 13 terms with scheme-dependent coefficients $c_{41}\dots c_{413}$. In a general scheme, the relation~\eq{superAD} yields
\begin{align*}
c_{41} &= 2\,\gamma_{\phi,10}  + 2\,\gamma_{\phi,27} + 4\,\gamma_{\phi,16} + 4\,\gamma_{\phi,3} + 4\,\gamma_{\phi,4} + 8\,\gamma_{\phi,1} + \gamma_{\phi,24} + \gamma_{\phi,37} + \gamma_{\phi,78}\\
       &= 2\,\gamma_{\psi,24} + 2\,\gamma_{\psi,6} + 2\,\gamma_{\psi,8} + \gamma_{\psi,112} + \gamma_{\psi,26} + \gamma_{\psi,54} + \gamma_{\psi,96}\,,\\
c_{42} &= c_{43} =  2\,\gamma_{\phi,4}  + 2\,\gamma_{\phi,7} + \gamma_{\phi,17} + \gamma_{\phi,21} + \gamma_{\phi,25} + \gamma_{\phi,73}\\
&\phantom{\ = c_{43} } = \gamma_{\psi,20} + \gamma_{\psi,28} + \gamma_{\psi,63} + \gamma_{\psi,88} \,,\\
c_{44} &= 2\,\gamma_{\phi,20}  + 2\,\gamma_{\phi,3} + 2\,\gamma_{\phi,29} + 2\,\gamma_{\phi,49} + 4\,\gamma_{\phi,5} + 4\,\gamma_{\phi,7} + \gamma_{\phi,79} \\
 &= 2\,\gamma_{\psi,21} + 2\,\gamma_{\psi,98} + \gamma_{\psi,118} + \gamma_{\psi,29} + \gamma_{\psi,69} + \gamma_{\psi,7}\,,\\
c_{45} &= \gamma_{\phi,5}  + \gamma_{\phi,70} + \gamma_{\phi,8} = \gamma_{\psi,47} + \gamma_{\psi,59} \,,\\
c_{46} &= 2\,\gamma_{\phi,9}  + 2\,\gamma_{\phi,6} + 3\,\gamma_{\phi,2} + \gamma_{\phi,12} + \gamma_{\phi,47} + \gamma_{\phi,71}\\
&=  2\,\gamma_{\psi,17} + 3\,\gamma_{\psi,9} + \gamma_{\psi,43} + \gamma_{\psi,52} + \gamma_{\psi,57}\,,\\
 c_{47} &= 2\,\gamma_{\phi,3}  + \gamma_{\phi,11} + \gamma_{\phi,22} + \gamma_{\phi,55} = 2\,\gamma_{\psi,16} + 4\,\gamma_{\psi,1} + \gamma_{\psi,113} + \gamma_{\psi,22} + \gamma_{\psi,42}\,,\\
c_{48} &= \gamma_{\phi,9}  + \gamma_{\phi,39} = \gamma_{\psi,3} + \gamma_{\psi,40} + \gamma_{\psi,67}\,,\\
c_{49}  &= c_{410}  = \gamma_{\phi,42} + \gamma_{\phi,60} + \gamma_{\phi,7} = 2\,\gamma_{\psi,13} + 3\,\gamma_{\psi,5} + \gamma_{\psi,33} + \gamma_{\psi,72} + \gamma_{\psi,93}\,,\\
c_{411} &= \gamma_{\phi,1}  + \gamma_{\phi,58} = \gamma_{\psi,30} + \gamma_{\psi,87}\,,\\
c_{412} &= 2\,\gamma_{\phi,8}  + \gamma_{\phi,31} + \gamma_{\phi,57} = 2\,\gamma_{\psi,12} + \gamma_{\psi,1} + \gamma_{\psi,105} + \gamma_{\psi,35} + \gamma_{\psi,45}\,,\\
c_{413} &= 2\,\gamma_{\phi,12}  + 2\,\gamma_{\phi,18} + 3\,\gamma_{\phi,14} + 4\,\gamma_{\phi,6} + 6\,\gamma_{\phi,2} + \gamma_{\phi,33} + \gamma_{\phi,40} + \gamma_{\phi,50} \\
&= 2\,\gamma_{\psi,14} + 2\,\gamma_{\psi,18} + 2\,\gamma_{\psi,3} + 3\,\gamma_{\psi,10} + 4\,\gamma_{\psi,4} + 6\,\gamma_{\psi,2} + \gamma_{\psi,122} \\
&\phantom{=\ }+ \gamma_{\psi,31}+ \gamma_{\psi,36} + \gamma_{\psi,48} + \gamma_{\psi,74} \,. \yestag{}
\end{align*}
Moreover, non-renormalisation theorems for $\mathcal{N}=1$ state that superpotential parameters do not exhibit UV poles from vertex corrections~\cite{Salam:1974jj,Grisaru:1979wc}, which stipulates $\hat{\beta}^{(4) {ABC}}_Y = 0$ and implies a number of constraints on the Yukawa coefficients
\begin{align*}
0 &= \beta_{y,21} + \beta_{y,562} 
= \beta_{y,295} + \beta_{y,89}
= 2\,\beta_{y,291} + \beta_{y,155} + \beta_{y,55}\,,\\
0 &= 2\,\beta_{y,454} + \beta_{y,5} + \beta_{y,57}
= 2\,\beta_{y,87} + \beta_{y,212} + \beta_{y,556}\,,\\
0 &= \beta_{y,137} + \beta_{y,219} + \beta_{y,409} + \beta_{y,58}
= \beta_{y,112} + \beta_{y,132} + \beta_{y,410} + \beta_{y,52}\,,\\
0 &= \beta_{y,16} + \beta_{y,286} + \beta_{y,447} + \beta_{y,88}
= 2\,\beta_{y,9} + \beta_{y,174} + \beta_{y,568} + \beta_{y,60}\,,\\
0 &= \beta_{y,293} + \beta_{y,47} + \beta_{y,56} + \beta_{y,586}\,,\\
0 &= 2\,\beta_{y,165} + 2\,\beta_{y,444} + 4\,\beta_{y,15} + \beta_{y,180} + \beta_{y,34}\,,\\
0 &= 2\,\beta_{y,113} + 2\,\beta_{y,459} + 2\,\beta_{y,7} + \beta_{y,157} + \beta_{y,49} + \beta_{y,64}\,,\\
0 &= 2\,\beta_{y,17} + 2\,\beta_{y,307} + 2\,\beta_{y,478} + 3\,\beta_{y,3} + \beta_{y,19} + \beta_{y,70}\,,\\
0 &= \beta_{y,144} + \beta_{y,20} + \beta_{y,206} + \beta_{y,300} + \beta_{y,476} + \beta_{y,8} + \beta_{y,80} + \beta_{y,95}\,,\\
0 &= \beta_{y,10} + \beta_{y,119} + \beta_{y,20} + \beta_{y,201} + \beta_{y,38} + \beta_{y,452} + \beta_{y,591} + \beta_{y,81}\,,\\
0 &= 2\,\beta_{y,46} + 2\,\beta_{y,48} + \beta_{y,167} + \beta_{y,231} + \beta_{y,287} + \beta_{y,437} + \beta_{y,73} + \beta_{y,94}\,,\\
0 &= 2\,\beta_{y,14} + 2\,\beta_{y,22} + \beta_{y,121} + \beta_{y,140} + \beta_{y,183} + \beta_{y,304} + \beta_{y,584} + \beta_{y,79}\,,\\
0 &= 2\,\beta_{y,18} + 3\,\beta_{y,29} + 3\,\beta_{y,4} + \beta_{y,139} + \beta_{y,245} + \beta_{y,312} + \beta_{y,415} + \beta_{y,71} + \beta_{y,76}\,,\\
0 &= 2\,\beta_{y,117} + 2\,\beta_{y,145} + 2\,\beta_{y,237} + 2\,\beta_{y,33} + 2\,\beta_{y,35} + 2\,\beta_{y,457} + 2\,\beta_{y,50} \\
&\quad + 2\,\beta_{y,51} + \beta_{y,558} + \beta_{y,564}\,,\\
0 &= 2\,\beta_{y,10} + 2\,\beta_{y,39} + 2\,\beta_{y,40} + 2\,\beta_{y,8} + 4\,\beta_{y,6} + \beta_{y,110} + \beta_{y,115} + \beta_{y,151}  \\
&\quad + \beta_{y,161} + \beta_{y,197} + \beta_{y,223} + \beta_{y,442} + \beta_{y,54} + \beta_{y,597} + \beta_{y,63} + \beta_{y,67}\,. \yestag{}
\end{align*}
Furthermore, the identification of scalar quartic interaction with the Yukawas $\lambda^{AB}_{\phantom{AB}CD} = Y^{ABE}Y_{ECD}$ also holds under RG transformations, which gives rise to the condition~\eq{susy-quartic} and in consequence 
\begin{align*}
   2\,c_{41} &= 2\,\beta_{\lambda,1}   + 4\,\beta_{\lambda,2} + \beta_{\lambda,136} + \beta_{\lambda,57}\,,\\
    2\,c_{42} &= 2\,\beta_{\lambda,30}  + \beta_{\lambda,105}\,,\\
    2\,c_{44} &= 2\,\beta_{\lambda,27}  + \beta_{\lambda,1} + \beta_{\lambda,130}\,,\\
    2\,c_{45} &= \beta_{\lambda,97} \,,\\
    2\,c_{46} &= 3\,\beta_{\lambda,33}  + \beta_{\lambda,98}\,,\\
    2\,c_{47} &= 2\,\beta_{\lambda,55}  + 4\,\beta_{\lambda,3} + \beta_{\lambda,127}\,,\\
    2\,c_{48} &= \beta_{\lambda,126} + \beta_{\lambda,29} \,,\\
    2\,c_{49} &= 3\,\beta_{\lambda,32}  + \beta_{\lambda,100}\,,\\
    2\,c_{411} &= \beta_{\lambda,96} \,,\\
    2\,c_{412} &= \beta_{\lambda,121} + \beta_{\lambda,3} \,,\\
    2\,c_{413} &= 2\,\beta_{\lambda,29}  + \beta_{\lambda,102} + \beta_{\lambda,123}\,,\\
 0 &= \beta_{\lambda,28} = \beta_{\lambda,31} = \beta_{\lambda,48} = \beta_{\lambda,49} = \beta_{\lambda,20} = 2\,\beta_{\lambda,36} + \beta_{\lambda,118} = 4\,\beta_{\lambda,34} + \beta_{\lambda,120}\,,\\
 0 &= 2\,\beta_{\lambda,52} + \beta_{\lambda,205} = 4\,\beta_{\lambda,54} + \beta_{\lambda,207} = 2\,\beta_{\lambda,38} + \beta_{\lambda,292} = \beta_{\lambda,349} + \beta_{\lambda,40} = 2\,\beta_{\lambda,22} + \beta_{\lambda,85}\,,\\
 0 &= 3\,\beta_{\lambda,24} + \beta_{\lambda,61} = 2\,\beta_{\lambda,25} + \beta_{\lambda,58} = \beta_{\lambda,47}  = \beta_{\lambda,247} + \beta_{\lambda,39} + \beta_{\lambda,51}\,,\\
 0 &= \beta_{\lambda,248} + \beta_{\lambda,41} + \beta_{\lambda,51} = 4\,\beta_{\lambda,116} + 8\,\beta_{\lambda,45} + \beta_{\lambda,473} = \beta_{\lambda,42} + \beta_{\lambda,500} + \beta_{\lambda,54}\,,\\
 0 &= \beta_{\lambda,45} + \beta_{\lambda,52} + \beta_{\lambda,617} = \beta_{\lambda,474} + \beta_{\lambda,7} + \beta_{\lambda,96} = 2\,\beta_{\lambda,50} + \beta_{\lambda,343} + \beta_{\lambda,9}\,,\\
 0 &= \beta_{\lambda,12} + \beta_{\lambda,286} + \beta_{\lambda,53} = \beta_{\lambda,15} + \beta_{\lambda,591} + \beta_{\lambda,97} = 4\,\beta_{\lambda,21} + 4\,\beta_{\lambda,90} + \beta_{\lambda,226}\,,\\
 0 &= \beta_{\lambda,184} + \beta_{\lambda,26} + \beta_{\lambda,35}  = \beta_{\lambda,126} + \beta_{\lambda,29} + \beta_{\lambda,439} + \beta_{\lambda,50}\,,\\
 0 &= 2\,\beta_{\lambda,238} + 2\,\beta_{\lambda,39} + \beta_{\lambda,47} + \beta_{\lambda,532} = 2\,\beta_{\lambda,239} + 2\,\beta_{\lambda,41} + \beta_{\lambda,47} + \beta_{\lambda,630}\,,\\
 0 &= 2\,\beta_{\lambda,2} + 2\,\beta_{\lambda,271} + 2\,\beta_{\lambda,53} + \beta_{\lambda,403} = 2\,\beta_{\lambda,66} + 2\,\beta_{\lambda,84} + 4\,\beta_{\lambda,4} + \beta_{\lambda,211}\,,\\
 0 &= 2\,\beta_{\lambda,146} + 2\,\beta_{\lambda,37} + 2\,\beta_{\lambda,6} + \beta_{\lambda,232} = 12\,\beta_{\lambda,10} + 3\,\beta_{\lambda,86} + 4\,\beta_{\lambda,71} + \beta_{\lambda,214}\,,\\
 0 &= 2\,\beta_{\lambda,65} + 4\,\beta_{\lambda,13} + \beta_{\lambda,191} + \beta_{\lambda,293} = 2\,\beta_{\lambda,14} + \beta_{\lambda,140} + \beta_{\lambda,350} + \beta_{\lambda,72}\,,\\
 0 &= 2\,\beta_{\lambda,662} + 4\,\beta_{\lambda,174} + 4\,\beta_{\lambda,176} + 5\,\beta_{\lambda,19} = 2\,\beta_{\lambda,167} + 2\,\beta_{\lambda,46} + \beta_{\lambda,21} + \beta_{\lambda,661}\,,\\
 0 &= 2\,\beta_{\lambda,76} + 4\,\beta_{\lambda,23} + 4\,\beta_{\lambda,26} + \beta_{\lambda,183} =  3\,\beta_{\lambda,32} + \beta_{\lambda,100} + \beta_{\lambda,43} + \beta_{\lambda,438} + \beta_{\lambda,476}\,,\\
 0 &= 3\,\beta_{\lambda,33} + \beta_{\lambda,458} + \beta_{\lambda,46} + \beta_{\lambda,593} + \beta_{\lambda,98} = \beta_{\lambda,121} + \beta_{\lambda,3} + \beta_{\lambda,401} + \beta_{\lambda,479} + \beta_{\lambda,53}\,,\\
 0 &= 3\,\beta_{\lambda,5} + \beta_{\lambda,233} + \beta_{\lambda,288} + \beta_{\lambda,37} + \beta_{\lambda,70} = 3\,\beta_{\lambda,10} + \beta_{\lambda,448} + \beta_{\lambda,502} + \beta_{\lambda,51} + \beta_{\lambda,71}\,,\\
 0 &= 2\,\beta_{\lambda,39} + 2\,\beta_{\lambda,41} + 4\,\beta_{\lambda,42} + \beta_{\lambda,110} + \beta_{\lambda,111} + \beta_{\lambda,246}\,,\\
 0 &= \beta_{\lambda,241} + \beta_{\lambda,265} + \beta_{\lambda,30} + \beta_{\lambda,38} + \beta_{\lambda,467} + \beta_{\lambda,51}\,,\\
 0 &= 2\,\beta_{\lambda,34} + 2\,\beta_{\lambda,45} + \beta_{\lambda,115} + \beta_{\lambda,326} + \beta_{\lambda,40} + \beta_{\lambda,558}\,,\\
 0 &= 2\,\beta_{\lambda,50} + \beta_{\lambda,161} + \beta_{\lambda,328} + \beta_{\lambda,4} + \beta_{\lambda,46} + \beta_{\lambda,466}\,,\\
 0 &= 2\,\beta_{\lambda,52} + 4\,\beta_{\lambda,48} + \beta_{\lambda,204} + \beta_{\lambda,283} + \beta_{\lambda,628} + \beta_{\lambda,8}\,,\\
 0 &= 2\,\beta_{\lambda,11} + 2\,\beta_{\lambda,39} + \beta_{\lambda,106} + \beta_{\lambda,355} + \beta_{\lambda,503} + \beta_{\lambda,74}\,,\\
 0 &= 2\,\beta_{\lambda,11} + 2\,\beta_{\lambda,8} + \beta_{\lambda,155} + \beta_{\lambda,298} + \beta_{\lambda,37} + \beta_{\lambda,80}\,,\\
 0 &= 2\,\beta_{\lambda,49} + 2\,\beta_{\lambda,54} + \beta_{\lambda,12} + \beta_{\lambda,206} + \beta_{\lambda,259} + \beta_{\lambda,526}\,,\\
 0 &= 2\,\beta_{\lambda,38} + 2\,\beta_{\lambda,43} + \beta_{\lambda,109} + \beta_{\lambda,13} + \beta_{\lambda,147} + \beta_{\lambda,304}\,,\\
 0 &= 2\,\beta_{\lambda,40} + 2\,\beta_{\lambda,43} + \beta_{\lambda,112} + \beta_{\lambda,14} + \beta_{\lambda,143} + \beta_{\lambda,341}\,,\\
 0 &= 2\,\beta_{\lambda,14} + 2\,\beta_{\lambda,152} + 2\,\beta_{\lambda,37} + 2\,\beta_{\lambda,513} + \beta_{\lambda,198} + \beta_{\lambda,253}\,,\\
 0 &= \beta_{\lambda,159} + \beta_{\lambda,18} + \beta_{\lambda,250} + \beta_{\lambda,36} + \beta_{\lambda,51} + \beta_{\lambda,556}\,,\\
 0 &= 2\,\beta_{\lambda,18} + 2\,\beta_{\lambda,41} + \beta_{\lambda,107} + \beta_{\lambda,371} + \beta_{\lambda,618} + \beta_{\lambda,83}\,,\\
 0 &= 4\,\beta_{\lambda,165} + 4\,\beta_{\lambda,18} + 4\,\beta_{\lambda,46} + \beta_{\lambda,114} + \beta_{\lambda,419} + \beta_{\lambda,92}\,,\\
 0 &= 2\,\beta_{\lambda,15} + 2\,\beta_{\lambda,23} + \beta_{\lambda,311} + \beta_{\lambda,53} + \beta_{\lambda,585} + \beta_{\lambda,95}\,,\\
 0 &= 2\,\beta_{\lambda,228} + 2\,\beta_{\lambda,27} + 2\,\beta_{\lambda,272} + 2\,\beta_{\lambda,70} + 3\,\beta_{\lambda,203} + 6\,\beta_{\lambda,5} + \beta_{\lambda,408}\,,\\
 0 &= 2\,\beta_{\lambda,146} + 2\,\beta_{\lambda,254} + 2\,\beta_{\lambda,37} + 2\,\beta_{\lambda,6} + \beta_{\lambda,236} + \beta_{\lambda,44} + \beta_{\lambda,589}\,,\\
 0 &= 2\,\beta_{\lambda,55} + 2\,\beta_{\lambda,6} + 4\,\beta_{\lambda,3} + \beta_{\lambda,127} + \beta_{\lambda,393} + \beta_{\lambda,480} + \beta_{\lambda,62}\,,\\
 0 &= 2\,\beta_{\lambda,387} + 2\,\beta_{\lambda,82} + 4\,\beta_{\lambda,15} + 4\,\beta_{\lambda,18} + \beta_{\lambda,168} + \beta_{\lambda,44} + \beta_{\lambda,659}\,,\\
 0 &= 2\,\beta_{\lambda,1} + 2\,\beta_{\lambda,189} + 2\,\beta_{\lambda,266} + 2\,\beta_{\lambda,65} + 4\,\beta_{\lambda,13} + \beta_{\lambda,396} + \beta_{\lambda,533} + \beta_{\lambda,56}\,,\\
 0 &= 16\,\beta_{\lambda,15} + 16\,\beta_{\lambda,18} + 2\,\beta_{\lambda,221} + 4\,\beta_{\lambda,164} + 4\,\beta_{\lambda,92} + 4\,\beta_{\lambda,94} + 8\,\beta_{\lambda,82} + \beta_{\lambda,425}\,,\\
 0 &= 2\,\beta_{\lambda,173} + 2\,\beta_{\lambda,388} + 2\,\beta_{\lambda,392} + 2\,\beta_{\lambda,81} + 8\,\beta_{\lambda,16} + \beta_{\lambda,175} + \beta_{\lambda,19} + \beta_{\lambda,660}\,,\\
 0 &= 2\,\beta_{\lambda,29} + 3\,\beta_{\lambda,201} + 3\,\beta_{\lambda,5} + \beta_{\lambda,102} + \beta_{\lambda,123} + \beta_{\lambda,404} + \beta_{\lambda,440} + \beta_{\lambda,484} + \beta_{\lambda,70}\,,\\
 0 &= 2\,\beta_{\lambda,35} + 2\,\beta_{\lambda,46} + 3\,\beta_{\lambda,5} + \beta_{\lambda,113} + \beta_{\lambda,235} + \beta_{\lambda,313} + \beta_{\lambda,465} + \beta_{\lambda,590} + \beta_{\lambda,70}\,,\\
 0 &= 2\,\beta_{\lambda,375} + 4\,\beta_{\lambda,78} + 4\,\beta_{\lambda,8} + 4\,\beta_{\lambda,9} + \beta_{\lambda,170} + \beta_{\lambda,177} + \beta_{\lambda,19} + \beta_{\lambda,218} + \beta_{\lambda,620}\,,\\
 0 &= 2\,\beta_{\lambda,276} + 4\,\beta_{\lambda,115} + 4\,\beta_{\lambda,119} + 4\,\beta_{\lambda,28} + 4\,\beta_{\lambda,45} + 8\,\beta_{\lambda,34} + \beta_{\lambda,1} + \beta_{\lambda,129} \\
 &\quad+ \beta_{\lambda,471} + \beta_{\lambda,599}\,,\\
 0 &= 2\,\beta_{\lambda,11} + 2\,\beta_{\lambda,9} + \beta_{\lambda,155} + \beta_{\lambda,255} + \beta_{\lambda,322} + \beta_{\lambda,37} + \beta_{\lambda,380} + \beta_{\lambda,44} + \beta_{\lambda,555} + \beta_{\lambda,79}\,,\\
 0 &= 2\,\beta_{\lambda,145} + 2\,\beta_{\lambda,63} + 4\,\beta_{\lambda,6} + 4\,\beta_{\lambda,75} + 4\,\beta_{\lambda,79} + 4\,\beta_{\lambda,9} + 8\,\beta_{\lambda,11} + 8\,\beta_{\lambda,7} + \beta_{\lambda,217} + \beta_{\lambda,345}\,,\\
 0 &= 2\,\beta_{\lambda,40} + 2\,\beta_{\lambda,43} + \beta_{\lambda,112} + \beta_{\lambda,13} + \beta_{\lambda,147} + \beta_{\lambda,258} + \beta_{\lambda,316} + \beta_{\lambda,368} + \beta_{\lambda,44} + \beta_{\lambda,464}\,,\\
 0 &= 2\,\beta_{\lambda,38} + 2\,\beta_{\lambda,40} + \beta_{\lambda,108} + \beta_{\lambda,13} + \beta_{\lambda,154} + \beta_{\lambda,256} + \beta_{\lambda,359} + \beta_{\lambda,37} + \beta_{\lambda,451} + \beta_{\lambda,506}\,,\\
 0 &= 2\,\beta_{\lambda,39} + 2\,\beta_{\lambda,41} + 4\,\beta_{\lambda,42} + \beta_{\lambda,110} + \beta_{\lambda,111} + \beta_{\lambda,16} + \beta_{\lambda,252} + \beta_{\lambda,370} + \beta_{\lambda,378} + \beta_{\lambda,553}\,,\\
 0 &= 2\,\beta_{\lambda,38} + 2\,\beta_{\lambda,43} + \beta_{\lambda,109} + \beta_{\lambda,14} + \beta_{\lambda,143} + \beta_{\lambda,16} + \beta_{\lambda,196} + \beta_{\lambda,314} + \beta_{\lambda,365} + \beta_{\lambda,584}\,,\\
 0 &= 2\,\beta_{\lambda,13} + 2\,\beta_{\lambda,14} + 2\,\beta_{\lambda,17} + 4\,\beta_{\lambda,12} + \beta_{\lambda,138} + \beta_{\lambda,156} + \beta_{\lambda,192} + \beta_{\lambda,299} + \beta_{\lambda,64} + \beta_{\lambda,73}\,,\\
 0 &= 2\,\beta_{\lambda,150} + 2\,\beta_{\lambda,357} + 2\,\beta_{\lambda,511} + 2\,\beta_{\lambda,64} + 2\,\beta_{\lambda,73} + 4\,\beta_{\lambda,13} + 4\,\beta_{\lambda,14} + 4\,\beta_{\lambda,195} + 8\,\beta_{\lambda,12}\\
 &\quad + \beta_{\lambda,395} + \beta_{\lambda,433}\,,\\
 0 &= 2\,\beta_{\lambda,61} + 2\,\beta_{\lambda,71} + 3\,\beta_{\lambda,87} + 6\,\beta_{\lambda,10} + 6\,\beta_{\lambda,24} + \beta_{\lambda,212} + \beta_{\lambda,227} + \beta_{\lambda,260} + \beta_{\lambda,37} \\
 &\quad+ \beta_{\lambda,469} + \beta_{\lambda,528}\,,\\
 0 &= 2\,\beta_{\lambda,135} + 2\,\beta_{\lambda,171} + 2\,\beta_{\lambda,281} + 2\,\beta_{\lambda,57} + 2\,\beta_{\lambda,600} + 4\,\beta_{\lambda,1} + 4\,\beta_{\lambda,17} + 4\,\beta_{\lambda,190} + 8\,\beta_{\lambda,2} \\
 &\quad+ \beta_{\lambda,19} + \beta_{\lambda,397} + \beta_{\lambda,632}\,,\\
 0 &= 2\,\beta_{\lambda,14} + 2\,\beta_{\lambda,77} + 2\,\beta_{\lambda,83} + 2\,\beta_{\lambda,93} + 4\,\beta_{\lambda,18} + 4\,\beta_{\lambda,26} + \beta_{\lambda,137} + \beta_{\lambda,222} + \beta_{\lambda,330} \\
 &\quad+ \beta_{\lambda,364} + \beta_{\lambda,562} + \beta_{\lambda,72}\,,\\
 0 &= 2\,\beta_{\lambda,167} + 2\,\beta_{\lambda,186} + 2\,\beta_{\lambda,21} + 2\,\beta_{\lambda,35} + 2\,\beta_{\lambda,4} + 2\,\beta_{\lambda,46} + 4\,\beta_{\lambda,22} + \beta_{\lambda,113} + \beta_{\lambda,162} \\
 &\quad+ \beta_{\lambda,329} + \beta_{\lambda,426} + \beta_{\lambda,627} + \beta_{\lambda,67} + \beta_{\lambda,91}\,,\\
 0 &= 2\,\beta_{\lambda,131} + 2\,\beta_{\lambda,161} + 2\,\beta_{\lambda,180} + 2\,\beta_{\lambda,27} + 2\,\beta_{\lambda,35} + 2\,\beta_{\lambda,4} + 2\,\beta_{\lambda,46} + 4\,\beta_{\lambda,25} + \beta_{\lambda,113} \\
 &\quad+ \beta_{\lambda,229} + \beta_{\lambda,282} + \beta_{\lambda,414} + \beta_{\lambda,460} + \beta_{\lambda,88}\,,\\
 0 &= 2\,\beta_{\lambda,159} + 2\,\beta_{\lambda,178} + 2\,\beta_{\lambda,18} + 2\,\beta_{\lambda,26} + 2\,\beta_{\lambda,30} + 2\,\beta_{\lambda,36} + 4\,\beta_{\lambda,31} + \beta_{\lambda,104} + \beta_{\lambda,117} \\
 &\quad+ \beta_{\lambda,240} + \beta_{\lambda,274} + \beta_{\lambda,410} + \beta_{\lambda,595} + \beta_{\lambda,93}\,,\\
 0 &= 2\,\beta_{\lambda,145} + 2\,\beta_{\lambda,362} + 2\,\beta_{\lambda,63} + 4\,\beta_{\lambda,6} + 4\,\beta_{\lambda,75} + 4\,\beta_{\lambda,8} + 4\,\beta_{\lambda,80} + 8\,\beta_{\lambda,11} + 8\,\beta_{\lambda,7} \\
 &\quad+ \beta_{\lambda,170} + \beta_{\lambda,19} + \beta_{\lambda,216} + \beta_{\lambda,335} + \beta_{\lambda,398} + \beta_{\lambda,583}\,,\\
 0 &= 2\,\beta_{\lambda,142} + 2\,\beta_{\lambda,168} + 2\,\beta_{\lambda,17} + 2\,\beta_{\lambda,187} + 2\,\beta_{\lambda,220} + 2\,\beta_{\lambda,377} + 4\,\beta_{\lambda,76} + 4\,\beta_{\lambda,82} \\
 &\quad+ 4\,\beta_{\lambda,93} + 4\,\beta_{\lambda,95} + 8\,\beta_{\lambda,15} + 8\,\beta_{\lambda,18} + 8\,\beta_{\lambda,23} + 8\,\beta_{\lambda,26} + \beta_{\lambda,336} + \beta_{\lambda,428} + \beta_{\lambda,625}\,,\\
 0 &= 16\,\beta_{\lambda,20} + 16\,\beta_{\lambda,22} + 16\,\beta_{\lambda,25} + 2\,\beta_{\lambda,134} + 2\,\beta_{\lambda,208} + 2\,\beta_{\lambda,285} + 2\,\beta_{\lambda,416} + 4\,\beta_{\lambda,162} + 4\,\beta_{\lambda,181} \\
 &\quad + 4\,\beta_{\lambda,21} + 4\,\beta_{\lambda,67} + 4\,\beta_{\lambda,88} + 4\,\beta_{\lambda,91} + 8\,\beta_{\lambda,4} + 8\,\beta_{\lambda,59} + 8\,\beta_{\lambda,89} + \beta_{\lambda,224} + \beta_{\lambda,635}\,,\\
 0 &= 2\,\beta_{\lambda,13} + 2\,\beta_{\lambda,14} + 2\,\beta_{\lambda,17} + 2\,\beta_{\lambda,197} + 2\,\beta_{\lambda,200} + 4\,\beta_{\lambda,12} + 4\,\beta_{\lambda,16} + \beta_{\lambda,139} + \beta_{\lambda,156} \\
 &\quad + \beta_{\lambda,172} + \beta_{\lambda,194} + \beta_{\lambda,321} + \beta_{\lambda,367} + \beta_{\lambda,373} + \beta_{\lambda,434} + \beta_{\lambda,560} + \beta_{\lambda,64} + \beta_{\lambda,73} + \beta_{\lambda,81}\,.
\tag{\stepcounter{equation}\theequation}\
\end{align*}

\clearpage
\bibliographystyle{jhep}
\bibliography{ref.bib}
\end{document}